\documentclass{sty/thesis}
\usepackage{sty/mythesis}
\usepackage{textcomp, gensymb, pifont}
\usepackage{url}
\usepackage{soul}
\usepackage{tikz}
\usepackage{setspace}
\usepackage{graphicx}
\usepackage{hhline}

\usepackage{mathptmx}

\usepackage[nocompress]{cite}

\usepackage[export]{adjustbox}

\usepackage{textcomp}

\usepackage{comment}
\usepackage{subfig}
\usepackage{wrapfig}
\usepackage{hyperref}
\usepackage{xspace}
\usepackage{placeins}

\newcommand{\ignore}[1]{}   

\let\oldcelsius\celsius
\renewcommand{\celsius}{~\oldcelsius\xspace}

\newcommand{\titleShortMeDiC}[0]{MeDiC\xspace}
\newcommand{\titleLongMeDiC}[0]{Memory Divergence Correction\xspace}
\newcommand{\titleShortSMS}[0]{SMS\xspace}
\newcommand{\titleLongSMS}[0]{Staged Memory Scheduler\xspace}
\newcommand{\titleShortMASK}[0]{MASK\xspace}
\newcommand{\titleLongMASK}[0]{Multi Address Space Concurrent Kernels\xspace}
\newcommand{\titleLongMASKEmph}[0]{\textbf{M}ulti-\textbf{A}ddress \textbf{S}pace Concurrent \textbf{K}ernels\xspace}

\newcommand{\titleShortMOSAIC}[0]{Mosaic\xspace}

\newcommand{\batch}[0]{batch\xspace}
\newcommand{\Batch}[0]{Batch\xspace}
\newcommand{\batches}[0]{batches\xspace}
\newcommand{\allocatorNameLong}[0]{\textsc{\textbf{Co}ntiguity-\textbf{co}nserving \textbf{A}llocation}\xspace}
\newcommand{\allocatorName}[0]{\textsc{\textbf{CocoA}\xspace}}
\newcommand{\policyName}[0]{\textsc{Lazy-Coalescer}\xspace}
\newcommand{\compactionName}[0]{\textsc{CaC}\xspace}
\newcommand{\compactionNameLong}[0]{\textsc{\textbf{C}ontiguity-\textbf{a}ware \textbf{C}ompaction}\xspace}

\newcommand*\mycirc[1]
{\tikz[baseline=(char.base)]{
  \node[shape=circle,draw,inner sep=0.5pt,fill=black,text=white,font=\sffamily] (char) {#1};}}

\newcommand*\mycirctwo[1]{%
  \tikz[baseline=(char.base)]{
  \node[shape=circle,draw,inner sep=0.0pt,fill=black,text=white,font=\bfseries] (char) {#1};}%
}%

\newlength{\realbaselineskip}
\newcommand{\name}{\textsc} 
\newcommand{\tlbtokenname}[0]{\name{TLB-Fill Tokens}\xspace}
\newcommand{\cachebypass}[0]{\name{TLB-request-aware L2 Bypass}\xspace}
\newcommand{\dramsched}[0]{\name{Address-space-aware DRAM Scheduler}\xspace}

\newcommand{\goldQ}[0]{\name{Golden Queue}\xspace}
\newcommand{\silverQ}[0]{\name{Silver Queue}\xspace}
\newcommand{\normalQ}[0]{\name{Normal Queue}\xspace}

\newboolean{publicversion}
\setboolean{publicversion}{true}

\ifthenelse{\boolean{publicversion}}{
  \newcommand{\grumbler}[2]{}
  \newcommand{\assign}[1]{}
  \newcommand{\respond}[3]{}
  \newcommand{\changesI}[0]{}
  \newcommand{\changesII}[0]{}
  \newcommand{\changesIII}[0]{}
  \newcommand{\changesIIII}[0]{}
  \newcommand{\changesIIIII}[0]{}
  \newcommand{\changesIIIIII}[0]{}
  \newcommand{\changesIIIIIII}[0]{}
  
  \newcommand{\changesIV}[0]{}
  \newcommand{\changesV}[0]{}
  \newcommand{\changesVI}[0]{}
  \newcommand{\changesVII}[0]{}
  \newcommand{\changesVIII}[0]{}
  \newcommand{\changesIX}[0]{}

}{
  \newcommand{\grumbler}[2]{\textcolor{red}{\bf #1: #2}}
  \newcommand{\changesI}[0]{}
  \newcommand{\changesII}[0]{}
  \newcommand{\changesIII}[0]{}
  \newcommand{\changesIIII}[0]{}
  \newcommand{\changesIIIII}[0]{}
  \newcommand{\changesIIIIII}[0]{}
  \newcommand{\changesIIIIIII}[0]{}
  
  \newcommand{\changesIV}[0]{}
  \newcommand{\changesV}[0]{}
  \newcommand{\changesVI}[0]{}
  \newcommand{\changesVII}[0]{}
  \newcommand{\changesVIII}[0]{}
  \newcommand{\changesIX}[0]{}

}

\newcommand{\cjr}[1]{\grumbler{CJR}{#1}}

\newcommand{\vm}[1]{\grumbler{VanceM}{#1}}
\newcommand{\rachata}[1]{\grumbler{Rachata}{#1}}

\newcommand{\red}[1]{#1}
\newcommand{\yellow}[1]{#1}
\newcommand{\green}[1]{#1}
\newcommand{\paragraphbe}[1]{\textbf{\textit{#1}} }
\newcommand{\para}[1]{\textbf{\textit{#1}} }

\usepackage{cleveref}
\crefformat{footnote}{#2\footnotemark[#1]#3}

\newcommand{\squishlist}{
	\begin{list}{$\bullet$}
		{ \setlength{\itemsep}{0pt}      \setlength{\parsep}{3pt}
			\setlength{\topsep}{3pt}       \setlength{\partopsep}{0pt}
			\setlength{\leftmargin}{1.5em} \setlength{\labelwidth}{1em}
			\setlength{\labelsep}{0.5em} } }
\newcommand{\squishend}{
    \end{list}  }

\title{{\bf {\Large Techniques for Shared Resource Management\\in Systems with Throughput Processors}}}

\begin{document}

\maketitle
\pagenumbering{roman}

\newpage
\begin{center}
\vspace*{\fill}
Copyright \textcopyright\ 2017 Rachata Ausavarungnirun
\end{center}

\newpage

\chapter*{Acknowledgements}

First and foremost, I would like to thank my parents, Khrieng and Ruchanee
Ausavarungnirun for their endless encouragement, love, and support. In
addition to my family, I would like to thank my advisor, Prof. Onur Mutlu, for
providing me with great research environment. He taught me many important
aspects of research and shaped me into the researcher I am today. 

I would like to thank all my committee members, Prof. James Hoe, Dr. Gabriel
Loh, Prof. Chris Rossbach and Prof. Kayvon Fatahalian, who provided me multiple
feedback on my research and spent a lot of their time and effort to help me
complete this dissertation. Special thank to Professor James Hoe, my
first mentor at CMU, who taught me all the basics since my sophomore year.
Professor Hoe introduced me to many interesting research projects within CALCM.
Thanks to Dr. Gabriel Loh for his guidance, which helped me tremendously during
the first four years of my PhD. Thanks to Prof. Chris
Rossbach for being a great mentor, providing me with guidance, feedback and
support for my research. Both Dr. Loh and Prof. Rossbach provided me with lots
of real-world knowledge from the industry, which further enhanced the
quality of my research. Lastly, thanks to Prof. Kayvon Fatahalian for his
knowledge and valuable comments on my GPU research.

All members of SAFARI have been like a family to me. This dissertation is done
thanks to lots of support and feedback from them.
Donghyuk Lee has always been a good friend and a great mentor. His work ethic
is something I always look up to. Thanks to Kevin Chang for all the valuable
feedback throughout my PhD. Thanks to Yoongu Kim and Lavanya Subramanian for
teaching me on several DRAM-related topics. Thanks to Samira Khan and Saugata
Ghose for their guidance. Thanks to Hongyi Xin and Yixin Luo for their positive
attitudes and their friendship. Thanks to Vivek Seshadri and Gennady Pekhimenko
for their research insights. Thanks to Chris Fallin and Justin Meza for all the
helps, especially during the early years of my PhD. They provided tremendous
help when I am preparing for my qualification exam. Thanks to Nandita
Vijaykumar for all GPU-related discussions. Thanks to Hanbin Yoon, Jamie
Liu, Ben Jaiyen, Chris Craik, Kevin Hsieh, Yang Li, Amirali Bouroumand, Jeremie
Kim, Damla Senol and Minesh Patel for all their interesting research
discussions.

In additional to people in the SAFARI group, I would like to thank Onur Kayiran
and Adwait Jog, who have been great colleagues and have been providing me with
valuable discussions on various GPU-related research topics. Thanks to Mohammad
Fattah for a great collaboration on Network-on-chip. Thanks to Prof. Reetu Das
for her inputs on my Network-on-chip research projects. 
Thanks to Eriko Nurvitadhi and Peter Milder, both of whom were my mentors
during my undergrad years. 
Thanks to John and Claire Bertucci for their fellowship support. Thanks to Dr.
Pattana Wangaryattawanich and Dr. Polakit Teekakirikul for their friendship and
mental support. Thanks to several members of the Thai Scholar community as well
as several members of the Thai community in Pittsburgh for their friendship.
Thanks to support from AMD, Facebook, Google, IBM, Intel, Microsoft, NVIDIA,
Qualcomm, VMware, Samsung, SRC, and support from NSF grants numbers 0953246,
1065112, 1147397, 1205618, 1212962, 1213052, 1302225, 1302557, 1317560,
1320478, 1320531, 1409095, 1409723, 1423172, 1439021 and 1439057.

Lastly, I would like to give a special thank to my wife, Chatchanan
Doungkamchan for her endless love, support and encouragement. She understands
me and helps me with every hurdle I have been through. 
Her work ethic and the care she gives to her research motivate me to work
harder to become a better researcher. She provides me with the perfect
environment that allows me to focus on improving myself and my work while
trying to make sure neither of us are burned-out from over working. I could not
have completed any of the works done in this dissertation without her support.


\newpage
\chapter*{Abstract}

The continued growth of the computational capability of throughput processors
has made throughput processors the platform of choice for a wide variety of
high performance computing applications. Graphics Processing Units (GPUs) are
a prime example of throughput processors that can deliver high performance for
applications ranging from typical graphics applications to general-purpose data
parallel (GPGPU) applications. However, this success has been accompanied by
new performance bottlenecks throughout the memory hierarchy of GPU-based
systems. This dissertation identifies and eliminates performance bottlenecks
caused by major sources of interference throughout the memory hierarchy.


Specifically, we provide an in-depth analysis of inter- and intra-application
as well as inter-address-space interference that 
significantly degrade the performance and efficiency of GPU-based systems.


To minimize such interference, we introduce changes to the memory
hierarchy for systems with GPUs that allow the memory hierarchy to be aware of
both CPU and GPU applications' characteristics. We introduce
mechanisms to dynamically analyze different applications'
characteristics and propose four major changes throughout the memory
hierarchy. 

First, we introduce \titleLongMeDiC
(\titleShortMeDiC), a cache management mechanism that mitigates
intra-application interference in GPGPU applications by allowing the shared L2
cache and the memory controller to be aware of the GPU's warp-level memory divergence characteristics.
\titleShortMeDiC uses this warp-level memory divergence information to give
more cache space and more memory bandwidth to warps that benefit most from utilizing such resources.
Our evaluations show that \titleShortMeDiC significantly outperforms multiple
state-of-the-art caching policies proposed for GPUs.

Second, we introduce the \titleLongSMS (\titleShortSMS), an
application-aware CPU-GPU memory request scheduler that mitigates inter-application
interference in heterogeneous CPU-GPU systems.  \titleShortSMS creates a
fundamentally new approach to memory controller design that decouples
the memory controller into three significantly simpler structures, each of
which has a separate task, These structures operate together to greatly improve both system
performance and fairness. Our three-stage memory controller first groups
requests based on row-buffer locality. This grouping allows the second stage to
focus on inter-application scheduling decisions. These two stages enforce
high-level policies regarding performance and fairness. As a result, the last
stage is simple logic that deals only with the low-level DRAM
commands and timing. SMS is also configurable: it allows the system software
to trade off between the quality of service provided to the CPU versus GPU applications.
Our evaluations show that \titleShortSMS not only reduces inter-application
interference caused by the GPU, thereby improving heterogeneous system
performance, but also provides better scalability and power efficiency compared
to multiple state-of-the-art memory schedulers.

%

Third, we redesign the GPU memory management unit to efficiently handle new
problems caused by the massive address translation parallelism present in GPU
computation units in multi-GPU-application environments. Running multiple GPGPU
applications concurrently induces significant inter-core thrashing on the
shared address translation/protection units; e.g., the shared Translation
Lookaside Buffer (TLB), a new phenomenon that we call \emph{inter-address-space
interference}. To reduce this interference, we introduce \emph{\titleLongMASK}
(\titleShortMASK).  \titleShortMASK introduces TLB-awareness throughout the GPU
memory hierarchy and introduces TLB- and cache-bypassing techniques to increase
the effectiveness of a shared TLB. 

Finally, we introduce \titleShortMOSAIC, a hardware-software cooperative
technique that further increases the effectiveness of TLB by modifying the
memory allocation policy in the system software. \titleShortMOSAIC introduces a
high-throughput method to support large pages in multi-GPU-application
environments. \yellow{The key idea is to ensure memory allocation preserve address space
contiguity to allow pages to be coalesced without any data movements.} Our
evaluations show that the \titleShortMASK-\titleShortMOSAIC combination
provides a simple mechanism that eliminates the performance overhead of address
translation in GPUs without significant changes to GPU hardware, thereby
greatly improving GPU system performance. 


The key conclusion of this dissertation is that a combination of GPU-aware
cache and memory management techniques can effectively mitigate the memory
interference on current and future GPU-based systems as well as other types
of throughput processors.

\rachata{KevinC: The last sentence might not be enough. Might want to add another sentence to say something.}
\rachata{KevinC: Might want to break the MASK and MOSAIC.}


\newpage
\listoffigures
\listoftables
\newpage
\tableofcontents
\newpage

\doublespacing

\setcounter{page}{1}
\pagenumbering{arabic}

\setcounter{secnumdepth}{3}


\chapter{Introduction}



\rachata{First introduce throughput processors and then refocus by saying this dissertation
focuses on one of the most popular throughput processor, which is the Graphics Processing Units.}

\rachata{First define throughput processor}

Throughput processor is a type of processors that consists of numerous simple
processing cores. 
Throughput processor allows applications to achieve very high throughput by executing a massive
number of compute operations on these processing cores in parallel within a
single
cycle~\cite{solomon62,senzig-afips65,crane-ec65,hellerman-ec66,cdc7600,cdcstar,illiac,cray1,cdc6600,cdc6600-2,hep,masa-fmt,april-fmt,tera-mta,amd-fusion,apu,kaveri,haswell,amdzen,skylake,powervr,arm-mali,tegra,tegrax1,fermi,kepler,maxwell,pascal,amdr9,radeon,vivante-gpgpu}.
These throughput processors incorporate a variety of processing paradigms, such as vector processors, which utilize a specific execution model called
Single Instruction Multiple Data (SIMD)
model that allows one instruction to be operated on multiple data~\cite{solomon62,senzig-afips65,crane-ec65,hellerman-ec66,cdc7600,cdcstar,illiac,cray1},
processors that utilize a technique called fine-grained multithreading, which allows the processor to issue
instructions from different threads after every cycle~\cite{cdc6600,cdc6600-2,hep,masa-fmt,april-fmt,tera-mta}, or
processors that utilize both techniques~\cite{amd-fusion,apu,kaveri,haswell,amdzen,skylake,powervr,arm-mali,tegra,tegrax1,fermi,kepler,maxwell,pascal,amdr9,radeon,vivante-gpgpu}.
One of the most prominent throughput processors available in modern day computing
systems that utilize both SIMD execution model and fine-grained multithreading is the Graphics 
Processing Units (GPUs). This dissertation uses GPUs as an example class of
throughput processors.


GPUs have enormous parallel processing power due to
the large number of computational units they provide. Modern GPU programming
models exploit this processing power using a large amount of thread-level
parallelism.  GPU applications can be broken down into thousands of threads,
allowing GPUs to use an execution model called SIMT (Single Instruction Multiple
Thread), which enables the GPU cores to tolerate dependencies and long
memory latencies.
The thousands of threads within a GPU application are clustered into \emph{work
groups} (or \emph{thread blocks}), where each thread block consists of a
collection of threads that are run concurrently. Within a thread block, threads
are further grouped into smaller units, called \emph{warps}~\cite{lindholm} or
\emph{wavefronts}~\cite{ati-wavefront}. Every cycle, each GPU core executes a
single warp. Each thread in a warp executes the same instruction (i.e., is at
the same program counter) in lockstep, which is an example of the \emph{SIMD} (Single Instruction,
Multiple Data)~\cite{flynn} execution model. This highly-parallel SIMT/SIMD execution model allows the GPU to
complete several hundreds to thousands of operations every cycle.


GPUs are present in many modern systems. These GPU-based systems range 
from traditional discrete GPUs~\cite{fermi,kepler,maxwell,pascal,amdr9,radeon,vivante-gpgpu,arm-mali,powervr} to heterogeneous CPU-GPU
architectures~\cite{amd-fusion,apu,kaveri,haswell,amdzen,skylake,powervr,arm-mali,tegra,tegrax1}.
In all of these systems with GPUs, resources throughout the memory hierarchy, e.g.,
core-private and shared caches, main memory, the interconnects, and the memory management units are
shared across multiple threads and processes that execute concurrently in both the CPUs and the GPUs.

\section{Resource Contention and Memory Interference Problem in Systems with GPUs}

Due to the limited shared resources in these systems, applications oftentimes
are not able to achieve their ideal throughput (as measured by, e.g., computed instructions per cycle). Shared resources become the
bottleneck and create inefficiency because accesses from one thread or
application can interfere with accesses from other threads or applications in any shared resources,
leading to both bandwidth and space contention, resulting in lower performance. \emph{The main goal of this dissertation is to analyze
and mitigate the major memory interference problems throughout shared resources in the
memory hierarchy of current and future systems with GPUs.} 

We focus on three major types of memory interference that occur in
systems with GPUs: 1) intra-application interference among different GPU threads,
2) inter-application interference that is caused by both CPU and GPU applications,
and 3) inter-address-space interference during address translation when
multiple GPGPU applications concurrently share the GPUs. 

\paragraphbe{Intra-application interference} is a type of interference that originates from
GPU threads within the \emph{same} GPU application. When a GPU executes a GPGPU
application, the threads that are scheduled to run on the GPU cores execute
concurrently. Even though these threads belong to the same kernel, they contend for
shared resources, causing interference to each other~\cite{medic,donglihpca15,chen-micro47,chen-mes14}.
This intra-application interference leads to the significant slowdown of threads
running on GPU cores and lowers the performance of the GPU.

\paragraphbe{Inter-application interference} is a type of interference that is caused by concurrently-executing CPU and GPU applications. It occurs in systems
where a CPU and a GPU share the main memory system. This type of interference is especially observed
in recent heterogeneous CPU-GPU systems~\cite{skylake,haswell,sandybridge,ivybridge,bobcat,kaveri,apu,tegra,amdzen,powervr,arm-mali,sms,cpugpu-micro,nmnl-pact13,jeong2012qos}, which introduce 
an integrated Graphics Processing Unit
(GPU) on the same die with the CPU cores. Due to the GPU's ability to execute a
very large number of parallel threads, GPU applications typically demand
significantly more memory bandwidth than typical CPU applications. Unlike GPU
applications that are designed to tolerate the long memory latency by employing
massive amounts of multithreading~\cite{sms,haswell,skylake,kaveri,apu,fermi,kepler,maxwell,pascal,amdr9,radeon,firepro,amdzen,powervr,arm-mali,vivante-gpgpu,tegra,tegrax1}, CPU
applications typically have much lower tolerance to
latency~\cite{sms,atlas,tcm,bliss,bliss-tpds,bliss-arxiv,mise,stfm,parbs,pam,pa-micro08}. The high bandwidth consumption of the GPU applications
heavily interferes with the progress of other CPU applications that share the same
hardware resources.

\paragraphbe{Inter-address-space interference} arises due to the address translation
process in an environment where multiple GPU applications share the same GPU,
e.g., a shared GPU in a cloud infrastructure. We discover that when multiple
GPGPU applications concurrently use the same GPU, the address translation
process creates additional contention at the shared memory hierarchy, including the
Translation Lookaside Buffers (TLBs), caches, and main memory. This particular
type of interference can cause a severe slowdown to all applications and the system when
multiple GPGPU applications are concurrently executed on a system with GPUs.

While previous works propose mechanisms to reduce interference and improve the
performance of GPUs (See Chapter~\ref{sec:dissertation-background} for a
detailed analyses of these previous works), these approaches 1) focus only on a
subset of the shared resources, such as the shared cache or the memory
controller and 2) generally do not take into account the characteristics of the applications executing on the GPUs. 

%

\section{Thesis Statement and Our Overarching Approach:\\Application Awareness}



With the understanding of the causes of memory interference, our thesis
statement is that \textbf{a combination of GPU-aware cache and memory
management techniques can mitigate memory interference caused by GPUs on current and
future systems with GPUs.}
To this end, we propose to mitigate memory interference in current and future
GPU-based systems via GPU-aware and GPU-application-aware resource management
techniques. We redesign the memory hierarchy such that
\emph{each component in the memory hierarchy is aware of the GPU applications'
characteristics.} The key idea of our approach is to extract important features
of different applications in the system and use them in managing memory hierarchy resources much more intelligently. These key features consist of, but are not
limited to, memory access characteristics, 
utilization of the shared cache, usage of shared main memory and demand for the
shared TLB. Exploiting these features, we introduce modifications to the shared
cache, the memory request scheduler, the shared TLB and the GPU memory
allocator to reduce the amount of inter-application, intra-application and
inter-address-space interference based on applications' characteristics.  We
give a brief overview of our major new mechanisms in the rest of this section.



\subsection{Minimizing Intra-application Interference}

Intra-application interference occurs when multiple threads in the GPU contend
for the shared cache and the shared main memory. Memory requests from one
thread can interfere with memory requests from other threads, leading to low
system performance. As a step to reduce this intra-application interference, we
introduce \titleLongMeDiC{} (\titleShortMeDiC{})~\cite{medic}, a cache and
memory controller management scheme that is designed to be aware of different
types of warps that access the shared cache, and selectively prioritize warps
that benefit the most from utilizing the cache. This new mechanism
first characterizes different types of warps based on how much benefit they
receive from the shared cache. 
To effectively characterize warp-type, \titleShortMeDiC uses the memory
divergence patterns, i.e., the diversity of how long each load and store
instructions in the warp takes. We observe that GPGPU applications exhibit
different levels of heterogeneity in their memory divergence behavior at the
shared L2 cache within the GPU. As a result, (1)~some warps benefit
significantly from the cache, while others make poor use of it; (2)~the
divergence behavior of a warp tends to remain stable for long periods of the
warp's execution; and (3)~the impact of memory divergence can be amplified by
the high queuing latencies at the L2 cache.

%
%

Based on the heterogeneity in warps' memory divergence behavior, we propose a
set of techniques, collectively called \emph{\titleLongMeDiC{}}
(\titleShortMeDiC{}), that reduce the negative performance impact of memory
divergence and cache queuing. \titleShortMeDiC{} uses warp divergence
characterization to guide three warp-aware components in the memory hierarchy:
(1)~a cache bypassing mechanism that exploits the latency tolerance of warps
that do not benefit from using the cache, to both alleviate queuing delay and
increase the hit rate for warps that benefit from using the cache, (2)~a cache
insertion policy that prevents data from warps that benefit from using the
cache from being prematurely evicted, and (3)~a memory controller that
prioritizes the few requests received from warps that benefit from using the
cache, to minimize stall time. Our evaluation shows that \titleShortMeDiC{} is
effective at exploiting inter-warp heterogeneity and delivers significant
performance and energy improvements over the state-of-the-art GPU cache
management technique~\cite{donglihpca15}.

\subsection{Minimizing Inter-application Interference}

Inter-application interference occurs when multiple processor cores (CPUs) and
a GPU integrated together on the same chip share the off-chip DRAM (and
perhaps some caches). In such as system, requests from the GPU can heavily
interfere with requests from the CPUs, leading to low system performance and
starvation of cores. Even though previously-proposed application-aware memory
scheduling policies designed for CPU-only
scenarios (e.g., ~\cite{fr-fcfs,parbs,atlas,tcm,stfm,pa-micro08,bliss,bliss-tpds,bliss-arxiv,mise,pam}) can be applied on a CPU-GPU
heterogeneous system, we observe that the GPU requests occupy a significant portion of request
buffer space and thus reduce the visibility of CPU cores' requests to the memory controller, leading to lower system
performance. 
Increasing the request buffer space requires complex logic to analyze
applications' characteristics, assign priorities for each memory request and
enforce these priorities when the GPU is present. As a result, these past
proposals for application-aware memory scheduling in CPU-only systems can
perform poorly on a CPU-GPU heterogeneous system at low \mbox{complexity}
(as we show in this dissertation).

%

To minimize the inter-application interference in CPU-GPU heterogeneous systems, we introduce a
new memory controller called the \titleLongSMS
(\titleShortSMS)~\cite{sms}, which is both application-aware and GPU-aware. 
Specifically, \titleShortSMS is designed to facilitate GPU applications' high bandwidth
demand, improving performance and fairness significantly. \titleShortSMS
introduces a fundamentally new approach that decouples the three primary tasks
of the memory controller into three significantly simpler structures that
together improve system performance and fairness. The three-stage memory
controller first groups requests based on row-buffer locality in its first
stage, called the \emph{Batch Formation stage}. This grouping allows the second
stage, called the \emph{Batch Scheduler} stage, to focus mainly on
inter-application scheduling decisions. These two stages collectively enforce
high-level policies regarding performance and fairness, and therefore the last
stage can get away with using simple per-bank FIFO queues (no further command
reordering within each bank) and straight forward logic that deals only with
the low-level DRAM commands and timing. This last stage is called the DRAM
Command Scheduler stage.



Our evaluation shows that \titleShortSMS is effective at reducing
inter-application interference. \titleShortSMS delivers superior performance
and fairness compared to state-of-the-art memory
schedulers~\cite{fr-fcfs,parbs,atlas,tcm}, while providing a design that is
significantly simpler to implement and that has significantly lower power consumption.

\subsection{Minimizing Inter-address-space Interference} 

Inter-address-space interference occurs 
when the GPU is \emph{shared} among multiple GPGPU applications in large-scale
computing environments~\cite{kepler,fermi,maxwell,pascal,firepro,arunkumar-isca17,vijay-hpca17,mafia}.
Much of the inter-address-space interference problem in a contemporary GPU lies
within the memory system, where multi-application execution requires virtual
memory support to manage the address spaces of each application and to provide
memory protection. We observe that when multiple GPGPU applications spatially share
the GPU, a significant amount of inter-core thrashing occurs on the shared
TLB within the GPU.  We observe that this contention at the shared TLB  is high
enough to prevent the GPU from successfully hiding memory latencies,
which causes TLB contention to become a first-order performance concern.

Based on our analysis of the TLB contention in a modern GPU system executing multiple applications, we introduce two mechanisms. First, we design
\titleLongMASK (\titleShortMASK{}). 
The key idea of \titleShortMASK{} is to 1) extend the GPU memory hierarchy to efficiently support
address translation via the use of multi-level TLBs, and 2) use
translation-aware memory and cache management techniques to maximize throughput in the
presence of inter-address-space contention. \titleShortMASK{} uses a novel
token-based approach to reduce TLB miss overheads by limiting the number of thread that can use the shared TLB, and its L2 cache bypassing
mechanisms and address-space-aware memory scheduling reduce the inter-address-space
interference. We show that \titleShortMASK{} restores much of
the thread-level parallelism that was previously lost due to address
translation.

Second, to further minimize the inter-address-space interference, we introduce
\titleShortMOSAIC. \titleShortMOSAIC significantly decreases
inter-address-space interference at the shared TLB by increasing TLB reach
via support for multiple page sizes, including very large pages. 
To enable multi-page-size support, we provide two \emph{key
observations}. First, we observe that the vast majority of memory allocations
and memory deallocations are performed \emph{en masse} by GPGPU applications in
phases, typically soon after kernel launch or before kernel exit. Second, long-lived
memory objects that usually increase fragmentation and induce complexity in CPU memory
management are largely absent in the GPU setting. These two observations make
it relatively easy to predict memory access patterns of GPGPU applications and simplify the task of
detecting when a memory region can benefit from using large pages. 

Based on the prediction of the memory access patterns, \titleShortMOSAIC{} 1) modifies GPGPU applications'
memory layout in system software to preserve address space contiguity, which allows the GPU
to splinter and coalesce pages very fast without moving data 
and 2) periodically performs memory compaction while still preserving address space
contiguity to avoid memory bloat and data fragmentation. Our prototype shows
that \titleShortMOSAIC is very effective at reducing inter-address-space
interference at the shared TLB and limits the number of shared TLB miss rate to
less than 1\% on average (down from 25.4\% in the baseline shared TLB).


In summary, \titleShortMASK incorporates TLB-awareness throughout the memory
hierarchy and introduces TLB- and cache-bypassing techniques to increase the
effectiveness of a shared TLB. \titleShortMOSAIC provides a hardware-software
cooperative technique that modifies the memory allocation policy in the system
software and introduces a high-throughput method to support large pages in
multi-GPU-application environments. The \titleShortMASK-\titleShortMOSAIC
combination provides a simple mechanism to eliminate the performance overhead
of address translation in GPUs without requiring significant changes in GPU
hardware. These techniques work together to significantly improve system
performance, IPC throughput, and fairness over the state-of-the-art memory
management technique~\cite{powers-hpca14}.




\section{Contributions}

We make the following major contributions:

\begin{itemize}

\item We provide an in-depth analyses of three different types of memory
interference in systems with GPUs. Each of these three types of interference
significantly degrades the performance and efficiency of the GPU-based systems.
To minimize memory interference, we introduce mechanisms to dynamically
analyze different applications' characteristics and propose four major changes
throughout the memory hierarchy of GPU-based systems. 

\item We introduce \titleLongMeDiC (\titleShortMeDiC). \titleShortMeDiC is a mechanism that minimizes
intra-application interference in systems with GPUs. \titleShortMeDiC is the
first work that observes that the different warps within a GPGPU application
exhibit heterogeneity in their memory divergence behavior at the shared L2
cache, and that some warps do not benefit from the few cache hits that they
have. We show that this memory divergence behavior tends to remain consistent throughout
long periods of execution for a warp, allowing for fast, online warp divergence
characterization and prediction. \titleShortMeDiC takes advantage of this
warp characterization via a combination of \emph{warp-aware} cache bypassing, cache
insertion and memory scheduling techniques. Chapter~\ref{sec:medic} provides
the detailed design and evaluation of \titleShortMeDiC.

\item We demonstrate how the GPU memory traffic in heterogeneous CPU-GPU
systems can cause severe inter-application interference, leading to poor
performance and fairness. We propose a new memory controller design, the
\titleLongSMS (\titleShortSMS), which delivers superior performance and
fairness compared to three state-of-the-art memory
schedulers~\cite{fr-fcfs,atlas,tcm}, while providing a design that is
significantly simpler to implement. The key insight behind \titleShortSMS's
scalability is that the primary functions of sophisticated memory controller
algorithms can be decoupled into different stages in a multi-level scheduler.
Chapter~\ref{sec:sms} provides the design and the evaluation of \titleShortSMS
in detail.

\item We perform a detailed analysis of the major problems in state-of-the-art
GPU virtual memory management that hinders high-performance multi-application execution. We
discover a new type of memory interference, which we call \emph{inter-address-space
interference}, that arises from a significant amount of inter-core thrashing on
the shared TLB within the GPU. We also discover that the TLB contention is high
enough to prevent the GPU from successfully hiding memory latencies, 
which causes TLB contention to become a first-order performance concern in GPU-based systems.
Based on our analysis, we introduce \titleLongMASK (\titleShortMASK). \titleShortMASK extends the GPU memory hierarchy to
efficiently support address translation through the use of multi-level TLBs,
and uses translation-aware memory and cache management to maximize IPC (instruction per cycle) throughput
in the presence of inter-application contention. \titleShortMASK{} restores
much of the thread-level parallelism that was previously lost due to address
translation. Chapter~\ref{sec:mask} analyzes the effect of inter-address-space
interference and provides the detailed design and evaluation of \titleShortMASK.

\item To further minimize the inter-address-space interference, we introduce
\titleShortMOSAIC. \titleShortMOSAIC further increases the effectiveness of TLB
by providing a hardware-software cooperative technique that modifies the memory
allocation policy in the system software. \titleShortMOSAIC introduces a low
overhead method to support large pages in multi-GPU-application environments.
The key idea of \titleShortMOSAIC is to ensure memory allocation preserve address space
contiguity to allow pages to be coalesced without any data movements.
Our prototype shows that \titleShortMOSAIC significantly increases the
effectiveness of the shared TLB in a GPU and further reduces inter-address-space interference.
Chapter~\ref{sec:mosaic} provides the detailed design and evaluation of
\titleShortMOSAIC.

\end{itemize}

\section{Dissertation Outline}

This dissertation is organized into eight Chapters.
Chapter~\ref{sec:baseline-main} presents background on modern GPU-based
systems. Chapter~\ref{sec:dissertation-background} discusses related prior
works on resource management, where techniques can potentially be applied to
reduce interference in GPU-based systems. Chapter~\ref{sec:medic} presents the
design and evaluation of \titleShortMeDiC.  \titleShortMeDiC is a mechanism
that minimizes \emph{intra-application interference} by redesigning the shared
cache and the memory controller to be aware of different types of warps.
Chapter~\ref{sec:sms} presents the design and evaluation of \titleShortSMS.
\titleShortSMS is a GPU-aware and application-aware memory controller design
that minimizes the \emph{inter-application interference}.
Chapter~\ref{sec:mask} presents a detailed analysis of the performance impact
of \emph{inter-address-space interference}. It then proposes \titleShortMASK, a
mechanism that minimizes \emph{inter-address-space interference} by introducing
TLB-awareness throughout the memory hierarchy.  Chapter~\ref{sec:mosaic}
presents the design for \titleShortMOSAIC.  \titleShortMOSAIC provides a
hardware-software cooperative technique that reduces \emph{inter-address-space
interference} by lowering contention at the shared TLB.
Chapter~\ref{sec:lesson} provides the summary of common principles and lessons
learned. Chapter~\ref{sec:conclusion} provides the summary of this dissertation
as well as future research directions that are enabled by this dissertation. 

\rachata{Done with most of Onur's comment up to here}



\chapter{The Memory Interference Problem \protect\linebreak in Systems with GPUs}
\label{sec:baseline-main}

We first provide background on the architecture of a modern GPU, and then we
discuss the bottlenecks that highly-multithreaded applications can face either
when executed alone on a GPU or when executing with other CPU or GPU applications.

\section{Modern Systems with GPUs}
\label{sec:main-gpu-baseline} 



In this section, we provide a detailed explanation of the GPU architecture that is
available on modern systems. Section~\ref{sec:main-gpu-baseline} discusses a
typical modern GPU
architecture~\cite{fermi,kepler,maxwell,pascal,amdr9,radeon,vivante-gpgpu,arm-mali,powervr,amd-fusion,apu,kaveri,haswell,amdzen,skylake,tegra,tegrax1} as well as its
memory hierarchy.  Section~\ref{sec:main-cpugpu-baseline} discusses the design
of a modern CPU-GPU heterogeneous
architecture~\cite{apu,kaveri,haswell,skylake} and its memory hierarchy.
Section~\ref{sec:main-gpucloud-baseline} discusses the memory management unit
and support for address translation.

\subsection{GPU Core Organization} 
\begin{figure}[h!]
	\centering
	\includegraphics[width=0.9\textwidth]{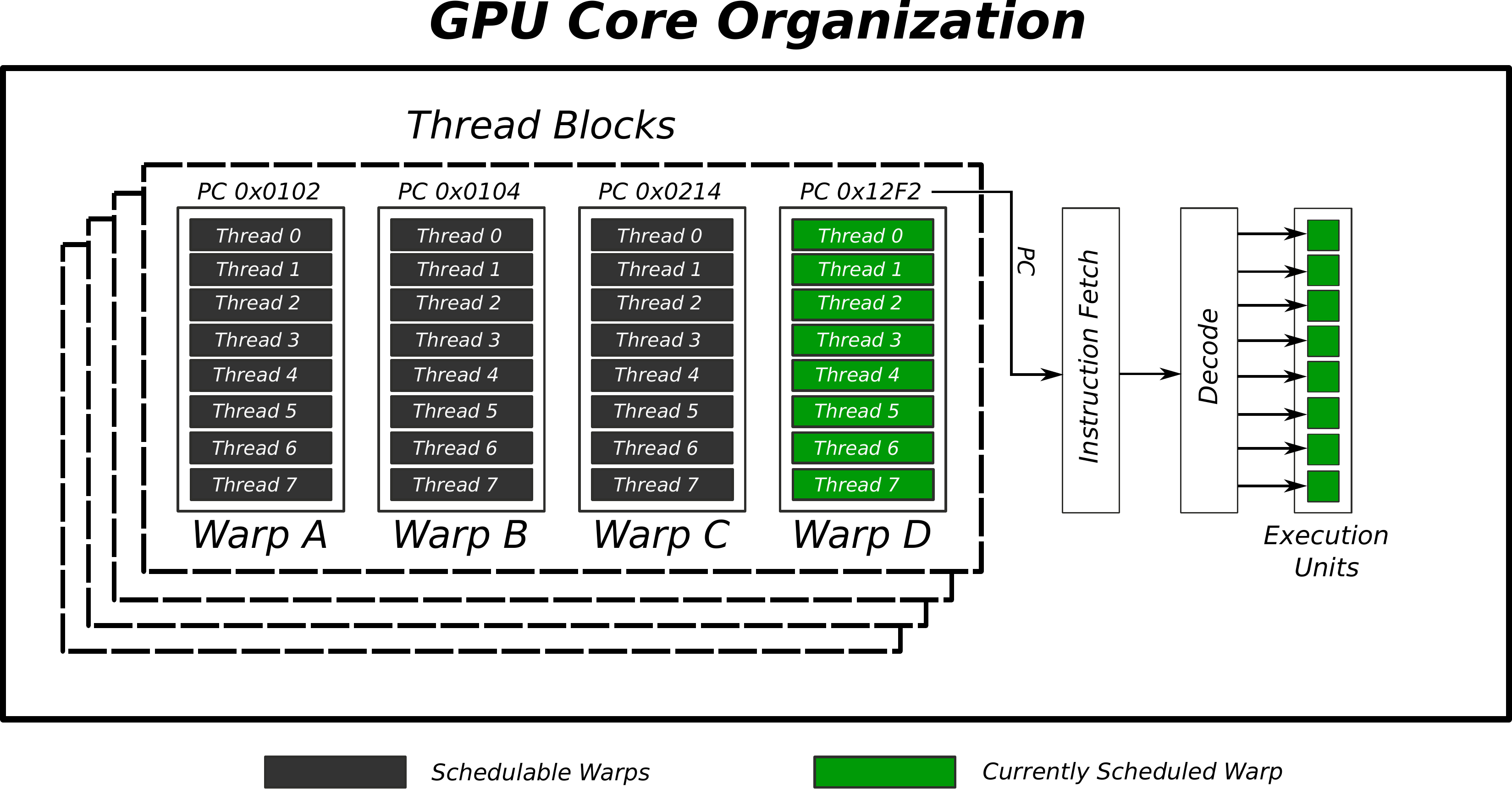}%
	\caption{Organization of threads, warps, and thread blocks.} 
	\label{fig:gpu-warp-baseline}
\vspace{-5pt}
\end{figure}

A typical GPU consists of several GPU cores called \emph{shader cores} (sometimes called
\emph{streaming multiprocessors}, or SMs). 
As shown in Figure~\ref{fig:gpu-warp-baseline}, a GPU core executes SIMD-like
instructions~\cite{flynn}. Each SIMD instruction can potentially operate on
multiple pieces of data in parallel. Each data piece operated on by a different
thread of control. Hence, the name SIMT (Single Instruction Multiple Thread).
Multiple threads that are the same are grouped into a warp. A warp is a
collection of threads that are executing the same instruction (i.e., are at the
same Program Counter). Multiple warps are grouped into a thread block. 
Every cycle, a GPU core fetches an available warp (a warp is available if none
of its threads are stalled), and issues an instruction associated with those
threads (in the example from Figure~\ref{fig:gpu-warp-baseline}, this instruction is
from \emph{Warp D} and the address of this instruction is $0x12F2$).  In this way, a GPU can
potentially retire as many instructions as the number of cores multiplied by
the number of threads per warp, enabling high instruction-per-cycle (IPC)
throughput. More detail on GPU core organization can be found
in~\cite{tbc,gebhart,warpsub,dwf,gpgpu-sim,warp-level-div,han-reducing-div,coherence-hpca13}.

\subsection{GPU Memory Hierarchy}
\label{sec:background-gpu-memory}

\begin{figure}[h!]
	\centering
	\vspace{2pt}
	\includegraphics[width=0.6\columnwidth]{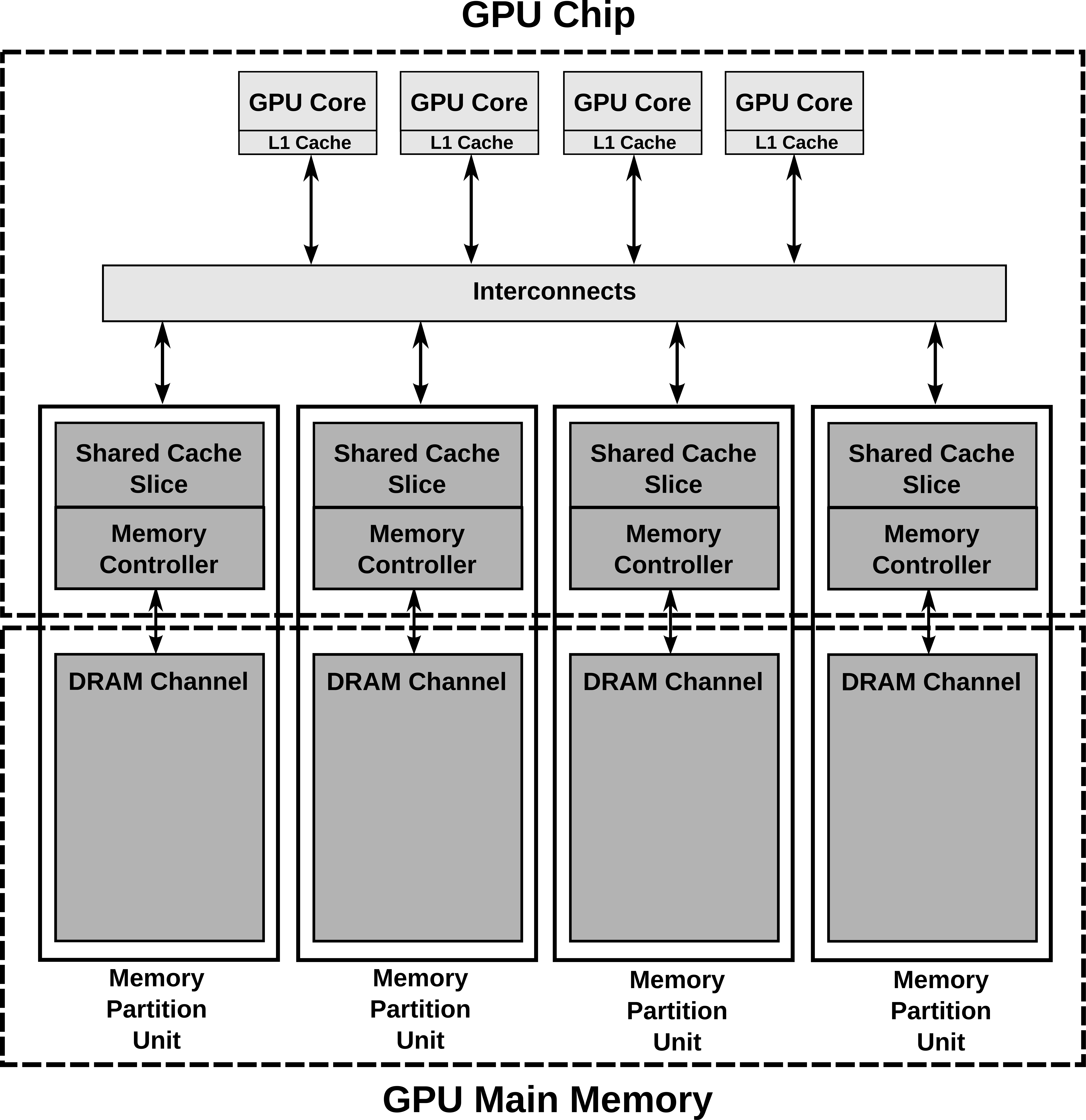}%
	\caption{Overview of a modern GPU architecture.} 
	\vspace{-6pt}
	\label{fig:gpu-spec-baseline}
\end{figure}

When there is a load or store instruction that needs to access data from the main
memory, the GPU core sends a memory request to the memory hierarchy, which is
shown in Figure~\ref{fig:gpu-spec-baseline}. This hierarchy typically contains
a private data cache, and an interconnect (typically a crossbar) that connects
all the cores with the shared data cache. If the target data is
present neither in the private nor the shared data cache, a memory request is
sent to the main memory in order to retrieve the data.

\paragraphbe{GPU Cache Organization and Key Assumptions.} Each core has its own
private L1 data, texture, and constant caches, as well as a software-managed
scratchpad
memory~\cite{kepler,fermi,lindholm,maxwell,pascal,radeon,ati-wavefront,zorua}.
In addition, the GPU also has several shared L2 cache slices and memory
controllers. 
Because there are several methods to design the GPU memory hierarchy, we assume
the baseline that decouples the memory channels into multiple memory partitions.
A \emph{memory partition unit} combines a single L2 cache slice (which is
banked) with a designated memory controller that connects the GPU to off-chip
main memory.  Figure~\ref{fig:gpu-spec-baseline} shows a simplified view of how
the cores (or SMs), caches, and memory partitions are organized in our baseline
GPU. 

\paragraphbe{GPU Main Memory Organization.}
Similar to systems with CPUs, a GPU uses DRAM (organized as hierarchical two-dimensional arrays of
bitcells) as main memory. Reading or writing data to DRAM requires that a row of bitcells from
the array first be read into a row buffer. This is required because the act of
reading the row destroys the row's contents, and so a copy of the bit values
must be kept (in the row buffer). Reads and writes operate directly on the row
buffer. Eventually, the row is ``closed'' whereby the data in the row buffer is
written back into the DRAM array. Accessing data already loaded in the row
buffer, also called a row buffer hit, incurs a shorter latency than when the
corresponding row must first be ``opened'' from the DRAM array.  A modern
memory controller (memory controller) must orchestrate the sequence of commands to open, read,
write and close rows. Servicing requests in an order that increases row-buffer
hit rate tends to improve overall throughput by reducing the average latency to
service requests. The memory controller is also responsible for enforcing a wide variety of
timing constraints imposed by modern DRAM standards (e.g., DDR3) such as
limiting the rate of page-open operations (t$_\text{FAW}$) and ensuring a
minimum amount of time between writes and reads (t$_\text{WTR}$). More detail on
timing constraints and DRAM operation can be found in~\cite{salp,al-dram,tl-dram,ava-dram,raidr,seshadri2013rowclone,lisa,donghyuk-stack,dsarp,chargecache,softmc}.

Each two-dimensional array of DRAM cells constitutes a bank, and a group of banks
forms a rank. All banks within a rank share a common set of command and data buses,
and the memory controller is responsible for scheduling commands such that each bus is used
by only one bank at a time. Operations on multiple banks may occur in parallel (e.g.,
opening a row in one bank while reading data from another bank's row buffer) so
long as the buses are properly scheduled and any other DRAM timing constraints
are honored. A memory controller can improve memory system throughput by scheduling requests
such that bank-level parallelism or BLP (i.e., the number of banks simultaneously
busy responding to commands) is higher~\cite{parbs,cjlee-micro09}.
A memory system implementation may support multiple independent memory channels (each
with its own ranks and banks)~\cite{mcp,rajeev-pact10} to further
increase the number of memory requests that can be serviced at the same time.
A key challenge in the implementation of modern, high-performance memory controllers is to
effectively improve system performance by maximizing both row-buffer hits and
BLP while simultaneously providing fairness among multiple CPUs and the GPU~\cite{sms}.

\paragraphbe{Key Assumptions.} We assume the memory controller consists of a
centralized memory request buffer. Additional details of the memory
controller design can be found in
Sections~\ref{sec:methodology-medic},~\ref{sec:meth-mask}
and~\ref{sec:meth-mosaic}.


\subsection{Intra-application Interference within GPU Applications}

While many GPGPU applications can tolerate a significant amount of memory
latency due to their parallelism through the SIMT execution model, many
previous works (e.g.,~\cite{caba,owl-asplos13,largewarps,osp-isca13,gpgpu-sim,warp-level-div,han-reducing-div,dwf,tbc,chatterjee-sc14,ccws,tor-micro13,warpsub}) observe
that GPU cores often stall for a significant fraction of time. One significant
source of these stalls is the contention at the shared GPU memory
hierarchy~\cite{largewarps,warpsub,ccws,caba,owl-asplos13,osp-isca13,nmnl-pact13,chatterjee-sc14,medic}. 
The large amount of parallelism in GPU-based systems creates a significant
amount of contention on the GPU's memory hierarchy. Even through all threads in
the GPU execute the codes from the same application, data accesses from one
warp can interfere with data accesses from other warps. This interference comes
in several forms such as additional cache thrashing and queuing delays
at both the shared cache and shared main memory. These combine to lower the
performance of GPU-based systems. We call this interference the
\emph{intra-application interference}.

\emph{Memory divergence}, where the threads of a warp reach a memory
instruction, and some of the threads' memory requests take longer to service
than the requests from other
threads~\cite{largewarps,warpsub,chatterjee-sc14,medic}, further exacerbates
the effect of \emph{intra-application interference}. Since all threads within a
warp operate in lockstep due to the SIMD execution model, the warp cannot
proceed to the next instruction until the \emph{slowest} request within the
warp completes, and \emph{all} threads are ready to continue execution.

Chapter~\ref{sec:medic} provides detailed analyses on how to reduce
\emph{intra-application interference} at the shared cache and the shared main
memory.

\rachata{Done with most of Onur's comment up to here}

%

\section{GPUs in CPU-GPU Heterogeneous Architectures}
\label{sec:main-cpugpu-baseline}

\begin{figure}[h!]
\centering
\includegraphics[width=0.8\textwidth]{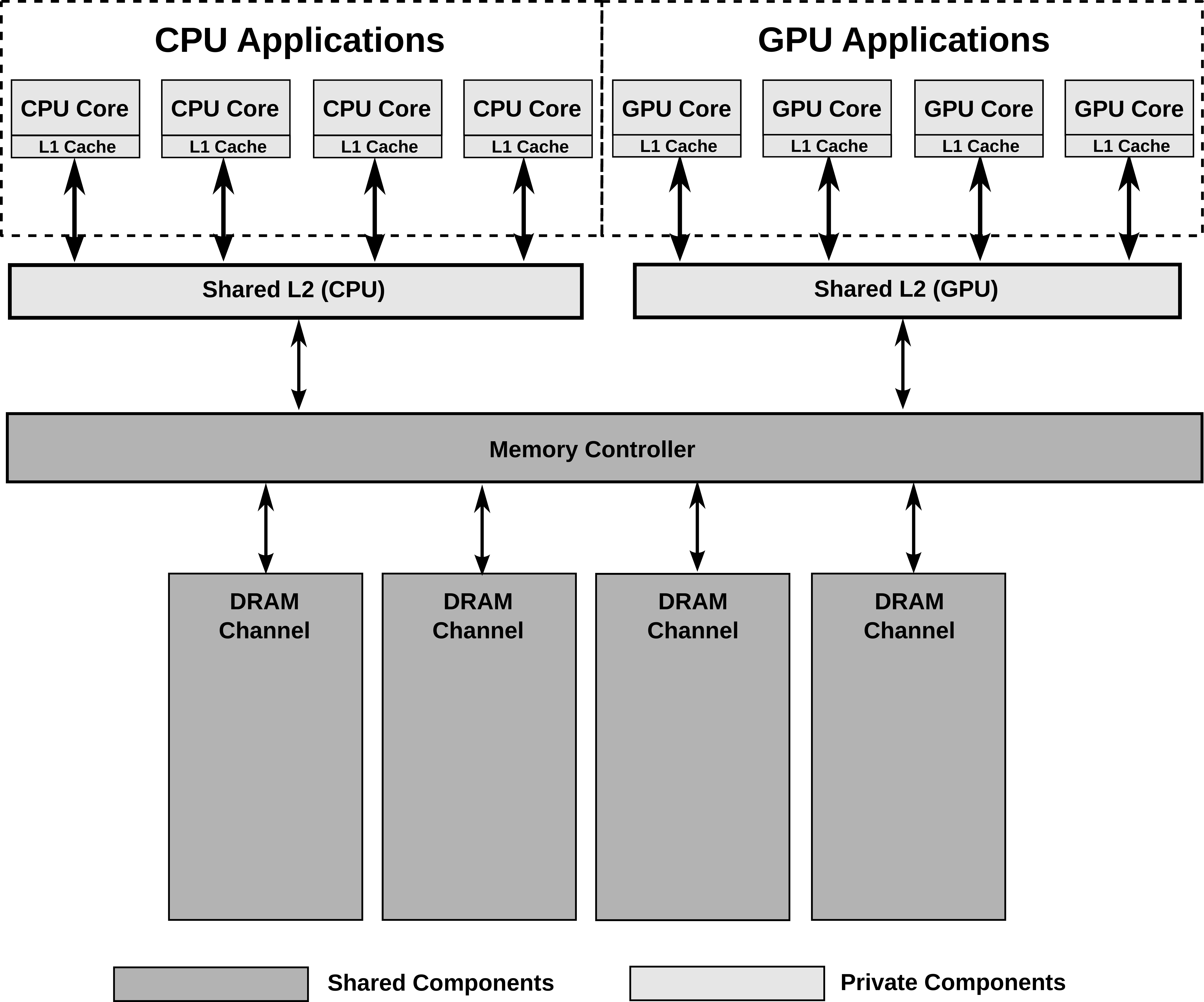}
\caption{The memory hierarchy of a heterogeneous CPU-GPU architecture.}
\label{fig:cpugpu-baseline}
\end{figure}

Aside from using off-chip discrete GPUs, modern architectures integrate
Graphics Processors integrate a GPU on the same chip as the CPU
cores~\cite{skylake,haswell,sandybridge,ivybridge,bobcat,kaveri,apu,tegra,amdzen,powervr,arm-mali,sms,cpugpu-micro,jeong2012qos}.
Figure~\ref{fig:cpugpu-baseline} shows the design of these recent heterogeneous
CPU-GPU architecture. As shown in Figure~\ref{fig:cpugpu-baseline}, parts of the
memory hierarchy are being shared across both CPU and GPU applications.

\paragraphbe{Key Assumptions.} We make two key assumptions for
the design of heterogeneous CPU-GPU systems. First, we assume that the GPUs and
the CPUs do not share the last level caches. Second, we assume that the memory
controller is the first point in the memory hierarchy that CPU applications and
GPU applications share resources. We applied multiple memory scheduler designs
as the baseline for our evaluations. Additional details of these baseline design
can be found in Sections~\ref{sec:meth-sms} and ~\ref{sec:sms-qual-eval}.

\subsection{Inter-application Interference across CPU and GPU Applications}
\label{sec:baseline-cpugpu-interfere}


As illustrated in Figure~\ref{fig:cpugpu-baseline}, the main memory is a major
shared resource among cores in modern chip multiprocessor (CMP) systems. Memory
requests from multiple cores interfere with each other at the main memory and
create \emph{inter-application interference}, which is a significant impediment to individual
application and system performance. Previous works on CPU-only application-aware memory
scheduling~\cite{stfm,parbs,atlas,tcm,bliss,pam,pa-micro08} have addressed the problem by
being aware of application characteristics at the memory controller and
prioritizing memory requests to improve system performance and fairness. This
approach of application-aware memory request scheduling has provided good
system performance and fairness in multicore systems.


As opposed to CPU applications, GPU applications are not very latency sensitive
as there are a large number of independent threads to cover long memory
latencies. However, the GPU requires a significant amount of bandwidth far
exceeding even the most memory-intensive CPU applications. As a result, a GPU
memory scheduler~\cite{lindholm,kepler,pascal} typically needs a large request
buffer that is capable of request coalescing (i.e., combining multiple requests
for the same block of memory into a single combined request~\cite{lindholm}).
Furthermore, since GPU applications are bandwidth intensive, often with
streaming access patterns, a policy that maximizes the number of row-buffer
hits is effective for GPUs to maximize overall throughput. Hence, a memory
scheduler that can improve the effective DRAM bandwidth such as the FR-FCFS
scheduler with a large request
buffer~\cite{fr-fcfs,frfcfs-patent,gpgpu-sim,complexity} tends to perform well
for GPUs. 

This conflicting preference between CPU applications and GPU applications (CPU
applications benefit from lower memory request latency while GPU applications
benefit from higher DRAM bandwidth) further complicates the design of memory
request scheduler for CPU-GPU heterogeneous systems. A design that favors
lowering the latency of CPU requests is undesirable for GPU applications while
a design that favors providing high bandwidth is undesirable for CPU
applications. 

In this dissertation, Chapter~\ref{sec:sms} provides an in-depth analysis of
this \emph{inter-application interference} and provides a method to mitigate the
interference in CPU-GPU heterogeneous architecture.

\section{GPUs in Multi-GPU-application Environments}
\label{sec:main-gpucloud-baseline} 

Recently, a newer set of analytic GPGPU applications, such as the Netflix movie
recommendation systems~\cite{netflix}, or a stock market
analyzer~\cite{tobias-stock}, require a closely connected, highly virtualized,
shared environment. These applications, which benefit from the amount of
parallelism GPU provides, do not need to use all resources in the GPU to
maximize their performance. Instead, these emerging applications benefit from
concurrency - by running a few of these applications together, each sharing
some resources on the GPU. NVIDIA GRID~\cite{kepler,pascal} and AMD
FirePro~\cite{firepro} are two examples of spatially share GPU resources across
multiple applications.

\begin{figure}
\centering
\includegraphics[width=0.8\textwidth]{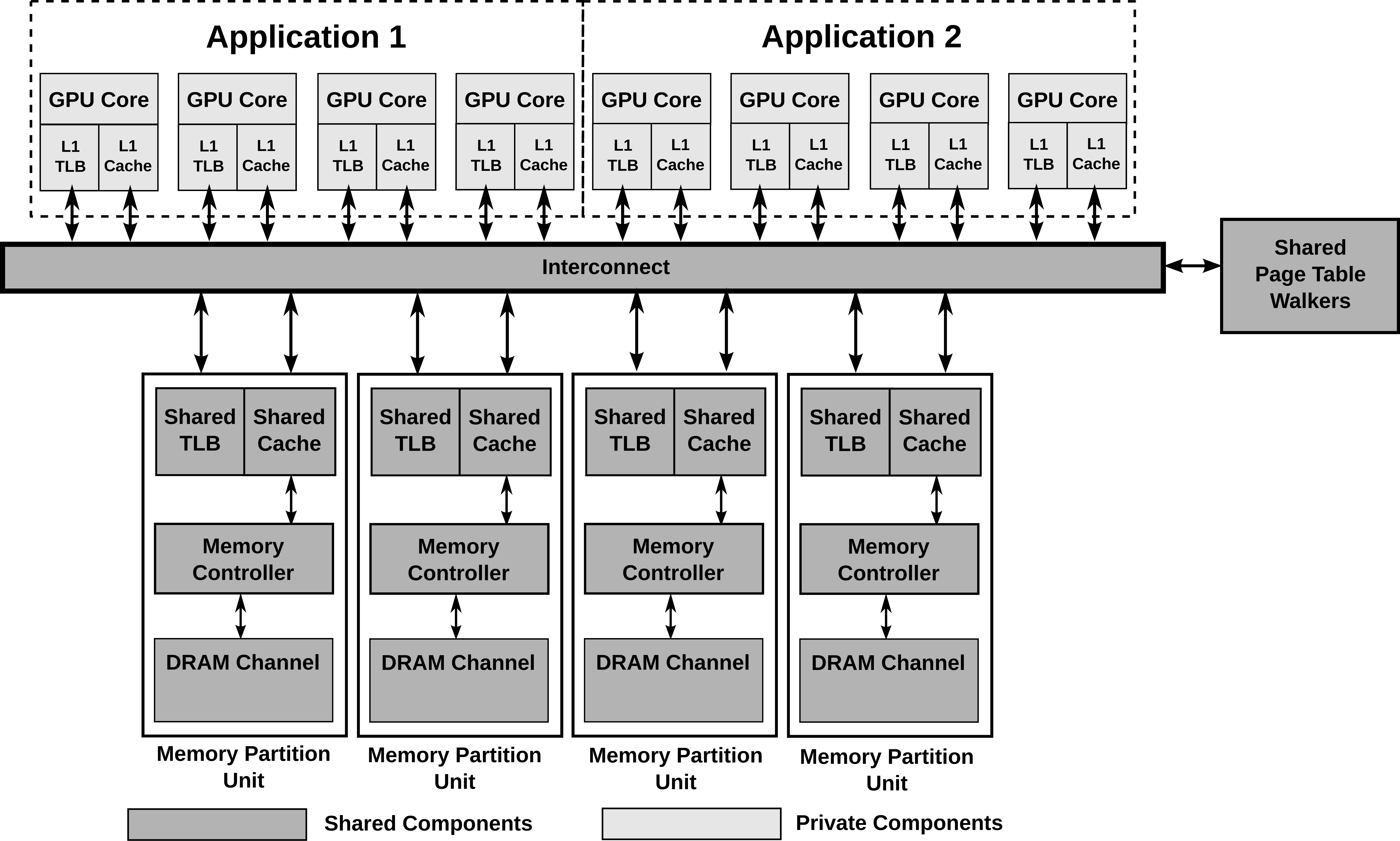}
\caption{A GPU design showing two concurrent GPGPU applications concurrently sharing the GPUs.}
\label{fig:GRID}
\end{figure}

Figure~\ref{fig:GRID} shows the high-level design of how a GPU can be spatially
shared across two GPGPU applications. In this example, the GPUs contain
multiple shared page table walkers, which are responsible for translating a virtual
address into a physical address. This design also contains two level of
translation lookaside buffers (TLBs), which cache the virtual-to-physical
translation. This design allows the GPU to co-schedule kernels, even
applications, concurrently because \emph{address translation enables memory
protection across multiple GPGPU applications.}

\paragraphbe{Key Assumptions.} The page table walker can be placed at different
locations in the GPU memory hierarchy. The GPU MMU design proposed by Power et
al. places a parallel page table walkers between the private L1 and the shared L2
caches~\cite{powers-hpca14}. Other alternative designs place the page table
walker at the Input-Output Memory Management Unit (IOMMU), which directly
connects to the main
memory~\cite{amd-fusion,apu,kaveri,haswell,amdzen,skylake,powervr,arm-mali,tegra,tegrax1,fermi,kepler,maxwell,pascal,amdr9,radeon,vivante-gpgpu,powervr},
and another GPU MMU design proposed by Cong et al. uses the CPU's page table
walker to perform GPU page walks~\cite{cong-hpca17}. We found that placing a parallel
page table walkers at the shared L2 cache provides the best performance. Hence,
we assume the baseline proposed by Power et al. that utilized the per-core
private TLB and place the page table walker at the shared L2
cache~\cite{powers-hpca14}.

\subsection{Inter-address-space Interference on Multiple GPU Applications}

While concurrently executing multiple GPGPU applications that have
complementary resource demand can improve GPU utilization, these applications
also share two critical resources: the shared address translation unit
and the shared TLB. We find that when multiple applications spatially share the
GPU, there is a significant amount of thrashing on the shared TLB within the
GPU because multiple applications from different address spaces are contending
at the shared TLB, the page table walker as well as the shared L2 data cache.
We define this phenomenon as the \emph{inter-address-space interference}. 

The amount of parallelism on GPUs further exacerbate the performance impact of
\emph{inter-address-space interference}. We found that an address translation
in response to a single TLB miss typically stalls tens of warps. As a result, a
small number of outstanding TLB misses can result in a significant number of
warps to become unschedulable, which in turn limits the GPU's most essential
latency-hiding capability. We observe that providing address translation in
GPUs reduce the GPU performance to 47.3\% of the ideal GPU with no address
translation, which is a significant performance overhead. As a result, it is
even more crucial to mitigate this \emph{inter-address-space interference}
throughout the GPU memory hierarchy in multi-GPU-application environments.
Chapters~\ref{sec:mask} and~\ref{sec:mosaic} provide detailed design
descriptions of the two mechanisms we propose that can be used to reduce this
\emph{inter-address-space interference}.

%
%
%


\chapter{Related Works on Resource Management \protect\linebreak in Systems with GPUs}
\label{sec:dissertation-background}

Several previous works have been proposed to address the memory interference problem
in systems with GPUs. These previous proposals address certain parts of the
main memory hierarchy. In this chapter, we first provide the background on the GPU's
execution model. Then, we provide breakdowns of previous works
on GPU resource management throughout the memory hierarchy as well as
differences between these previous works and techniques presented in this dissertation.

\section{Background on the Execution Model of GPUs}
\label{sec:disstertation-gpu-design}

Modern day GPUs employ two main techniques to enable their parallel processing
power: SIMD, which executes multiple data within a single instruction, and
fine-grain multithreading, which prevents the GPU cores from stalling by issuing
instructions from different threads every cycle. This section provides the
background on previous machines and processors that apply similar techniques.

\subsection{SIMD and Vector Processing}


The SIMD execution model, which includes vector processing, been used by
several machines in the past.  Slotnik et al. in the Solomon
Computer~\cite{solomon62}, Senzig and Smith~\cite{senzig-afips65}, Crane and
Guthens~\cite{crane-ec65}, Hellerman~\cite{hellerman-ec66}, CDC
7600~\cite{cdc7600}, CDC STAR-100~\cite{cdcstar}, Illiac IV~\cite{illiac} and
Cray I~\cite{cray1} are examples of machines that employ a vector processor. In
modern systems, Intel MMX~\cite{mmx,peleg-ieeemicro96} and Intel SSE~\cite{haswell} also apply
SIMD in order to improve performance. As an alternative of using one
instruction to execute multiple data, VLIW~\cite{vliw} generate codes for a
parallel machine that allows multiple instructions to operate on multiple
data concurrently in a single cycle. Intel i860~\cite{grimes1989intel}
and Intel Itanium~\cite{itanium} are examples of processors with the VLIW technology.

\subsection{Fine-grained Multithreading}

Fine-grain multithreading, which is a technique that allows the processor to
issue instructions from different threads every cycle, is the key component
that enables latency hiding capability in modern day GPUs. CDC
6600~\cite{cdc6600,cdc6600-2}, Denelcor HEP~\cite{hep}, MASA~\cite{masa-fmt},
APRIL~\cite{april-fmt} and Tera MTA~\cite{tera-mta} are examples of machines
that utilize fine-grain multithreading.

\section{Background on Techniques to Reduce Interference of Shared Resources}
\label{sec:dissertation-background-memory}

Several techniques to reduce interference at the shared cache, shared off-chip
main memory as well as the shared interconnect have been proposed.  In this
section, we provide a brief discussion of these works.

\subsection{Cache Bypassing Techniques}

\paragraphbe{Hardware-based Cache Bypassing Techniques.} 
Several hardware-based cache bypassing mechanisms have been proposed in both CPU
and GPU setups. Li et al. propose PCAL, a bypassing mechanism that
addresses the cache thrashing problem by throttling the number of threads that
time-share the cache at any given time~\cite{donglihpca15}. The key idea of
PCAL is to limit the number of threads that get to access the cache. 
Li et al.~\cite{li-ics15} propose a cache bypassing mechanism that allows only
threads with high reuse to utilize the cache. The key idea is to use locality
filtering based on the reuse characteristics of GPGPU applications, with only
high reuse threads having access to the cache. Xie et al.~\cite{xie-hpca15}
propose a bypassing mechanism at the thread block level. In their mechanism,
the compiler statically marks whether thread blocks prefer caching or
bypassing. At runtime, the mechanism dynamically selects a subset of thread
blocks to use the cache, to increase cache utilization.
Chen et al.~\cite{chen-micro47,chen-mes14} propose a combined warp throttling and
bypassing mechanism for the L1 cache based on the cache-conscious warp
scheduler~\cite{ccws}. The key idea is to bypass the cache when resource
contention is detected. This is done by embedding history information into the
L2 tag arrays. The L1 cache uses this information to perform bypassing
decisions, and only warps with high reuse are allowed to access the L1 cache. 
Jia et al. propose an L1 bypassing mechanism~\cite{mrpb}, whose
key idea is to bypass requests when there is an associativity stall.
Dai et al. propose a mechanism to bypass cache based on a model of a cache
miss rate~\cite{dai-dac16}.

There are also several other CPU-based cache bypassing techniques. These
techniques include using additional buffers track cache statistics to predict 
cache blocks that have high utility based on reuse count
~\cite{annex-cache,gaur-isca11,zhang-ispled14,kharbutli-ieeetran,etsion-tc,chaudhuri-pact12,xiang-ics09,liu-micro08}, 
reuse distance~\cite{doung-micro12,gupta-ipdps13,chaudhuri-pact12,feng-interact12,park-sc13,yu-dasc,gao2010dueling,youfeng-micro02},
behavior of the cache block~\cite{jalminger-iccp03} 
or miss rate~\cite{mct,tyson-micro95}

\rachata{I think that is all}



\paragraphbe{Software-based Cache Bypassing Techniques.} 
Because GPUs allow software to specify whether to utilize the cache or
not~\cite{nvidia-bypassing,maxwell-guide}.  Software based cache bypassing techniques have also
been proposed to improve system performance. Li et al.~\cite{li-sc15} propose
a compiler-based technique that performs cache bypassing using a method similar
to PCAL~\cite{donglihpca15}. Xie et al.~\cite{xie-iccad13} propose a mechanism
that allows the compiler to perform cache bypassing for global load
instructions. Both of these mechanisms apply bypassing to \emph{all} loads and
stores that utilize the shared cache, without requiring additional
characterization at the compiler level. Mekkat et al.~\cite{mekkat-pact13}
propose a bypassing mechanism for when a CPU and a GPU share the last level
cache. Their key idea is to bypass GPU cache accesses when CPU applications are
cache sensitive, which is not applicable to GPU-only execution.


\subsection{Cache Insertion and Replacement Policies} 

Many works have proposed different insertion policies for CPU systems
(e.g.,~\cite{bip,eaf-vivek,dip,rrip}). Dynamic Insertion Policy
(DIP)~\cite{dip} and Dynamic Re-Reference Interval Prediction
(DRRIP)~\cite{rrip} are insertion policies that account for cache thrashing.
The downside of these two policies is that they are unable to distinguish
between high-reuse and low-reuse blocks in the same thread~\cite{eaf-vivek}.
The Bi-modal Insertion Policy~\cite{bip} dynamically characterizes the cache
blocks being inserted. None of these works on cache insertion and replacement
policies~\cite{bip,eaf-vivek,dip,rrip} take warp type characteristics or memory
divergence behavior into account.


%

\subsection{Cache and Memory Partitioning Techniques}


Instead of mitigating the interference problem between applications by
scheduling requests at the memory controller, Awasthi et al. propose a
mechanism that spreads data in the same working set across memory channels in
order to increase memory level parallelism~\cite{rajeev-pact10}. Muralidhara et
al. propose memory channel partitioning (MCP) to map applications to different
memory channels based on their memory-intensities and row-buffer locality to
reduce inter-application interference~\cite{mcp}. Mao et al. propose to
partition GPU channels and only allow a subset of threads to access each memory
channel~\cite{mao-temp}. In addition to channel partitioning, several works
also propose to partition DRAM banks~\cite{xie-hpca14,ikeda-icpads13,liu-pact12} 
and the shared cache~\cite{lavanya-asm,ucp} to improve performance. These
partitioning techniques are orthogonal to our proposals and can be combined
to improve the performance of GPU-based systems.


\subsection{Memory Scheduling on CPUs} 

Memory scheduling algorithms improve system performance by reordering memory
requests to deal with the different constraints and behaviors of DRAM. 
The first-ready-first-come-first-serve (FR-FCFS)~\cite{fr-fcfs} algorithm
attempts to schedule requests that result in row-buffer hits (first-ready), and
otherwise prioritizes older requests (FCFS). FR-FCFS increases DRAM throughput,
but it can cause fairness problems by under-servicing applications with low
row-buffer locality. 
Ebrahimi et al.~\cite{pam} propose PAM, a memory scheduler that prioritizes
critical threads in order to improve the performance of multithreaded
applications. Ipek et al. propose a self-optimizing memory scheduling that
improve system performance with reinforcement
learning~\cite{reinforcement-learning}. Mukundan and Martinez propose MORSE, a
self-optimizing reconfigurable memory scheduler~\cite{morse-hpca12}. Lee et al.
propose two prefetch aware memory scheduling
designs~\cite{pa-micro08,cjlee-micro09}. Stuecheli et al.~\cite{vwq-isca10} and
Lee et al.~\cite{lee2010dram} propose memory schedulers that are aware of
writeback requests. Seshadri et al.~\cite{dbi} propose to simplify the
implementation of row-locality-aware write back by exploiting the dirty-block
index. Several application-aware memory scheduling
algorithms~\cite{mutlu-podc08,atlas,tcm,stfm,parbs,bliss,mise} have been proposed to balance
both performance and fairness. Parallelism-aware Batch Scheduling
(PAR-BS)~\cite{parbs} batches requests based on their arrival times (older
requests batched first). Within a batch, applications are ranked to preserve
bank-level parallelism (BLP) within an application's requests.  Kim et al.
propose ATLAS~\cite{atlas}, which prioritizes applications that have received
the least memory service. As a result, applications with low memory
intensities, which typically attain low memory service, are prioritized.
However, applications with high memory intensities are deprioritized and hence
slowed down significantly, resulting in unfairness. Kim et al. further propose
TCM~\cite{tcm}, which addresses the unfairness problem in ATLAS. TCM first
clusters applications into low and high memory-intensity clusters based on
their memory intensities. TCM always prioritizes applications in the low
memory-intensity cluster, however, among the high memory-intensity applications
it shuffles request priorities to prevent unfairness. Ghose et al. propose a
memory scheduler that takes into account of the criticality of each load and
prioritizes loads that are more critical to CPU performance~\cite{ghose2013}.
Subramanian et al. propose MISE~\cite{mise}, which is a memory scheduler that
estimates slowdowns of applications and prioritizes applications that are
likely to be slow down the most. Subramanian et al. also propose
BLISS~\cite{bliss,bliss-tpds}, which is a mechanism that separates applications into a
group that interferes with other applications and another group that does not,
and prioritizes the latter group to increase performance and fairness. Xiong et
al. propose DMPS, a ranking based on latency sensitivity~\cite{xiong-taco16}.
Liu et al. propose LAMS, a memory scheduler that prioritizes requests based on
the latency of servicing each memory request ~\cite{liu-ipccc16}.

\subsection{Memory Scheduling on GPUs} 

Since GPU applications are bandwidth intensive, often with streaming access
patterns, a policy that maximizes the number of row-buffer hits is effective
for GPUs to maximize overall throughput. As a result, FR-FCFS with a large
request buffer tends to perform well for GPUs~\cite{gpgpu-sim}. In view of this,
previous work~\cite{complexity} designed mechanisms to reduce the complexity of
row-hit first based (FR-FCFS) scheduling. Jeong et al. propose a QoS-aware
memory scheduler that guarantees the performance of GPU applications by
prioritizing Graphics applications over CPU applications until the system can
guarantee a frame can be rendered within its deadline, and prioritize CPU
applications afterward~\cite{jeong2012qos}. Jog et
al.~\cite{adwait-critical-memsched} propose CLAM, a memory scheduler that
identifies critical memory requests and prioritizes them in the main memory.

Aside from CPU-GPU heterogeneous systems, Usui et at. propose
SQUASH~\cite{usui-squash} and DASH~\cite{usui-dash}, which are
accelerator-aware memory controller designs that improve the performance of
systems with CPU and hardware accelerators. Zhao et al. propose FIRM, a memory
controller design that improves the performance of systems with persistent
memory~\cite{jishen-firm}.

\subsection{DRAM Designs}

Aside from memory scheduling and memory partitioning techniques, previous works
propose new designs that are capable of reducing memory latency in
conventional
DRAM~\cite{chang-sigmetric16,lisa,dsarp,al-dram,tl-dram,ava-dram,donghyuk-stack,salp,donghyuk-ddma,micron-rldram3,sato-vlsic1998,hart-compcon1994,hidaka-ieeemicro90,hsu-isca1993,kedem-1997,son-isca2013,luo-dsn2014,chatterjee-micro2012,phadke-date2011,shin-hpca2014,chandrasekar-date2014,o-isca2014,zheng-micro2008,ware-iccd2006,ahn-cal2009,ahn-taco2012,chang-sigmetric17,imw2013,superfri}
as well as non-volatile
memory~\cite{meza-weed2013,ku-ispass2013,meza-cal2012,yoon-iccd2012,qureshi-isca2009,qureshi-micro2009,lee-isca2009,lee-ieeemicro2010,lee-cacm2010,meza-iccd2012}.
Previous works on bulk data transfer~\cite{gschwind-cf2006, gummaraju-pact2007,
kahle-ibmjrd2005,carter-hpca1999,zhang-ieee2001,seo-patent,intelioat,zhao-iccd2005,jiang-pact2009,seshadri2013rowclone,lu-micro2015,lisa}
and in-memory
computation~\cite{ahn-isca2015,ahn-isca2015-2,7056040,7429299,guo-wondp14,592312,
seshadri-cal2015,mai-isca2000,draper-ics2002,
seshadri-micro2015,seshadri-arxiv2016,hsieh-iccd2016,tom-isca16,
amirali-cal2016, stone-1970, fraguela-2003,375174,808425,
4115697,694774,sura-2015,zhang-2014,akin-isca2015,kim.bmc18,pim-chapter,googlePIM,
babarinsa-2015,7446059,6844483,pattnaik-pact2016,ambit} can be used improve DRAM
bandwidth.  Techniques to reduce the overhead of DRAM refresh~\cite{raidr,
venkatesan-hpca2006,bhati-isca2015, lin-iccd2012, agrawal-memsys2016,
nair-isca2013,liu-isca2013,khan-dsn2016,khan-sigmetrics2014,khan-cal2017,khan-micro2017,patel-isca2017,qureshi-dsn2015,
kim-asic2001,baek-tc2014,agrawal-hpca2014,ohsawa-islped1998,qureshi-dsn2015}
can be applied to improve the performance of GPU-based systems.
Data compression techniques can also be used on the main memory to increase the
effective DRAM
bandwidth~\cite{bdi-pact12,lcp-micro13,toggle-hpca16,compress-reuse-hpca15,caba}.
These techniques can be used to mitigate the performance impact of memory
interference and improve the performance of GPU-based systems. They are
orthogonal and can be combined with techniques proposed in this dissertation.

Previous works on data prefetching can also be used to mitigate high DRAM
latency~\cite{pa-micro08,seshadri-taco2015,cjlee-micro09,nesbit-pact2004,srinath-hpca2007,lai-isca2001,alameldeen-hpca2007,baer-1995,cao-sigmetrics1995,dahlgren-1995,ebrahimi-micro09,jouppi-isca90,hur-micro2006,joseph-isca1997,cooksey-asplos2002,ebrahimi-hpca,ebrahimi-isca2011,mutlu-isca2005,
mutlu-hpca2003, mutlu-micro2005,
hashemi-isca2016,lee-tc2011,hashemi-micro2016}. However, these techniques generally
increase DRAM bandwidth, which lead to lower GPU performance.

Upcoming works~\cite{vijaykumar-isca18,vijaykumar-xmem-isca18} propose cross-layer abstractions to enable the programmer
to better manage GPU memory system resources by expressing semantic information
about high-level data structures.

\subsection{Interconnect Contention Management}

Aside from the shared cache and the shared off-chip main memory, on-chip interconnect is another shared
resources on the GPU memory hierarchy. While this dissertation does not focus on the contention
of shared on-chip interconnect, many previous works provide mechanisms
to reduce contention of the shared on-chip interconnect. These include works on
hierarchical on-chip network
designs~\cite{hird,hird-journal,ravindran97,xiangdong95,hr-model,ravindran98,numachine,das09,sci,ksr},
low cost router designs~\cite{kim09,hird,hird-journal,mullins04,rotary-router,Kodi08,grot-hpca2009}, bufferless
interconnect
designs~\cite{hotpotato,gomez08,scarab,chaosrouter,casebufferless,chipper,HAT-SBAC_PAD-2012,minbd,minbd-book,hotnets2010,sigcomm12,hep,tera-mta,cm}
and Quality-of-Service-aware interconnect
designs~\cite{pvc,grot2010,grot-isca2011,Reetu-MICRO2009,aergia,maze-routing,a2c,het-nocs-dac13}.

%

\section{Background on Memory Management Unit and Address Translation Designs}
\label{sec:dissertation-background-mmu}
\label{sec:virt-back}

\rachata{Why this is needed for GPU virtualization and why these matters.}

\rachata{This section might more work compared to other.}

Aside from the caches and the main memory, the memory management unit (MMU) is
another important component in the memory hierarchy. The MMU provides address
translation for applications running on the GPU. When multiple GPGPU
applications are concurrently running, the MMU is also provides memory
protection across different virtual address spaces that are concurrently using
the GPU memory. This section first introduces previous works on concurrent
GPGPU application. Then, we provide background on previous works on TLB
designs that aids address translation.


\subsection{Background on Concurrent Execution of GPGPU Applications}

\paragraphbe{Concurrent Kernels and GPU Multiprogramming.} 
The opportunity to improve utilization 
with concurrency is well-recognized but previous 
proposals~\cite{asplos-sree,wang-hpca16,li2014symbiotic},
do not support memory protection.
Adriaens et al.~\cite{gpu-multitasking} 
observe the need for spatial sharing across protection domains
but do not propose or evaluate a design.
NVIDIA GRID~\cite{grid} and
AMD FirePro~\cite{firepro} support static
partitioning of hardware to allow kernels from different VMs
to run concurrently---partitions are determined at startup,
causing fragmentation and under-utilization.
The goal of our proposal, \titleShortMASK, is a flexible dynamic partitioning of shared resources. 
NVIDIA's Multi Process Service (MPS)~\cite{mps} allows multiple processes to 
launch kernels on the GPU: the service provides no memory protection or error
containment. 
Xu et al~\cite{warp-slicer} propose Warped-Slicer,
which is a mechanism for multiple applications to spatially share a GPU core.
Similar to MPS, Warped-Slicer provides no memory protection, and is not suitable
for supporting multi-application in a multi-tenant cloud setting.

\paragraphbe{Preemption and Context Switching.}
Preemptive context switching
is an active research area~\cite{isca-2014-preemptive, gebhart,wang-hpca16}.
Current architectural support~\cite{lindholm,pascal} will likely improve in
future GPUs. Preemption and spatial multiplexing are complementary to the goal
of this dissertation, and exploring techniques to combine them is future work.


\paragraphbe{GPU Virtualization.} 
Most current hypervisor-based full virtualization techniques for
GPGPUs~\cite{gdev,gpuvm,gVirt} must support a virtual device abstraction
without dedicated hardware support for VDI found  in GRID~\cite{grid} and
FirePro~\cite{firepro} . Key components missing from these proposals includes
support for dynamic partitioning of hardware resources and efficient techniques
for handling over-subscription. Performance overheads incurred by some of these designs
argue strongly for hardware assists such as those we propose. By contrast,
API-remoting solutions such as vmCUDA~\cite{vmCUDA} and rCUDA~\cite{rcuda}
provide near native performance but require modifications to guest software and
sacrifice both isolation and compatibility.

\paragraphbe{Other Methods to Enable Virtual Memory.}
Vesely et al.\ analyze support for virtual memory in heterogeneous
systems~\cite{abhishek-ispass16}, finding that the cost of address translation
in GPUs is an order of magnitude higher than in CPUs and that high latency
address translations limit the GPU's latency hiding capability and hurts
performance (an observation in-line with our own findings.
We show additionally that thrashing due to
interference further slows applications sharing the GPU. Our proposal, \titleShortMASK, is capable
not only of reducing interference between multiple applications,
but of reducing the TLB miss rate in
single-application scenarios as well. We expect that our techniques are
applicable to CPU-GPU heterogeneous system. 

Direct segments~\cite{direct-segment}
and redundant memory mappings~\cite{rmm} reduce address translation overheads by mapping large contiguous
virtual memory to contiguous physical address space which reduces address translation
overheads by increasing the reach of TLB entries. These techniques are complementary
to those in \titleShortMASK, and may eventually become relevant in GPU settings as well.

\paragraphbe{Demand Paging in GPUs.}
Demand paging is an important functionality for memory virtualization that 
is challenging for GPUs~\cite{abhishek-ispass16}. 
Recent works~\cite{tianhao-hpca16}, AMD's hUMA~\cite{huma}, as well as
NVIDIA's PASCAL architecture~\cite{tianhao-hpca16,pascal}
support for demand paging in GPUs. As identified in MOSAIC, these techniques
can be costly in GPU environment.

\subsection{TLB Designs}

\paragraphbe{GPU TLB Designs.} Previous works have explored 
the design space for TLBs in heterogeneous systems with
GPUs~\cite{cong-hpca17,powers-hpca14,pichai-asplos14,abhishek-ispass16}, and the adaptation of 
x86-like TLBs to a heterogeneous CPU-GPU setting~\cite{powers-hpca14}. 
Key elements in these designs include probing the TLB after L1
coalescing to reduce the number of parallel TLB requests, shared
concurrent page table walks, and translation caches to 
reduce main memory accesses. Our proposal, \titleShortMASK, owes
much to these designs, but we show empirically that contention patterns 
at the shared L2 layer require additional support to accommodate cross-context 
contention.
Cong et al. propose a TLB design similar to our baseline GPU-MMU design~\cite{cong-hpca17}. However,
this design utilizes the host (CPU) MMU to perform page walks, which is
inapplicable in the context of multi-application GPUs.
Pichai et al.~\cite{pichai-asplos14} explore TLB design 
for heterogeneous CPU-GPU systems, and
add TLB awareness to the existing CCWS GPU warp scheduler~\cite{ccws},
which enables parallel TLB access on the L1 cache level,
similar in concept to the Powers design~\cite{powers-hpca14}. 
Warp scheduling is orthogonal to our work: 
incorporating a TLB-aware CCWS warp scheduler 
to \titleShortMASK could further improve performance.

\paragraphbe{CPU TLB Designs.} 
Bhattacharjee et al. examine shared last-level TLB
designs~\cite{inter-core-tlb} as well as page walk cache
designs~\cite{large-reach}, proposing a mechanism that can accelerate
multithreaded applications by sharing translations between cores. However,
these proposals are likely to be less effective for multiple concurrent GPGPU
applications because translations are not shared between virtual address
spaces.  Barr et al.\ propose SpecTLB~\cite{spectlb}, which speculatively
predicts address translations to avoid the TLB miss latency.  Speculatively predicting
address translation can be complicated and costly in GPU because there can be
multiple concurrent TLB misses to many different TLB entries in the GPU.

\paragraphbe{Mechanisms to Support Multiple Page Sizes.}
TLB miss overheads can be reduced by accelerating page table
walks~\cite{barr-isca10,large-reach} or reducing their
frequency~\cite{jayneel-isca16}; by reducing the number of TLB misses (e.g.
through prefetching~\cite{bhattacharjee-pact09, kandiraju-isca02,saulsbury-isca00}, 
prediction~\cite{prediction-tlb}, or structural change to
the TLB~\cite{talluri-asplos94, binh-colt, binh-hpca14} or TLB
hierarchy~\cite{bhattacharjee-hpca11, lustig-13, srikantaiah-micro10,ahn-tocs15, ahn-isca12, rmm, direct-segment, jayneel-micro14}).
Multipage mapping techniques~\cite{talluri-asplos94, binh-colt, binh-hpca14}
map multiple pages with a single TLB entry, improving TLB reach by a small
factor (e.g., to 8 or 16); much greater improvements to TLB reach are needed to
deal with modern memory sizes.  Direct segments~\cite{direct-segment,jayneel-micro14} 
extend standard paging with a large segment to map the
majority of an address space to a contiguous physical memory region, but
require application modifications and are limited to workloads able to a single
large segment.  Redundant memory mappings (RMM)~\cite{rmm} extend TLB
reach by mapping {\it ranges} of virtually and physically contiguous pages in a
range TLB. 

A number of related works propose hardware support to recover and expose
address space contiguity. GLUE~\cite{binh-micro15} groups contiguous, aligned
small page translations under a single speculative large page translation in
the TLB.  Speculative translations (similar to SpecTLB~\cite{spectlb}) can be
verified by off-critical-path page table walks, reducing effective page-table
walk latency. GTSM~\cite{gtsm} provides hardware support to leverage the
address space contiguity of physical memory even when pages have been retired
due to bit errors.  Were such features to become available, hardware mechanisms
for preserving address space contiguity could reduce the overheads induced by proactive
compaction, which is a feature we introduce in our proposal, \titleShortMOSAIC.

The policies and mechanisms used to implement transparent large page
support in \titleShortMOSAIC{} are informed by a wealth of previous research
on operating system support for large pages for CPUs. 
Navarro et al.~\cite{superpage} identify
contiguity-awareness and fragmentation reduction as primary concerns for large page management,
proposing reservation-based allocation and deferred promotion of base pages to large pages.
These ideas are widely used in modern operating systems~\cite{thp}.
Ingens~\cite{ingens} eschews reservation-based allocation in favor of
the utilization-based promotion based on a bit vector which tracks
spatial and temporal utilization of base pages, implementing promotion
and demotion asynchronously, rather than in a page fault handler. 
These basic ideas heavily inform \titleShortMOSAIC{}'s design, which
attempts to emulate these same policies in hardware. In contrast
to Ingens, \titleShortMOSAIC{} can rely on dedicated hardware to provide
access frequency and distribution, and need not infer it by sampling
access bits whose granularity may be a poor fit for the page size.

Gorman et al.~\cite{gorman-ismm08} propose a placement policy for an operating system's 
physical page allocator that mitigates fragmentation and promotes address space contiguity 
by grouping pages according to relocatability. Subsequent work~\cite{gorman-wiosca10}
proposes a software-exposed interface for applications to explicitly
request large pages like {\tt libhugetlbfs}~\cite{libhugetlbfs}.
These ideas are complementary to ideas presented in this thesis.
\titleShortMOSAIC can plausibly benefit from similar policies simplified
to be hardware-implementable, and we leave that investigation as future work.

%





\chapter{Reducing Intra-application Interference \protect\linebreak with Memory Divergence Correction}
\label{sec:medic-introduction} 
\label{sec:medic} 

%
%

Graphics Processing Units (GPUs) have enormous parallel processing power to
leverage thread-level parallelism.  GPU applications can be broken down into
thousands of threads, allowing GPUs to use \emph{fine-grained multithreading}~\cite{cdc6600,smith-hep} to
prevent GPU cores from stalling due to dependencies and long memory latencies.
Ideally, there should always be available threads for GPU cores to continue
execution, preventing stalls within the core. GPUs also take advantage of the
\emph{SIMD} (Single Instruction, Multiple Data) execution model~\cite{flynn}.  The thousands of threads within a GPU application are
clustered into \emph{work groups} (or \emph{thread blocks}), with each thread
block consisting of multiple smaller bundles of threads that are run
concurrently.  Each such thread bundle is called a \emph{wavefront}~\cite{ati-wavefront} or
\emph{warp}~\cite{lindholm}.  In each cycle, each GPU core executes a single warp. Each thread
in a warp executes the same instruction (i.e., is at the same program counter).
Combining SIMD execution with fine-grained multithreading allows a GPU to complete
several hundreds of operations every cycle in the ideal case.

In the past, GPUs strictly executed graphics applications, which naturally
exhibit large amounts of concurrency. 
In recent years, with tools such as
CUDA~\cite{programmingguide} and OpenCL~\cite{khronos2008opencl}, programmers
have been able to adapt non-graphics applications to GPUs, writing these
applications to have thousands of threads that can be run on a SIMD computation
engine. Such adapted non-graphics programs are known as general-purpose GPU
(GPGPU) applications. Prior work has demonstrated that many scientific and data
analysis applications can be executed significantly faster when programmed to
run on GPUs~\cite{rodinia,parboil,mars,lonestar}.



\begin{figure}[h]
	\centering
        \vspace{3pt}
	\includegraphics[width=0.8\columnwidth]{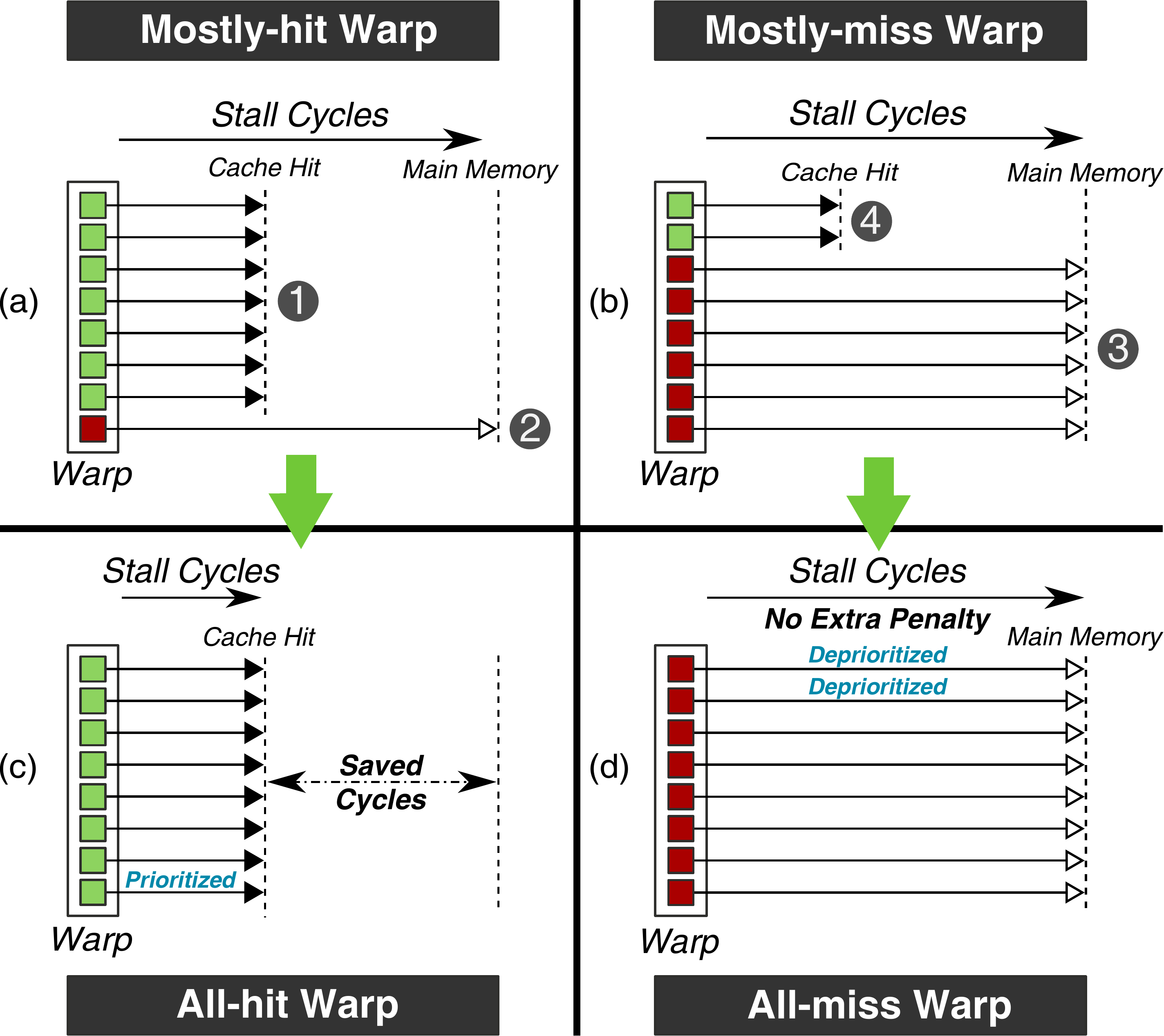}	
    \caption{Memory divergence within a warp. (a) and (b) show the heterogeneity between
    \emph{mostly-hit} and \emph{mostly-miss} warps, respectively.  (c) and (d) show the change in stall time from 
    converting \emph{mostly-hit warps into all-hit warps}, and \emph{mostly-miss warps into all-miss
    warps}, respectively.}
	\label{fig:mem-div}
\end{figure}

While many GPGPU applications can tolerate a significant amount of memory
latency due to their parallelism and the use of fine-grained multithreading, many
previous works (e.g.,~\cite{caba,owl-asplos13,largewarps,osp-isca13}) observe that GPU cores still stall for
a significant fraction of time when running many other GPGPU applications.  One significant source of these stalls
is \emph{memory divergence}, where the threads of a warp reach a memory
instruction, and some of the threads' memory requests take longer to service
than the requests from other threads~\cite{largewarps,warpsub,chatterjee-sc14}.
Since all threads within a warp operate in lockstep due to the SIMD execution
model, the warp cannot proceed to the next instruction until the
\emph{slowest} request within the warp completes, and \emph{all} threads are
ready to continue execution.  Figures~\ref{fig:mem-div}a and~\ref{fig:mem-div}b show
\red{examples of memory divergence within a warp, which we will explain in more detail soon.}

\FloatBarrier


In this work, we make three new key observations about the memory 
divergence behavior of GPGPU warps:

\vspace{3pt}
\paragraphbe{Observation 1:} There is \emph{heterogeneity across warps} in the degree
of memory divergence experienced by each warp at the shared L2 cache (i.e., the
percentage of threads within a warp that miss in the cache varies widely).  
Figure~\ref{fig:mem-div} shows examples of two different \emph{types of warps}, with
eight threads each, that exhibit different degrees of memory divergence:
\begin{itemize}
    \item Figure~\ref{fig:mem-div}a shows a \emph{mostly-hit warp}, where most
        of the warp's memory accesses hit in the cache (\mycirc{1}). However, a
        single access misses in the cache and must go to main memory
        (\mycirc{2}). As a result, the \emph{entire warp} is stalled until the
        much longer cache miss completes. 
    \item Figure~\ref{fig:mem-div}b shows a \emph{mostly-miss warp}, where 
        most of the warp's memory requests miss in the cache (\mycirc{3}), resulting in many
        accesses to main memory.  Even though some requests are cache hits (\mycirc{4}),
        these do not benefit the execution time of the warp.
\end{itemize}

\paragraphbe{Observation 2:} \emph{A warp tends to retain its memory 
divergence behavior} (e.g., whether or not it is mostly-hit or mostly-miss)
\emph{for long periods of execution}, and is thus predictable.  As we 
show in Section~\ref{sec:mech-medic}, this predictability enables us to perform
history-based warp divergence characterization.



\vspace{3pt}
\paragraphbe{Observation 3:} Due to the amount of thread parallelism
within a GPU, \emph{a large number of memory requests can arrive at the L2
cache in a small window of execution time, leading to significant queuing
delays}.  Prior work observes high access
latencies for the shared L2 cache within a GPU~\cite{sisoftware,
coherence-hpca13,demystify}, but does not identify \emph{why}
these latencies are so high.  We show that when a large number of requests arrive
at the L2 cache, both the limited number of read/write ports and backpressure
from cache bank conflicts force many of these requests to queue up for long
periods of time.  We observe that this queuing latency can sometimes add
\emph{hundreds} of cycles to the cache access latency, and that non-uniform
queuing across the different cache banks exacerbates memory divergence.

Based on these three observations, we aim to devise a mechanism that has two major goals: (1)~convert mostly-hit warps into
\emph{all-hit warps} (warps where \emph{all} requests hit in the cache, as shown in
Figure~\ref{fig:mem-div}c), and (2)~convert mostly-miss warps into
\emph{all-miss warps} (warps where \emph{none} of the requests hit in the cache,
as shown in Figure~\ref{fig:mem-div}d).  As we can see in
Figure~\ref{fig:mem-div}a, the stall time due to memory divergence for the 
mostly-hit warp can be eliminated by converting only the single cache miss (\mycirc{2}) into
a hit.  Doing so requires additional cache space.  If we convert the two cache
hits of the mostly-miss warp (Figure~\ref{fig:mem-div}b, \mycirc{4}) into cache misses, 
we can cede the cache space previously used by these hits to the mostly-hit warp, thus converting the
mostly-hit warp into an all-hit warp.  Though the mostly-miss warp is now an
all-miss warp (Figure~\ref{fig:mem-div}d), it incurs no extra stall penalty, as the warp was already waiting on the other six
cache misses to complete.  Additionally, now that it is an all-miss warp, 
we predict that its future memory requests will also not be in the L2 cache,
so we can simply have these
requests \emph{bypass the cache}.  In doing so, the requests from the all-miss
warp can completely avoid unnecessary L2 access and queuing delays.  This decreases the total
number of requests going to the L2 cache, thus reducing the queuing latencies for
requests from mostly-hit and all-hit warps, as there is less contention.

We introduce \emph{\titleLongMeDiC{}} (\emph{\titleShortMeDiC{}}), a GPU-specific mechanism that
exploits \emph{memory divergence heterogeneity} across warps at the shared cache and
at main memory to improve the overall performance of GPGPU applications.
\titleShortMeDiC{} consists of three different components, which work together to
achieve our goals of converting mostly-hit warps into all-hit warps and
mostly-miss warps into all-miss warps: (1)~a warp-type-aware \emph{cache bypassing mechanism},
which prevents requests from mostly-miss and all-miss warps from accessing the
shared L2 cache (Section~\ref{sec:bypass}); (2)~a warp-type-aware \emph{cache
insertion policy}, which prioritizes requests from mostly-hit and all-hit warps
to ensure that they all become cache hits (Section~\ref{sec:insertion}); and
(3)~a warp-type-aware \emph{memory scheduling mechanism}, which 
prioritizes requests from mostly-hit warps that were not successfully
converted to all-hit warps, in order to minimize the stall time due to
divergence (Section~\ref{sec:mem-controller}).  These three components are all
driven by an online mechanism that can identify the expected memory divergence
behavior of each warp (Section~\ref{sec:identify}).


This dissertation makes the following contributions:
\begin{itemize}
\item \red{We observe that the different warps within a GPGPU application
    exhibit heterogeneity in their memory divergence behavior at the shared L2
    cache, and that some warps do not benefit from the few cache hits that they
    have. This memory divergence behavior tends to remain consistent throughout
    long periods of execution for a warp, allowing for fast, online warp divergence characterization and prediction.}

\item We identify a new performance bottleneck in GPGPU application execution that
    can contribute significantly to memory divergence: due to the very large number of memory requests
    issued by warps in GPGPU applications that contend at the shared L2 cache, many of these requests
    experience \emph{high cache queuing latencies}.

\item Based on our observations, we propose \emph{\titleLongMeDiC{}}, \red{a new mechanism that
    exploits the stable memory divergence behavior of warps  
    to (1)~improve the effectiveness of the cache by favoring warps that take the most
    advantage of the cache, (2)~address the cache queuing problem, and (3)~improve the effectiveness
    of the memory scheduler by favoring warps that benefit most from prioritization.}
    We compare \titleShortMeDiC{} to four different cache management mechanisms,
    and show that it improves performance by 21.8\% and energy efficiency by
    20.1\% across a wide variety of GPGPU workloads compared to a
    a state-of-the-art GPU cache management mechanism~\cite{donglihpca15}.

\end{itemize}

\section{Background}
\label{sec:background-medic} 

We first provide background on the architecture of a modern GPU, and then
we discuss the bottlenecks that highly-multithreaded applications 
can face when executed on a GPU. These
applications can be compiled using OpenCL~\cite{khronos2008opencl} or
CUDA~\cite{programmingguide}, either of which converts a general purpose application into a GPGPU
program that can execute on a GPU.

\subsection{Baseline GPU Architecture}

\begin{figure}[h]
	\centering
	\vspace{3pt}
	\includegraphics[width=0.45\columnwidth]{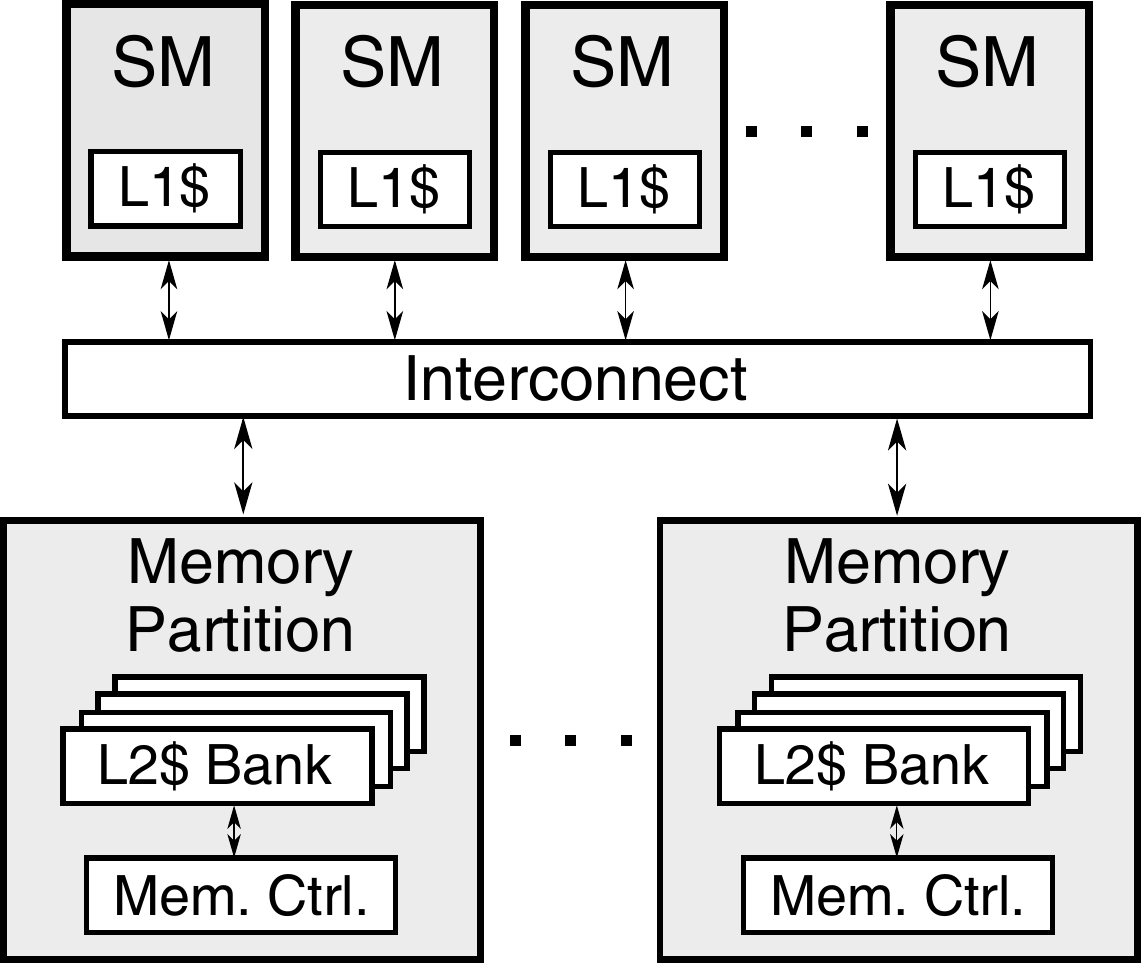}%
	\caption{Overview of the baseline GPU architecture.} 
	\label{fig:gpu-spec}
\end{figure}

A typical GPU consists of several \emph{shader cores} (sometimes called
\emph{streaming multiprocessors}, or SMs). In this work, we set the number of
shader cores to 15, with 32 threads per warp in each core, corresponding to the
NVIDIA GTX480 GPU based on the Fermi architecture~\cite{fermi}. The GPU we
evaluate can issue up to 480 concurrent memory accesses per
cycle~\cite{gtx480-config}. Each core has its own private L1 data, texture, and
constant caches, as well as a scratchpad
memory~\cite{kepler,fermi,lindholm}. In addition, the GPU also has several
shared L2 cache slices and memory controllers. A \emph{memory partition unit} combines
a single L2 cache slice (which is banked) with a designated memory controller
that connects to off-chip main memory.
Figure~\ref{fig:gpu-spec} shows a simplified view of how the cores (or SMs),
caches, and memory partitions are organized in our baseline GPU. 

\FloatBarrier

\subsection{Bottlenecks in GPGPU Applications}

Several previous works have analyzed the benefits and limitations 
of using a GPU for general purpose workloads (other than graphics purposes), 
including characterizing the impact of microarchitectural changes on
applications~\cite{gpgpu-sim} or developing performance
models that break down performance bottlenecks in GPGPU
applications~\cite{carole-wu, perf-model,hong-isca09,liu-icpp07,ma-asap12,govindaraju-sc06}. 
\red{All of these works show benefits from using a
throughput-oriented GPU.} However, a significant number of applications are unable to fully
utilize all of the available parallelism within the GPU, leading to periods of
execution where no warps are available for execution~\cite{caba}.

\red{When there are no available warps, the GPU cores stall, and the
application stops making progress until a warp becomes available. Prior
work has investigated two problems that can delay some warps from becoming available
for execution: (1) \emph{branch divergence}, which occurs when a branch in the same SIMD
instruction resolves into multiple different
paths~\cite{gpgpu-sim,warp-level-div,han-reducing-div,dwf,largewarps}, and (2)
\emph{memory divergence}, which occurs when the simultaneous memory requests
from a single warp spend different amounts of time retrieving their
associated data from memory~\cite{largewarps,warpsub,chatterjee-sc14}. In this
work, we focus on the memory divergence problem; prior work on branch divergence
is complementary to our work.}

\section{Motivation and Key Observations}
\label{sec:medic-motiv}

We make three new key observations about memory divergence (at the shared L2 cache).
First, we observe that the degree of memory divergence can differ across warps. 
This inter-warp heterogeneity affects how well each warp takes advantage of
the shared cache.  Second, we observe that a warp's memory divergence behavior tends
to remain stable for long periods of execution, making it
predictable.  Third, we observe that requests to the shared cache experience
long queuing delays due to the large amount of parallelism in GPGPU programs,
which exacerbates the memory divergence problem and slows down GPU execution.
Next, we describe each of these observations in detail and motivate our solutions.

\subsection{Exploiting Heterogeneity Across Warps}
\label{sec:warp-util}

\begin{figure}[h]
	\centering
	\includegraphics[width=0.7\columnwidth]{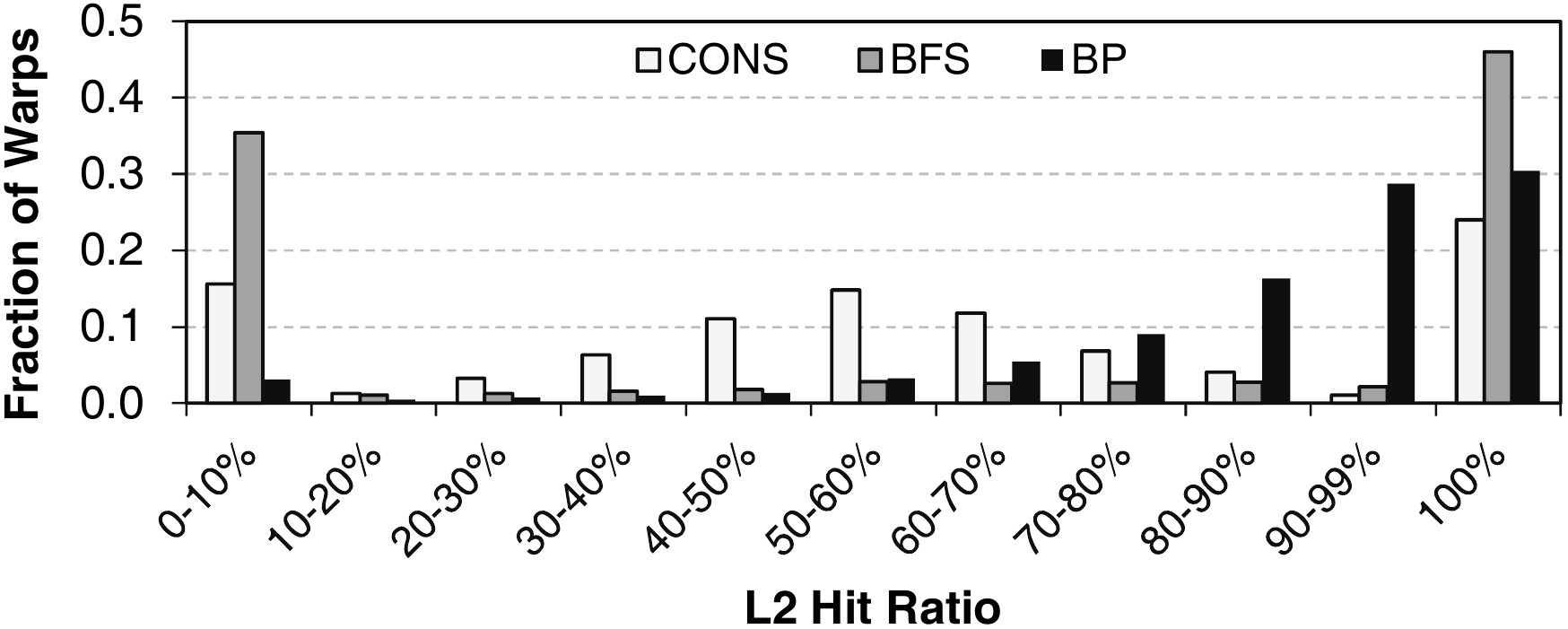}%
        \vspace{-1pt}
	\caption{L2 cache hit ratio of different warps in three representative GPGPU applications (see Section~\ref{sec:methodology-medic} for methods).} 
	\label{fig:warp-spread}
\end{figure}

\begin{figure}[h]
	\centering
	\includegraphics[width=0.7\textwidth]{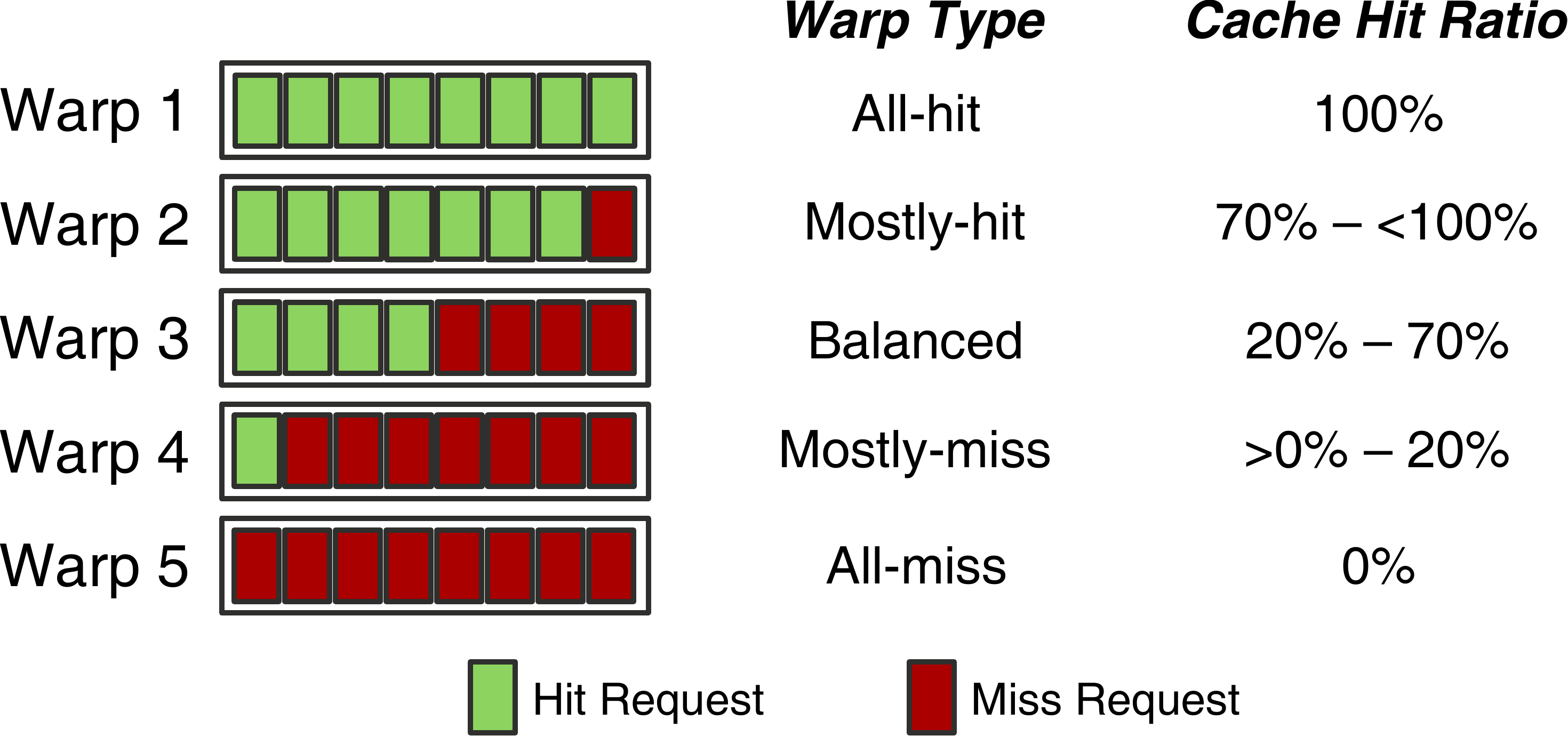}%
	\caption{Warp type categorization based on the shared cache hit ratios.
    Hit ratio values are empirically chosen.} 
	\label{fig:warp-types}
\end{figure}

We observe that different warps have different amounts of sensitivity to memory latency and
cache utilization.  We study the cache utilization of a warp by determining
its \emph{hit ratio}, the percentage of memory requests that hit in the cache when the warp issues a single
memory instruction.  As Figure~\ref{fig:warp-spread} shows, the warps from each
of our three representative GPGPU applications are distributed across all
possible ranges of \emph{hit ratio}, exhibiting significant heterogeneity. 
To better characterize warp behavior,
we break the warps down into the five types shown in
Figure~\ref{fig:warp-types} based on their hit ratios: \emph{all-hit},
\emph{mostly-hit}, \emph{balanced}, \emph{mostly-miss}, and \emph{all-miss}.

\FloatBarrier

\begin{figure}[h]
	\centering
	\includegraphics[width=\columnwidth]{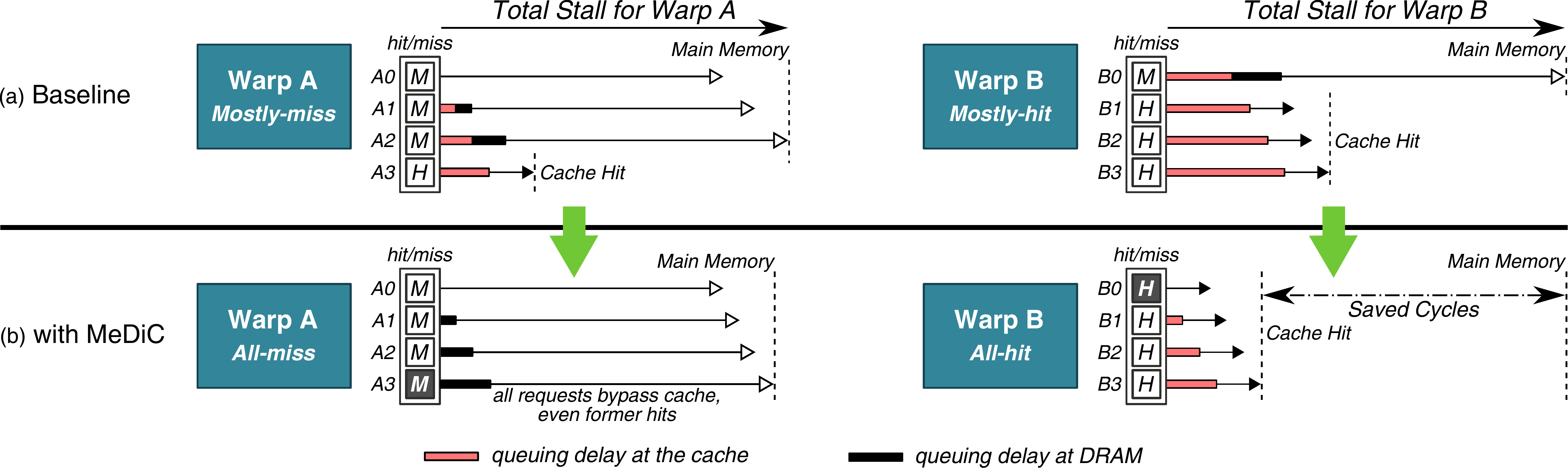}%
	\vspace{-2pt}
    \caption{(a) Existing inter-warp heterogeneity, (b) exploiting the heterogeneity with \titleShortMeDiC{} to improve performance.} 
	\label{fig:mem-div-motiv}
	\vspace{-3pt}
\end{figure}

This inter-warp heterogeneity in cache utilization provides new opportunities for
performance improvement. We illustrate two such opportunities by walking through a
simplified example, shown in Figure~\ref{fig:mem-div-motiv}. Here, we have two
warps, \emph{A} and~\emph{B}, where~\emph{A} is a mostly-miss warp (with
three of its four memory requests being L2 cache misses) and~\emph{B} is a
mostly-hit warp with only a single L2 cache miss (request~\emph{B0}).  Let us
assume that warp~\emph{A} is scheduled first.

As we can see in Figure~\ref{fig:mem-div-motiv}a, the mostly-miss
warp \emph{A} does not benefit at all from the cache: even though one of its requests (\emph{A3}) hits in the cache, warp~\emph{A}
cannot continue executing until \emph{all} of its memory requests are serviced.
As the figure shows, using the cache to speed up only request~\emph{A3} has no
material impact on warp \emph{A}'s stall time.  In addition, while
requests~\emph{A1} and~\emph{A2} do not hit in the cache, they still incur
a queuing latency at the cache while they wait to be looked up in the cache tag array.

On the other hand, the mostly-hit warp \emph{B} can be penalized
significantly.  First, since warp~\emph{B} is scheduled after the mostly-miss warp~\emph{A},
all four of warp~\emph{B}'s requests incur a large L2 queuing delay, \emph{even
though the cache was not useful to speed up warp~\emph{A}}.  On top of this unproductive
delay, since request~\emph{B0} misses in the cache, it holds up execution
of the entire warp while it gets serviced by main memory.  The overall
effect is that despite having \emph{many more} cache hits (and thus much better cache
utility) than warp~\emph{A}, warp~\emph{B} ends up stalling for as long as or even longer than the mostly-miss warp~\emph{A} stalled for.

\FloatBarrier

To remedy this problem, we set two goals (Figure~\ref{fig:mem-div-motiv}b):

\noindent\emph{1) Convert the mostly-hit warp~\emph{B} into an all-hit warp.}  
By converting \emph{B0} into a hit,
warp~\emph{B} no longer has to stall on any memory misses, which enables the warp to
become ready to execute much earlier.  This requires a little additional space in the
cache to store the data for \emph{B0}.

\noindent\emph{2) Convert the mostly-miss warp~\emph{A} into an all-miss warp.} 
Since a single cache hit is of no effect to
warp~\emph{A}'s execution, we convert \emph{A0} into a cache miss.  This frees up the cache
space \emph{A0} was using, and thus creates cache space for storing \emph{B0}.  In addition, warp~\emph{A}'s requests
can now skip accessing the cache and go straight to main memory, which has two
benefits: \emph{A0--A2} complete faster because they no longer experience the
cache queuing delay that they incurred in Figure~\ref{fig:mem-div-motiv}a, and
\emph{B0--B3} also complete faster because they must queue behind a smaller
number of cache requests. Thus, bypassing the cache for warp~\emph{A}'s request allows \emph{both} warps to stall
for less time, improving GPU core utilization.

\red{To realize these benefits, we propose to 
(1)~develop a mechanism that can identify mostly-hit and
mostly-miss warps; 
(2)~design a mechanism that allows mostly-miss warps
to yield their ineffective cache space to mostly-hit warps, similar to
how the mostly-miss warp~\emph{A} in Figure~\ref{fig:mem-div-motiv}a turns into
an all-miss warp in Figure~\ref{fig:mem-div-motiv}b, so that warps such
as the mostly-hit warp~\emph{B} can become all-hit warps; 
(3)~design a mechanism that bypasses the cache for requests from mostly-miss
and all-miss warps such as warp~\emph{A}, to decrease warp stall time and
reduce lengthy cache queuing latencies; and 
(4)~prioritize requests from mostly-hit warps across the memory hierarchy, at both the shared L2 cache and at the memory controller, to
minimize their stall time as much as possible, similar to how the mostly-hit
warp~\emph{B} in Figure~\ref{fig:mem-div-motiv}a turns into an all-hit
warp in Figure~\ref{fig:mem-div-motiv}b.}

\begin{figure}[h!!!]
	\centering
	\includegraphics[width=\columnwidth]{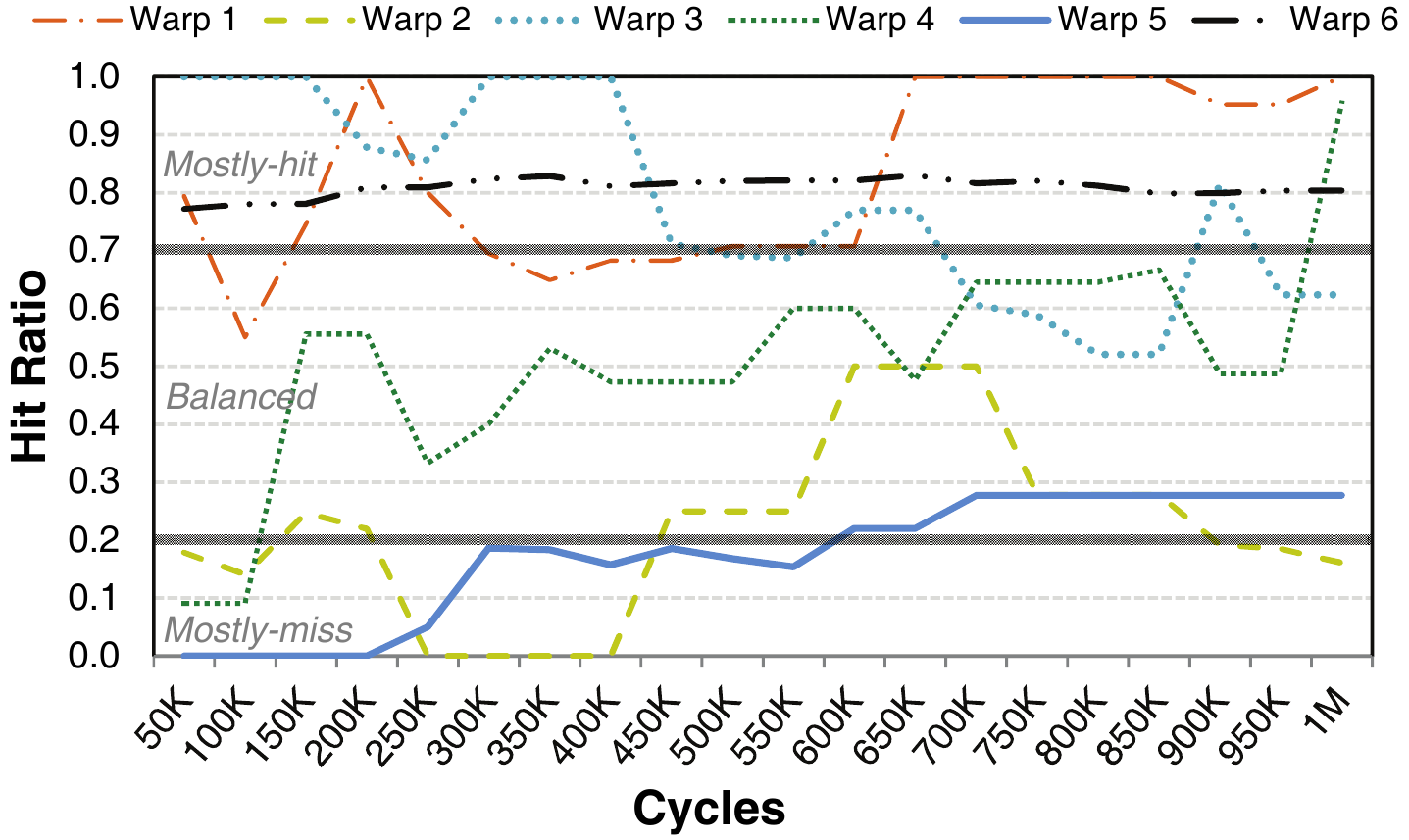}%
	\vspace{-1.7pt}
	\caption{Hit ratio of randomly selected warps over time.} 
	\label{fig:warp-ratio-trend}
\end{figure}

\green{A key challenge is how to group warps into different
warp types. In this work, we observe that warps tend to exhibit stable cache hit
behavior over long periods of execution. A warp consists of several threads that
repeatedly loop over the same instruction sequences. This secs/medic/results in similar
hit/miss behavior at the cache level across different instances of the same warp. As
a result, a warp measured to have a particular hit ratio is likely to maintain a
similar hit ratio throughout a lengthy phase of execution. We observe that most CUDA
applications exhibit this trend.

Figure~\ref{fig:warp-ratio-trend} shows
the hit ratio over a duration of one million cycles, for six randomly selected warps from our CUDA
applications. We also plot horizontal lines to illustrate the hit ratio cutoffs
that we set in Figure~\ref{fig:warp-types} for our mostly-hit~($\geq$70\%) and
mostly-miss ($\leq$20\%) warp types.  Warps~1, 3, and~6 spend the majority of their time
with high hit ratios, and are classified as mostly-hit warps.  Warps~1 and~3 do,
however, exhibit some long-term (i.e., 100k+ cycles) shifts to the balanced
warp type.  Warps~2 and~5 spend a long time as mostly-miss warps, though they both experience a single long-term
shift into balanced warp behavior.  As we can see, \emph{warps
tend to remain in the same warp type} at least for hundreds of thousands of cycles.  }

As a result of this relatively
stable behavior, our mechanism, \titleShortMeDiC{} (described in detail in
Section~\ref{sec:mech-medic}), samples the hit ratio of each warp and uses this data
for warp characterization.  To account for the long-term hit ratio
shifts, \titleShortMeDiC{} resamples the hit ratio every 100k~cycles.

\FloatBarrier


\subsection{Reducing the Effects of L2 Queuing Latency}
\label{sec:cache-latency}

Unlike CPU applications, GPGPU applications can issue as many as hundreds of
memory instructions per cycle.  All of these memory
requests can arrive concurrently at the L2 cache, which is the first shared level of
the memory hierarchy, creating a bottleneck. Previous
works~\cite{gpgpu-sim,sisoftware,demystify,coherence-hpca13} point out that
the latency for accessing the L2 cache can take hundreds of cycles, even though
the nominal cache lookup latency is significantly lower (only tens of cycles).
While they identify this disparity, these earlier efforts do not identify or analyze the
source of these long delays.

\begin{figure}[h!!!]
	\centering
    \includegraphics[width=\columnwidth]{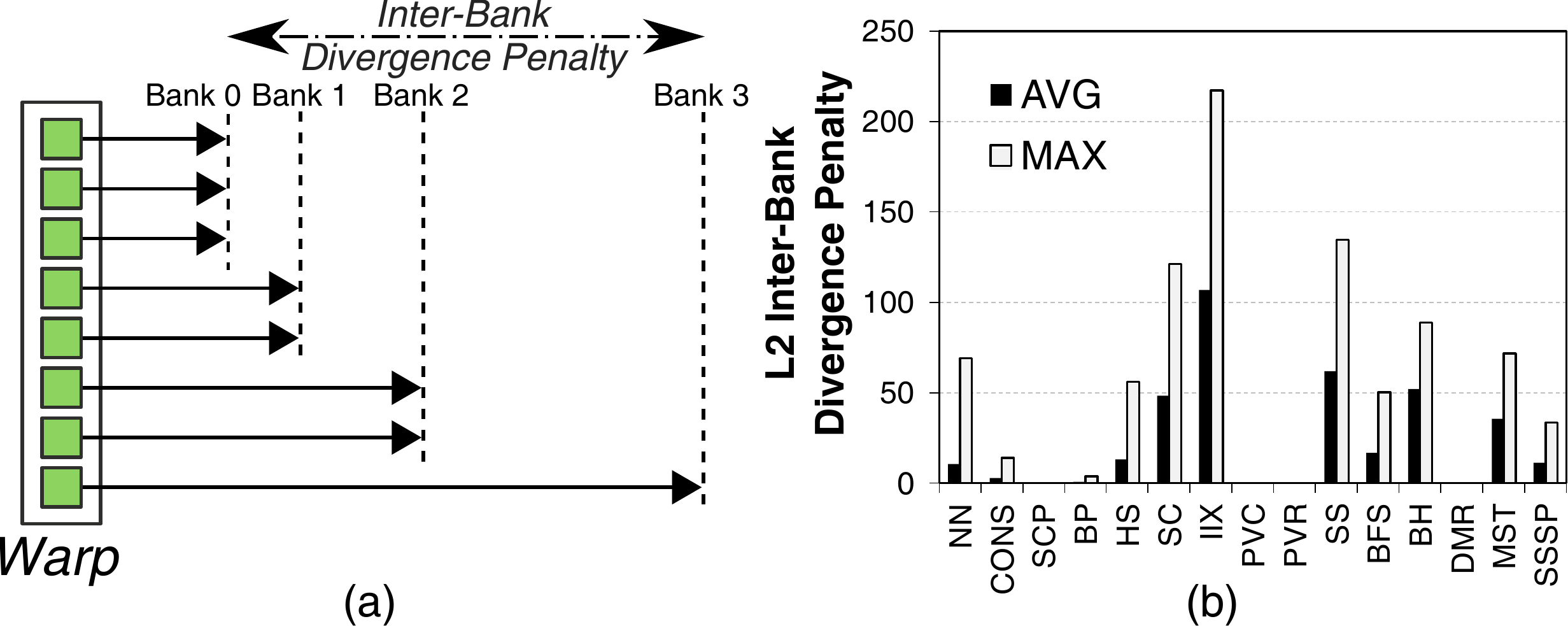}%
	\caption{\red{Effect of bank queuing latency divergence in the L2 cache: (a) example of the impact on stall time of skewed queuing latencies, (b) inter-bank divergence penalty due to skewed queuing for all-hit warps, in cycles.}} 
	\label{fig:skew-latency-effect}
\end{figure}

We make a new observation that identifies an important source of the long L2 cache
access delays in GPGPU systems. L2 bank conflicts can cause queuing delay,
which can differ from one bank to another and lead to the disparity of cache
access latencies across different banks. As Figure~\ref{fig:skew-latency-effect}a shows, even if every cache
access within a warp hits in the L2 cache, each access can incur a different
cache latency due to non-uniform queuing, and the warp has to stall until the \emph{slowest} cache
access retrieves its data (i.e., memory divergence can occur). 
For each set of simultaneous requests issued by an all-hit warp, we define its
\emph{inter-bank divergence penalty} to be the difference between the fastest
cache hit and the slowest cache hit, as depicted in
Figure~\ref{fig:skew-latency-effect}a.

In order to confirm this behavior, we modify GPGPU-Sim~\cite{gpgpu-sim} to
accurately model L2 bank conflicts and queuing delays (see
Section~\ref{sec:methodology-medic} for details).  We then measure the average and
maximum inter-bank divergence penalty observed \emph{only for all-hit warps}
in our different CUDA applications, shown in
Figure~\ref{fig:skew-latency-effect}b.  We find that \emph{on average}, an all-hit warp
has to stall for an additional 24.0~cycles
because some of its requests go to cache banks with high
access contention.

\FloatBarrier

\begin{figure}[h!!!]
	\centering
	\includegraphics[width=\columnwidth]{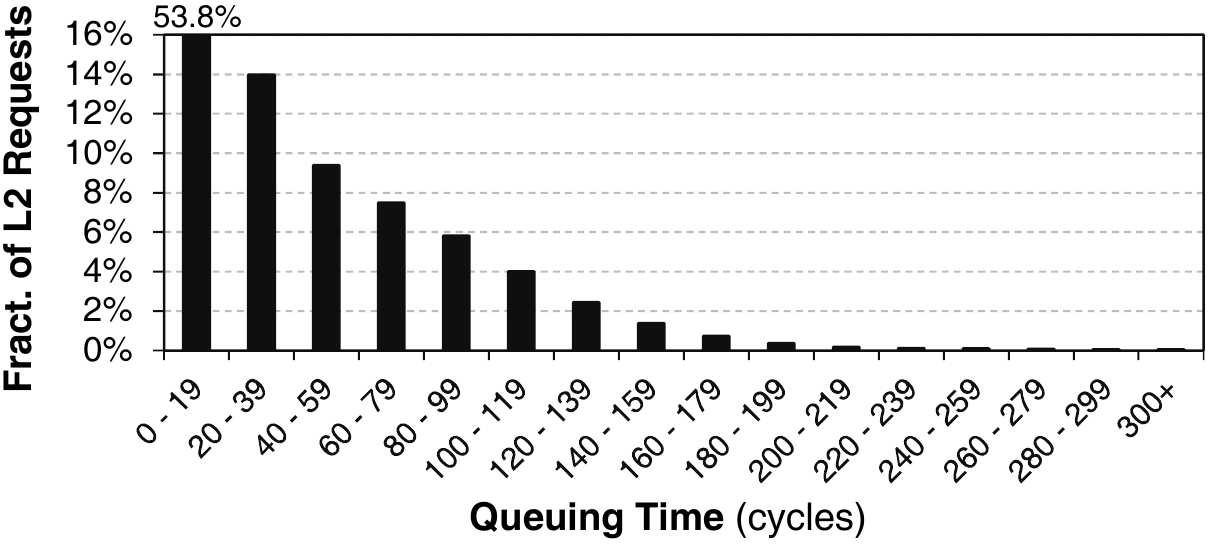}	
    \caption{Distribution of per-request queuing latencies for L2 cache requests from BFS.}
	\label{fig:l2-queue-latency}
\end{figure}

To quantify the magnitude of queue contention, we analyze the queuing delays for
a two-bank L2 cache where the tag lookup latency is set to one cycle.
We find that even with such a small cache lookup latency, a significant number of
requests experience tens, if not hundreds, of cycles of queuing delay.
Figure~\ref{fig:l2-queue-latency} shows the distribution of these delays for
BFS~\cite{lonestar}, across all of its individual L2 cache requests. BFS contains one compute-intensive
kernel and two memory-intensive kernels. We observe that requests generated by
the compute-intensive kernel do not incur high queuing latencies, while
requests from the memory-intensive kernels suffer from significant queuing
delays. On average, across all three kernels, cache requests spend 34.8~cycles
in the queue waiting to be serviced, which is quite high considering the
idealized one-cycle cache lookup latency. 

\begin{figure}[h!!!]
	\centering
	\includegraphics[width=\columnwidth]{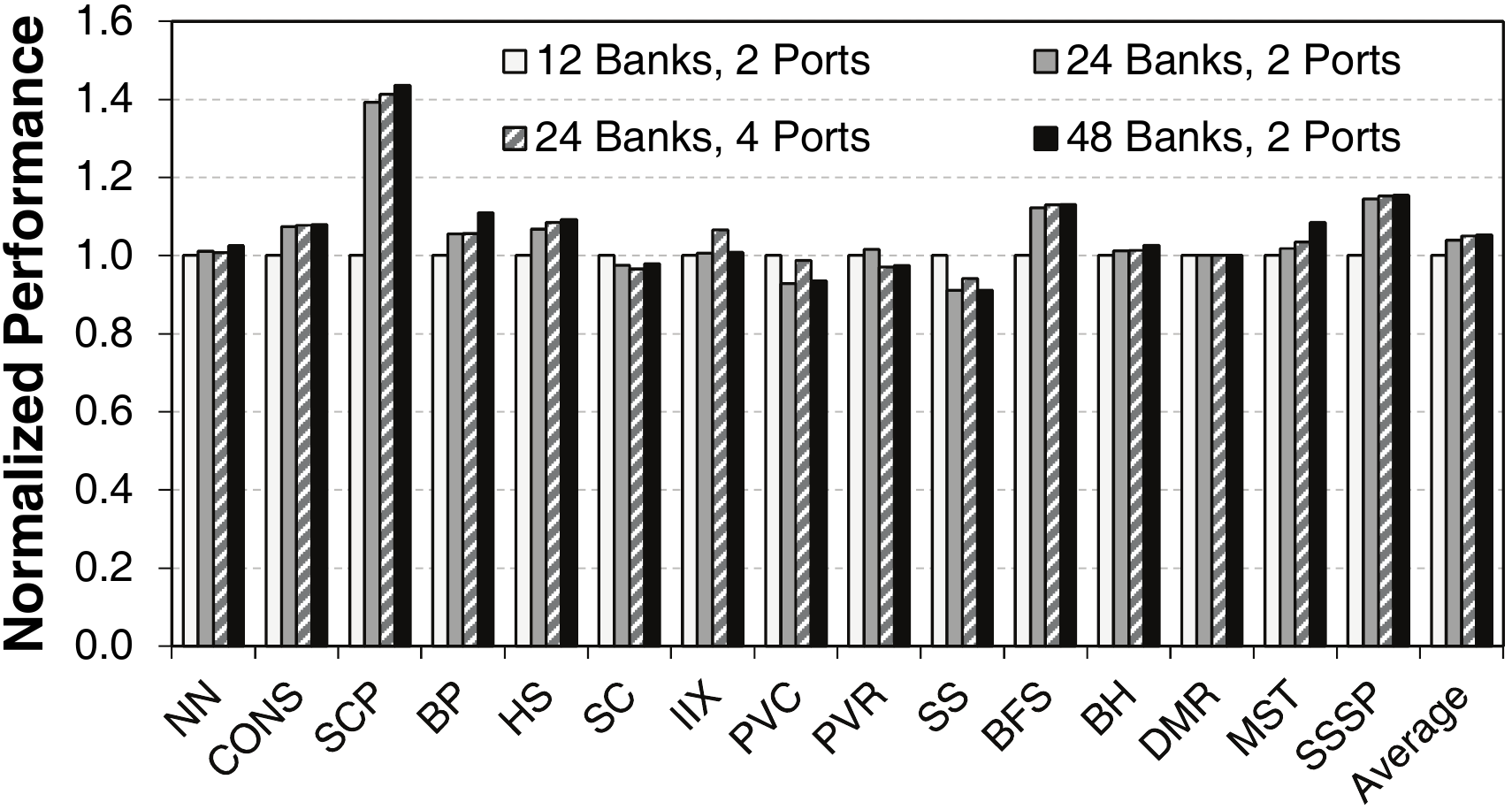}%
	\caption{Performance of GPGPU applications with different number of banks and ports per bank, normalized to a 12-bank cache with 2~ports per bank.} 
	\label{fig:motiv-bank}
\end{figure}

One naive solution to the L2 cache queuing problem is to increase the number of banks, without reducing the
number of physical ports per bank and without increasing the size of the shared
cache. However, as shown in Figure~\ref{fig:motiv-bank}, 
\red{the average performance improvement from doubling the number of
banks to 24 (i.e., 4 banks per memory partition) is less than 4\%, while
the improvement from quadrupling the banks is less than 6\%.
There are two key reasons for this minimal
performance gain. First, while more cache banks can help to distribute the
queued requests, these extra banks do not change the memory
divergence behavior of the warp (i.e., the warp hit ratios remain unchanged).
 Second, non-uniform bank access patterns still remain, causing cache requests to queue up
unevenly at a few banks.\footnote{Similar problems have been observed for bank conflicts in
main memory~\cite{rau-isca91,salp}.}}

\subsection{Our Goal}
\label{sec:bypass-motiv}

Our goal of \titleShortMeDiC is to improve cache utilization and reduce cache queuing
latency by taking advantage of heterogeneity between different types of warps. To this end, we create a mechanism that (1)~tries to eliminate
mostly-hit and mostly-miss warps by converting as many of them as possible to
all-hit and all-miss warps, respectively; (2)~\red{reduces the queuing delay at the L2 
cache by bypassing requests from mostly-miss and all-miss warps,
such that each L2 cache hit experiences a much lower overall L2 cache latency;
and (3)~prioritizes mostly-hit warps in the memory scheduler to minimize the amount of
time they stall due to a cache miss.}

\FloatBarrier

\section{\titleShortMeDiC{}: \titleLongMeDiC}
\label{sec:mech-medic}


In this section, we introduce \titleLongMeDiC (\titleShortMeDiC{}), a set of techniques that
take advantage of the memory divergence heterogeneity across warps, as discussed in
Section~\ref{sec:medic-motiv}.  These techniques work independently of each other, but act
synergistically to provide a substantial performance improvement. 
In Section~\ref{sec:identify}, we propose a mechanism that identifies and groups
warps into different warp types based on their degree of memory divergence,
as shown in Figure~\ref{fig:warp-types}.

\begin{figure*}[h]
        \vspace{3pt}
	\centering
	\includegraphics[width=\textwidth]{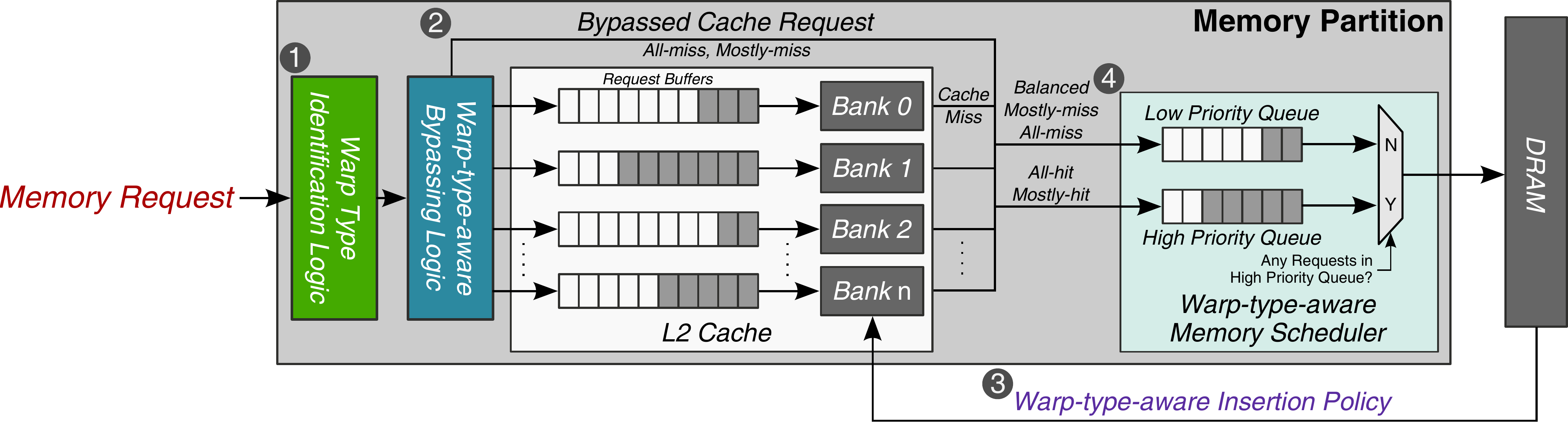}	
	\vspace{-7pt}
	\caption{Overview of \titleShortMeDiC{}: \protect\mycirc{1} warp type identification logic, \protect\mycirc{2} warp-type-aware cache bypassing, \protect\mycirc{3} warp-type-aware cache insertion policy, \protect\mycirc{4} warp-type-aware memory scheduler.} 
	\label{fig:main-mechanism}
	\vspace{6pt}
\end{figure*}

As depicted in
Figure~\ref{fig:main-mechanism}, \titleShortMeDiC{} uses \mycirc{1} warp
type identification to drive three different components: \mycirc{2} a \emph{warp-type-aware cache bypass
mechanism} (Section~\ref{sec:bypass}), which bypasses requests from all-miss and
mostly-miss warps to reduce the L2 queuing delay; \mycirc{3} a \emph{warp-type-aware
cache insertion policy} (Section~\ref{sec:insertion}), which
works to keep cache lines from mostly-hit warps while demoting lines from
mostly-miss warps; and \mycirc{4} a \emph{warp-type-aware memory scheduler}
(Section~\ref{sec:mem-controller}), which prioritizes DRAM requests from
mostly-hit warps as they are highly latency sensitive. We analyze the hardware cost of \titleShortMeDiC{}
in Section~\ref{sec:hw-cost}.

\subsection{Warp Type Identification}
\label{sec:identify}

In order to take advantage of the memory divergence heterogeneity across
warps, we must first add hardware that can identify the divergence behavior of
each warp.  The key idea is to periodically sample the hit ratio of a warp,
and to classify the warp's divergence behavior as one of the five types in
Figure~\ref{fig:warp-types} based on the observed hit ratio
(see Section~\ref{sec:warp-util}).  This information can then be used to drive
the warp-type-aware components of \titleShortMeDiC{}.  In general, warps tend to retain the same memory divergence behavior for long periods of execution. However, as we observed in Section~\ref{sec:warp-util}, there
can be some long-term shifts in warp divergence behavior, requiring periodic
resampling of the hit ratio to potentially adjust the warp type.

Warp type identification through hit ratio sampling
requires hardware within the cache to periodically count the number of hits
and misses each warp incurs.  We append two counters to the metadata stored
for each warp, which represent the total number of cache hits and
cache accesses for the warp.
We reset these counters periodically, and set the
bypass logic to operate in a profiling phase for each warp after this reset.\footnote{In this work, we reset the hit ratio every
100k cycles for each warp.}
During profiling, which lasts for the first 30 cache accesses of each warp, the bypass logic (which we explain in
Section~\ref{sec:bypass}) does not make any cache bypassing decisions, to allow the counters to accurately
characterize the current memory divergence behavior of the warp.  At the end of 
profiling, the warp type is determined and stored in the metadata.

%

\subsection{Warp-type-aware Shared Cache Bypassing}
\label{sec:bypass}

Once the warp type is known and a warp generates a request to the L2 cache, our
mechanism first decides whether to bypass the cache based on the warp type.
The key idea behind \emph{warp-type-aware cache bypassing}, as discussed in Section~\ref{sec:warp-util}, 
is to convert mostly-miss warps into all-miss warps, as they do not benefit
greatly from the few cache hits that they get.  By bypassing these requests, we
achieve three benefits: (1)~bypassed requests can avoid L2 queuing latencies
entirely, (2)~other requests that do hit in the L2 cache experience
shorter queuing delays due to the reduced contention, and (3)~space is created
in the L2 cache for mostly-hit warps.

The cache bypassing logic must make a simple decision: if an incoming memory
request was generated by a mostly-miss or all-miss warp, the request is bypassed
directly to DRAM.  This is determined by reading the warp type stored in the
warp metadata from the warp type identification mechanism.  A simple 2-bit
demultiplexer can be used to determine whether a request is sent to the L2 bank
arbiter, or directly to the DRAM request queue.

\vspace{5pt}
\paragraphbe{Dynamically Tuning the Cache Bypassing Rate.} While cache bypassing alleviates queuing pressure at the L2
cache banks, it can have a negative impact on other portions of the memory
partition.  For example, bypassed requests that were originally cache hits
now consume extra off-chip memory bandwidth, and can increase queuing delays at
the DRAM queue.  If we lower the number of bypassed requests (i.e., reduce
the number of warps classified as mostly-miss), we can reduce 
DRAM utilization.  After examining a random selection of kernels from three
applications (BFS, BP, and CONS), we find that the ideal number of warps classified as mostly-miss differs
for each kernel. Therefore, we add a mechanism that \emph{dynamically} tunes 
the hit ratio boundary between mostly-miss warps and balanced warps (nominally
set at 20\%; see Figure~\ref{fig:warp-types}).  If the cache miss rate increases
significantly, the hit ratio boundary is lowered.\footnote{In our evaluation,
we reduce the threshold value between mostly-miss warps and balanced warps by 5\% for every 5\% increase in cache miss rate.}

\paragraphbe{Cache Write Policy.} Recent GPUs support
multiple options for the L2 cache write policy~\cite{fermi}. In this work, we assume that
the L2 cache is write-through~\cite{coherence-hpca13}, so our bypassing logic can
always assume that DRAM contains an up-to-date copy of the data. For write-back
caches, previously-proposed
mechanisms~\cite{gupta-ipdps13,jaewoong-micro12,mekkat-pact13} can be used in
conjunction with our bypassing technique to ensure that bypassed requests
get the correct data. For correctness, fences and
atomic instructions from bypassed warps still access the L2 for cache lookup, 
but are not allowed to store data in the cache.

\subsection{Warp-type-aware Cache Insertion Policy}
\label{sec:insertion}

Our cache bypassing mechanism frees up space within the L2 cache, which we want
to use for the cache misses from mostly-hit warps (to convert these memory 
requests into cache hits).  However, even with the new bypassing mechanism, other warps (e.g., balanced, mostly-miss)
still insert some data into the cache. In order to aid the
conversion of mostly-hit warps into all-hit warps, we develop a \emph{warp-type-aware
cache insertion policy}, whose key idea is to ensure that for a given cache set,
data from mostly-miss warps are evicted first, while data from mostly-hit
warps and all-hit warps are evicted last.


To ensure that a cache block from a mostly-hit warp stays in the cache 
for as long as possible, we insert the block closer to the MRU position.
A cache block requested by a mostly-miss warp is inserted closer to the LRU position,
making it more likely to be evicted.  To track the status of these cache 
blocks, we add two bits of metadata to each cache block, indicating the warp
type.\footnote{Note that cache blocks
from the all-miss category share the same 2-bit value as the mostly-miss category because they always get bypassed (see Section~\ref{sec:bypass}).}
These bits are then appended to the replacement policy bits. As a
result, a cache block from a mostly-miss warp is more likely to get
evicted than a block from a balanced warp.  Similarly, a cache block from a 
balanced warp is more likely to be evicted than a block from a mostly-hit
or all-hit warp.

%

\subsection{Warp-type-aware Memory Scheduler}
\label{sec:mem-controller}


Our cache bypassing mechanism and cache insertion policy work to increase the
likelihood that \emph{all} requests from a mostly-hit warp become cache hits,
converting the warp into an all-hit warp.  However, due to cache conflicts, or
due to poor locality, there may still be cases when a mostly-hit warp cannot be
fully converted into an all-hit warp, and is therefore unable to avoid stalling
due to memory divergence as at least one of its requests has to go to
DRAM. In such a case, we want to minimize the amount of
time that this warp stalls. To this end, we propose a \emph{warp-type-aware memory
scheduler} that prioritizes the occasional DRAM request from a mostly-hit
warp.


The design of our memory scheduler is very simple. Each memory request is
tagged with a single bit, which is set if the memory request comes from a
mostly-hit warp (or an all-hit warp, in case the warp was mischaracterized). 
We modify the request queue at the memory controller to
contain two different queues (\mycirc{4} in Figure~\ref{fig:main-mechanism}),
where a \emph{high-priority queue} contains all requests
that have their mostly-hit bit set to one.  The \emph{low-priority queue} contains all other
requests, whose mostly-hit bits are set to zero. Each queue uses
FR-FCFS~\cite{fr-fcfs,frfcfs-patent} as the scheduling policy; however, the scheduler
always selects requests from the high priority queue over requests
in the low priority queue.\footnote{Using two queues 
ensures that high-priority requests are not blocked
by low-priority requests even when the low-priority queue is full. Two-queue
priority also uses simpler logic design than comparator-based priority~\cite{bliss,bliss-arxiv}.}

%


\section{Methodology}
\label{sec:methodology-medic}

We model our mechanism using GPGPU-Sim 3.2.1~\cite{gpgpu-sim}.
Table~\ref{table:config} shows the configuration of the GPU. We modified
GPGPU-Sim to accurately model cache bank conflicts, and added the cache
bypassing, cache insertion, and memory scheduling mechanisms needed to support
\titleShortMeDiC{}. We use GPUWattch~\cite{gpuwattch} to evaluate power
consumption.

\begin{table*}[h]
  \centering
\vspace{3pt}
    \begin{tabular}{ll}
        \toprule
\textbf{System Overview}           &  15 cores, 6 memory partitions\\
        \cmidrule(rl){1-2}
\textbf{Shader Core Config.}           &  1400 MHz, 9-stage pipeline, GTO scheduler~\cite{ccws}\\
        \cmidrule(rl){1-2}
\textbf{Private L1 Cache}    &  16KB, 4-way associative, LRU, L1 misses are coalesced before \\
                             &  ccessing L2, 1 cycle latency \\
        \cmidrule(rl){1-2} 
\textbf{Shared L2 Cache}   &  \red{768KB total, 16-way associative, LRU, 2 cache banks} \\
                           &  \red{2 interconnect ports per memory partition, 10 cycle latency} \\
        \cmidrule(rl){1-2} 
\textbf{DRAM}   & GDDR5 1674 MHz, 6 channels (one per memory partition)\\
                & FR-FCFS scheduler~\cite{fr-fcfs,frfcfs-patent} 8 banks per rank, burst length 8\\
        \bottomrule
    \end{tabular}%
  \caption{Configuration of the simulated system.}
  \label{table:config}%
\end{table*}%

\begin{table*}[h]
\centering
\begin{tabular}{|c|c|c|c|c|c|c|}
\hline
\textbf{\#} & \textbf{Application} & \textbf{AH} & \textbf{MH} &\textbf{BL} &\textbf{MM} &\textbf{AM} \\ \hline\hline 
1           & Nearest Neighbor (NN)~\cite{cuda-sdk} & 19\% & \textbf{79\%} & 1\% & 0.9\% & 0.1\% \\
\hline
2           & Convolution Separable (CONS)~\cite{cuda-sdk} & 9\% & 1\% & \textbf{82\%} & 1\% & 7\% \\
\hline
3           & Scalar Product (SCP)~\cite{cuda-sdk}  & 0.1\% & 0.1\% & 0.1\% & 0.7\% & \textbf{99\%} \\
\hline
4           & Back Propagation (BP)~\cite{rodinia}  & 10\% & 27\% & \textbf{48\%} & 6\% & 9\% \\
\hline
5           & Hotspot (HS)~\cite{rodinia}  & 1\% & 29\% & \textbf{69\%} & 0.5\% & 0.5\% \\
\hline
6           & Streamcluster (SC)~\cite{rodinia} & 6\% & 0.2\% & 0.5\% & 0.3\% & \textbf{93\%} \\
\hline
7           & Inverted Index (IIX)~\cite{mars} & \textbf{71\%} & 5\% & 8\% & 1\% & 15\% \\
\hline
8           & Page View Count (PVC)~\cite{mars} & 4\% & 1\% & \textbf{42\%} & 20\% & 33\% \\
\hline
9           & Page View Rank (PVR)~\cite{mars} & 18\% & 3\% & 28\% & 4\% & \textbf{47\%} \\
\hline
10          & Similarity Score (SS)~\cite{mars} & \textbf{67\%} & 1\% & 11\% & 1\% & 20\% \\
\hline
11          & Breadth-First Search (BFS)~\cite{lonestar}& \textbf{40\%} & 1\% & 20\% & 13\% & 26\% \\
\hline
12          & Barnes-Hut N-body Simulation (BH)~\cite{lonestar} & \textbf{84\%} & 0\% & 0\% & 1\% & 15\% \\
\hline
13          & Delaunay Mesh Refinement (DMR)~\cite{lonestar} & \textbf{81\%} & 3\% & 3\% & 1\% & 12\% \\
\hline
14          & Minimum Spanning Tree (MST)~\cite{lonestar} & \textbf{53\%} & 12\% & 18\% & 2\% & 15\%  \\
\hline
15          & Survey Propagation (SP)~\cite{lonestar} & \textbf{41\%} & 1\% & 20\% & 14\% & 24\% \\
\hline
\hline
\end{tabular}%
\caption{{Evaluated GPGPU applications and the characteristics of their warps.}}%
\vspace{-2pt}
\label{table:apps}
\end{table*}

\vspace{5pt}
\paragraphbe{Modeling L2 Bank Conflicts.}\green{ In order to analyze the
detailed caching behavior of applications in modern GPGPU architectures, we
modified GPGPU-Sim to accurately model banked caches.\footnote{We
validate that the performance values reported for our applications before and
after our modifications to GPGPU-Sim are equivalent.} Within each memory
partition, we divide the shared L2 cache into two banks. When a memory
request misses in the L1 cache, it is sent to the memory partition
through the shared interconnect. However, it can only be sent if there is a
free port available at the memory partition (we dual-port each memory
partition). Once a request arrives at the port, a unified bank arbiter
dispatches the request to the request queue for the appropriate cache bank
(which is determined statically using some of the memory address bits).  If the
bank request queue is full, the request remains at the incoming port until
the queue is freed up. Traveling through the port and arbiter consumes an extra
cycle per request. In order to prevent a bias towards any one port or any one
cache bank, the simulator rotates which port and which bank are first examined
every cycle.}

\red{When a request misses in the L2 cache, it is sent to the DRAM request
queue, which is shared across all L2 banks as previously implemented in
GPGPU-Sim.  When a request returns from DRAM, it is inserted into one of
the per-bank DRAM-to-L2 queues. Requests returning from the L2 cache
to the L1 cache go through a unified memory-partition-to-interconnect queue (where round-robin priority is used 
to insert requests from different banks into the queue).}

\vspace{3pt}
\paragraphbe{GPGPU Applications.} We evaluate our system across multiple
GPGPU applications from the CUDA SDK~\cite{cuda-sdk}, Rodinia~\cite{rodinia},
MARS~\cite{mars}, and Lonestar~\cite{lonestar} benchmark suites.\footnote{\green{We 
use default tuning parameters for all applications.}}  These
applications are listed in Table~\ref{table:apps}, along with the breakdown of
warp characterization. The dominant warp type for each application is marked in
\emph{bold} (AH:~all-hit, MH:~mostly-hit, BL:~balanced, MM:~mostly-miss, 
AM:~all-miss; see Figure~\ref{fig:warp-types}). We simulate 500 million
instructions for each kernel of our application, though some kernels complete
before reaching this instruction count. 

\vspace{3pt}
\paragraphbe{Comparisons.} In addition to the baseline secs/medic/results, we compare
each individual component of \titleShortMeDiC{} with state-of-the-art policies.
\green{We compare our bypassing mechanism with three different
cache management policies. First, we compare to PCAL~\cite{donglihpca15}, a
token-based cache management mechanism. 
PCAL limits the number of threads that get to access the cache by using tokens. 
If a cache request is a miss, it causes a replacement only if the warp has a token. PCAL, as
modeled in this work, first grants tokens to the warp that recently used the cache,
then grants any remaining tokens to warps that access the cache in order of
their arrival.  Unlike the original proposal~\cite{donglihpca15}, which applies PCAL to the L1 caches,
we apply PCAL to the shared L2 cache. We sweep the number of tokens per epoch
and use the configuration that gives the best average performance. Second, we
compare \titleShortMeDiC{} against a random bypassing policy (\textbf{Rand}), where a percentage of 
randomly-chosen warps bypass the cache every 100k cycles. For every workload, we
statically configure the percentage of warps that bypass the cache such that Rand yields the best
performance. This comparison point is designed to show the value of warp type information
in bypassing decisions.} Third, we compare to a program counter (PC) based bypassing policy (\textbf{PC-Byp}). 
This mechanism bypasses requests from \emph{static instructions} that 
mostly miss (as opposed to requests from mostly-miss \emph{warps}). This comparison point is designed
to distinguish the value of tracking hit ratios at the warp level instead of at the instruction
level.

We compare our memory scheduling mechanism with
the baseline first-ready, first-come first-serve (FR-FCFS) memory scheduler~\cite{fr-fcfs,frfcfs-patent}, which is reported to
provide good performance on GPU and GPGPU
workloads~\cite{complexity,chatterjee-sc14,sms}.
We compare our cache
insertion with the Evicted-Address Filter~\cite{eaf-vivek}, a state-of-the-art CPU cache insertion policy. 

\vspace{5pt}
\paragraphbe{Evaluation Metrics.} \green{We report performance secs/medic/results using
the harmonic average of the IPC speedup (over the baseline GPU) of each kernel
of each application.\footnote{We confirm that for each application, 
all kernels have similar speedup values, and that aside from SS
and PVC, there are no outliers (i.e., no kernel has a much higher
speedup than the other kernels).  To verify that harmonic speedup is not swayed greatly by these few
outliers, we recompute it for SS and PVC \emph{without} these
outliers, and find that the outlier-free speedup is within 1\% of the
harmonic speedup we report in the dissertation.}  Harmonic speedup was shown to reflect the average
normalized execution time in multiprogrammed
workloads~\cite{harmonic_speedup}.
We calculate energy efficiency for each workload by dividing the IPC by the
energy consumed.}

\vspace{5pt}


\section{Evaluation}
\label{sec:eval}

\vspace{3pt}


\subsection{Performance Improvement of \titleShortMeDiC{}}
\vspace{3pt}

\begin{figure}[h!!!]
	\centering
	\includegraphics[width=\columnwidth]{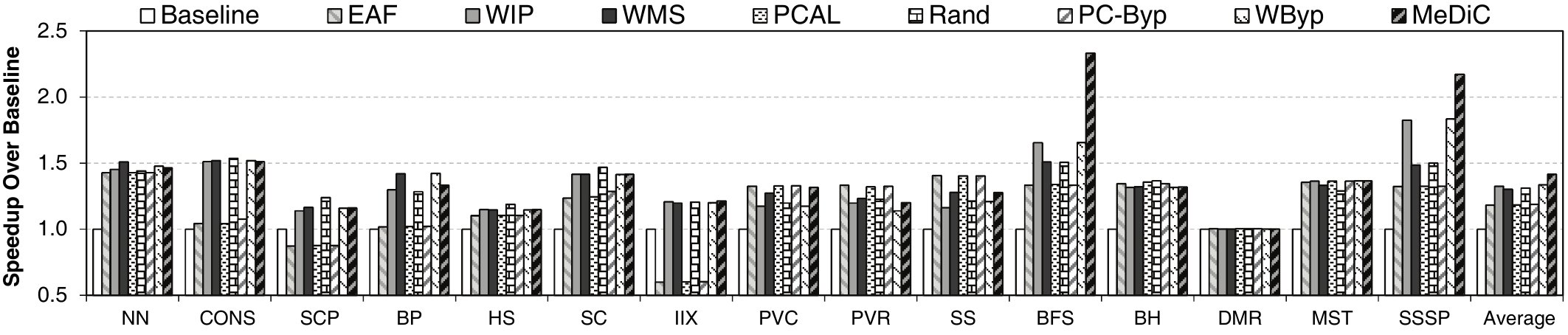}%
	\caption{\red{Performance of \titleShortMeDiC{}.}} 
	\label{fig:main-res}
\end{figure}

Figure~\ref{fig:main-res} shows the performance of \titleShortMeDiC{} compared to
the various state-of-the-art mechanisms 
(EAF~\cite{eaf-vivek}, PCAL~\cite{donglihpca15}, Rand, PC-Byp) from
Section~\ref{sec:methodology-medic},\footnote{We tune the configuration of each of
these previously-proposed mechanisms such that those mechanisms achieve the
highest performance secs/medic/results.} as well as the performance of each individual
component in \titleShortMeDiC{}.

\textbf{Baseline} shows the
performance of the unmodified GPU using FR-FCFS as the memory
scheduler~\cite{fr-fcfs,frfcfs-patent}. 
\textbf{EAF} shows the performance of the Evicted-Address Filter~\cite{eaf-vivek}.
\textbf{WIP} shows the performance of our
warp-type-aware insertion policy by itself. 
\textbf{WMS} shows the performance of our
warp-type-aware memory scheduling policy by itself. 
\red{\textbf{PCAL} shows the performance of the PCAL bypassing mechanism
proposed by Li et al.~\cite{donglihpca15}.
\textbf{Rand} shows the performance
of a cache bypassing mechanism that performs bypassing decisions randomly on
a fixed percentage of warps.}
\textbf{PC-Byp} shows the performance of the bypassing mechanism that uses the
PC as the criterion for bypassing instead of the warp-type.
\textbf{WByp} shows
the performance of our warp-type-aware bypassing policy by itself.


From these secs/medic/results, we draw the following conclusions:
\vspace{-1pt}
\begin{itemize}
\item Each component of \titleShortMeDiC{} individually provides significant
performance improvement: WIP (32.5\%), WMS (30.2\%), and WByp (33.6\%).
\titleShortMeDiC{}, which combines all three mechanisms, provides a 41.5\% performance
improvement over Baseline, on average.  \titleShortMeDiC{} matches or outperforms its
individual components for all benchmarks except BP, where \titleShortMeDiC{} has a 
higher L2 miss rate and lower row buffer locality than WMS and WByp.


\item WIP outperforms EAF~\cite{eaf-vivek} by 12.2\%. We observe that the
key benefit of WIP is that cache blocks from
mostly-miss warps are much more likely to be evicted. In addition, WIP
reduces the cache miss rate of several
applications (see Section~\ref{sec:cache-miss}). 

\item WMS provides significant performance gains (30.2\%) over Baseline,
because the memory scheduler prioritizes requests from warps that have
a high hit ratio, allowing these warps to become active much sooner than
they do in Baseline. 

\item WByp
provides an average 33.6\% performance improvement over Baseline, because
it is effective at reducing the L2 queuing latency.  We show the change in
queuing latency and provide a more detailed
analysis in Section~\ref{sec:cache-miss}.

\item \green{Compared to PCAL~\cite{donglihpca15}, WByp provides 12.8\% better performance, and
full \titleShortMeDiC{} provides 21.8\% better performance. We observe that
while PCAL reduces the amount of cache thrashing, the reduction in thrashing
does not directly translate into better performance. We observe that
warps in the mostly-miss category sometimes have high reuse, and acquire tokens to
access the cache. This causes less cache space to become available for mostly-hit warps,
limiting how many of these warps become all-hit. However, when high-reuse warps
that possess tokens are mainly in the mostly-hit category (PVC, PVR, SS, and BH), we find that
PCAL performs better than WByp.}


\item Compared to Rand,\footnote{\green{Note that our evaluation uses an ideal random
bypassing mechanism, where we manually select the best individual
percentage of requests to bypass the cache for each workload.  As a result, the
performance shown for Rand is better than can be practically realized.}} \titleShortMeDiC{} performs
6.8\% better, because \titleShortMeDiC{} is able to make bypassing decisions
that do not increase the miss rate significantly. This leads to lower off-chip
bandwidth usage under \titleShortMeDiC{} than under Rand. 
Rand increases the cache miss rate by 10\% for the kernels of
several applications (BP, PVC, PVR, BFS, and MST). \red{We observe that in many
cases, \titleShortMeDiC{} improves the performance of applications that tend to
generate a large number of memory requests, and thus experience substantial
queuing latencies. We further analyze the effect of \titleShortMeDiC{} on queuing delay in
Section~\ref{sec:cache-miss}.}

\item Compared to PC-Byp, \titleShortMeDiC performs 12.4\% better. We observe
that the overhead of tracking the PC becomes significant, and that thrashing
occurs as two PCs can hash to the same index, leading to inaccuracies in the
bypassing decisions.

\end{itemize}

\green{We conclude that each component of \titleShortMeDiC{}, and the full
\titleShortMeDiC{} framework, are effective. Note that each component of
\titleShortMeDiC{} addresses the same problem (i.e., memory divergence of threads
within a warp) using different techniques on different parts of the memory
hierarchy. For the majority of workloads, one optimization is enough. However,
we see that for certain high-intensity workloads (BFS and SSSP), the congestion is
so high that we need to attack divergence on multiple fronts. Thus,
\titleShortMeDiC{} provides better average performance than all of its individual
components, especially for such memory-intensive workloads. }

\subsection{Energy Efficiency of \titleShortMeDiC{}}

\begin{figure}[h!!!]
	\centering
	\includegraphics[width=\columnwidth]{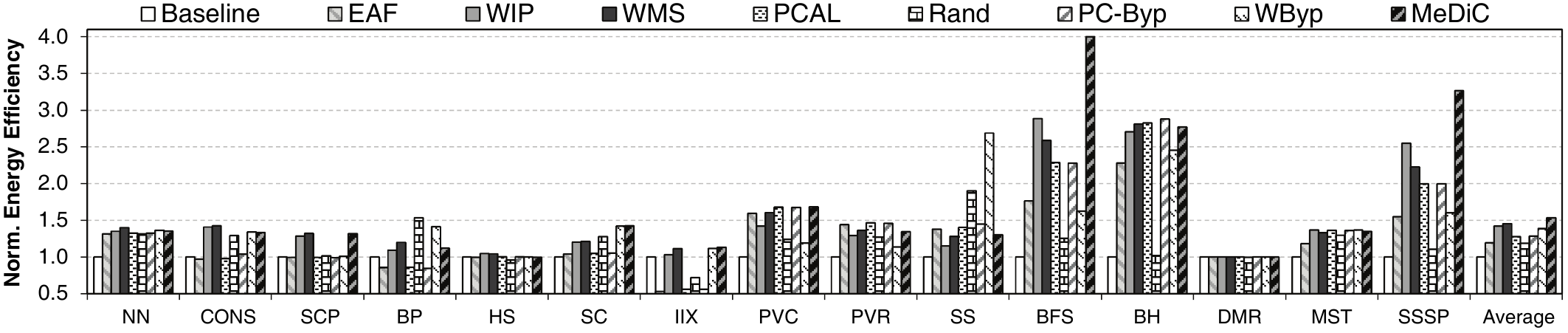}%
	\caption{\red{Energy efficiency of \titleShortMeDiC{}.}}
	\label{fig:energy-eff}
\end{figure}

\titleShortMeDiC{} provides significant GPU energy efficiency improvements, 
as shown in Figure~\ref{fig:energy-eff}. All three components of \titleShortMeDiC{},
as well as the full \titleShortMeDiC{} framework, are more energy efficient than
all of the other works we compare against.  \titleShortMeDiC{} is 53.5\% more
energy efficient than Baseline.  WIP itself is 19.3\% more energy efficient
than EAF.  WMS is 45.2\% more energy efficient than Baseline, which uses an FR-FCFS
memory scheduler~\cite{fr-fcfs, frfcfs-patent}.  WByp and \titleShortMeDiC{} are more energy efficient
than all of the other evaluated bypassing mechanisms, with 8.3\% and 20.1\% more 
efficiency than PCAL~\cite{donglihpca15}, respectively.

For all of our applications, the energy
efficiency of \titleShortMeDiC{} is better than or equal to Baseline, because even though our
bypassing logic sometimes increases energy consumption by sending more memory
requests to DRAM, the resulting performance improvement outweighs this additional energy. We
also observe that our insertion policy reduces the L2 cache miss rate,
allowing \titleShortMeDiC{} to be even more energy efficient by not wasting energy
on cache lookups for requests of all-miss warps.

\subsection{Analysis of Benefits}
\label{sec:cache-miss}

\begin{figure}[h!!!]
	\centering
	\vspace{3pt}
	\includegraphics[width=0.7\columnwidth]{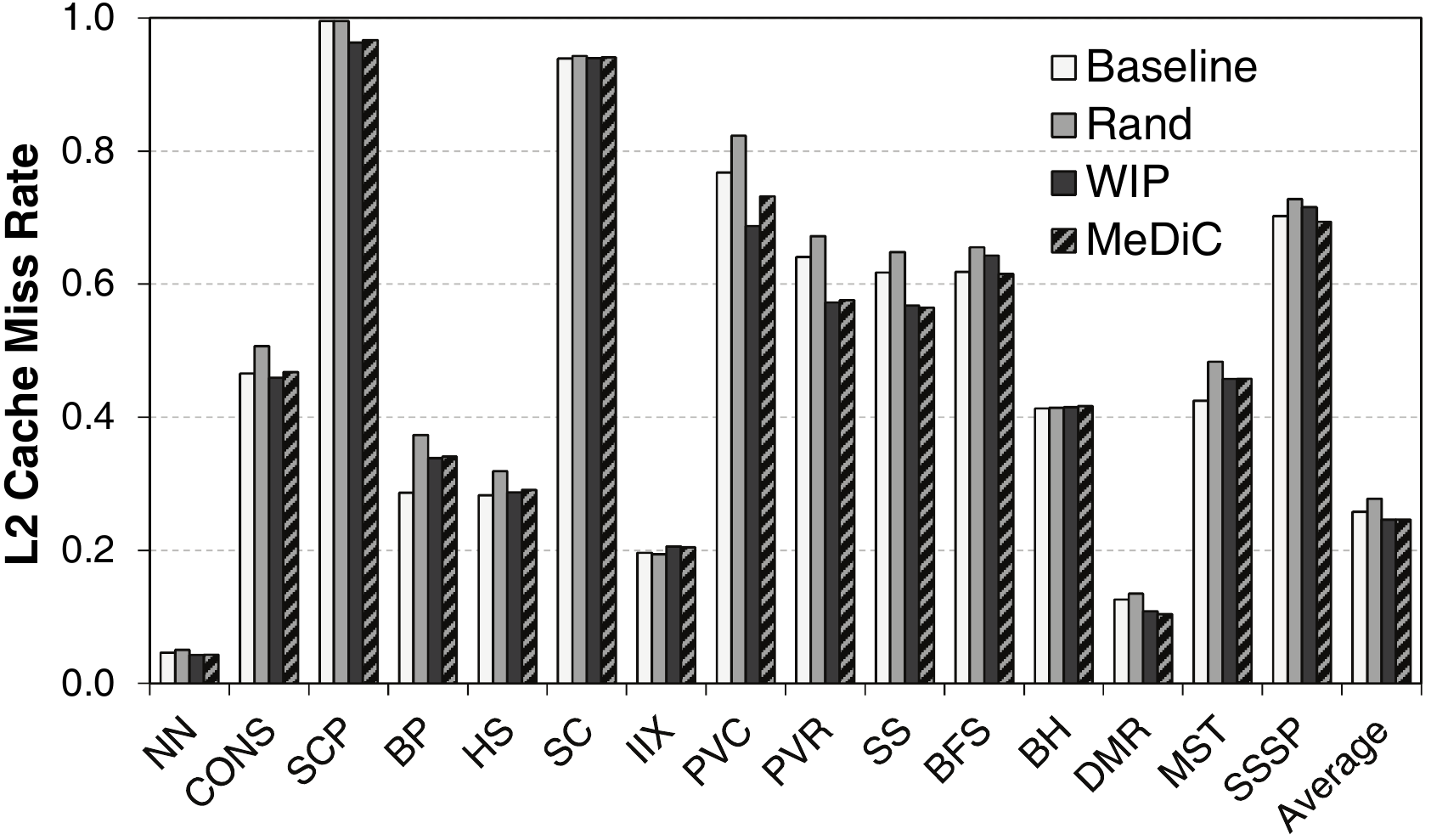}%
	\caption{L2 Cache miss rate of \titleShortMeDiC{}.} 
	\label{fig:miss-rate}
\end{figure}

\paragraphbe{Impact of \titleShortMeDiC{} on Cache Miss Rate.} One possible
downside of cache bypassing is that the bypassed requests
can introduce extra cache misses. Figure~\ref{fig:miss-rate} shows the cache miss rate
for Baseline, Rand, WIP, and \titleShortMeDiC{}.

Unlike Rand, \titleShortMeDiC{} does not increase the cache miss rate over Baseline for most of our applications.
The key factor behind this is WIP, the insertion policy in \titleShortMeDiC{}. We
observe that WIP on its own provides significant cache miss rate reductions for
several workloads (SCP, PVC, PVR, SS, and DMR). For the two workloads (BP and
BFS) where WIP increases the miss rate (5\% for
BP, and 2.5\% for BFS), the bypassing mechanism in \titleShortMeDiC{} is able to
contain the negative effects of WIP by dynamically tuning how aggressively
bypassing is performed based on the change in cache miss rate (see
Section~\ref{sec:bypass}). We conclude that \titleShortMeDiC{} does not hurt the
overall L2 cache miss rate.

\begin{figure}[h!!!]
	\centering
	\includegraphics[width=0.7\columnwidth]{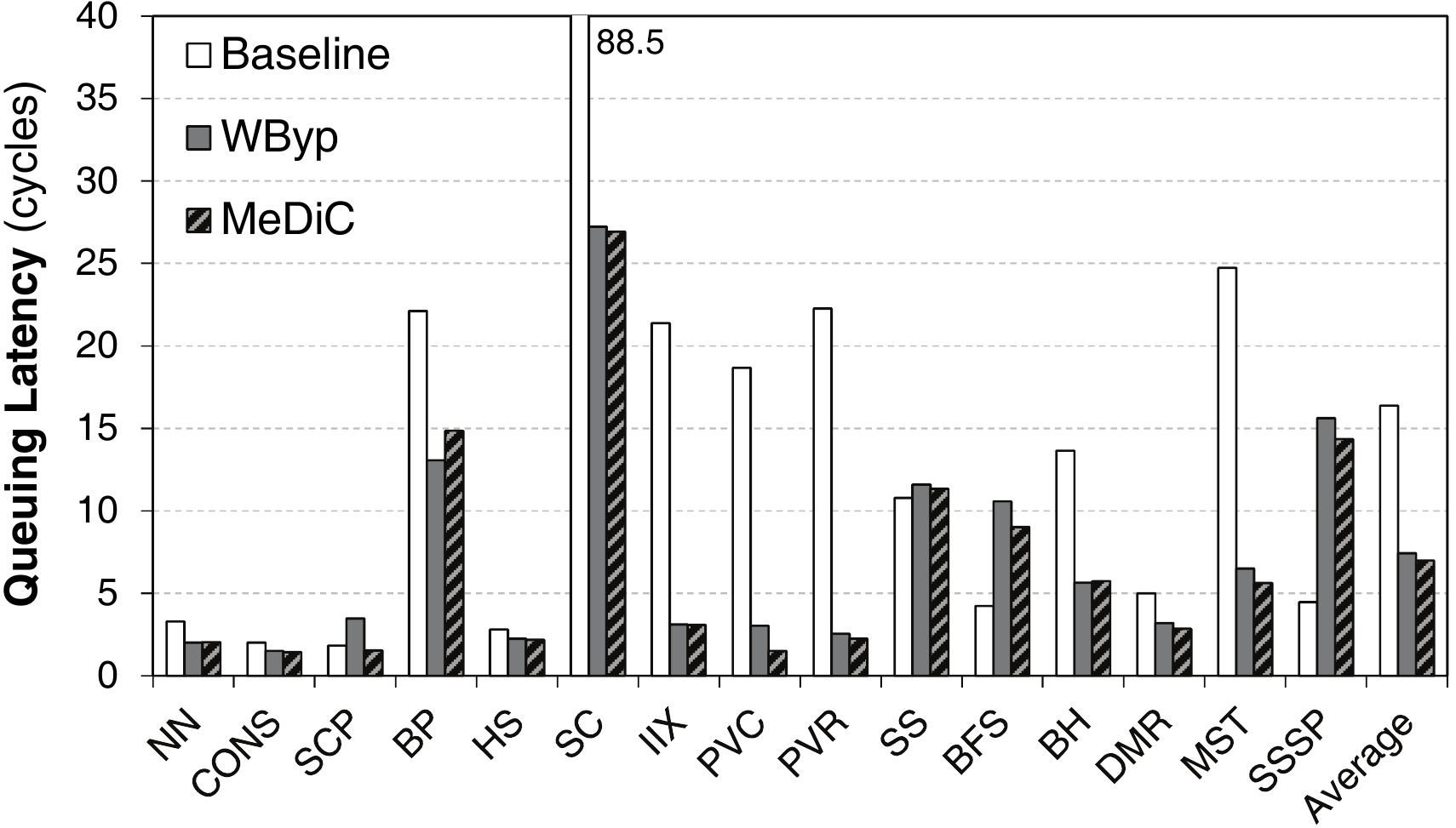}%
	\caption{L2 queuing latency for warp-type-aware bypassing and \titleShortMeDiC{}, compared to Baseline L2 queuing latency.} 
	\label{fig:queue-latency}
\end{figure}

\vspace{5pt}
\paragraphbe{Impact of \titleShortMeDiC{} on Queuing Latency.}
Figure~\ref{fig:queue-latency} shows the average L2 cache queuing latency
for WByp and \titleShortMeDiC{}, compared to Baseline queuing latency. For most workloads,
WByp reduces the queuing latency significantly (up to 8.7x in the case of PVR).
This reduction secs/medic/results in significant performance
gains for both WByp and \titleShortMeDiC{}.

There are two applications where the
queuing latency increases significantly: BFS and SSSP. We observe that
when cache bypassing is applied, the GPU cores retire instructions at a
much faster rate (2.33x for BFS, and 2.17x for SSSP). This increases the
pressure at each shared resource, including a sharp increase in the rate of cache
requests arriving at the L2 cache. This additional backpressure
secs/medic/results in higher L2 cache queuing latencies for both applications.

When all three mechanisms in \titleShortMeDiC{} (bypassing, cache insertion,
and memory scheduling) are combined, we observe that the queuing latency reduces even further. 
This additional reduction occurs because
the cache insertion mechanism in \titleShortMeDiC{} reduces the cache miss rate.
We conclude that in general, 
\titleShortMeDiC{} significantly alleviates the L2 queuing bottleneck.

\begin{figure}[h!!!]
	\centering
	\includegraphics[width=0.7\columnwidth]{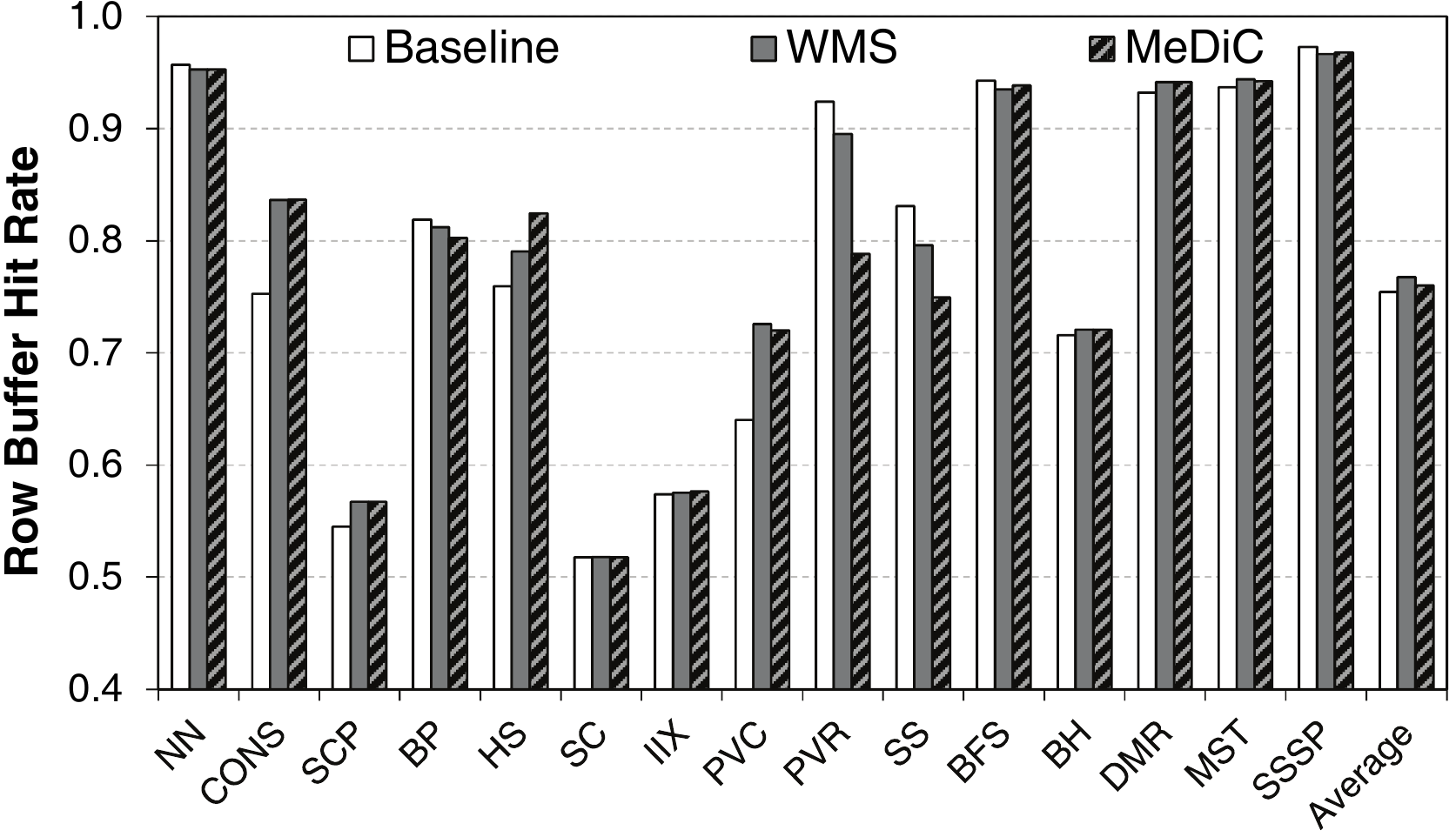}%
	\caption{Row buffer hit rate of warp-type-aware memory scheduling and \titleShortMeDiC{}, compared to Baseline.} 
	\label{fig:rowbufferlocality}
\end{figure}

\vspace{5pt}
\paragraphbe{Impact of \titleShortMeDiC{} on Row Buffer Locality.} Another
possible downside of cache bypassing is that
it may increase the number of requests serviced by DRAM,
which in turn can affect DRAM row buffer locality. Figure~\ref{fig:rowbufferlocality}
shows the row buffer hit rate for WMS and \titleShortMeDiC{}, compared to the Baseline
hit rate.

Compared to Baseline, WMS has a
negative effect on the row buffer locality of six applications (NN, BP, PVR, SS, BFS, and SSSP), and
a positive effect on seven applications (CONS, SCP, HS, PVC, BH, DMR, and MST).  We
observe that even though the row buffer locality of some applications decreases, the
overall performance improves, as the memory scheduler prioritizes requests from
warps that are more sensitive to long memory latencies. Additionally,
prioritizing requests from warps that send a small number of memory requests
(mostly-hit warps) over warps that send a large number of memory requests (mostly-miss
warps) allows more time for mostly-miss warps to batch requests together, improving their
row buffer locality.  Prior work on GPU memory scheduling~\cite{sms} has 
observed similar behavior, where batching requests together allows GPU requests
to benefit more from row buffer locality.

\subsection{Identifying Reuse in GPGPU Applications}
\label{sec:reuse}

While WByp bypasses warps that have low cache
utility, it is possible that some cache blocks fetched by these bypassed warps
get accessed frequently. Such a frequently-accessed cache block may
be needed later by a mostly-hit warp, and thus leads to an extra cache
miss (as the block bypasses the cache).
To remedy this,
we add a mechanism to \titleShortMeDiC{} that ensures all high-reuse cache blocks still get to
access the cache.  The key idea, building upon the state-of-the-art mechanism
for block-level reuse~\cite{eaf-vivek}, is to use a
Bloom filter to track the high-reuse cache blocks, and to use this filter to
override bypassing decisions.  We call this combined design
\textbf{\titleShortMeDiC{}-reuse}.

\begin{figure}[h!!!]
	\centering
	\includegraphics[width=0.7\columnwidth]{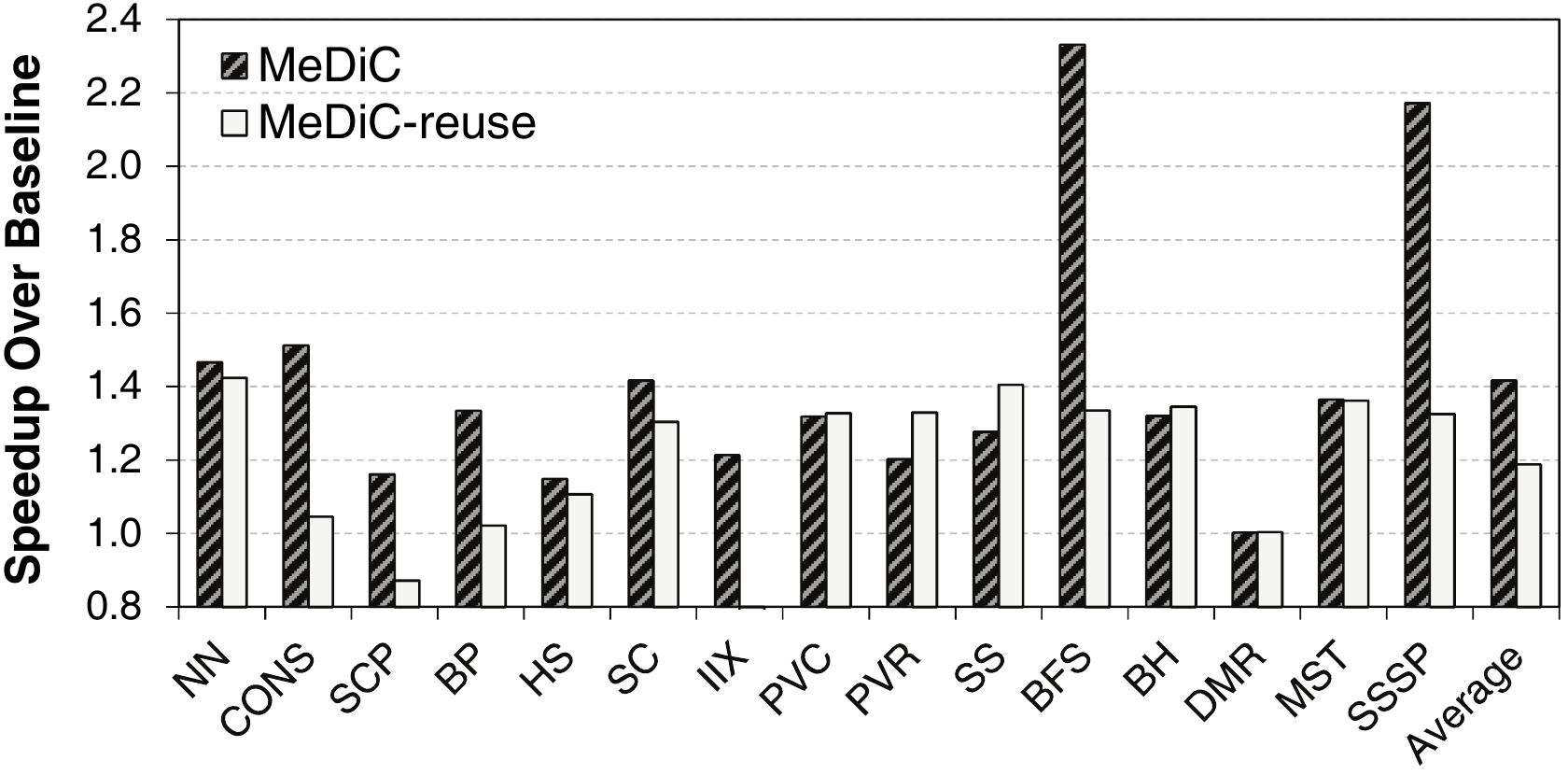}%
	\caption{Performance of \titleShortMeDiC{} with Bloom filter based reuse detection mechanism from the EAF cache~\cite{eaf-vivek}.} 
	\label{fig:reuse}
\end{figure}

Figure~\ref{fig:reuse} shows that \titleShortMeDiC{}-reuse suffers 16.1\% performance
degradation over \titleShortMeDiC{}. There are two reasons behind this degradation.
First, we observe that \titleShortMeDiC{} likely implicitly captures blocks with high reuse, as
these blocks tend to belong to all-hit and mostly-hit warps. 
Second, we observe that several GPGPU applications
contain access patterns that cause severe false positive aliasing within the
Bloom filter used to implement EAF and \titleShortMeDiC{}-reuse.  This leads to
some low reuse cache accesses from mostly-miss and all-miss warps taking up
cache space unnecessarily, resulting in cache thrashing.  We conclude that
\titleShortMeDiC{} likely implicitly captures the high reuse cache blocks that are relevant
to improving memory divergence (and thus performance).  However, there may
still be room for other mechanisms that make the best of block-level cache
reuse and warp-level heterogeneity in making caching decisions.



\subsection{Hardware Cost} 
\label{sec:hw-cost}
\titleShortMeDiC{} requires additional metadata storage in two locations.
First, each warp needs to maintain its own hit
ratio.  This can be done by adding 22~bits to the metadata of each warp: two
10-bit counters to track the number of L2 cache hits and the number of L2 cache
accesses, and 2~bits to store the warp type.\footnote{We
combine the mostly-miss and all-miss categories into a single warp type value,
because we perform the same actions on both types of warps.} To efficiently account for
overflow, the two counters that track L2 hits and L2 accesses
are shifted right when the most significant bit of the latter counter is set.
Additionally, the metadata for each cache line contains two bits, in
order to annotate the warp type for the cache insertion policy. The total
storage needed in the cache is $2~\times$~\emph{NumCacheLines} bits. In all, \titleShortMeDiC{}
comes at a cost of 5.1~kB, or less than 1\% of the L2 cache size.

To evaluate the trade-off of storage overhead, we evaluate a GPU where this
overhead is converted into additional L2 cache space for the baseline GPU. 
We conservatively increase the L2 capacity by 5\%, and find that this additional
cache capacity does not improve the performance of any of our workloads by more than 1\%.
As we discuss in the chapter, contention due to warp interference and divergence, and not due 
to cache capacity, is the root cause behind the performance bottlenecks that \titleShortMeDiC{} alleviates.
We conclude that \titleShortMeDiC{} can deliver significant performance improvements
with very low overhead.

%
%
%
%
%

%
%
%
%
%
%



\section{MeDiC: Conclusion}

Warps from GPGPU applications exhibit heterogeneity in their memory divergence
behavior at the shared L2 cache within the GPU.  We find that (1)~some warps 
benefit significantly from the cache, while others make poor use of it;  
(2)~such divergence behavior for a warp tends to remain stable for long periods of the
warp's execution; and (3)~the impact of memory divergence can be
amplified by the high queuing latencies at the L2 cache.


We propose \emph{\titleLongMeDiC} (\titleShortMeDiC{}), whose key idea is to
identify memory divergence heterogeneity in hardware and use this information to drive cache
management and memory scheduling, by prioritizing warps that take the greatest
advantage of the shared cache. To achieve this, \titleShortMeDiC{} consists of three
\emph{warp-type-aware} components for (1)~cache bypassing, (2)~cache insertion,
and (3)~memory scheduling.  \titleShortMeDiC{} delivers significant performance and
energy improvements over multiple previously proposed policies, and over a
state-of-the-art GPU cache management technique.  We conclude that exploiting inter-warp
heterogeneity is effective, and hope future works explore other ways of
improving systems based on this key observation.

\chapter{Reducing Inter-application Interference with Staged Memory Scheduling}
\label{sec:sms}

As the number of cores continues to increase in modern chip multiprocessor (CMP)
systems, the DRAM memory system is becoming a critical shared resource.
Memory requests from multiple cores interfere with each other, and this
inter-application interference is a significant impediment to individual
application and overall system performance. Previous work on application-aware
memory scheduling \cite{atlas,tcm,stfm, parbs} has addressed the problem by
making the memory controller aware of application characteristics and
appropriately prioritizing memory requests to improve system performance and
fairness.

Recent systems~\cite{bobcat,sandybridge,tegra} present an additional
challenge by introducing integrated graphics processing units (GPUs) on
the same die with CPU cores. GPU applications typically demand significantly
more memory bandwidth than CPU applications due to the GPU's capability of
executing a large number of parallel threads. GPUs use
single-instruction multiple-data (SIMD) pipelines to concurrently execute multiple threads,
where a batch of threads running the same instruction is called a
wavefront or warp. When a wavefront stalls on a memory instruction, the GPU core
hides this memory access latency by switching to another wavefront to avoid stalling
the pipeline. Therefore, there can be thousands of outstanding memory requests
from across all of the wavefronts. This is fundamentally more memory intensive than CPU
memory traffic, where each CPU application has a much
smaller number of outstanding requests due to the sequential execution model
of CPUs.

\begin{figure}
\centering
\includegraphics[width=0.7\columnwidth]{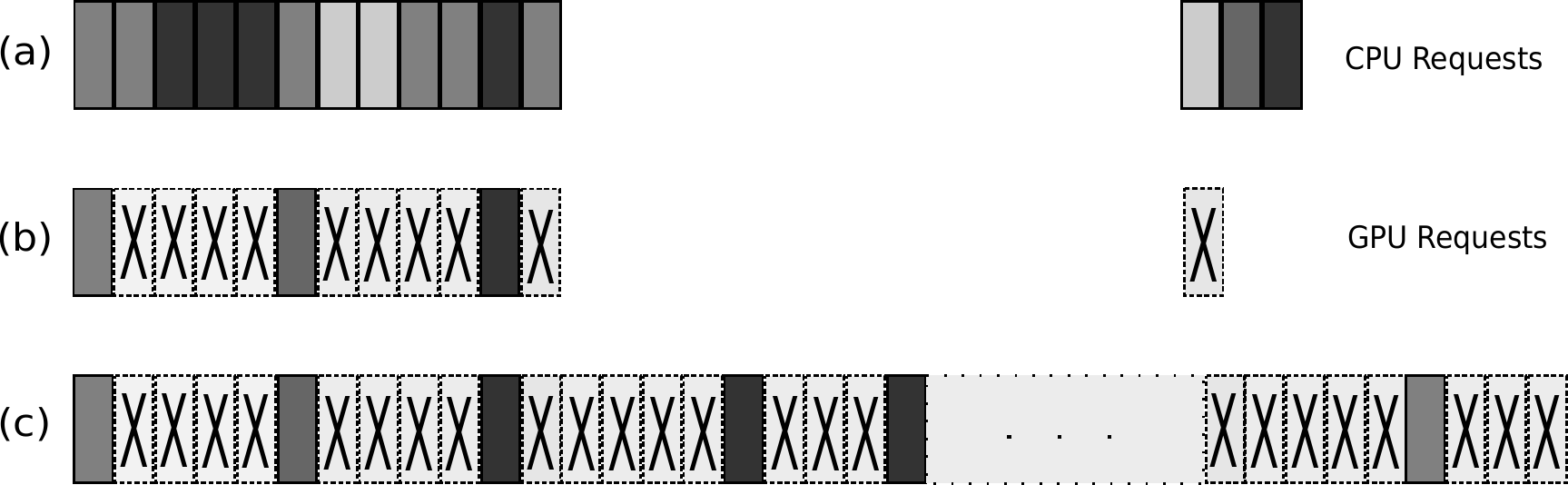}
\caption{Limited visibility example. (a) CPU-only information, (b) Memory controller's visibility, (c) Improved visibility}
\label{fig:visibility}
\end{figure}

Recent memory scheduling research has focused on
memory interference between applications in CPU-only scenarios. These
past proposals are built around a single centralized request buffer at each memory
controller (MC). The scheduling algorithm implemented in the memory controller
analyzes the stream of requests in the centralized request buffer to
determine application memory characteristics, decides on a priority for each
core, and then enforces these
priorities. Observable memory characteristics may include the number of requests
that result in row-buffer hits, the bank-level parallelism of each core, memory request
rates, overall fairness metrics, and other information.
Figure~\ref{fig:visibility}(a) shows the CPU-only scenario where the request
buffer only holds requests from the CPUs. In this case, the memory controller sees a number of
requests from the CPUs and has visibility into their memory behavior. On the
other hand, when the request buffer is shared between the CPUs and the GPU, as
shown in Figure~\ref{fig:visibility}(b), the large volume of requests from the
GPU occupies a significant fraction of the memory controller's request buffer, thereby limiting the
memory controller's visibility of the CPU applications' memory behaviors.

One approach to increasing the memory controller's visibility across a larger
window of memory requests is to increase the size of its request buffer. This
allows the memory controller to observe more requests from the CPUs to better
characterize their memory behavior, as shown in
Figure~\ref{fig:visibility}(c). For instance, with a large request buffer, the
memory controller can identify and service multiple requests from one CPU core to the same row such that
they become row-buffer hits, however, with a small request buffer as shown in
Figure~\ref{fig:visibility}(b), the memory controller may not even see these requests at the same time
because the GPU's requests have occupied the majority of the entries.

Unfortunately, very large request buffers impose
significant implementation challenges including the die area for the larger
structures and the additional circuit complexity for analyzing so many requests,
along with the logic needed for assignment and enforcement of priorities. Therefore, while building
a very large, centralized memory controller request buffer could lead to good memory
scheduling decisions, the approach is unattractive due to the resulting area,
power, timing and complexity costs.

In this work, we propose the \titleLongSMS (\titleShortSMS), a decentralized
architecture for memory scheduling in the context of integrated multi-core CPU-GPU systems. The key idea in \titleShortSMS is to decouple the various functional requirements
of memory controllers and partition these tasks across several simpler hardware
structures which operate in a staged fashion.
The three primary functions of the memory controller, which map to the
three stages of our proposed memory controller architecture, are:

\begin{enumerate}
\item Detection of basic within-application memory characteristics (e.g.,
row-buffer locality).
\item Prioritization across applications (CPUs and GPU) and enforcement
of policies to reflect the priorities.
\item Low-level command scheduling (e.g., activate, precharge, read/write), enforcement
of device timing constraints (e.g., t$_\text{RAS}$, t$_\text{FAW}$, etc.), and
resolving resource conflicts (e.g., data bus arbitration).
\end{enumerate}

Our specific \titleShortSMS implementation makes widespread use of distributed FIFO structures
to maintain a very simple implementation, but at the same time \titleShortSMS can provide
fast service to low memory-intensity (likely latency sensitive) applications and effectively
exploit row-buffer locality and bank-level parallelism for high memory-intensity (bandwidth
demanding) applications. While \titleShortSMS provides a specific implementation, our
staged approach for memory controller organization provides a general framework for exploring scalable
memory scheduling algorithms capable of handling the diverse memory needs of
integrated CPU-GPU systems of the future.


This work makes the following contributions:

\begin{itemize} 
\item We identify and present the challenges posed to existing memory
scheduling algorithms due to the highly memory-bandwidth-intensive
characteristics of GPU applications.
\item We propose a new decentralized, \emph{multi-stage} approach to
memory scheduling that effectively handles the interference caused by
bandwidth-intensive applications, while simplifying the hardware
implementation.
\item We evaluate our approach against four previous memory scheduling
algorithms~\cite{fr-fcfs, atlas, parbs, tcm} across a wide variety workloads
and CPU-GPU systems and show that it provides better performance and fairness.
As an example, our evaluations on a CPU-GPU
system show that \titleShortSMS improves system performance by 41.2\% and
fairness by 4.8$\times$ across 105 multi-programmed workloads on a 16-CPU/1-GPU, four
memory controller system, compared to the best previous memory scheduler
TCM~\cite{tcm}.

\end{itemize}

\section{Background}
\label{sec:background-sms}

\noindent In this section, we re-iterate DRAM organization
and discuss how past research attempted to
deal with the challenges of providing performance and fairness
for modern memory systems.

\clearpage

\subsection{Main Memory Organization}

\noindent DRAM is organized as two-dimensional arrays of bitcells. Reading or writing
data to DRAM requires that a row of bitcells from the array first be read into
a row buffer. This is required because the act of reading the row destroys the
row's contents, and so a copy of the bit values must be kept (in the row
buffer). Reads and writes operate directly on the row buffer. Eventually the
row is ``closed'' whereby the data in the row buffer are written back into the
DRAM array. Accessing data already loaded in the row buffer, also called a row
buffer hit, incurs a shorter latency than when the corresponding row must first
be ``opened'' from the DRAM array.
A modern memory controller (MC) must orchestrate the
sequence of commands to open, read, write and close rows. Servicing requests
in an order that increases row-buffer hits tends to improve overall throughput
by reducing the average latency to service requests. The MC is also
responsible for enforcing a wide variety of timing constraints imposed by
modern DRAM standards (e.g., DDR3) such as limiting the rate of page-open
operations (t$_\text{FAW}$) and ensuring a minimum amount of time between
writes and reads (t$_\text{WTR}$).

Each two dimensional array of DRAM cells constitutes a bank, and a group of banks
form a rank. All banks within a rank share a common set of command and data buses,
and the memory controller is responsible for scheduling commands such that each bus is used
by only one bank at a time. Operations on multiple banks may occur in parallel (e.g.,
opening a row in one bank while reading data from another bank's row buffer) so
long as the buses are properly scheduled and any other DRAM timing constraints
are honored. A memory controller can improve memory system throughput by scheduling requests
such that bank-level parallelism or BLP (i.e., the number of banks simultaneously
busy responding to commands) is increased.
A memory system implementation may support multiple independent memory channels (each
with its own ranks and banks) to further
increase the number of memory requests that can be serviced at the same time.
A key challenge in the implementation of modern, high-performance memory controllers is to
effectively improve system performance by maximizing both row-buffer hits and
BLP while simultaneously providing fairness among multiple CPUs and the GPU.

\subsection{Memory Scheduling}
\noindent Accessing off-chip memory is one of the major bottlenecks in microprocessors.
Requests that miss in the last level cache incur long latencies, and as
multi-core processors increase the number of CPUs, the problem gets worse
because all of the cores must share the limited off-chip memory bandwidth. The
large number of requests greatly increases contention for the memory data and command
buses. Since a bank can only process one command at a time, the large number
of requests also increases bank contention where requests must wait for busy
banks to finish servicing other requests. A request from one core can
also cause a row buffer containing data for another core to be closed, thereby
reducing the row-buffer hit rate of that other core (and vice-versa). All of
these effects increase the latency of memory requests by both increasing
queuing delays (time spent waiting for the memory controller to start servicing a request) and
DRAM device access delays (due to decreased row-buffer hit rates and bus
contention).

The memory controller is responsible for buffering and servicing memory requests from the
different cores and the GPU. Typical implementations make use of a memory
request buffer to hold and keep track of all in-flight requests. Scheduling
logic then decides which requests should be serviced, and issues the corresponding
commands to the DRAM devices. Different memory scheduling algorithms may attempt
to service memory requests in an order different than the order in which the requests
arrived at the memory controller, in order to increase row-buffer hit rates, bank level parallelism,
fairness, or achieve other goals.

\subsection{Memory Scheduling in CPU-only Systems}
\noindent Memory scheduling algorithms improve system performance by reordering
memory requests to deal with the different constraints and behaviors of DRAM.
The first-ready-first-come-first-serve (FR-FCFS)~\cite{fr-fcfs} algorithm attempts
to schedule requests that result in row-buffer hits (first-ready), and otherwise prioritizes
older requests (FCFS). FR-FCFS increases DRAM throughput, but it can cause fairness problems
by under-servicing applications with low row-buffer locality.
Several application-aware memory scheduling
algorithms~\cite{atlas,tcm,stfm,parbs} have been proposed to balance both
performance and fairness. Parallelism-aware Batch Scheduling
(PAR-BS)~\cite{parbs} batches requests based on their arrival times
(older requests batched first). Within a batch,
applications are ranked to preserve bank-level parallelism (BLP)
within an application's requests. More recently, ATLAS~\cite{atlas}
proposes prioritizing applications that have received the least memory service.
As a result, applications with low memory intensities, which typically
attain low memory service, are prioritized. However, applications with high
memory intensities are deprioritized and hence slowed down significantly,
resulting in unfairness. The most recent work on application-aware memory
scheduling, Thread Cluster Memory scheduling (TCM)~\cite{tcm}, addresses this
unfairness problem. TCM first clusters applications into low and high memory-intensity
clusters based on their memory intensities. TCM always prioritizes applications in
the low memory-intensity cluster, however, among the high memory-intensity
applications it shuffles request priorities to prevent unfairness.

\subsection{Characteristics of Memory Accesses from GPUs}

A typical CPU application only has a relatively small number of outstanding
memory requests at any time. The size of a processor's instruction window
bounds the number of misses that can be simultaneously exposed to the memory
system. Branch prediction accuracy limits how large the instruction window can
be usefully increased. In contrast, GPU applications have very
different access characteristics, generating many more memory requests
than CPU applications. A GPU application can consist of many thousands
of parallel threads, where memory stalls on one group of threads can be hidden
by switching execution to one of the many other groups of threads.

\begin{figure}
\centering
\includegraphics[width=1.0\textwidth]{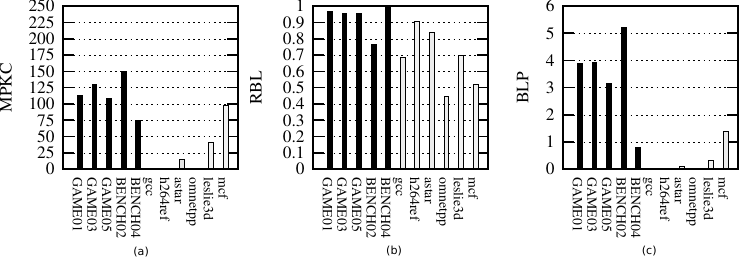}
\caption{GPU memory characteristic. (a) Memory-intensity, measured by memory requests
per thousand cycles, (b) Row buffer locality, measured by the fraction
of accesses that hit in the row buffer, and (c) Bank-level parallelism.}
\label{fig:sms-cpugpu-intensity}
\end{figure}

Figure~\ref{fig:sms-cpugpu-intensity} (a) shows the memory request rates for a representative
subset of our GPU
applications and the most memory-intensive SPEC2006 (CPU) applications, as
measured by memory requests per thousand cycles (see Section~\ref{sec:meth-sms} for
simulation methodology descriptions) when each application runs alone on the system. The raw
bandwidth demands of the GPU applications are often multiple times
higher than the SPEC benchmarks. Figure~\ref{fig:sms-cpugpu-intensity} (b) shows the
row-buffer hit rates (also called row-buffer locality or RBL). The GPU applications
show consistently high levels of RBL, whereas the SPEC benchmarks exhibit more
variability. The GPU programs have high levels of spatial locality, often due
to access patterns related to large sequential memory accesses (e.g., frame buffer
updates). Figure~\ref{fig:sms-cpugpu-intensity}(c) shows the BLP for each application,
with the GPU programs consistently making use of far banks at the same time.

In addition to the high-intensity memory traffic of GPU applications, there are
other properties that distinguish GPU applications from CPU applications.
The TCM~\cite{tcm} study observed that CPU applications with streaming access
patterns typically exhibit high RBL but low BLP,
while applications with less uniform access patterns typically
have low RBL but high BLP.
In contrast, GPU applications have {\em both} high RBL and
high BLP. The combination of high memory intensity, high RBL and high BLP means
that the GPU will cause significant interference to other applications
across all banks, especially when using a memory scheduling algorithm
that preferentially favors requests that result in row-buffer hits.

\subsection{What Has Been Done in the GPU?}

\noindent As opposed to CPU applications, GPU applications are not very latency sensitive
as there are a large number of independent threads to cover long memory
latencies. However, the GPU requires a significant amount of bandwidth far
exceeding even the most memory-intensive CPU applications. As a result, a GPU
memory scheduler~\cite{lindholm} typically needs a large request buffer that is
capable of request coalescing (i.e., combining multiple requests for the same
block of memory into a single combined request~\cite{programmingguide}).
Furthermore, since GPU
applications are bandwidth intensive, often with streaming access patterns, a
policy that maximizes the number of row-buffer hits is effective for GPUs to
maximize overall throughput. As a result, FR-FCFS with a large request buffer
tends to perform well for GPUs~\cite{gpgpu-sim}. In view of this, previous
work~\cite{complexity} designed mechanisms to reduce the complexity of
row-hit first based (FR-FCFS) scheduling.

\section{Challenges with Existing Memory Controllers}
\label{sec:sms-motivation}

\subsection{The Need for Request Buffer Capacity}
\begin{figure}
\centering
\includegraphics[width=0.7\columnwidth]{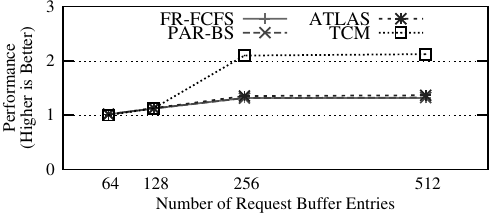}
\caption{Performance at different request buffer sizes}
\label{fig:reqQ}
\end{figure}

The results from Figure~\ref{fig:sms-cpugpu-intensity} showed that GPU applications have
very high memory intensities. As discussed in Section~\ref{sec:baseline-cpugpu-interfere}, the
large number of GPU memory requests occupy many of the memory controller's request buffer
entries, thereby making it very difficult for the memory controller to properly determine the
memory access characteristics of each of the CPU applications.
Figure~\ref{fig:reqQ} shows the performance impact of increasing the memory controller's
request buffer size for a variety of memory scheduling algorithms (full
methodology details can be found in Section~\ref{sec:meth-sms}) for a 16-CPU/1-GPU
system. By increasing the size of the request buffer from 64 entries to 256
entries,\footnote{For all sizes, half of the entries are reserved for the CPU requests.} previously proposed memory controller algorithms can gain up to 63.6\% better
performance due to this improved visibility.

\subsection{Implementation Challenges in Providing Request Buffer Capacity}
The results above show that when the memory controller has enough visibility across the global
memory request stream to properly characterize the behaviors of each core, a
sophisticated algorithm like TCM can be effective at making good scheduling
decisions. Unfortunately, implementing a sophisticated algorithm like TCM over
such a large scheduler introduces very significant implementation challenges.
For all algorithms that use a centralized request buffer and prioritize requests
that result in row-buffer hits (FR-FCFS, PAR-BS, ATLAS, TCM), associative logic
(CAMs) will be needed for each entry to compare its requested row against
currently open rows in the DRAM banks. For all algorithms that prioritize
requests based on rank/age (FR-FCFS, PAR-BS, ATLAS, TCM), a large comparison tree is
needed to select the highest ranked/oldest request from all request buffer
entries. The size of this comparison tree grows with request buffer size.
Furthermore, in addition to this logic for reordering requests and enforcing
ranking/age, TCM also requires additional logic to continually monitor each
CPU's last-level cache MPKI rate (note that a CPU's instruction count is not
typically available at the memory controller), each core's RBL which requires additional {\em
shadow row buffer index} tracking~\cite{fst,fst-tocs}, and each core's BLP.

Apart from the logic required to implement the policies of the specific memory
scheduling algorithms, all of these memory controller designs need additional logic to enforce
DDR timing constraints. Note that different timing constraints will apply
depending on the {\em state} of each memory request. For example, if a memory
request's target bank currently has a different row loaded in its row buffer,
then the memory controller must ensure that a precharge (row close) command is allowed to
issue to that bank (e.g., has t$_\text{RAS}$ elapsed since the row was
opened?), but if the row is already closed, then different timing constraints
will apply. For each request buffer entry, the memory controller will determine whether or
not the request can issue a command to the DRAM based on the current state of
the request and the current state of the DRAM system. That is, {\em every}
request buffer entry (i.e., all 256) needs an independent instantiation of the
DDR compliance-checking logic (including data and command bus availability
tracking).
This type of monolithic memory controller effectively implements a large out-of-order scheduler; note that typical
instruction schedulers in modern out-of-order processors only have about 32-64
entries~\cite{microarch}. Even after accounting for the clock speed differences
between CPU core and DRAM command frequencies, it is very difficult to implement
a fully-associative\footnote{Fully associative in the sense that a request in
{\em any} one of the request buffer entries could be eligible to issue in a
given cycle.}, age-ordered/prioritized, out-of-order scheduler with 256-512
entries~\cite{palacharla}.

\section{The \titleLongSMS}
\label{sec:mechanism-sms}

The proposed \titleLongSMS (\titleShortSMS) is structured to reflect the primary functional
tasks of the memory scheduler. Below, we first describe the overall \titleShortSMS
algorithm, explain additional implementation details, step through the rationale
for the design, and then walk through the hardware implementation.

\subsection{The \titleShortSMS Algorithm}
\paragraphbe{\Batch Formation.}
The first stage of \titleShortSMS consists of several simple FIFO structures, one
per {\em source} (i.e., a CPU core or the GPU).  Each request from a given
source is initially inserted into its respective FIFO upon arrival at the
memory controller.  A {\em\batch} is simply one or more memory requests from
the same source that access the same DRAM row.  That is, all requests within a
\batch, except perhaps for the first one, would be row-buffer hits if scheduled
consecutively.  A \batch is complete or {\em ready} when an incoming request
accesses a different row, when the oldest request in the \batch has exceeded a
threshold age, or when the FIFO is full.  Ready \batches may then be considered
by the second stage of the \titleShortSMS.

\ignore{
\begin{figure*}[h]
\centering
\includegraphics[width=4in,height=2in]{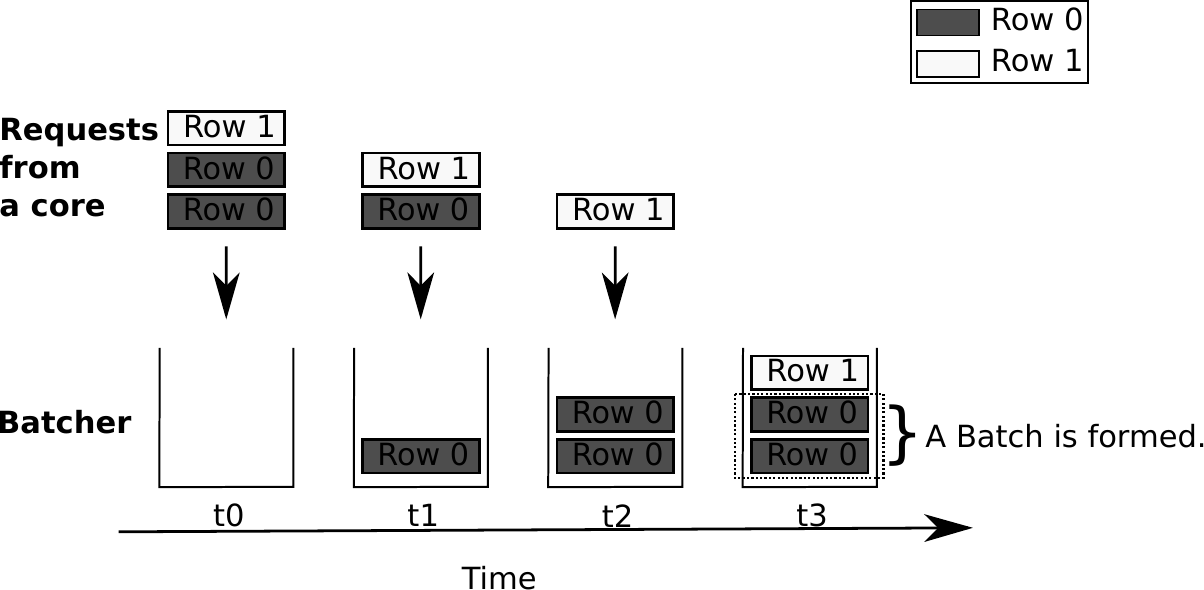}
\caption{A timeline showing how a batch of requests is formed for an application}
\label{fig:batcher}
\end{figure*}
}

\paragraphbe{\Batch Scheduler.}
The \batch formation stage has combined individual memory requests into \batches of
row-buffer hitting requests.  The next stage, the \batch scheduler, deals directly
with \batches, and therefore need not worry about scheduling to optimize for
row-buffer locality.  Instead, the \batch scheduler can focus on higher-level policies
regarding inter-application interference and fairness.  The goal of the \batch
scheduler is to prioritize \batches from applications that are latency critical,
while making sure that bandwidth-intensive applications (e.g., the GPU) still make
reasonable progress.

The \batch scheduler operates in two states: pick and drain.  In the pick state, the
\batch scheduler considers each FIFO from the \batch formation stage.  For each
FIFO that contains a {\em ready} \batch, the \batch scheduler picks one \batch based
on a balance of shortest-job first (SJF) and round-robin principles.  For SJF, the
\batch scheduler chooses the core (or GPU) with the fewest total memory requests
across all three stages of the \titleShortSMS. SJF prioritization reduces average
request service latency, and it tends to favor latency-sensitive applications,
which tend to have fewer total requests. The other component of the \batch
scheduler is a round-robin policy that simply cycles through each of the
per-source FIFOs ensuring that high memory-intensity applications receive
adequate service. Overall, the \batch scheduler chooses the SJF policy with a
probability of $p$, and the round-robin policy otherwise.

After picking a \batch, the \batch scheduler enters a drain state where it
forwards the requests from the selected \batch to the final stage of the \titleShortSMS.
The \batch scheduler simply dequeues one request per cycle until all requests
from the \batch have been removed from the selected \batch formation FIFO.  At this point,
the \batch scheduler re-enters the pick state to select the next \batch.

\ignore{
\floatname{algorithm}{Ruleset}
\begin{algorithm}
\caption{Batch prioritization in the batch scheduler}
\label{rules:schedalgo}
\footnotesize{
\begin{algorithmic}
\STATE \textbf{1a) Probabilistic SRPT (a)}: with probability $p$, prioritize
batch of application with smallest outstanding number of requests.
\STATE \textbf{1a) Probabilistic SRPT (b)}: with probability $p$, prioritize
each application's batch in a round-robin manner.
\STATE \textbf{2) Within each application: FCFS} Older batch first
\end{algorithmic}
}
\end{algorithm}
}
 
\paragraphbe{DRAM Command Scheduler.}
The last stage of the \titleShortSMS is the DRAM command scheduler (DCS).  The DCS
consists of one FIFO queue per DRAM bank (e.g., eight banks/FIFOs for DDR3).
The drain phase of the \batch scheduler places the memory requests directly
into these FIFOs.  Note that because \batches are moved into the DCS FIFOs one
\batch at a time, any row-buffer locality within a \batch is preserved within a
DCS FIFO.  At this point, any higher-level policy decisions have already been
made by the \batch scheduler, therefore, the DCS can simply focus on issuing
low-level DRAM commands and ensuring DDR protocol compliance.

On any given cycle, the DCS only considers the requests at the {\em head} of
each of the per-bank FIFOs.  For each request, the DCS determines whether that
request can issue a command based on the request's current row-buffer state
(i.e., is the row buffer already open with the requested row, closed, or open
with the wrong row?) and the current DRAM state (e.g., time elapsed since a 
row was opened in a bank, data bus availability).  If more than one request is
eligible to issue a command, the DCS simply arbitrates in a round-robin fashion.

\subsection{Additional Algorithm Details}
\label{sec:additionalalgodetails}
\paragraphbe{\Batch Formation Thresholds.}
The \batch formation stage holds requests in the per-source FIFOs until a
complete \batch is ready.  This could unnecessarily delay requests as the
\batch will not be marked ready until a request to a {\em different} row
arrives at the memory controller, or the FIFO size has been reached.  This
additional queuing delay can be particularly devastating for low-intensity,
latency-sensitive applications.

\titleShortSMS considers an application's memory intensity in forming \batches.
For applications with low memory-intensity ($<$1 MPKC), \titleShortSMS completely
bypasses the \batch formation and \batch scheduler, and forwards requests directly to
the DCS per-bank FIFOs.  For these highly sensitive applications, such a bypass
policy minimizes the delay to service their requests.  Note that this bypass
operation will not interrupt an on-going drain from the \batch scheduler, which
ensures that any separately scheduled \batches maintain their row-buffer
locality.

For medium memory-intensity (1-10 MPKC) and high memory-intensity ($>$10 MPKC) applications,
the \batch formation stage uses age thresholds of 50 and 200 cycles,
respectively.  That is, regardless of how many requests are in the current
\batch, when the oldest request's age exceeds the threshold, the entire \batch
is marked ready (and consequently, any new requests that arrive, even if
accessing the same row, will be grouped into a new \batch).  Note that while
TCM uses the MPKI metric to classify memory intensity, \titleShortSMS uses misses per
thousand {\em cycles} (MPKC) since the per-application instruction counts are not
typically available in the memory controller.  While it would not be overly difficult to
expose this information, this is just one less implementation overhead that
\titleShortSMS can avoid.

\paragraphbe{Global Bypass.}
As described above, low memory-intensity applications can bypass the entire \batch
formation and scheduling process and proceed directly to the DCS.  Even for high
memory-intensity applications, if the memory system is lightly loaded (e.g.,
if this is the only application running on the system right now), then the \titleShortSMS
will allow all requests to proceed directly to the DCS.  This bypass is enabled
whenever the total number of in-flight requests (across {\em all} sources) in
the memory controller is less than sixteen requests.

\paragraphbe{Round-Robin Probability.}
As described above, the \batch scheduler uses a probability of $p$ to schedule
batches with the SJF policy and the round-robin policy otherwise. Scheduling
\batches in a round-robin order can ensure fair progress from high-memory
intensity applications. Our experimental results show that setting $p$ to 90\%
(10\% using the round-robin policy) provides a good performance-fairness
trade-off for \titleShortSMS. 

\subsection{\titleShortSMS Rationale}
\paragraphbe{In-Order \Batch Formation.}
It is important to note that \batch formation occurs in the order of request
arrival.  This potentially sacrifices some row-buffer locality as requests to
the same row may be interleaved with requests to other rows.  We considered
many variations of \batch formation that allowed out-of-order grouping of
requests to maximize the length of a run of row-buffer hitting requests, but
the overall performance benefit was not significant.  First, constructing very
large \batches of row-buffer hitting requests can introduce significant
unfairness as other requests may need to wait a long time for a bank to
complete its processing of a long run of row-buffer hitting
requests~\cite{minimalist}.  Second, row-buffer locality across \batches may
still be exploited by the DCS.  For example, consider a core that has three
\batches accessing row X, row Y, and then row X again.  If X and Y map to
different DRAM banks, say banks A and B, then the \batch scheduler will send the
first and third \batches (row X) to bank A, and the second \batch (row Y) to
bank B.  Within the DCS's FIFO for bank A, the requests for the first and third
\batches will all be one after the other, thereby exposing the row-buffer
locality across \batches despite the requests appearing ``out-of-order'' in
the original \batch formation FIFOs.

\paragraphbe{In-Order \Batch Scheduling.}
Due to contention and back-pressure in the system, it is possible that a FIFO
in the \batch formation stage contains more than one valid \batch.  In such a
case, it could be desirable for the \batch scheduler to pick one of the
\batches not currently at the head of the FIFO.  For example, the bank
corresponding to the head \batch may be busy while the bank for another \batch
is idle.  Scheduling \batches out of order could decrease the service latency
for the later \batches, but in practice it does not make a big difference and
adds significant implementation complexity.
It is important to note that even though \batches are dequeued from the \batch
formation stage in arrival order per FIFO, the request order {\em between}
the FIFOs may still slip relative to each other.  For example, the \batch
scheduler may choose a recently arrived (and formed) \batch from a high-priority
(i.e., latency-sensitive) source even though an older, larger \batch from a different
source is ready.

\paragraphbe{In-Order DRAM Command Scheduling.}
For each of the per-bank FIFOs in the DCS, the requests are already grouped
by row-buffer locality (because the \batch scheduler drains an entire \batch
at a time), and globally ordered to reflect per-source priorities.  Further
reordering at the DCS would likely just undo the prioritization decisions made
by the \batch scheduler.  Like the \batch scheduler, the in-order nature of
each of the DCS per-bank FIFOs does not prevent out-of-order scheduling at
the global level.  A CPU's requests may be scheduled to the DCS in arrival
order, but the requests may get scattered across different banks, and the
issue order among banks may slip relative to each other.

\subsection{Hardware Implementation}
\begin{figure*}[h!]
\centering
\includegraphics[width=0.7\textwidth]{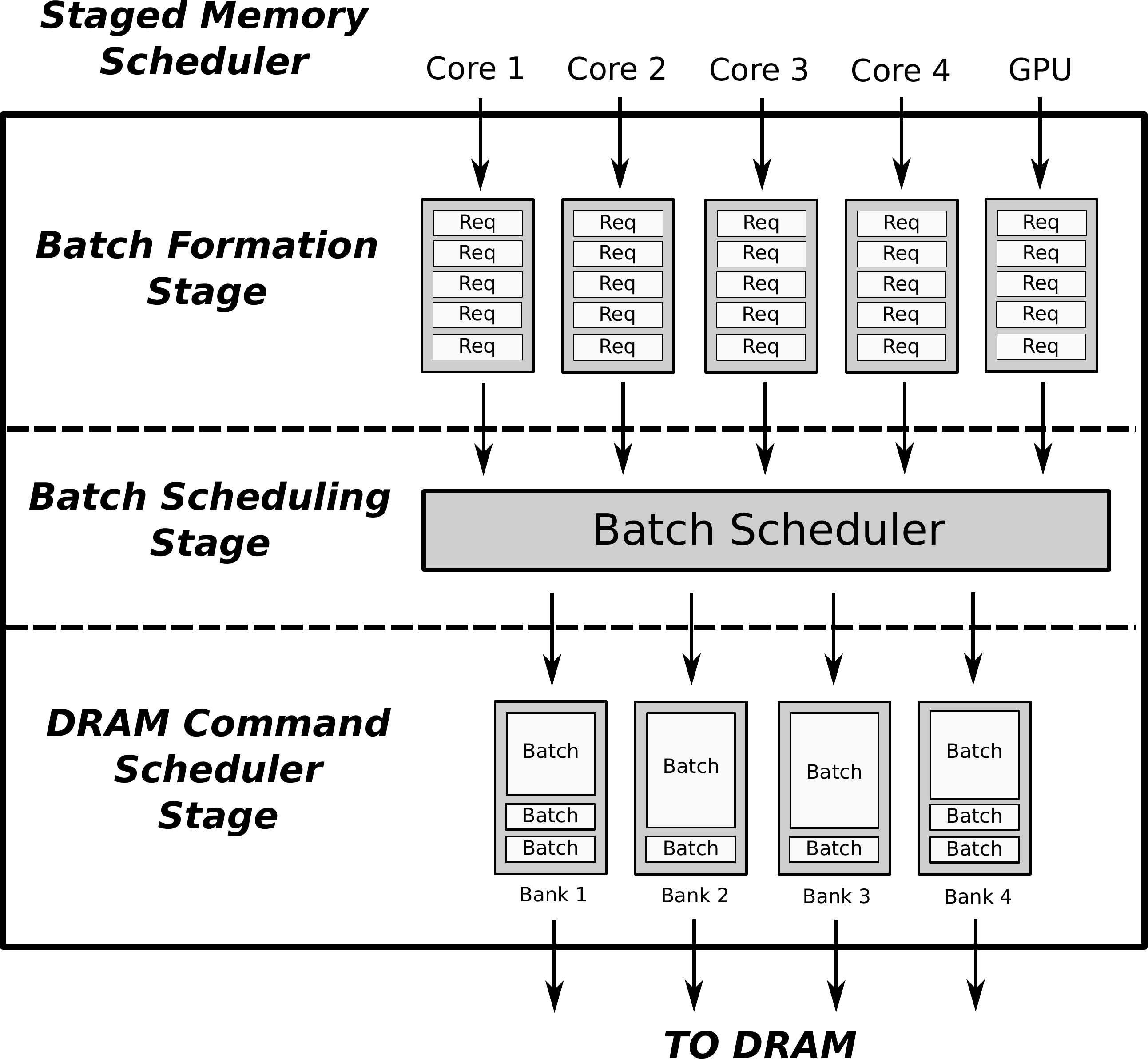}
\caption{The design of \titleShortSMS}
\label{fig:overall}
\end{figure*}

The staged architecture of \titleShortSMS lends directly to a low-complexity
hardware implementation. Figure~\ref{fig:overall} illustrates the overall
hardware organization of \titleShortSMS.\\[-0.1in]

\paragraphbe{\Batch Formation.} The \batch formation stage consists of
little more than one FIFO per source (CPU or GPU). Each FIFO maintains
an extra register that records the row index of the last request, so
that any incoming request's row index can be compared to determine if
the request can be added to the existing \batch. Note that this requires
only a single comparator (used only once at insertion) per FIFO. Contrast
this to a conventional monolithic request buffer where comparisons on
{\em every} request buffer entry (which is much larger than the number of FIFOs that \titleShortSMS
uses) must be made, potentially against all currently open rows across all
banks.\\[-0.1in]

\paragraphbe{\Batch Scheduler.} The \batch scheduling stage consists primarily
of combinatorial logic to implement the \batch picking rules. When using the
SJF policy, the \batch scheduler only needs to pick the \batch corresponding to
the source with the fewest in-flight requests, which can be easily performed
with a tree of MIN operators. Note that this tree is relatively shallow since
it only grows as a function of the number of FIFOs. Contrast this to the
monolithic scheduler where the various ranking trees grow as a function of the
total number of entries.\\[-0.1in]

\paragraphbe{DRAM Command Scheduler.} The DCS stage consists of the per-bank
FIFOs. The logic to track and enforce the various DDR timing and power
constraints is identical to the case of the monolithic scheduler, but the scale
is drastically different. The DCS's DDR command-processing logic only considers the
requests at the head of each of the per-bank FIFOs (eight total for DDR3),
whereas the monolithic scheduler requires logic to consider {\em every} request buffer entry (hundreds).\\[-0.1in]

\paragraphbe{Overall Configuration and Hardware Cost.}
The final configuration of \titleShortSMS that we use in this dissertation consists of
the following hardware structures. The \batch formation stage uses ten-entry
FIFOs for each of the CPU cores, and a twenty-entry FIFO for the GPU. The DCS
uses a fifteen-entry FIFO for each of the eight DDR3 banks. For sixteen cores
and a GPU, the aggregate capacity of all of these FIFOs is 300 requests,
although at any point in time, the \titleShortSMS logic can only consider or act
on a small subset of the entries (i.e., the seventeen at the heads of the
\batch formation FIFOs and the eight at the heads of the DCS FIFOs). In
addition to these primary structures, there are a small handful of bookkeeping
counters. One counter per source is needed to track the number of in-flight
requests; each counter is trivially managed as it only needs to be incremented
when a request arrives at the memory controller, and then decremented when the request is
complete. Counters are also needed to track per-source MPKC rates for
memory-intensity classification, which are incremented when a request arrives,
and then periodically reset.
Table~\ref{table:hardware} summarizes the amount of hardware overhead required for
each stage of \titleShortSMS.

\begin{table*}[ht]
\centering
\scriptsize
\begin{tabular}{| l | l | l |}
\hline \bf{Storage} & \bf{Description} & \bf{Size} \\
\hline \hline \multicolumn{3}{|l|} {\textbf{\textit{Storage Overhead of Stage 1: \Batch formation stage}} }\\
\hline CPU FIFO queues & A CPU core's FIFO queue & $N_{core} \times Queue\_Size_{core} = 160$ entries \\
\hline GPU FIFO queues & A GPU's FIFO queue & $N_{GPU} \times Queue\_Size_{GPU} = 20$ entries \\
\hline MPKC counters & Counts per-core MPKC & $N_{core} \times log_2 MPKC_{max} = 160$ bits \\
\hline Last request's row index & Stores the row index of & $(N_{core} + N_{GPU}) \times log_2 Row\_Index\_Size = 204$ bits \\ & the last request to the FIFO & \\
\hline \multicolumn{3}{|l|} {\textbf{\textit{Storage Overhead of Stage 2: \Batch Scheduler}} }\\
\hline CPU memory request counters & Counts the number of outstanding & $N_{core} \times log_2 Count_{max\_CPU} = 80$ bits\\ & memory requests of a CPU core &  \\
\hline GPU memory request counter & Counts the number of outstanding & $N_{GPU} \times log_2 Count_{max\_GPU} = 10$ bits\\ & memory requests of the GPU &  \\
\hline \hline \multicolumn{3}{|l|} {\textbf{\textit{Storage Overhead of Stage 3: DRAM Command Scheduler}} }\\
\hline Per-Bank FIFO queues & Contains a FIFO queue per bank & $N_{banks} \times Queue\_Size_{bank} = 120$ entries\\
\hline
\end{tabular}
\caption{\small Hardware storage required for \titleShortSMS \label{table:hardware}}
\end{table*}

\subsection{Experimental Methodology}
\label{sec:meth-sms}

We use an in-house cycle-accurate simulator to perform our evaluations. For our
performance evaluations, we model a system with sixteen x86 CPU cores and a
GPU.  For the CPUs, we model three-wide out-of-order processors with a cache hierarchy including per-core L1 caches and
a shared, distributed L2 cache.  The GPU does not share the CPU caches.
Table~\ref{table:sysparams} shows the detailed system
parameters for the CPU and GPU cores. The parameters for the main memory system
are listed in Table~\ref{table:sysparams}. Unless stated otherwise, we use four
memory controllers (one channel per memory controller) for all experiments. In order to prevent the GPU from
taking the majority of request buffer entries, we reserve half of the
request buffer entries for the CPUs. To model the memory
bandwidth of the GPU accurately, we perform coalescing on
GPU memory requests before they are sent to the memory controller~\cite{lindholm}.

\begin{table*}[th!]
\centering
\begin{tabular}{|l|l|}
\hline
\textbf{Parameter} & \textbf{Setting}\\
\hline
\hline
CPU Clock Speed & 3.2GHz \\ 
\hline
CPU ROB & 128 entries \\ 
\hline
CPU L1 cache & 32KB Private, 4-way \\ 
\hline
CPU L2 cache & 8MB Shared, 16-way \\ 
\hline
CPU Cache Rep. Policy & LRU \\ 
\hline
\hline
GPU SIMD Width & 800 \\ 
\hline
GPU Texture units & 40 \\ 
\hline
GPU Z units & 64 \\ 
\hline
GPU Color units & 16 \\ 
\hline
\hline
Memory Controller Entries & 300 \\
\hline
Channels/Ranks/Banks & 4/1/8 \\ 
\hline
DRAM Row buffer size & 2KB \\
\hline
DRAM Bus & 128 bits/channel \\ 
\hline
tRCD/tCAS/tRP & 8/8/8 ns \\
\hline
tRAS/tRC/tRRD & 20/27/4 ns \\ 
\hline
tWTR/tRTP/tWR & 4/4/6 ns \\
\hline
\hline
\end{tabular}
\caption{Simulation parameters.}
\label{table:sysparams}
\end{table*}

\paragraphbe{Workloads.} 
We evaluate our system with a set of 105 multiprogrammed workloads, each
simulated for 500~million cycles. Each workload consists of sixteen SPEC
CPU2006 benchmarks and one GPU application selected from a mix of video games and graphics
performance benchmarks. For each CPU benchmark, we use PIN~\cite{reddi2004pin,pin} with
PinPoints~\cite{pinpoint} to select the representative phase.  For the GPU
application, we use an industrial GPU simulator to collect memory requests
with detailed timing information.  These requests are collected after having
first been filtered through the GPU's internal cache hierarchy,
therefore we do not further model any caches for the GPU in our final hybrid
CPU-GPU simulation framework.

We classify CPU benchmarks into three categories (Low,
Medium, and High) based on their memory intensities, measured
as last-level cache misses per thousand instructions (MPKI). Table~\ref{table:mpki} shows
the MPKI for each CPU benchmark. Benchmarks with less than 1 MPKI are low
memory-intensive, between 1 and 25 MPKI are medium memory-intensive, and
greater than 25 are high memory-intensive. Based on these three categories, we
randomly choose a number of benchmarks from each category to form
workloads consisting of seven intensity mixes: L (All low), ML (Low/Medium), M
(All medium), HL (High/Low), HML (High/Medium/Low), HM (High/Medium) and H(All
high). The GPU benchmark is randomly selected for each workload
without any classification.

\ignore{
\textbf{Trace collection:} For the CPU benchmarks, we use PIN~\cite{pin}
with Pin-Point~\cite{pinpoint} to select representative samples, each consist
of 200 million instructions. For the GPU application, we use an industrial GPU
simulator and collect a trace of all memory requests that are filtered by the
GPU cache and goes into the memory controller. Because of this, our simulator
does not model the GPU cache hierarchy. The GPU traces have a time stamp that
specifies the time in which the memory request is generated.  For each
workloads, we simulate 500 million cycles of execution, when we reach the end
of a CPU or GPU traces, we rewind and fetch from the beginning of the trace.}

\begin{table}[h!] 
\centering 
\begin{tabular}{|c|c||c|c||c|c|} 
\hline Name & MPKI & Name & MPKI & Name & MPKI \\
\hline
\hline
tonto & 0.01 & sjeng & 1.08 & omnetpp & 21.85 \\ 
\hline
povray & 0.01 & gobmk & 1.19 & milc & 21.93 \\ 
\hline
calculix & 0.06 & gromacs & 1.67 & xalancbmk & 22.32 \\
\hline
perlbench & 0.11 & h264ref & 1.86 & libquantum & 26.27 \\
\hline
namd & 0.11 & bzip2 & 6.08 & leslie3d & 38.13 \\ 
\hline
dealII & 0.14 & astar & 7.6 & soplex & 52.45 \\
\hline
wrf & 0.21 & hmmer & 8.65 & GemsFDTD & 63.61 \\
\hline
gcc & 0.33 & cactusADM & 14.99 & lbm & 69.63 \\
\hline
& & sphinx3 & 17.24 & mcf & 155.30 \\
\hline
\end{tabular} 
\caption{L2 Cache Misses Per Kilo-Instruction (MPKI) of 26 SPEC 2006
  benchmarks.}
\label{table:mpki}
\end{table}

\paragraphbe{Performance Metrics.}
In an integrated CPUs and GPU system like the one we evalute, 
To measure system performance, we use \textit{CPU+GPU Weighted Speedup}
(Eqn.~\ref{eq:ws}), which is a sum of the CPU weighted
speedup~\cite{ws-metric2,harmonic_speedup} and the GPU speedup
multiply by the weight of the GPU. In addition, we measure
\textit{Unfairness}~\cite{Reetu-MICRO2009,atlas,tcm,vandierendonck}
using maximum slowdown for all the CPU cores. We
report the harmonic mean instead of arithmetic mean for \textit{Unfairness} in
our evaluations since slowdown is an inverse metric of speedup.

\ignore{
To measure system performance, we use \textit{Weighted
Speedup}~\cite{weighted_speedup,harmonic_speedup} (Eqn.~\ref{eq:ws}), which
has been shown to be equivalent to system
throughput~\cite{multiprogram_metrics} for all the CPU cores. We report the
GPU performance based on its relative memory throughput (Eqn.~\ref{eq:memTP}). In addition, we measure
\textit{Unfairness}~\cite{stc,atlas,tcm,vandierendonck} (Eqn.~\ref{eq:unfairness}) using
maximum slowdown for all the CPU cores. We report the harmonic mean instead of
arithmetic mean for \textit{Unfairness} in our evaluations since slowdown in an
inverse metric of speedup.  
}

\begin{equation}
CPU+GPU Weighted Speedup = \sum_{i=1}^{NCPU}{\frac{IPC_{i}^{shared}}{IPC_{i}^{alone}}} + WEIGHT*\frac{GPU_{FrameRate}^{shared}}{GPU_{FrameRate}^{alone}}
\label{eq:ws}
\end{equation}

\begin{equation}
Unfairness = \max_i{\frac{IPC_i^{alone}}{IPC_i^{shared}}}
\label{eq:unfairness}
\end{equation}



\section{Qualitative Comparison with Previous Scheduling Algorithms}
\label{sec:sms-qual-eval}
In this section, we compare \titleShortSMS qualitatively to previously proposed
scheduling policies and analyze the basic differences between \titleShortSMS and
these policies. The fundamental difference between \titleShortSMS and previously
proposed memory scheduling policies for CPU only scenarios is that the latter
are designed around a single, centralized request buffer which has poor
scalability and complex scheduling logic, while \titleShortSMS is built around a
decentralized, scalable framework.

\subsection{First-Ready FCFS (FR-FCFS)}
FR-FCFS~\cite{fr-fcfs} is a commonly used scheduling
policy in commodity DRAM systems. A FR-FCFS scheduler prioritizes requests
that result in row-buffer hits over row-buffer misses and otherwise prioritizes
older requests. Since FR-FCFS unfairly prioritizes applications with high
row-buffer locality to maximize DRAM throughput, prior
work~\cite{atlas,parbs,tcm,stfm,memattack} have observed that it has low system
performance and high unfairness.

\subsection{Parallelism-aware Batch Scheduling (PAR-BS)}
PAR-BS~\cite{parbs} aims to improve
fairness and system performance. In order to prevent unfairness, it forms
batches of outstanding memory requests and prioritizes the oldest batch, to
avoid request starvation. To improve system throughput, it prioritizes
applications with smaller number of outstanding memory requests within a batch.
However, PAR-BS has two major shortcomings. First, batching could cause older
GPU requests and requests of other memory-intensive CPU applications to be
prioritized over latency-sensitive CPU applications. Second, as previous
work~\cite{atlas} has also observed, PAR-BS does not take into account an
application's long term memory-intensity characteristics when it assigns
application priorities within a batch. This could cause memory-intensive
applications' requests to be prioritized over latency-sensitive applications'
requests within a batch.

\subsection{Adaptive per-Thread Least-Attained-Serviced Memory Scheduling (ATLAS)}
ATLAS~\cite{atlas} aims to improve system performance by prioritizing
requests of applications with lower attained memory service. This improves the
performance of low memory-intensity applications as they tend to have low
attained service. However, ATLAS has the disadvantage of not preserving
fairness. Previous work~\cite{atlas,tcm} have shown that simply prioritizing low memory
intensity applications leads to significant slowdown of memory-intensive
applications.

\subsection{Thread Cluster Memory Scheduling (TCM)}
TCM~\cite{tcm} is the best
state-of-the-art application-aware memory scheduler providing both system
throughput and fairness. It groups applications into either latency- or
bandwidth-sensitive clusters based on their memory intensities. In order to
achieve high system throughput and low unfairness, TCM employs different
prioritization policy for each cluster. To improve system throughput, a
fraction of total memory bandwidth is dedicated to latency-sensitive cluster
and applications within the cluster are then ranked based on memory intensity
with least memory-intensive application receiving the highest priority. On the
other hand, TCM minimizes unfairness by periodically shuffling applications
within a bandwidth-sensitive cluster to avoid starvation. This approach
provides both high system performance and fairness in CPU-only systems. In an
integrated CPU-GPU system, GPU generates a significantly larger amount of
memory requests compared to CPUs and fills up the centralized request buffer. 
As a result, the memory controller lacks the visibility of CPU memory requests
to accurately determine each application's memory access behavior. Without the
visibility, TCM makes incorrect and non-robust clustering decisions, which 
classify some applications with high memory intensity into the
latency-sensitive cluster. These misclassified applications cause
interference not only to low memory intensity applications, but also to each
other. Therefore, TCM causes some degradation in both system
performance and fairness in an integrated CPU-GPU system. As described in
Section~\ref{sec:sms-motivation}, increasing the request buffer size is a simple and
straightforward way to gain more visibility into CPU applications' memory
access behaviors. However, this approach is not scalable as we show in our
evaluations (Section~\ref{sec:eval}). In contrast, \titleShortSMS provides much
better system performance and fairness than TCM with the same number of
request buffer entries and lower hardware cost.

\section{Experimental Evaluation of SMS}
\label{sec:eval-sms}

We present the performance of five memory scheduler configurations:
FR-FCFS, ATLAS, PAR-BS, TCM, and SMS on the 16-CPU/1-GPU four-memory-controller system
described in Section~\ref{sec:meth-sms}. All memory schedulers use 300 request
buffer entries per memory controller; this size was chosen based on the results in
Figure~\ref{fig:reqQ} which showed that performance does not appreciably
increase for larger request buffer sizes. Results are presented in the
workload categories as described in Section~\ref{sec:meth-sms}, with
workload memory intensities increasing from left to right.

\begin{figure*}
\centering
\includegraphics[width=\textwidth]{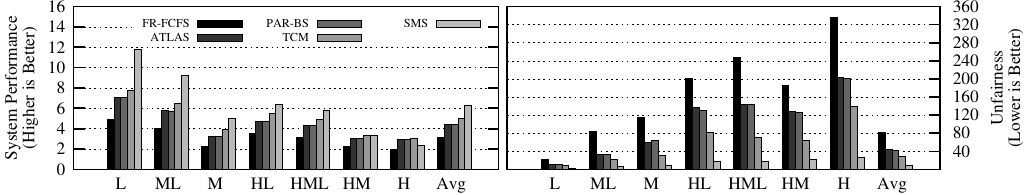}
\caption{System performance, and fairness for 7 categories of workloads (total
of 105 workloads)}
\label{fig:mainres}
\end{figure*}

Figure~\ref{fig:mainres} shows the system performance (measured as weighted
speedup) and fairness of the previously proposed algorithms and \titleShortSMS,
averaged across 15 workloads for each of the seven categories (105 workloads in total).
Compared to TCM, which
is the best previous algorithm for both system performance and fairness,
\titleShortSMS provides 41.2\% system performance improvement and 4.8$\times$ fairness
improvement. Therefore, we conclude that \titleShortSMS provides better system
performance and fairness than all previously proposed scheduling policies,
while incurring much lower hardware cost and simpler scheduling logic.

Based on the results for each workload category, we make the following
major observations: First, \titleShortSMS consistently outperforms previously
proposed algorithms (given the same number of request buffer entries), in terms
of both system performance and fairness across most of the workload categories.
Second, in the ``H'' category with only high memory-intensity workloads, \titleShortSMS
underperforms by 21.2\%/20.7\%/22.3\% compared to
ATLAS/PAR-BS/TCM, but \titleShortSMS still provides 16.3\% higher system
performance compared to FR-FCFS. The main reason for this behavior is that
ATLAS/PAR-BS/TCM improve performance by unfairly prioritizing certain applications over others,
which is reflected by their poor fairness results.  For instance, we observe
that TCM misclassifies some of these high memory-intensity applications into
the low memory-intensity cluster, which starves requests of applications in the
high memory-intensity cluster. On the other hand, \titleShortSMS preserves
fairness in all workload categories by using its probabilistic round-robin policy
as described in Section~\ref{sec:mechanism-sms}. As a result, \titleShortSMS provides
7.6$\times$/7.5$\times$/5.2$\times$ better fairness relative to
ATLAS/PAR-BS/TCM respectively, for the high memory-intensity category.

\subsection{Analysis of CPU and GPU Performance}

\begin{figure*}
\centering
\includegraphics[width=\textwidth]{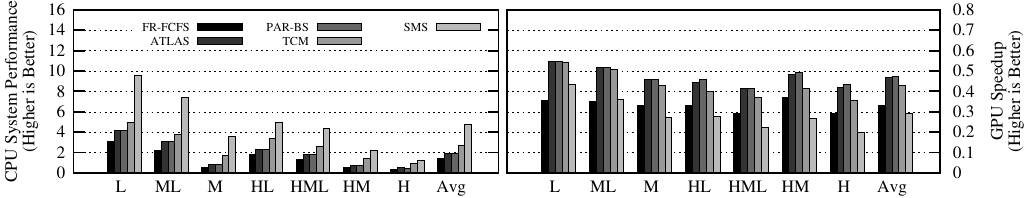}
\caption{CPUs and GPU Speedup for 7 categories of workloads (total
of 105 workloads)}
\label{fig:spdUp}
\end{figure*}

In this section, we study the performance of the CPU system and the GPU system
separately. Figure~\ref{fig:spdUp} shows CPU-only weighted speedup and GPU speedup.
Two major observations are in order. First, \titleShortSMS gains 1.76$\times$
improvement in CPU system performance over TCM. Second, \titleShortSMS achieves this
1.76$\times$ CPU performance improvement while delivering similar GPU performance
as the FR-FCFS baseline.\footnote{Note that our GPU Speedup metric is defined with
respect to the performance of the GPU benchmark running on the system by itself.
In all cases, the relative speedup reported is much less than 1.0 because the GPU
must now share memory bandwidth with 16 CPUs.}  The results show that TCM (and the other algorithms) end
up allocating far more bandwidth to the GPU, at significant performance and fairness
cost to the CPU applications.
\titleShortSMS appropriately deprioritizes the memory bandwidth
intensive GPU application in order to enable higher CPU performance and overall
system performance, while preserving fairness. Previously proposed scheduling
algorithms, on the other hand, allow the GPU to hog memory bandwidth and
significantly degrade system performance and fairness (Figure~\ref{fig:mainres}).

%

\subsection{Scalability with Cores and Memory Controllers}

\begin{figure}
\centering
\includegraphics[width=\textwidth]{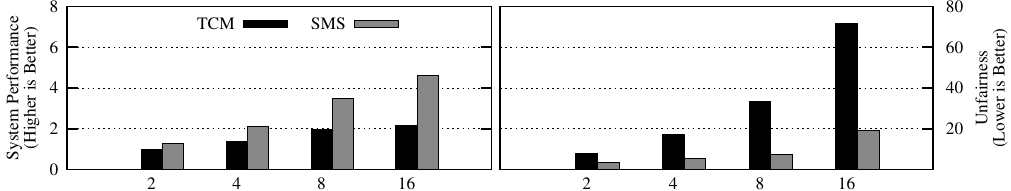}
\caption{\titleShortSMS vs TCM on a 16 CPU/1 GPU, 4 memory controller system with varying the number of cores}
\label{fig:coreSweep}
\end{figure}

\begin{figure}
\centering
\includegraphics[width=\textwidth]{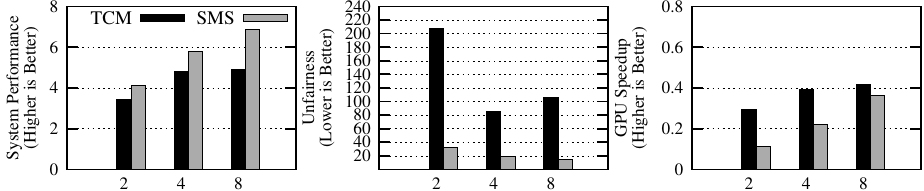}
\caption{\titleShortSMS vs TCM on a 16 CPU/1 GPU system with varying the number of channels}
\label{fig:chanSweep}
\end{figure}

Figure~\ref{fig:coreSweep} compares the performance and fairness of \titleShortSMS
against TCM (averaged over 75 workloads\footnote{We use 75 randomly selected workloads
per core count.  We could not use the same workloads/categorizations as earlier because
those were for 16-core systems, whereas we are now varying the number of cores.}) with the same number of
request buffers, as the number of cores is varied. We make the following
observations: First, \titleShortSMS continues to provide better system performance
and fairness than TCM. Second, the system performance gains and
fairness gains increase significantly as the number of cores and
hence, memory pressure is increased. \titleShortSMS's performance and
fairness benefits are likely to become more significant as core counts in future
technology nodes increase.

Figure~\ref{fig:chanSweep} shows the system performance and fairness of
\titleShortSMS compared against TCM  as the number of memory channels is varied.
For this, and all subsequent results, we perform our evaluations on 60 workloads
from categories that contain high memory-intensity applications. We observe that
\titleShortSMS scales better as the number of memory channels increases. As the
performance gain of TCM diminishes when the number of memory channels increases
from 4 to 8 channels, \titleShortSMS continues to provide performance improvement
for both CPU and GPU.

\ignore{

Figure~\ref{fig:coreSweep} compares the performance and unfairness of
\titleShortSMS against TCM with the same number (300) of request buffers. As the
number of cores and memory controllers is varied, we make several major
conclusions. First, \titleShortSMS consistently provides better system performance
and fairness than TCM. Second, the system performance and
fairness gains (194\% to 379\%) increase as the number of cores is increased.
This performance gain widens because \titleShortSMS is capable of throttling GPU
requests while allowing latency sensitive applications to progress faster.  As
a result, there is a cost of minor GPU bandwidth reduction (from 14.7\% to
39.5\%)
}

\subsection{Sensitivity to \titleShortSMS Design Parameters}

\ignore{
\begin{figure}
\centering
\includegraphics[width=\textwidth]{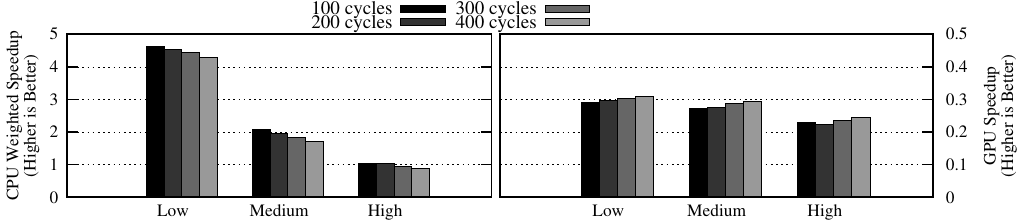}
\caption{\titleShortSMS sensitivity to \batch-age threshold}
\label{fig:batch-age}
\end{figure}
}
\paragraphbe{Effect of \Batch Formation.} 
Figure~\ref{fig:batchsize} shows the system performance and fairness of
\titleShortSMS as the maximum \batch size varies. When the \batch scheduler can forward
individual requests to the DCS, the
system performance and fairness drops significantly by 12.3\% and 1.9$\times$
compared to when it uses a maximum \batch size of ten. The reasons are twofold: First,
intra-application row-buffer locality is not preserved without forming
requests into \batches and this causes performance degradation due to longer
average service latencies. Second, GPU and high memory-intensity applications' requests
generate a lot of interference by destroying each other's and most
importantly latency-sensitive applications' row-buffer locality. With
a reasonable maximum \batch size (starting from ten onwards), intra-application
row-buffer locality is well-preserved with reduced interference to provide
good system performance and fairness. We have also observed that most CPU
applications rarely form \batches that exceed ten requests.  This is because
the in-order request stream rarely has such a long sequence of requests all
to the same row, and the timeout threshold also prevents the \batches from
becoming too large.  As a result, increasing the \batch size beyond ten
requests does not provide any extra benefit, as shown in
Figure~\ref{fig:batchsize}.

\begin{figure}
\centering
\begin{minipage}{3in}
\includegraphics[width=\textwidth]{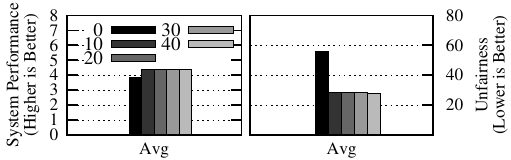}
\caption{\titleShortSMS sensitivity to \batch Size}
\label{fig:batchsize}
\end{minipage}
\begin{minipage}{3in}
\includegraphics[width=\textwidth]{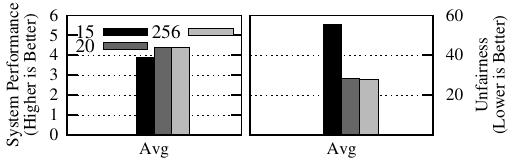}
\caption{\titleShortSMS sensitivity to DCS FIFO Size}
\label{fig:fifosize}
\end{minipage}
\end{figure}

\paragraphbe{DCS FIFO Size.}
Figure~\ref{fig:fifosize} shows the sensitivity of \titleShortSMS to the size of the
per-bank FIFOs in the DRAM Command Scheduler (DCS). Fairness degrades as the
size of the DCS FIFOs is increased. As the size of the per-bank DCS FIFOs
increases, the \batch scheduler tends to move more \batches from the \batch
formation stage to the DCS FIFOs. Once \batches are moved to the DCS FIFOs, they
cannot be reordered anymore.  So even if a higher-priority \batch were to
become ready, the \batch scheduler cannot move it ahead of any \batches
already in the DCS.  On the other hand, if these \batches were left in the
\batch formation stage, the \batch scheduler could still reorder them. Overall,
it is better to employ smaller per-bank DCS FIFOs that leave more \batches in
the \batch formation stage, enabling the \batch scheduler to see more \batches
and make better \batch scheduling decisions, thereby reducing starvation and improving
fairness.  The FIFOs only need to be large enough to keep the DRAM banks busy.


\subsection{Case Studies}

In this section, we study some additional workload setups and design choices.
In view of simulation bandwidth and time constraints, we reduce the simulation
time to 200M cycles for these studies.

\begin{figure}[h!!!]
\centering
\includegraphics[width=\textwidth]{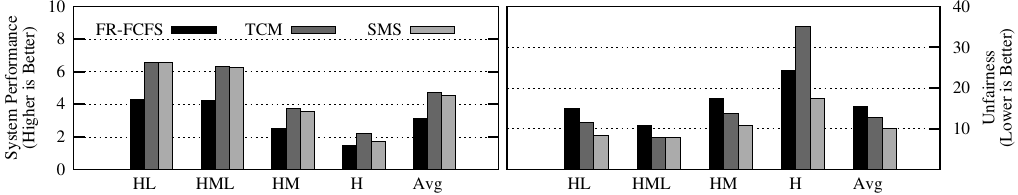}
\caption{System performance and fairness on a 16 CPU-only system.}
\label{fig:cpu}
\end{figure}

\paragraphbe{Case study 1: CPU-only Results.}
In the previous sections, we showed that \titleShortSMS effectively
mitigates inter-application interference in a CPU-GPU integrated system. In this
case study, we evaluate the performance of \titleShortSMS in a CPU-only scenario.
Figure~\ref{fig:cpu} shows the system performance and fairness of \titleShortSMS on
a 16-CPU system with exactly the same system parameters as described in
Section~\ref{sec:meth-sms}, except that the system does not have a GPU. We present
results only for workload categories with at least some high memory-intensity applications, as
the performance/fairness of the other workload categories are quite similar to TCM.
We observe that \titleShortSMS degrades performance by only 4\% compared to TCM,
while it improves fairness by 25.7\% compared to TCM on average across
workloads in the ``H'' category. \titleShortSMS's performance degradation mainly
comes from the ``H'' workload category (only high memory-intensity
applications); as discussed in our main evaluations in Section~\ref{sec:eval},
TCM mis-classifies some high memory-intensity applications into the low
memory-intensity cluster, starving requests of applications classified into the
high memory-intensity cluster. Therefore, TCM gains performance at the cost of
fairness. On the other hand, \titleShortSMS prevents this starvation/unfairness
with its probabilistic round-robin policy, while still maintaining good system
performance.

\begin{figure}[h!!!]
\centering
\includegraphics[width=0.8\textwidth]{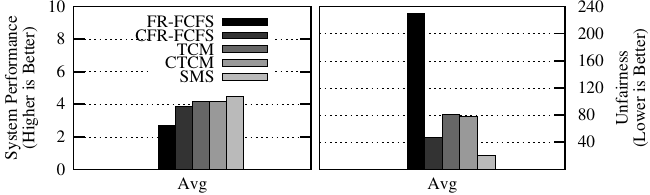}
\caption{Performance and Fairness when always prioritizing CPU requests over GPU requests}
\label{fig:gfrfcfs}
\end{figure}

\paragraphbe{Case study 2: Always Prioritizing CPU Requests over GPU Requests.}
Our results in the previous sections show that \titleShortSMS achieves its
system performance and fairness gains by appropriately managing the GPU request
stream. In this case study, we consider modifying previously proposed policies
by always deprioritizing the GPU. Specifically, we implement variants of the
FR-FCFS and TCM scheduling policies, CFR-FCFS and CTCM, where the CPU
applications' requests are always selected over the GPU's requests.
Figure~\ref{fig:gfrfcfs} shows the performance and fairness of FR-FCFS,
CFR-FCFS, TCM, CTCM and \titleShortSMS scheduling policies, averaged across
workload categories containing high-intensity applications. Several conclusions
are in order. First, by protecting the CPU applications' requests from the
GPU's interference, CFR-FCFS improves system performance by 42.8\% and fairness
by 4.82x as compared to FR-FCFS. This is because the baseline FR-FCFS is
completely application-unaware and it always prioritizes the row-buffer hitting
requests of the GPU, starving other applications' requests. Second, CTCM does
not improve system performance and fairness much compared to TCM, because
baseline TCM is already application-aware.  Finally, \titleShortSMS still
provides much better system performance and fairness than CFR-FCFS and CTCM
because it deprioritizes the GPU appropriately, but not completely, while
preserving the row-buffer locality within the GPU's request stream.  Therefore,
we conclude that \titleShortSMS provides better system performance and fairness
than merely prioritizing CPU requests over GPU requests.

\section{SMS: Conclusion}
\label{sec:conclusion}
\noindent While many advancements in memory scheduling policies have been made to deal
with multi-core processors, the integration of GPUs on the same chip as the
CPUs has created new system design challenges. This work has demonstrated
how the inclusion of GPU memory traffic can cause severe difficulties for
existing memory controller designs in terms of performance and especially fairness. 
In this dissertation, we propose a new approach, \titleLongSMS, that delivers superior
performance and fairness compared to state-of-the-art memory schedulers,
while providing a design that is significantly simpler to implement. The key
insight behind \titleShortSMS's scalability is that the primary functions of
sophisticated memory controller algorithms can be decoupled, leading to our multi-stage
architecture.  This research attacks a critical component of a fused CPU-GPU
system's memory hierarchy design, but there remain many other problems that
warrant further research.  For the memory controller specifically, additional explorations
will be needed to consider interactions with GPGPU workloads.  Co-design and
concerted optimization of the cache hierarchy organization, cache
partitioning, prefetching algorithms, memory channel partitioning, and the
memory controller  are likely needed to fully exploit future heterogeneous
computing systems, but significant research effort will be needed to find
effective, practical, and innovative solutions.


\chapter{Reducing Inter-address-space Interference with a TLB-aware Memory Hierarchy}
\label{sec:mask}

Graphics Processing Units (GPUs) provide high throughput by exploiting a high
degree of \changesI{\emph{thread-level parallelism}. A GPU executes hundreds of
threads concurrently, where \changesII{the threads are grouped into multiple
\emph{warps}}.}  
\changesI{The} GPU executes each warp in \emph{lockstep} (i.e., each thread in the warp executes the \emph{same}
instruction concurrently). When \changesI{one or more \changesII{threads of} a} warp stall, the GPU hides the latency of this stall 
by scheduling and executing another warp. 
\changesI{This high throughput provided by \changesIII{a GPU} creates an opportunity to
accelerate applications from \changesII{a wide range of} domains \changesVIII{(e.g.,} \changesIV{\cite{netflix, rodinia, parboil, mars, lonestar,tensorflow,lulesh,shoc,bowtie2,cudaprotien2014,molecular-gpgpu,mashimo2013molecular,nobile2016graphics}\changesVIII{)}.}}

\changesIII{GPU compute} density continues to increase \changesII{to support
\changesIV{demanding} applications.  For example,} emerging \changesIII{GPU} architectures are expected to provide 
as many as 128 \changesII{\emph{streaming multiprocessors} (i.e., \changesIV{GPU} cores)} per chip in the near 
future~\cite{arunkumar-isca17, vijay-hpca17}. \changesI{While the} increased 
compute density can help many \changesIII{individual general-purpose GPU (GPGPU)} applications, it exacerbates a growing
need to \emph{share} the GPU cores \changesIV{\emph{across multiple
applications} in order to fully utilize the large amount of GPU resources}. This is especially true in large-scale computing environments,
such as cloud servers, where \changesI{\changesII{diverse demands} for compute and memory
exist across different applications.} 
To enable efficient GPU 
utilization in the \changesIII{presence} of application heterogeneity, these large-scale
environments rely on the ability to \changesII{\emph{virtualize}} the \changesIII{GPU} compute 
resources and execute multiple applications \changesIV{\emph{concurrently}} \changesIII{on a single GPU}~\cite{ept, npt, 
amd-io-virt, intel-io-virt}.  

The adoption of GPUs in large-scale computing environments is hindered
by the primitive virtualization support in contemporary
GPUs~\cite{amd-fusion,apu,kaveri,haswell,amdzen,skylake,powervr,arm-mali,tegra,tegrax1,fermi,kepler,maxwell,pascal,amdr9,radeon,vivante-gpgpu}.
While hardware virtualization support \changesII{has improved for
integrated GPUs~\cite{amd-fusion,apu,kaveri,haswell,amdzen,skylake,powervr,
arm-mali,tegra,tegrax1}, where the GPU cores and CPU cores are \changesIV{on} the same chip and
share the same off-chip memory, virtualization support for}
discrete \changesIII{GPUs~\cite{fermi,kepler,maxwell,pascal,amdr9,radeon,vivante-gpgpu,arm-mali,powervr}}, \changesI{where the GPU is on a different chip than the CPU and has its own
memory,} is insufficient.
\changesII{Despite poor existing support for virtualization, discrete GPUs are \changesIII{likely to be}
more attractive than integrated GPUs for large-scale computing environments,
as they provide the highest-available compute density} and remain
the platform of choice in many
domains~\cite{tensorflow,netflix,mars,lonestar,rodinia}.

\sloppypar{Two alternatives for
virtualizating discrete GPUs are \changesIII{\emph{time} multiplexing and \emph{spatial}} multiplexing.
\changesIII{Modern} GPU architectures support time multiplexing using application
preemption~\cite{lindholm,pascal,isca-2014-preemptive,gebhart,kepler,wang-hpca16},
but this support currently does \changesIV{\emph{not}} scale  well because each additional
application increases contention for \changesII{the limited} GPU resources
(Section~\ref{s:time-multiplex}). Spatial multiplexing allows us to share a GPU
among \changesIII{concurrently-executing} applications much as we currently share \changesIV{multi-core} CPUs, by providing support for
\emph{multi-address-space concurrency} (i.e., the concurrent execution of \changesI{application}
kernels from different processes or guest VMs).  By efficiently and dynamically
managing \changesI{application} kernels that execute concurrently on the GPU, spatial multiplexing
\emph{avoids} the scaling issues of time multiplexing.  To support spatial
multiplexing, GPUs must provide architectural support for \emph{both} memory
virtualization and memory protection.}

\sloppypar{\changesIII{We find that existing techniques for spatial multiplexing 
in modern GPUs (e.g., \cite{asplos-sree, kepler,pascal,cudastream})
have two major shortcomings. \changesIV{They} either 
(1)~require significant programmer intervention to adapt
existing programs for spatial multiplexing; or 
(2)~sacrifice memory protection, which is a key requirement for virtualized 
systems.
To overcome these shortcomings, GPUs must utilize \changesIV{\emph{memory virtualization}~\cite{vm-contemporary}},
which \changesIV{enables} multiple applications to run concurrently while
providing memory protection.
While memory virtualization support in modern GPUs is also primitive, in large
part due to the poor performance of address translation,
several recent efforts have worked to improve address translation} within
GPUs~\cite{powers-hpca14, pichai-asplos14, tianhao-hpca16, abhishek-ispass16,cong-hpca17}.
\changesIII{These efforts introduce \emph{translation lookaside buffer} (TLB)
designs that improve performance significantly when a single application
executes on a GPU.}
\changesIII{Unfortunately, as we show in Section~\ref{sec:baseline},
even these improved address translation mechanisms suffer from high
performance overheads during spatial multiplexing,
as the limited capacities of the TLBs become a source of significant contention
within the GPU.}}

\changesII{In this \changesIII{chapter}, we} perform a thorough \changesIII{experimental} 
analysis of concurrent multi-application execution when
state-of-the-art address translation techniques are employed
\changesII{in a discrete GPU} (Section~\ref{sec:motiv}).
We make \changesIII{three} \emph{key observations} \changesI{from our analysis.}
\changesIV{First, a single TLB miss frequently stalls \emph{multiple} warps at once,
and incurs a very high latency, as each miss must walk through multiple levels
of a page table to find the desired address translation.
Second, due to high contention for shared address translation structures among
the multiple applications, the TLB miss rate increases significantly.
As a result, the GPU often does \changesV{\emph{not}} have enough warps that are ready to execute,
leaving GPU cores idle and defeating the GPU's latency hiding properties.}
\changesIII{Third}, \changesIII{contention} between applications induces significant thrashing on the 
shared L2 TLB and significant interference between TLB misses and data requests 
throughout the \changesIV{entire} GPU memory system.
With only a few simultaneous \changesIII{TLB miss requests}, it becomes 
difficult for the GPU to find a warp that can be scheduled for execution, 
\changesVII{which defeats the} GPU's basic \changesIII{fine-grained multithreading techniques~\cite{cdc6600,cdc6600-2,smith-hep,hep}}~\changesVII{that are essential} for hiding the latency of stalls.

Based on \changesI{our} extensive \changesIII{experimental} analysis, we conclude that
\emph{address translation \changesIII{is} a first-order performance concern} in GPUs
when multiple applications are executed concurrently.
\emph{\bf{Our goal}} in this work is to develop new techniques that can alleviate 
the severe address translation bottleneck \changesIII{in} state-of-the-art GPUs.

To this end, we propose \titleLongMASKEmph{} (\titleShortMASK{}), a new 
\changesII{GPU framework}
that minimizes inter-application interference 
and \changesII{address} translation overheads
\changesII{during concurrent application execution}.
The \changesII{\changesIV{overarching} idea of \titleShortMASK{}} is to make the entire memory
hierarchy \changesI{\emph{aware of TLB requests}}.
\titleShortMASK{} takes advantage of 
locality across \changesII{GPU} cores to reduce TLB misses, and relies on
three novel \changesII{mechanisms} to minimize \changesIII{address} translation overheads. 
First, \tlbtokenname provide \changesI{\emph{a contention-aware} mechanism 
to reduce thrashing in the shared L2 TLB},
including a bypass cache to increase the TLB hit rate.
Second, \changesIII{our \cachebypass mechanism provides \emph{contention-aware}
cache bypassing to reduce interference at the L2 cache between
address translation requests and data demand requests.}
Third, \changesIII{our \dramsched provides a \emph{contention-aware} 
memory controller policy that prioritizes address translation requests over
data demand requests to mitigate high address translation overheads.} 
\changesII{Working together, these three mechanisms are} highly effective at alleviating 
the address translation bottleneck, \changesIII{as our secs/mask-micro17/results show \changesIV{(Section~\ref{sec:design})}}.

\changesIV{Our comprehensive experimental evaluation shows that, via} the use of TLB-request-aware
policies throughout the memory hierarchy, \titleShortMASK{} 
\changesII{significantly reduces \changesIII{(1)~the} number of TLB misses that occur during
multi-application execution; and
\changesIII{(2)~the} overall latency of the remaining TLB misses, by
ensuring that page table walks are serviced quickly.}
\changesII{As a result, \titleShortMASK{} greatly increases the average number of 
threads that can be scheduled during long-latency stalls, which in turn} 
improves system throughput \changesIII{(weighted speedup~\cite{harmonic_speedup,ws-metric2})}
by 57.8\%, improves IPC throughput by 43.4\%, and reduces unfairness by
22.4\% over a state-of-the-art GPU memory management unit (MMU) design~\cite{powers-hpca14}.
\titleShortMASK{} provides performance within only 23.2\% of \changesIII{an ideal} TLB that
always hits.

This chapter makes the following \changesIV{major} contributions: 
\begin{itemize}[topsep=0.3em, leftmargin=1em, labelwidth=*, align=left, itemsep=0.3em] 

\item To our knowledge, this is the first work to \changesII{(1)~}provide a thorough
analysis of GPU memory virtualization under multi-address-space concurrency,
\changesII{(2)~}show the large impact of address translation on latency hiding
within a GPU, and
\changesII{(3)~demonstrate} the need for new techniques to alleviate
\changesIV{contention caused by address translation due to multi-application execution} in a GPU.


\item We propose \titleShortMASK{}~\cite{mask,mask-arxiv,mask-tech-report}, \changesIII{a new GPU framework that mitigates address translation overheads
in the \changesVIII{presence} of multi-address-space concurrency. \titleShortMASK consists} of three novel techniques that 
\changesI{work} together to increase TLB request awareness across the entire \changesIV{GPU} memory hierarchy.
\changesVIII{\titleShortMASK (1)}~significantly improves system performance,
IPC throughput, and fairness over a state-of-the-art GPU address
translation mechanism; \changesIV{and (2)~provides} practical support for spatially partitioning a
GPU across multiple address spaces.


\end{itemize}

\section{Background}
\label{sec:background}


There \changesIII{is an increasingly pressing} need to \emph{share} the GPU
hardware among multiple applications \changesIII{to improve GPU resource utilization}.  As a result, 
recent work\changesIV{~\cite{mosaic,asplos-sree,kepler,pascal,cudastream,gpu-multitasking,lindholm} 
enables} support for GPU virtualization, where a single physical GPU can be
shared transparently across multiple applications, with each application having
its own address space.\footnote{In this thesis, we use the term \emph{address space} 
to refer to distinct \emph{memory protection domains}, whose access to 
resources must be isolated and protected \changesIII{to enable} GPU virtualization.}  
Much of this work \changesIV{relies} on traditional time and spatial multiplexing techniques that
\changesIV{are} employed by CPUs, and state-of-the-art \changesVIII{GPUs contain} elements
of both types of techniques~\cite{gpuvm,gVirt,vmCUDA}.
Unfortunately, as we discuss in this section, existing GPU virtualization
implementations are \changesI{too \changesIV{coarse-grained: they} employ fixed 
hardware policies that leave system software \emph{without} 
\changesIV{mechanisms that can \emph{dynamically} 
reallocate GPU resources to different applications, which are required for
true application-transparent GPU virtualization.}

\subsection{Time Multiplexing}
\label{s:time-multiplex}

\sloppypar{Most modern systems time-share \changesII{(i.e., time multiplex) the GPU by
running kernels from multiple applications back-to-back}~\cite{kepler,lindholm}. 
These designs are optimized for the case
where \emph{no concurrency exists} between kernels from
different address spaces. This simplifies memory protection and scheduling at 
the cost of two fundamental \changesII{trade-offs}. \changesI{First, kernels from a single address space \changesIV{usually \emph{cannot}}
fully utilize all of the GPU's resources, leading to \changesIV{significant} resource \changesIII{underutilization~\cite{nmnl-pact13,cpugpu-micro,asplos-sree,wang-hpca16,mafia,caba}.}}
Second, \changesII{time multiplexing} limits the ability of a \changesI{GPU kernel} scheduler to provide
forward-progress or QoS guarantees, which can lead to 
unfairness and starvation~\cite{ptask}.}

While \changesIV{kernel preemption~\cite{isca-2014-preemptive,gebhart,wang-hpca16,kepler,pascal}} could allow a time-sharing scheduler to avoid 
\changesIII{a case where one GPU kernel unfairly uses \changesVIII{up} most of the execution time}
(e.g., by context switching at a fine granularity), \changesIV{such} preemption support 
remains an active research area \changesIV{in GPUs~\cite{isca-2014-preemptive, gebhart}.} Software
approaches~\cite{wang-hpca16} sacrifice memory protection. 
NVIDIA's Kepler~\cite{kepler} and Pascal~\cite{pascal} architectures 
support preemption at \changesII{the} thread block and instruction granularity, respectively.
We \changesII{empirically} find that neither \changesII{granularity is \changesV{\changesIV{effective at} minimizing
inter-application} interference.}

\changesI{To illustrate the performance overhead of time multiplexing,}
\changesIV{Figure~\ref{fig:context-switch-nvidia} shows how the execution time
increases when we use \changesVII{time multiplexing} to switch between multiple 
concurrently-executing processes, as opposed to executing the processes
back-to-back without any concurrent execution.
We perform these experiments on real NVIDIA K40~\cite{k40,kepler} and 
NVIDIA GTX 1080~\cite{gtx1080} GPUs.}
Each process runs a GPU kernel that interleaves basic
arithmetic operations with loads and stores into shared and global memory. 
\changesIV{We observe that as more processes execute concurrently, the overhead 
of time multiplexing grows significantly.  For example, on the NVIDIA GTX 1080,
\changesVII{time multiplexing} between two processes increases the total execution time by
12\%, as opposed to executing one process immediately after the other process
finishes.  When we increase the number of processes to 10, the overhead of
\changesVII{time multiplexing} increases to 91\%.}
On top of this high performance overhead, we \changesII{find that} inter-application
interference pathologies \changesI{(e.g., the starvation of one or more 
concurrently-executing \changesII{application} kernels)} \changesII{are easy to induce}: 
\changesII{an application} kernel from one
process consuming the majority of shared memory can easily cause \changesII{application} kernels from
other processes \changesVIII{to} \changesIII{never get scheduled \changesVIII{for execution} \changesIV{on} the GPU}. 
While we expect preemption support to improve in future hardware, we seek a 
\changesIII{multi-application concurrency} solution that does \changesIII{\emph{not}} depend on it. 

\begin{figure}[h!]
\centering
\vspace{-0.5em}
  \includegraphics[width=\columnwidth]{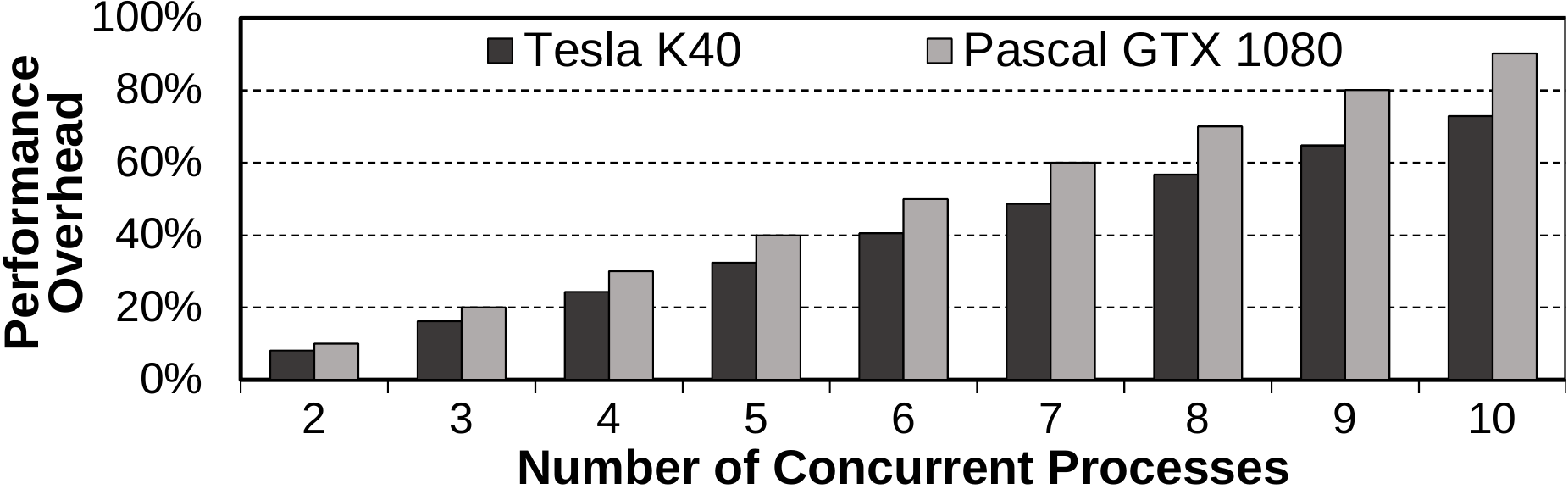}
\vspace{-1.5em}
  \caption{\changesIV{Increase in execution time when \changesVII{time multiplexing} is used
to execute processes concurrently} on real GPUs.}
\vspace{-1em}
  \label{fig:context-switch-nvidia}
\end{figure}


\subsection{Spatial Multiplexing}
\label{sec:spatial-multiplex}

Resource utilization can be improved with {\em spatial
multiplexing}~\cite{gpu-multitasking}, as the ability to execute multiple \changesII{application} kernels
\emph{concurrently} \changesIII{(1)~enables} the system to co-schedule kernels that have complementary
resource demands, and \changesIII{(2)~can} enable independent progress guarantees for different kernels.
\changesIII{Examples of spatial multiplexing support in modern GPUs include
(1)~}application-specific \changesII{\emph{software}}
scheduling of multiple kernels~\cite{asplos-sree}; \changesII{and
(2)~}\changesI{NVIDIA's \changesIII{CUDAstream} support~\cite{kepler,pascal,cudastream},} which co-schedules kernels from independent ``streams'' 
\changesII{by \changesIII{merging} them into} a single address space. 
\changesIII{\changesIV{Unfortunately,} these spatial multiplexing mechanisms have significant shortcomings.}
Software approaches (e.g., Elastic Kernels~\cite{asplos-sree}) require programmers to \emph{manually} time-slice kernels
to enable \changesIV{their mapping} onto CUDA streams for concurrency. While \changesIII{CUDAstream
supports} \changesII{the flexible partitioning of resources at runtime}, 
\changesIII{merging kernels into a single address 
space sacrifices memory protection, \changesIII{which is} a key requirement in virtualized settings.}

\changesII{True GPU support for multiple concurrent address spaces} can address these
shortcomings by \emph{enabling} \changesIII{hardware virtualization}. 
\changesIII{Hardware virtualization allows the system to \changesIV{(1)~}adapt to
changes in \changesIV{application resource utilization or (2)~mitigate interference at runtime, by dynamically
allocating} hardware resources to multiple concurrently-executing applications.}
NVIDIA and AMD both offer products~\cite{grid,firepro} with \changesIII{partial}
hardware virtualization support. \changesIII{However, these products simplify
memory protection by \emph{statically} partitioning the hardware resources prior to
program execution.} 
\changesIII{As a result,} these \changesV{systems} \changesIV{\emph{cannot}} adapt to changes in demand 
\changesIII{at runtime, \changesVIII{and, thus,} can still leave GPU resources
underutilized}. \changesI{To efficiently support \changesVIII{the} \changesIV{\emph{dynamic}}
sharing of GPU resources, GPUs must provide memory
virtualization \changesII{\emph{and}} memory protection, both of which \changesII{require}
\changesIV{efficient mechanisms} for virtual-to-physical address translation.}


\section{Baseline Design}
\label{sec:background-spatial}
\label{sec:baseline}
\label{sec:memory-protection}
\label{sec:pwc}

\changesIV{We} \changesIII{describe \changesV{(1)~the}} state-of-the-art 
address translation \changesV{mechanisms} for  GPUs, and \changesIII{(2)~the} overhead 
of these translation mechanisms when multiple applications share the GPU~\cite{powers-hpca14}.  
We analyze the shortcomings of \changesIV{state-of-the-art address translation mechanisms
for GPUs} \changesIII{in the presence of} multi-application concurrency in Section~\ref{sec:motiv},
which motivates the need for \titleShortMASK{}.}



\changesII{State-of-the-art GPUs extend the GPU memory 
hierarchy with translation lookaside buffers (TLBs)~\cite{powers-hpca14}.
TLBs \changesIV{(1)~greatly} reduce the overhead of address translation by caching recently-used
virtual-to-physical address mappings \changesIV{from a page table, and (2)~help} ensure that memory accesses from
application kernels running in different address spaces \changesIII{are} isolated \changesIII{from each other}.
Recent works~\cite{powers-hpca14, pichai-asplos14} propose optimized TLB designs
that improve \changesIV{address} translation performance for \changesIII{GPUs.}} 

\begin{figure}[h!]%
\centering
	\includegraphics[width=\columnwidth]{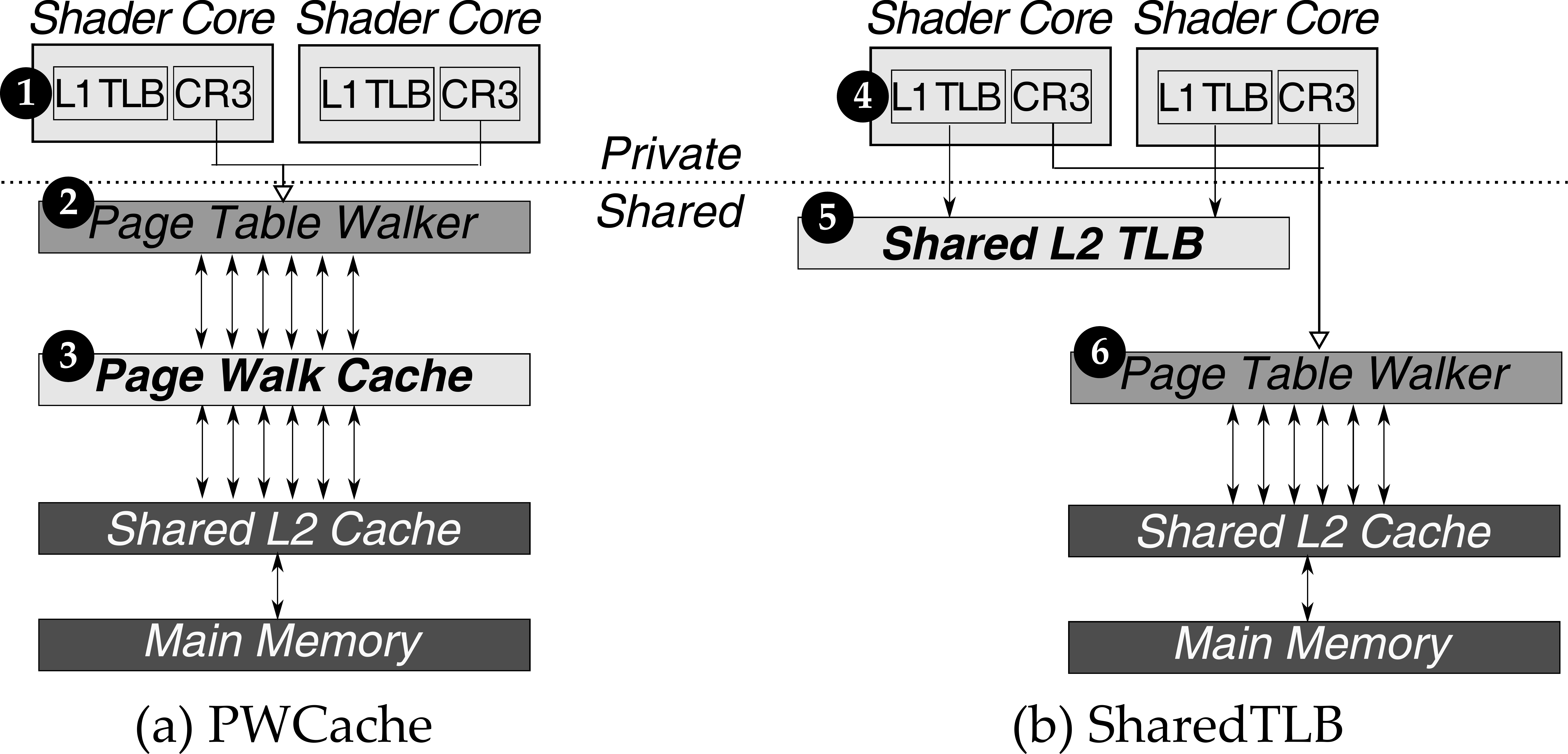}%
\caption{\changesII{Two variants of baseline GPU design.}}%
\label{fig:tlb-baseline}
\end{figure}

We adopt \changesIII{a baseline based on these state-of-the-art TLB designs,} whose
memory hierarchy makes use of one of two variants for address translation:
(1)~\changesIV{\emph{PWCache},} a \changesII{previously-proposed} design 
that \changesII{utilizes a \emph{shared page walk cache}} \changesIII{after the L1 TLB}~\cite{powers-hpca14} \changesIII{(Figure~\ref{fig:tlb-baseline}a)}; and 
(2)~\changesIV{\emph{SharedTLB},} \changesII{a} design that \changesII{utilizes a \emph{shared \changesIII{L2} TLB} \changesIII{after the L1 TLB (Figure~\ref{fig:tlb-baseline}b)}}. 
\changesII{The TLB caches translations that are stored in a multi-level
page table (we assume a four-level page table in this work).}
We extend \changesII{both TLB designs} to handle multi-address-space concurrency.
\changesII{Both variants}
\changesIV{incorporate private per-core L1 TLBs, and all cores share a 
highly-threaded page table walker.  
For PWCache, on a miss in the L1 TLB (\mycirc{1} in Figure~\ref{fig:tlb-baseline}a),
the GPU initializes a page table walk (\mycirc{2}), which probes \changesVIII{the} shared
page walk cache (\mycirc{3}).  Any page walk requests that miss in the page
walk cache go to the shared L2 cache and (if needed) main memory.
For \emph{SharedTLB}, on a miss in the L1 TLB (\mycirc{4} in Figure~\ref{fig:tlb-baseline}b),
the GPU checks whether the translation is available in \changesVIII{the} shared L2 TLB
(\mycirc{5}).  If the translation misses in the shared L2 TLB, the GPU
initiates a page table walk (\mycirc{6}), whose requests go to the shared L2
cache and (if needed) main memory.\footnote{In our
evaluation, we \changesIII{use an 8KB} page walk cache. The
shared L2 TLB is located next to the shared \changesIII{L2} cache.  \changesII{L1 and L2 TLBs
use \changesIV{the} LRU replacement policy.}}}

\begin{figure}[h!]
\centering
\includegraphics[width=\columnwidth]{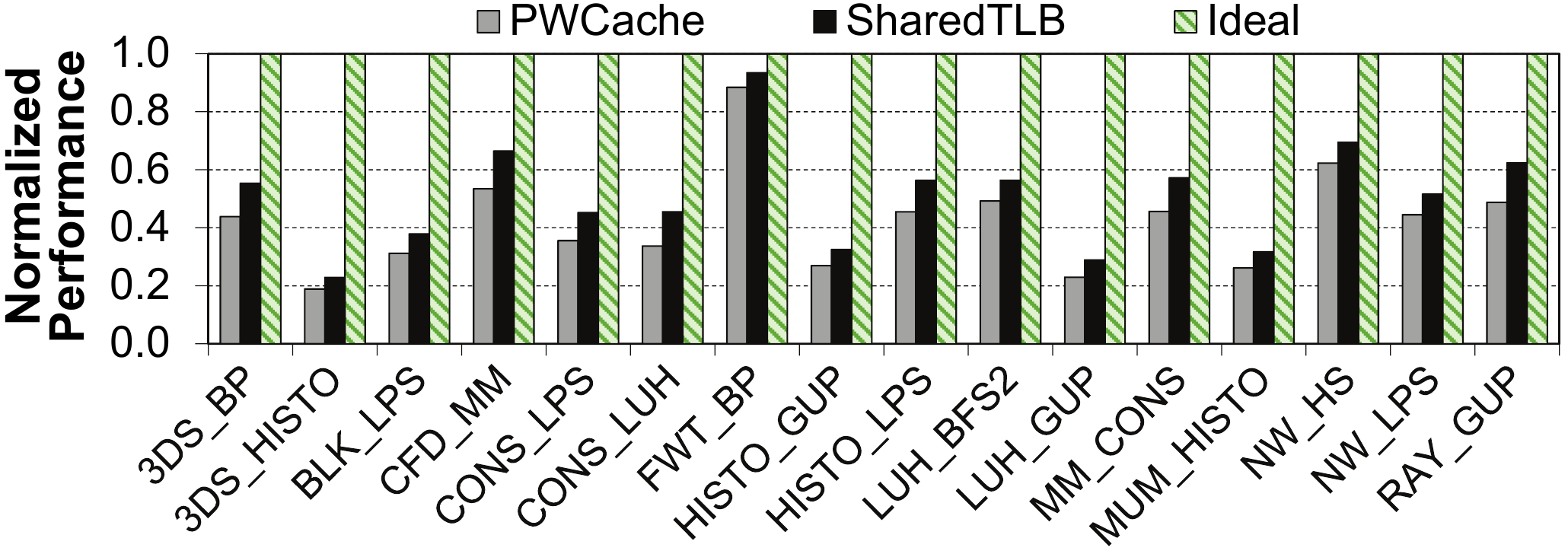}%
\caption{Baseline designs vs. ideal performance.}
\label{fig:pwcache}
\end{figure}

\changesIII{Figure~\ref{fig:pwcache} compares the performance} 
of \changesII{both \changesIII{baseline variants \changesIV{(\changesVIII{\emph{PWCache}},
depicted in Figure~\ref{fig:tlb-baseline}a, and \changesVIII{\emph{SharedTLB}}, depicted in Figure~\ref{fig:tlb-baseline}b)}, running two separate applications concurrently,} to an} ideal
scenario where every TLB access is a hit (see \changesVII{Table~\ref{table:config} for our
simulation configuration, and} Section~\ref{sec:meth-mask} for our
methodology).  We find that \emph{both} \changesII{variants} incur a significant performance
overhead \changesIII{(45.0\% and 40.6\% on average)} compared to the ideal case.\footnote{\changesI{\changesIV{We see} discrepancies between the performance of
our two baseline \changesII{variants} compared to the \changesII{secs/mask-micro17/results reported} by Power et
al.~\cite{powers-hpca14}. \changesV{These discrepancies} occur because \changesVIII{Power et al.\ assume} a \changesIV{much} higher L2 data cache access
latency (130 ns vs. \changesIII{our} 10 ns \changesVIII{latency}) and a \changesIV{much} higher shared L2 TLB access latency (130 ns vs. 
\changesIII{our} 10 ns \changesVIII{latency}).  Our cache latency model, with a 10 ns access latency plus queuing latency
(\changesVII{see Table~\ref{table:config} \changesVIII{in}} Section~\ref{sec:meth-mask}), accurately captures modern GPU parameters~\cite{maxwell}.}}
\changesIII{In order to retain the benefits of sharing a GPU across
multiple applications, we first} analyze the shortcomings of our baseline
design, and then use this analysis to develop \changesIV{our new mechanisms} that improve TLB
performance \changesIII{to make it \changesVIII{approach the ideal performance}}.

\section{Design Space Analysis}
\label{sec:motiv}

\changesI{To improve the performance of address
translation in GPUs,
we \changesII{first} analyze and characterize \changesII{the translation 
overhead in \changesIV{a state-of-the-art baseline (see Section~\ref{sec:baseline}),
taking into account especially the} performance challenges induced by multi-address-space concurrency and contention.
We first analyze how TLB misses can \changesII{limit the GPU's ability to hide
long-latency stalls, which directly impacts} 
performance (Section~\ref{sec:tlb-bottleneck}). \changesIV{Next,} we discuss two types
of memory interference that impact GPU performance: (1)~interference introduced 
by sharing GPU resources among multiple \changesIV{concurrent} applications (Section~\ref{sec:motiv-inter-thrashing}),
and (2)~interference introduced by sharing \changesII{the GPU memory hierarchy \changesIV{between}}  \changesIII{address translation requests} and 
\changesIII{data demand requests} (Section~\ref{sec:dram-interference}).
}

\subsection{Effect of TLB Misses on GPU Performance}
\label{sec:tlb-bottleneck}

GPU throughput relies on \emph{fine-grained
multithreading}\changesIV{~\cite{cdc6600,cdc6600-2,smith-hep,hep}} to hide memory latency.\footnote{\changesI{More detailed information about the GPU execution
model and its memory hierarchy can be found in \cite{mosaic,osp-isca13,owl-asplos13,caba,zorua,
medic,toggle-hpca16, cpugpu-micro, adwait-critical-memsched,rachata-thesis,
trogers-thesis,adwait-thesis, largewarps}.}}
\changesIII{We} observe a fundamental tension between address translation and fine-grained
multithreading.
The need to cache address translations at a page granularity,
combined with application-level spatial locality, \changesII{increase} the likelihood that \changesIII{address}
translations fetched in response to a TLB miss \changesII{are} needed by \changesIV{\emph{more than one}
\changesIII{warp} (i.e., many threads)}. Even with the massive levels of parallelism supported by GPUs, we
observe that a small number of outstanding TLB misses can result in the \changesIV{warp}
scheduler not having enough ready \changesIV{warps} to schedule, which in turn limits the
GPU's \changesII{essential} latency-hiding mechanism.

\changesIII{Figure~\ref{fig:stall-tlb} illustrates a scenario for an application
with four warps, where all four warps execute on the same GPU core.}
Figure~\ref{fig:stall-tlb}a shows how the GPU behaves when 
no virtual-to-physical address translation is required. When Warp~A 
\changesIII{performs a high-latency memory access (\mycirc{1} in Figure~\ref{fig:stall-tlb})}, 
the GPU core does \emph{not} stall \changesIV{since} other warps have schedulable 
\changesIII{instructions (Warps~B--D)}. In this case, the GPU core selects \changesIV{an active warp} (Warp~B) \changesIV{in} the next cycle \changesIV{(\mycirc{2})}, and continues 
issuing instructions. 
\changesIII{Even though Warps~B--D also perform memory accesses some time later, 
the accesses are independent of each other, and the GPU \changesIV{avoids stalling by switching to a warp that is \changesV{\emph{not}} waiting for a memory access \changesV{(\mycirc{3},~\mycirc{4})}}.}
Figure~\ref{fig:stall-tlb}b \changesIV{depicts} the same \changesIV{4 warps} \emph{when address translation
is required}. Warp~A misses in the TLB (indicated in red), and stalls (\changesV{\mycirc{5}})
until the \changesI{virtual-to-physical} translation \emph{finishes}. 
\changesIII{In Figure~\ref{fig:stall-tlb}b, due to spatial locality within the
application, the other warps (Warps~B--D) need the same address translation
as Warp~A.  As a result, they too stall \changesV{(\mycirc{6}, \mycirc{7}, \mycirc{8})}. \changesIV{At this point, the GPU no longer has any warps
that it can schedule, and the GPU core stalls until the address translation 
request completes.}
Once the address translation request completes \changesV{(\mycirc{9})}, the data demand
requests of the warps are issued to memory. \changesIV{Depending on the available
memory bandwidth and the parallelism of these data demand requests,} the data demand 
requests from Warps~B--D \changesIV{can incur additional} queuing
latency \changesV{(\mycirc{10},~\mycirc{11},~\mycirc{12})}. The GPU core can resume execution \changesV{\emph{only after}} the data demand request for
Warp~A is \changesIV{complete}~\changesV{(\mycirc{13})}.}

\changesIII{Three} phenomena harm performance in
this scenario. First, warps stalled on TLB misses reduce the availability of
schedulable warps, \changesI{which lowers \changesII{GPU}} utilization. \changesIV{In Figure~\ref{fig:stall-tlb}, no available
warp exists while \changesV{the} address translation request is pending, so the GPU utilization goes down to 0\% \changesV{for a long time}.} 
Second, \changesIII{address translation requests, which are a series of \emph{dependent} memory 
requests generated by a page walk,} must
complete before \changesIII{\changesIV{a} pending data demand \changesIV{request that requires the physical address} can be issued}, which reduces the \changesV{GPU's ability
to hide} latency by keeping \changesII{many} memory requests in flight. 
\changesIII{Third, when the address translation data becomes available, \emph{all} stalled 
warps \changesIV{that were waiting for the translation consecutively execute and} send their data demand requests \changesIV{to memory,} resulting in additional queuing delay for data demand
requests throughout the memory hierarchy.}

\begin{figure*}[h!]

\centering
	\includegraphics[width=\textwidth]{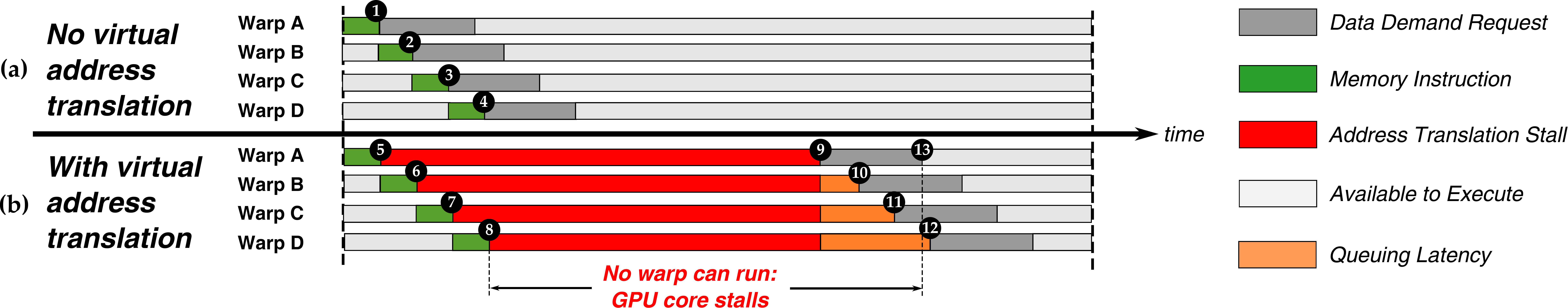}%
\caption{Example bottlenecks created by TLB misses.} 
\label{fig:stall-tlb}
\end{figure*}



\changesI{To illustrate how TLB misses \changesIV{significantly} reduce the number of
\changesIV{ready-to-schedule} warps in GPU applications,}
\changesIII{Figure~\ref{fig:concurrent-pw} shows the \changesIV{average number of} concurrent page 
table walks} (sampled every 10K cycles) for a range of applications,
\changesIII{and Figure~\ref{fig:schedulable-warp} shows
the average number of stalled warps per active \changesVI{TLB \changesVII{miss, in the}~\emph{SharedTLB}~\changesVII{baseline design}}}. \changesV{Error bars indicate the minimum and maximum values.}
\changesIII{We observe from Figure~\ref{fig:concurrent-pw} that}~\changesVII{more than 20~outstanding} TLB misses \changesIII{can} \changesIII{perform page
walks at the same time, all of which contend for access to address translation
structures.  From Figure~\ref{fig:schedulable-warp}, we observe that each} TLB miss 
\changesIII{can stall} \changesVII{more than} 30~warps \changesIV{out of the 64 warps in the core}.
\changesIII{The combined effect of these observations is that TLB misses in a
GPU can quickly stall a large number of warps \changesIV{within a \changesVII{GPU} core.
The GPU core}} 
must wait \changesII{for the misses to be 
resolved before issuing \changesIII{data demand}} requests \changesIII{and
resuming execution}. \changesII{Hence,} minimizing TLB 
misses and \changesII{the} page table walk latency is critical.

\begin{figure}[h!]
\centering
\includegraphics[width=\columnwidth]{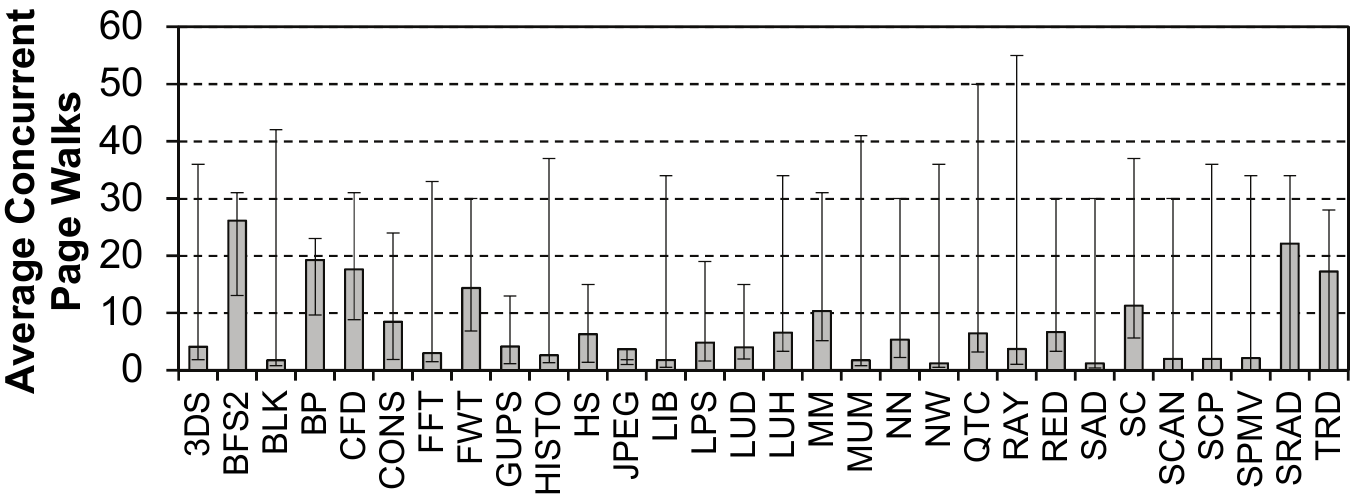}
\caption{Average number of concurrent page walks.}
\label{fig:concurrent-pw}
\end{figure}

\begin{figure}[h!]
\centering
\includegraphics[width=\columnwidth]{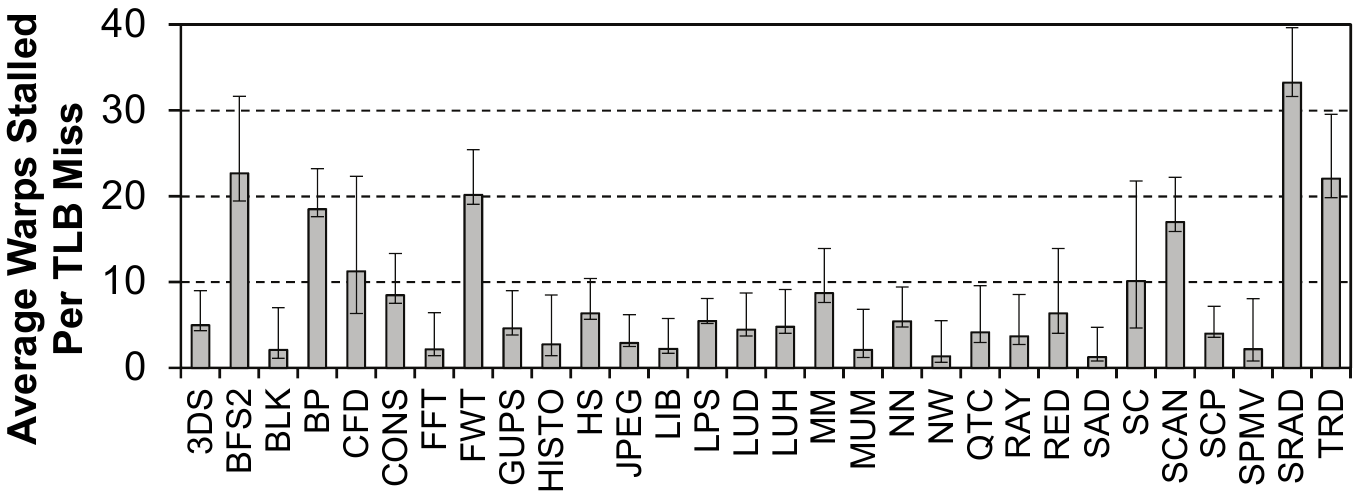}
\caption{Average number of warps stalled per TLB miss.}
\label{fig:schedulable-warp}
\end{figure}

\para{Impact of Large Pages.} \changesI{A large page} size can significantly improve
the coverage of the TLB~\cite{mosaic}. 
\changesIII{However, a TLB miss on a large page \changesII{stalls many more warps than
a TLB miss on a small page. \changesIV{We find that with a 2MB page size, the average number of stalled
warps increases \changesIV{to close} to 100\%~\cite{mosaic}, 
even though the average number of concurrent page \changesV{table walks}~\changesIV{never exceeds 5~misses per GPU core}}.  
Regardless of the page size, there is a strong need for mechanisms that mitigate the high 
cost of TLB misses.}} 

\subsection{Interference at the Shared TLB}
\label{sec:motiv-inter-thrashing}

\changesI{When multiple applications are concurrently executed, the address translation
overheads discussed in Section~\ref{sec:tlb-bottleneck} are exacerbated \changesIV{due to} inter-address-space interference.}
%
\changesIV{To study the impact of this interference, we measure how
the TLB miss rates change once another application is introduced.}
Figure~\ref{fig:interference-real} compares the \changesIII{\changesIV{512-entry} L2} TLB miss rate
\changesII{of four representative workloads when each application in the workload
runs in \changesIII{\changesVIII{isolation to} the miss rate when the \changesV{two applications run concurrently \changesVIII{and share}} the L2 TLB}}. 
\changesII{We observe from the figure} that inter-address-space interference 
\changesII{increases} \changesI{the TLB miss rate} \changesII{significantly for
most applications}.
\changesIII{This occurs because when the applications share the TLB, address
translation requests often induce \changesIV{TLB thrashing.} The resulting thrashing 
\changesIV{(1)~hurts} performance, and \changesIV{(2)~leads} to unfairness and starvation when applications generate 
TLB misses at different rates \changesIV{in the TLB (not shown)}.}

\begin{figure}[h!]
\centering
\includegraphics[width=0.85\columnwidth]{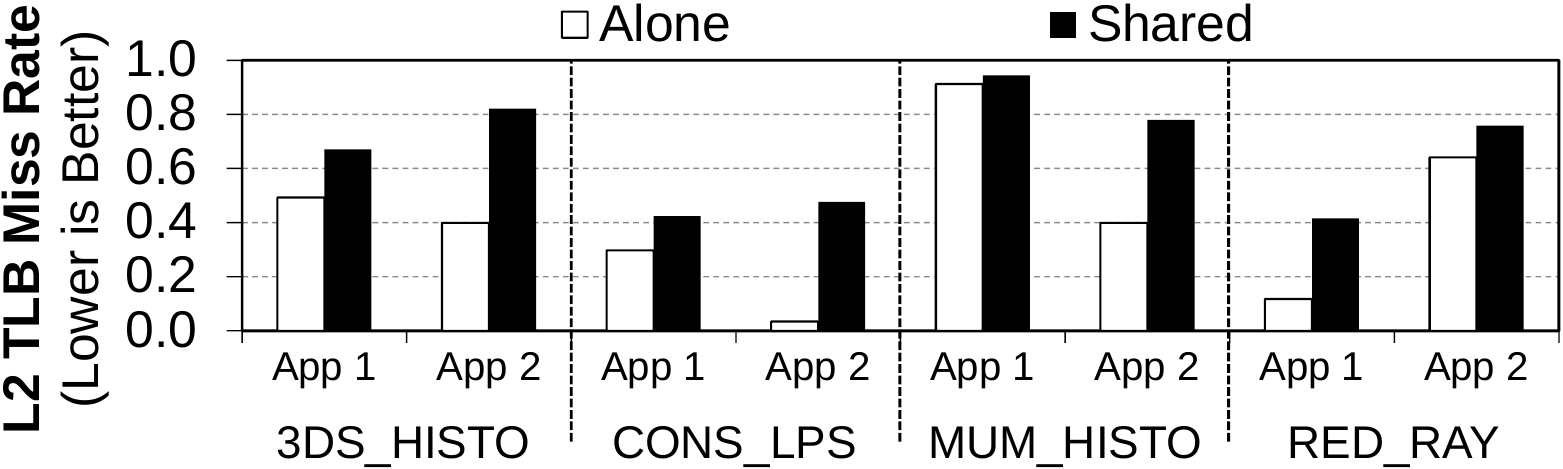}
\caption{\changesVII{Effect of} interference \changesVII{on the shared L2 TLB miss rate}. 
Each set of bars corresponds to a pair of \changesVII{co-running} applications \changesVIII{(e.g., ``3DS\_HISTO'' 
		denotes that the 3DS and HISTO benchmarks are run concurrently)}.} 
\label{fig:interference-real}
\end{figure}

\subsection{Interference Throughout the Memory Hierarchy}
\label{sec:bypassl2cache-motiv}
\label{sec:dram-interference}

\para{Interference at the Shared Data Cache.} 
Prior work~\cite{medic} \changesII{demonstrates} that while cache hits in GPUs reduce the consumption
of off-chip memory bandwidth, \changesVIII{the} \changesIV{cache hits} result in a lower load/store 
instruction latency only when \emph{every thread in the warp} hits in the cache. 
In contrast, when a page table walk hits in the shared L2
cache, the cache hit has the potential to help reduce the latency of \emph{other warps}
that have threads which access the same page in memory. 
However, TLB-related data can interfere with \changesIV{and displace cache
entries housing regular application data,} which can hurt the overall \changesIV{GPU} performance.



Hence, a trade-off exists between prioritizing \changesIII{address translation} requests \changesIV{vs.}
data \changesI{demand} requests in the GPU memory hierarchy. 
\changesI{\changesIV{Based on \changesVIII{an} empirical analysis of our workloads, we} \changesII{find} that} translation data
from \changesII{page table} levels closer to the page table
root are more likely to be \changesIV{\emph{shared}} across warps, and \changesII{typically hit in the cache.}
\yellow{We \changesII{observe that, for a 4-level page table,} the data cache hit rates of \changesIII{address translation}
requests across all workloads are 99.8\%, 98.8\%, 68.7\%, and \changesVIII{1.0\%} for the root,
first, second, and third levels \changesII{of the page table}, respectively.} \changesI{This means that \changesIII{address translation}
requests for the deepest \changesIV{page table} levels often do \changesIV{\emph{not}} utilize the cache well.} Allowing shared structures to
cache \changesIV{page table entries} from only the \changesV{page table} levels closer to the root could alleviate
the interference between low-hit-rate \changesIV{address translation data and regular application data}.




\para{Interference at Main Memory.}
Figure~\ref{fig:dram-util} characterizes the DRAM bandwidth \changesII{used by
\changesIII{address translation and data demand} requests,} \changesI{normalized to the maximum bandwidth available},
\changesII{for our workloads where two applications concurrently share the GPU.}
Figure~\ref{fig:dram-latency} compares the average latency \changesII{of
\changesIII{address translation \changesVIII{requests} and data demand} requests}. We see that even though \changesIII{address translation} requests consume
only 13.8\% of the \changesI{total} utilized DRAM bandwidth (2.4\% of the maximum \changesI{available} bandwidth),
their \changesVIII{average} DRAM latency is \changesIV{\emph{higher}} than that of data \changesIII{demand} requests. \changesIV{This is undesirable because} 
\changesI{\changesIII{address translation} requests \changesIV{usually stall} multiple warps, while data \changesIII{demand}
requests \changesIV{usually stall} only one warp (not shown).}
The \changesII{higher latency for \changesIII{address translation} requests} is caused by \changesIV{the} \changesVIII{FR-FCFS memory} \changesII{scheduling \changesIV{policy}}~\cite{fr-fcfs,frfcfs-patent}, which \changesII{prioritizes} accesses that hit in the
row buffer. Data \changesIII{demand} requests from GPGPU applications generally have very
high row buffer locality~\cite{sms,demystify,nmnl-pact13,complexity}, so a scheduler that cannot distinguish 
\changesIII{address translation} requests \changesIII{from data demand requests} effectively de-prioritizes \changesIII{the address translation requests}, increasing their latency,
\changesIV{and thus exacerbating the effect on stalled warps}.

\begin{figure}[h!]
\centering
\includegraphics[width=\columnwidth]{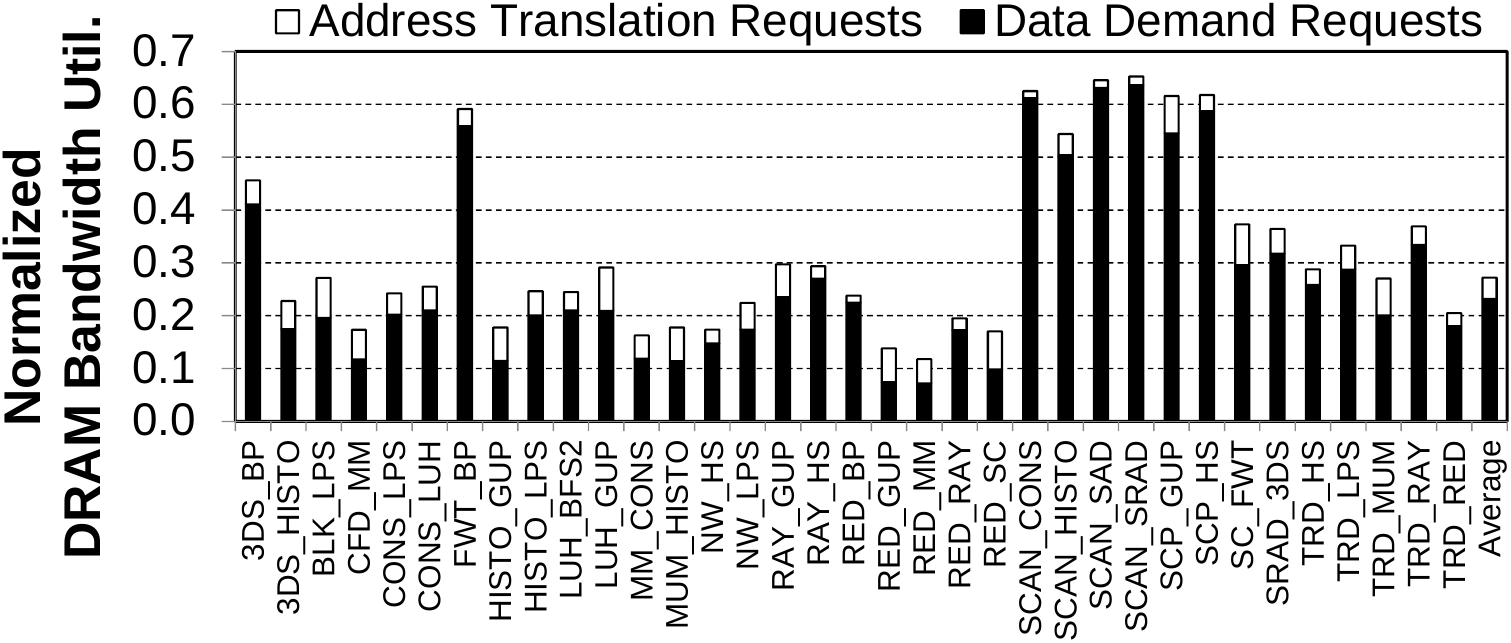}%
\caption{\changesIV{DRAM bandwidth utilization of address translation requests and data demand requests for} \changesII{two-application workloads}.}
\label{fig:dram-util}
\end{figure}

\begin{figure}[h!]
\centering
\includegraphics[width=\columnwidth]{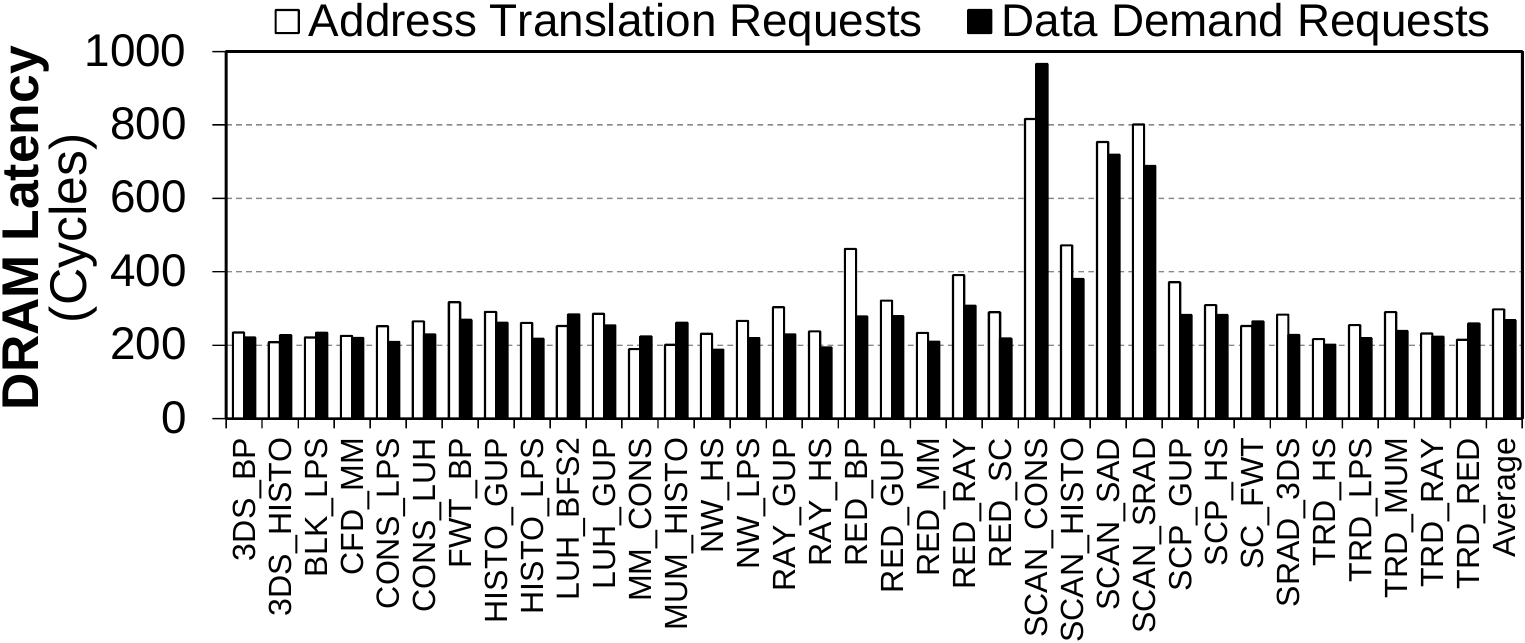}%
\caption{Latency of \changesIV{address translation requests and data demand requests for two-application workloads}.} 
\label{fig:dram-latency}
\end{figure}

\subsection{Summary and Our Goal}
\label{sec:our-goal}




\changesIV{We} make two important observations about address translation
in GPUs. First, address translation \changesIV{can greatly hinder a} GPU's ability to hide latency \changesV{by exploiting}
thread-level parallelism, \changesIV{since one single TLB miss can stall \emph{multiple} warps.} 
Second, \changesIV{during concurrent execution, multiple applications generate} 
\changesIV{inter-address-space interference throughout the GPU memory hierarchy,} 
which \changesIV{further increases the TLB miss latency and memory latency}. In light of these observations, 
\changesII{\emph{our goal}} is to 
\changesV{alleviate} the \changesIV{address} translation overhead \changesIV{in GPUs}
\changesV{in three ways:} (1)~increasing the TLB hit rate by \changesIV{reducing} TLB thrashing,
(2)~decreasing interference between \changesIII{address translation
requests and data demand requests} in the shared L2 cache, \changesVIII{and} (3)~decreasing \changesII{the} TLB miss
latency by prioritizing \changesIII{address translation} requests in DRAM \changesIV{without sacrificing DRAM
bandwidth utilization}. 


\section{Design of MASK}
\label{sec:design}

\changesII{To improve support for multi-application concurrency in state-of-the-art
GPUs, we introduce \titleShortMASK.}
\changesII{\titleShortMASK is a framework that provides memory protection support and} employs three \changesII{mechanisms} in the memory hierarchy to
reduce address translation overheads while requiring minimal \changesI{hardware changes, as}
illustrated in Figure~\ref{fig:overall-design}.
First, we introduce \tlbtokenname, \changesI{which regulate the number of warps that
can fill \changesII{(i.e., insert entries)} into the shared TLB in order}
to \changesII{reduce TLB thrashing}, and utilize a \changesII{small} \changesIV{TLB} bypass cache 
\changesII{to hold TLB entries from warps \changesIV{that are} not allowed to fill the shared TLB \changesIV{due to not having enough tokens}}
(\mycirc{1}). Second, we design \changesV{an}~\cachebypass mechanism,
\changesI{which significantly increases the shared L2 data cache
utilization \changesIV{and hit rate}} by reducing interference from \changesII{the TLB-related data 
that does not have high temporal locality} (\mycirc{2}).
Third, we design an \dramsched to further reduce
interference between \changesII{address translation requests and data demand} \changesVI{requests 
 (\mycirc{3}). }
 \changesII{In this section, we describe the detailed \changesIV{design and} implementation of \titleShortMASK.}
 We analyze 
the hardware cost of \titleShortMASK in Section~\ref{sec:overhead}.

\begin{figure*}[h!]
\centering
\includegraphics[width=\columnwidth]{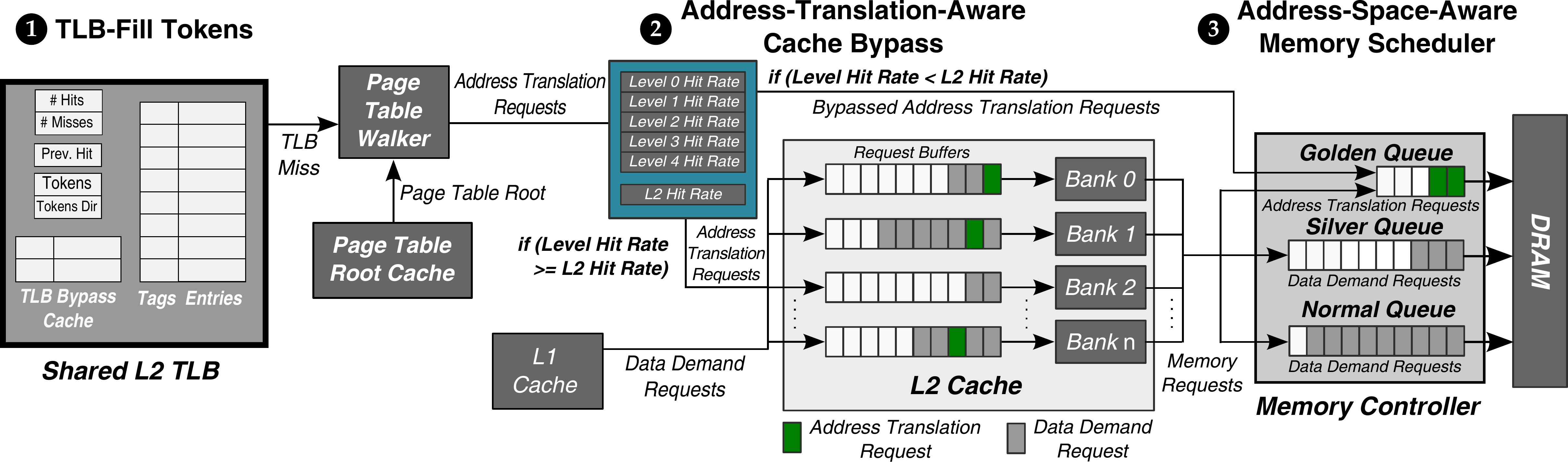}%
\caption{\titleShortMASK design overview.} 
\label{fig:overall-design}
\end{figure*}

\subsection{\changesIV{Enforcing} Memory Protection}

\changesII{\changesIV{Unlike} previously-proposed GPU sharing techniques that do \changesIV{\emph{not}} enable
memory protection~\cite{mafia,wang-hpca16,kepler,lindholm,pascal,isca-2014-preemptive,gebhart}, 
\titleShortMASK provides memory protection by 
allowing different \changesIV{GPU} cores to be assigned to
different address spaces.}
\titleShortMASK uses per-core page table \changesIV{\emph{root}} registers (similar to \changesI{the \changesVIII{CR3 register in x86} systems~\cite{intelx86}})
to set the current address space on each core. The \changesII{page table root \changesIV{register}} value 
\changesIV{from each GPU core} is also stored
in a page table root cache \changesIV{for use by the page table walker}. 
\changesIV{If a GPU core's page table root register value changes, the GPU core \changesV{conservatively} drains all in-flight
memory requests in order to \changesVII{ensure correctness}.} 
\changesIV{We extend each L2 TLB entry with an \changesVIII{address space} identifier \changesVIII{(ASID)}}. TLB flush operations
target a single \changesIV{GPU} core, flushing the core's L1 TLB,
and all entries in the L2 TLB \changesIV{that contain the matching \changesVIII{address space} identifier.}


\subsection{Reducing L2 TLB Interference}
\label{sec:fill-bypassing}

Sections~\ref{sec:tlb-bottleneck} and~\ref{sec:motiv-inter-thrashing}
demonstrate the need to minimize \changesII{TLB misses, which induce
long-latency stalls}. \titleShortMASK{} addresses this need 
with a new mechanism called \tlbtokenname  (\mycirc{1} in
Figure~\ref{fig:overall-design}).
To reduce \changesIV{inter-address-space interference} at the shared L2 TLB, we use an epoch- and
token-based scheme to limit the number of warps from each \changesIV{GPU} core that can
fill (and therefore contend for) the L2 TLB. While every warp can probe
the shared L2 TLB, only warps with tokens can \changesV{\emph{fill}} the
shared L2 TLB. 
\changesII{\changesV{Page table entries (PTEs) requested by} warps without tokens are only \changesIV{buffered in} a small
\changesIV{\emph{TLB bypass cache}}.}
This token-based mechanism 
requires two components: \yellow{(1)~a component to determine the number of tokens allocated to
each application, and (2)~a component that implements a policy for assigning tokens to warps \changesIV{within an application}.}


When a TLB \changesIV{request} arrives \changesIV{at the L2 TLB controller}, 
the GPU probes tags for both the shared L2 TLB 
and \changesIV{the TLB} bypass cache 
in parallel. A hit \changesI{in} either the TLB or the \changesIV{TLB} bypass cache yields a TLB hit.

\para{Determining the Number of Tokens.} 
\changesII{Every epoch,\footnote{\label{foot:epoch}We empirically select an epoch length of 100K cycles.}  \titleShortMASK tracks \changesI{(1)~the} \changesIV{L2 TLB} miss rate for each
application and \changesI{(2)~the} total number of \changesIV{all} warps in each core.}
After the first epoch,\footnote{\changesIV{Note that during the first epoch, \titleShortMASK does \changesV{\emph{not}} perform TLB bypassing.}}
the initial number of tokens for each application is set to a \changesII{predetermined} fraction of the total number of warps per application.

At the end of any subsequent epoch, for each application, \titleShortMASK
compares \changesII{the application's} shared L2 TLB miss rate \changesVIII{during} the current epoch to 
\changesII{its} miss rate from
the previous epoch. 
\changesII{If the miss rate \emph{increases} by more than 2\%, this indicates
that shared TLB contention is \emph{high} at the current token count, so
\titleShortMASK \emph{decreases} the number of tokens \changesIV{allocated to the application}.
If the miss rate \emph{decreases} by more than 2\%, this indicates
that shared TLB contention is \emph{low} at the current token count, so
\titleShortMASK \emph{increases} the number of tokens \changesIV{allocated to the application}.
If the miss rate change is within 2\%, the TLB contention has not 
changed significantly, and the token count remains unchanged.}

\para{Assigning Tokens to Warps.} 
Empirically, we observe that
(1)~\changesII{the different warps of an application tend to} have \changesI{similar TLB miss rates;} 
and (2)~it is beneficial for warps that already have
tokens to retain them, as it is likely that their TLB entries are already
in the shared L2 TLB. We leverage these two observations to
simplify the token assignment logic: \changesIV{our mechanism assigns tokens
to warps, one token per warp,} \changesI{\changesIV{in an order} based on the warp ID \changesII{(i.e., if there are
$n$~tokens, the $n$~warps with the lowest warp ID values receive tokens)}}. 
\changesII{This simple} heuristic is effective at reducing \changesIV{TLB} thrashing,
as contention at the shared L2 TLB is reduced based on the number of
tokens, and highly-used TLB entries that \changesIV{are requested by warps} \changesVIII{without} tokens can still
fill the \changesIV{TLB} bypass cache \changesIV{and thus \changesVIII{still take advantage of} locality}.

\para{\changesV{TLB} Bypass Cache.} While \tlbtokenname can reduce thrashing in the shared L2
TLB, a handful of highly-reused \changesIV{\changesV{PTEs} may be requested by} warps with no tokens, \changesVIII{which cannot insert the PTEs into}
the shared L2 TLB. To address this, we add \changesI{a} \emph{\changesIV{TLB} bypass cache},
which is a small 32-entry \changesI{fully-associative} cache.  Only warps \changesIV{\emph{without}} tokens
can fill the \changesIV{TLB} bypass cache \changesIV{in our evaluation}. To preserve consistency and correctness, \titleShortMASK
flushes \changesVIII{\emph{all} contents} of the TLB and the \changesIV{TLB} bypass cache when \changesII{a \changesV{PTE} is}
modified. \changesII{Like the L1 and L2 TLBs, the \changesIV{TLB} bypass cache uses \changesIV{the} LRU replacement 
policy.}



\subsection{Minimizing Shared L2 \changesIV{Cache} Interference}
\label{sec:data-bypassing}

We find that a TLB miss generates shared \changesV{L2}
cache accesses with varying degrees of locality. Translating addresses through
a multi-level page table (\changesII{e.g., the four-level table used in} \titleShortMASK{}) can generate dependent
memory requests \changesII{at} each level.
This causes significant queuing latency at the shared L2 cache, 
corroborating observations from previous work~\cite{medic}.
Page table entries in levels
closer to the root are more likely to be shared \changesIV{and thus reused} across threads than
entries near the leaves. 

\changesI{To address both interference and queuing delays \changesIV{due to address translation requests} at the shared L2 cache,} we
introduce \changesII{\changesV{an} \cachebypass mechanism (\mycirc{2} in
Figure~\ref{fig:overall-design})}. \yellow{To
determine which \changesII{address translation requests} \changesIV{should bypass \changesV{(i.e., skip \changesVII{probing and} filling the L2 cache)}}, we leverage our insights from
Section~\ref{sec:bypassl2cache-motiv}. 
\changesII{Recall that page table entries closer to the leaves have poor
\changesIV{cache} hit rates \changesIV{(i.e., the number of cache hits over all cache accesses)}.  
We make two observations from our \changesV{detailed study} on the page
table hit rates at each \changesIV{page table} level \changesVII{(see our technical report~\cite{mask-tech-report})}.
First, not all page table levels have the same hit rate across workloads (e.g.,
the level~3 hit rate for the \changesVIII{MM\_CONS} workload is only 58.3\%, but is
94.5\% for \changesVIII{RED\_RAY}).
Second, the}
hit rate behavior can change over time.  This means that \changesIV{a scheme that statically \changesV{bypasses} address translation requests \changesVIII{for} a certain page table level}
is \changesIV{\emph{not}} effective, as \changesV{such a scheme} cannot adapt to \changesV{dynamic hit rate behavior} changes.}
Because of the sharp drop-off in \changesVIII{the} L2 cache hit rate \changesIV{of address translation requests} after the first few
levels, we can simplify the \changesIV{mechanism to determine when address translation requests should \changesV{bypass the} L2 cache by comparing} the L2
cache hit rate of each page \changesI{table} level for \changesII{address translation} requests
to the L2 cache hit rate \changesI{of} \changesII{data demand} requests. We impose L2 cache bypassing \changesI{for \changesII{address translation} requests
from a particular page table level} when the hit rate \changesI{of} \changesII{address translation} requests \changesI{to that
page table level} falls below the hit rate \changesI{of}
\changesII{data demand} requests. 
\changesIV{The shared \changesV{L2 TLB has} counters to \changesVII{track} the cache hit rate of each page table level.}
\yellow{\changesV{Each memory request is} tagged with a three-bit
\changesV{value} that indicates \changesV{its} page walk depth, allowing \titleShortMASK{} to differentiate between
\changesVII{request types}. These bits are set to
\emph{zero} for data \changesII{demand} requests, and to 7 for any depth higher than
6.}
\cjr{Our current page table depth is 4 right?}~\cjr{Don't we need 
separate hit/miss counters for TLB vs non-TLB requests to make this possible?
If so, should fix Figure 12.}





\subsection{Minimizing Interference at Main Memory}
\label{sec:dram-sched}

\changesIV{There are two types of interference that occur at main memory:
(1)~data demand requests can interfere with address translation requests, as we 
saw in Section~\ref{sec:dram-interference}; and
(2)~data demand requests from multiple applications can interfere with each
other.}
\titleShortMASK{}'s memory controller design mitigates both forms of interference using 
an \dramsched{} \changesVIII{(\mycirc{3} in Figure~\ref{fig:overall-design})}.

The \dramsched breaks the traditional DRAM request buffer into three
separate queues. The first queue,
called the \goldQ{}, \changesIV{is} a small FIFO queue.\footnote{We observe
that \changesII{address translation} requests have low row \changesIV{buffer} locality. Thus, \changesIV{there is no
significant performance benefit if the memory controller reorders address 
translation requests within the \goldQ to exploit row buffer locality.}} 
\changesII{Address translation} requests always go to the
\goldQ{}, while \changesII{data demand} requests go \changesV{to}~\changesII{one of the two other}
queues (\changesI{\changesV{the size of each queue is}} similar to the size of a typical DRAM request \changesV{buffer}).  The second
queue, called the \silverQ{}, contains \changesI{data \changesII{demand} requests} from \emph{one}
selected application. The last queue, called the \normalQ, contains
data \changesII{demand} requests from all \changesIV{\emph{other}} applications. The \goldQ is used to prioritize TLB
misses over data \changesII{demand} requests. \changesI{The \silverQ allows the GPU to (1)~avoid
starvation when one \changesIV{or more applications hog} memory bandwidth, and
(2)~improve fairness when multiple applications execute concurrently~\cite{memattack,sms}. When
one application unfairly hogs DRAM bandwidth in the \normalQ, the \silverQ can
process \changesII{data demand} requests from another application that would otherwise be starved or
unfairly delayed.}


Our \dramsched always prioritizes requests in 
the \goldQ over requests in the \silverQ, which are \changesI{always} prioritized over requests in the \normalQ. 
\changesI{To provide higher priority to applications that are likely to be stalled due to concurrent 
TLB misses, and to \changesII{minimize the time that bandwidth-heavy applications have access to the 
silver queue,}}
each application takes turns being assigned to the \silverQ based on two \changesIV{per-application} metrics: 
(1)~the number of concurrent page walks, and (2)~the number of warps stalled per active TLB miss.
The number of \changesII{data demand} requests each application can add to the \silverQ, \changesIV{when the
application gets its turn,} is
shown \changesV{as $thresh_i$} in Equation~\ref{eq:silver}.
\changesIV{After application}~$i$ ($App_i$) \changesV{\changesVII{reaches} its quota},
the next application ($App_{i+1}$) is \changesV{then} allowed to send \changesV{its} requests to the \silverQ, \changesIV{and so on.}  Within \changesII{both the \silverQ and \normalQ},
FR-FCFS~\cite{fr-fcfs,frfcfs-patent} is used to
schedule requests.


\begin{equation}
\hspace{-5pt}
thresh_i = thresh_{max} x \frac{\hfill ConPTW_i * WarpsStalled_i}{\sum_{j=1}^{numApps}{ConPTW_j*WarpsStalled_j}}
\label{eq:silver}
\end{equation}
\vm{Is there any way to make this equation easier to read? Should it be numbered if it's the only one?}

To track the number of outstanding concurrent page walks \changesI{($ConPTW$ in Equation~\ref{eq:silver})}, 
we add a 6-bit counter per application to the shared L2 TLB.\footnote{We leave techniques to virtualize this counter
for more than 64 applications as future work.} \changesIV{This counter tracks the number of concurrent
TLB misses}.} 
To track the number of warps stalled per \changesIV{active TLB miss} \changesI{($WarpsStalled$ in Equation~\ref{eq:silver})}, we add a 6-bit counter to \changesI{each} TLB
MSHR entry, \changesI{\changesVIII{which} tracks the maximum number of warps that hit in
the entry.}~\changesV{The}~\dramsched resets all of these counters every epoch \changesI{(see Section~\ref{sec:fill-bypassing}).}

We find that the number of concurrent \changesII{address translation} requests that go to each
memory channel is small, so our design has an additional benefit of lowering \changesI{the} page table
walk latency \changesIV{(because it prioritizes address translation requests)} while minimizing interference.

\subsection{Page Faults and TLB Shootdowns}
\label{sec:mech-other}

Address translation inevitably introduces page faults.  Our design can
be extended to use techniques from previous works, such as performing
copy-on-write for \changesIV{handling page} faults~\cite{powers-hpca14}, and either exception
support~\cite{igpu} or demand paging techniques~\cite{tianhao-hpca16,pascal,cc-numa-gpu-hpca15}
for major faults. 
\changesVI{We leave this as future work.} 

\yellow{Similarly, TLB shootdowns are required when \changesIV{a GPU core
changes its address space or when a page table entry is updated.} Techniques to
reduce TLB shootdown overhead}~\cite{unitd,tlb-consistency,banshee} are
well-explored and can be \changesV{used with}~\titleShortMASK{}.



\section{Methodology}
\label{sec:meth-mask}

\changesII{To evaluate \titleShortMASK, we} model the NVIDIA Maxwell architecture~\cite{maxwell}, \changesI{and the TLB-fill} bypassing,
\changesVIII{cache bypassing}, and memory scheduling mechanisms in \titleShortMASK{}, using the Mosaic
simulator~\cite{mosaic}, which is based on GPGPU-Sim 3.2.2~\cite{gpgpu-sim}. We
heavily modify the simulator to accurately model the behavior of CUDA Unified
Virtual Addressing~\cite{maxwell,pascal} as described below.
Table~\ref{table:config} provides \changesIV{the details of} our baseline GPU configuration.
\changesI{Our baseline uses the FR-FCFS memory scheduling policy}~\cite{fr-fcfs,frfcfs-patent},
\yellow{based on findings from previous
works}~\cite{sms,complexity,nvidia-hpca17} \changesI{which show that \changesV{FR-FCFS} provides
good performance for GPGPU applications compared to other, more sophisticated
schedulers}~\cite{atlas,tcm}. \changesI{We have open-sourced our modified simulator online~\cite{mosaic.github}.}

\begin{table}[h!]
\begin{footnotesize}
\centering
\setlength{\tabcolsep}{0.3em}
\begin{tabular}{ll}
        \toprule
\multicolumn{2}{c}{\textbf{GPU Core \changesIX{Configuration}}} \\
	\midrule
\textbf{System Overview}           &  30 cores, 64 execution \changesIV{units} per core.\\
\textbf{Shader Core}           &  1020 MHz, 9-stage pipeline, 64 threads per warp, \\  & \yellow{GTO scheduler}~\cite{ccws}.\\
\textbf{Page Table Walker}   & \changesIV{Shared page table walker, traversing 4-level page tables.} \\
        \midrule
\multicolumn{2}{c}{\textbf{Cache and Memory \changesIX{Configuration}}} \\
        \midrule
\textbf{Private L1 Cache}    &  16KB, 4-way associative, LRU, L1 misses are \\ & coalesced before accessing L2, 1-cycle latency.\\
\textbf{Private L1 TLB}    &  64 entries per core, fully associative, LRU, 1-cycle latency.\\
\textbf{Shared L2 Cache}   &  2MB total, 16-way associative, LRU, 16 cache banks, \\ & 2 ports per cache bank, 10-cycle latency \\
\textbf{Shared L2 TLB}   &  512 entries total, 16-way associative, LRU, 2 ports, \\ & 10-cycle latency \\
\textbf{Page Walk Cache}    &  16-way 8KB, \changesV{10-cycle} latency \\
\textbf{DRAM}   & \changesIV{GDDR5 1674 MHz~\cite{gddr5}, 8 channels,} 8 banks per rank, \changesV{1 rank,} \\ & FR-FCFS scheduler~\cite{fr-fcfs,frfcfs-patent}, burst length 8\\
        \bottomrule
\end{tabular}%
\vspace{.5em}
\caption{Configuration of the simulated system.}
\vspace{-2em}
\label{table:config}%
\end{footnotesize}%
\end{table}%

\para{TLB and Page Table \changesV{Walker} Model.} \changesI{We accurately model \changesII{both TLB design variants} discussed in Section~\ref{sec:pwc}}. 
We employ the non-blocking TLB implementation used \changesIV{by} Pichai et al.~\cite{pichai-asplos14}.
Each core has a private L1 TLB. The page table
walker is shared \changesV{\changesVII{across} threads}, and admits up to 64 concurrent threads for walks. 
On a TLB miss, a page table walker generates a series
of dependent requests that probe the \changesV{L2 cache} and main memory as
needed. We \changesI{faithfully} model 
\changesIV{the multi-level page walks.}

\para{Workloads.}
We randomly select 27 applications from the CUDA SDK~\cite{cuda-sdk},
Rodinia~\cite{rodinia}, Parboil~\cite{parboil},
LULESH~\cite{lulesh,lulesh-origin}, and SHOC~\cite{shoc} suites. We classify
these benchmarks based on their L1 and L2 TLB miss rates into one of four
groups, \changesI{as} shown in Table~\ref{table:bench}.
For our multi-application secs/mask-micro17/results, we randomly select 35 pairs of applications,
avoiding \changesI{pairs where both applications} have \changesVIII{a} \changesIV{low L1 TLB miss rate (i.e., $<$20\%) and low L2 TLB miss rate (i.e., $<$20\%)}, since
these applications are relatively insensitive to \changesIV{address translation} overheads. 
The application that finishes first is relaunched to keep the
GPU core \changesV{busy} and \changesIV{\changesVIII{maintain} memory} contention.

\begin{table}[h!]
\begin{footnotesize}
\centering
\begin{tabular}{|c|c||l|}
        \hline
\textbf{L1 TLB Miss Rate} & \textbf{L2 TLB Miss Rate}   & \textbf{Benchmark Name} \\
\hhline{|=|=#=|}
Low & Low   & LUD, NN \\
        \hline
Low & High   & BFS2, FFT, HISTO, NW, \\
& & QTC, RAY, SAD, SCP \\
        \hline
High & Low & BP, GUP, HS, LPS \\
        \hline
High & High  & 3DS, BLK, CFD, CONS,\\
& & FWT, LUH, MM, MUM, RED, SC, \\
& & SCAN, SRAD, TRD \\
        \hline
\end{tabular}%
\vspace{.2em}
\caption{Categorization of \changesIV{workloads}.}
\vspace{-2em}
\label{table:bench}%
\end{footnotesize}%
\end{table}%

We divide \changesIV{35 application-pairs} into three workload categories based on the number of
applications that have both high L1 and L2 TLB miss rates, \changesII{as high 
TLB miss rates at both levels indicate a high amount of pressure on the
limited TLB resources}. \changesIV{\emph{n-HMR}} contains
\changesV{application-pairs} where \emph{n} applications in the \changesV{workload} have \changesIV{both} high L1 and
L2 TLB miss rates.


%

\para{Evaluation Metrics.} 
We report performance using \changesIV{\emph{weighted
speedup}~\cite{harmonic_speedup,ws-metric2}}, \changesI{a commonly-used metric to evaluate the performance of a
multi-application workload~\cite{sms,cpugpu-micro,tcm,atlas,parbs,stfm,mcp,aergia,Reetu-MICRO2009,mise,lavanya-asm,usui-dash}}.
\changesI{Weighted speedup is} defined as
$\sum{\frac{IPC_{Shared}}{IPC_{Alone}}}$, \changesI{where}  $IPC_{alone}$ is the
IPC of an application that runs on the same number of \changesIV{GPU} cores, but does \changesIV{\emph{not}}
share GPU resources with any other application, and $IPC_{shared}$ is
the IPC of an application \changesIV{when it runs} concurrently with other
applications. We report the unfairness of each design using
\changesIV{\emph{maximum slowdown}}, defined as $Max{\frac{IPC_{Alone}}{IPC_{Shared}}}$~\cite{ebrahimi-micro09,sms,Reetu-MICRO2009,
atlas,tcm,vandierendonck,bliss,bliss-tpds,mise,lavanya-asm,usui-dash}.

\para{Scheduling and Partitioning of Cores.} 
\changesIV{We assume an oracle GPU scheduler that finds the \emph{best} partitioning of the GPU cores for each pair
of applications. For each pair of applications that are concurrently executed, the scheduler partitions
the cores according to the best weighted speedup for that pair \changesV{found by}
an exhaustive search over all possible static \changesIV{core} partitionings.
}
\changesIV{\emph{Neither the L2 cache
  nor main memory are partitioned}. All applications 
  can use all of the shared L2 cache and the main memory.}



\para{Design Parameters.} \titleShortMASK exposes two configurable parameters:
$InitialTokens$ for \tlbtokenname, \changesVI{and $thresh_{max}$ for} the \dramsched. A sweep
over the range of possible $InitialTokens$ values reveals less than 1\% performance
variance, as \changesVIII{\tlbtokenname are} effective at reconfiguring the total number
of tokens to a steady-state value (Section~\ref{sec:fill-bypassing}).
In our evaluation, we set $InitialTokens$ to 80\%. We \changesVI{set $thresh_{max}$} to $500$ empirically.


\section{Evaluation}
\label{sec:eval}

\cjr{
\begin{compactitem}
\item Do GPU-MMU and Static enjoy the same oracular schedule? Would be more accurate at least for 
	Static to partition based on the available schemes for GRID. I'll bet that would leave some cores
	unused. \rachata{Yes. All of them enjoys the oracle. Note that we detach ourselves from GRID because
        of comments from MICRO reviewers. Essentially it is hard to claim what we model represents GRID due to
        limited informations.}
\item Might want to group plots by their category to save space.
\item We need to find a way to make plots bigger... \rachata{Grouping them by category might do the trick}
\item When using figure Xa, Xb, etc, we need to use the subcaption command instead of adding letters by hand--the colors aren't coming out right.
\item Of course, the real baseline that neither we nor our competitors include is a design with no memory protection, or one where address translation and TLBs are in the memory controller. Too late for ISCA probably, given that we have another paper to deal with as well.
\end{compactitem}
}

We compare the performance of \titleShortMASK against \changesIV{four} GPU designs.  The
first, called \emph{Static}, uses a static spatial partitioning of resources, where an
oracle is used to partition GPU cores, but the shared L2 cache and memory
channels are partitioned equally \changesV{across applications}. This design is intended
to capture key design aspects of NVIDIA GRID~\cite{grid} and AMD FirePro~\cite{firepro},
\changesII{\changesVI{based on} publicly-available information}.
The second design, called \emph{PWCache}, models the 
\changesIV{page walk cache baseline design we discuss in Section~\ref{sec:baseline}. 
The third design, called \emph{SharedTLB}, models the 
\changesV{shared L2 TLB} baseline design we discuss in Section~\ref{sec:baseline}.}
The \changesIV{fourth} \changesV{design, \emph{Ideal}, represents} a hypothetical GPU
where every single TLB access is a TLB hit. 
In addition to these designs, we report \changesI{the performance of the} individual
components of \titleShortMASK{}: \tlbtokenname(\titleShortMASK{}-TLB),
\cachebypass(\titleShortMASK{}-Cache), and \dramsched(\titleShortMASK{}-DRAM).

\subsection{Multiprogrammed Performance}
\label{sec:eval-multi}

\changesI{Figure~\ref{fig:multi-ws} compares \changesII{the \changesV{average}
performance by workload category} of \emph{Static}, \changesIV{\emph{PWCache}, \emph{SharedTLB},} and 
\emph{Ideal} to \titleShortMASK and \changesIV{the three individual components of \titleShortMASK}}. 
\changesI{We make \changesIV{two} observations from 
\changesII{Figure~\ref{fig:multi-ws}. 
First, compared} to \changesIV{\emph{SharedTLB}, which is the best-performing baseline}, 
\titleShortMASK{} \changesI{improves the weighted speedup by} 57.8\% \changesI{on average. Second, we find that}
\titleShortMASK performs only 23.2\% worse than \emph{Ideal} \changesIV{(where all accesses to the L1 TLB are hits)}.
\changesII{This demonstrates that \titleShortMASK reduces a large portion of the TLB miss overhead.}

\begin{figure}[h!]
\centering
\includegraphics[width=\columnwidth]{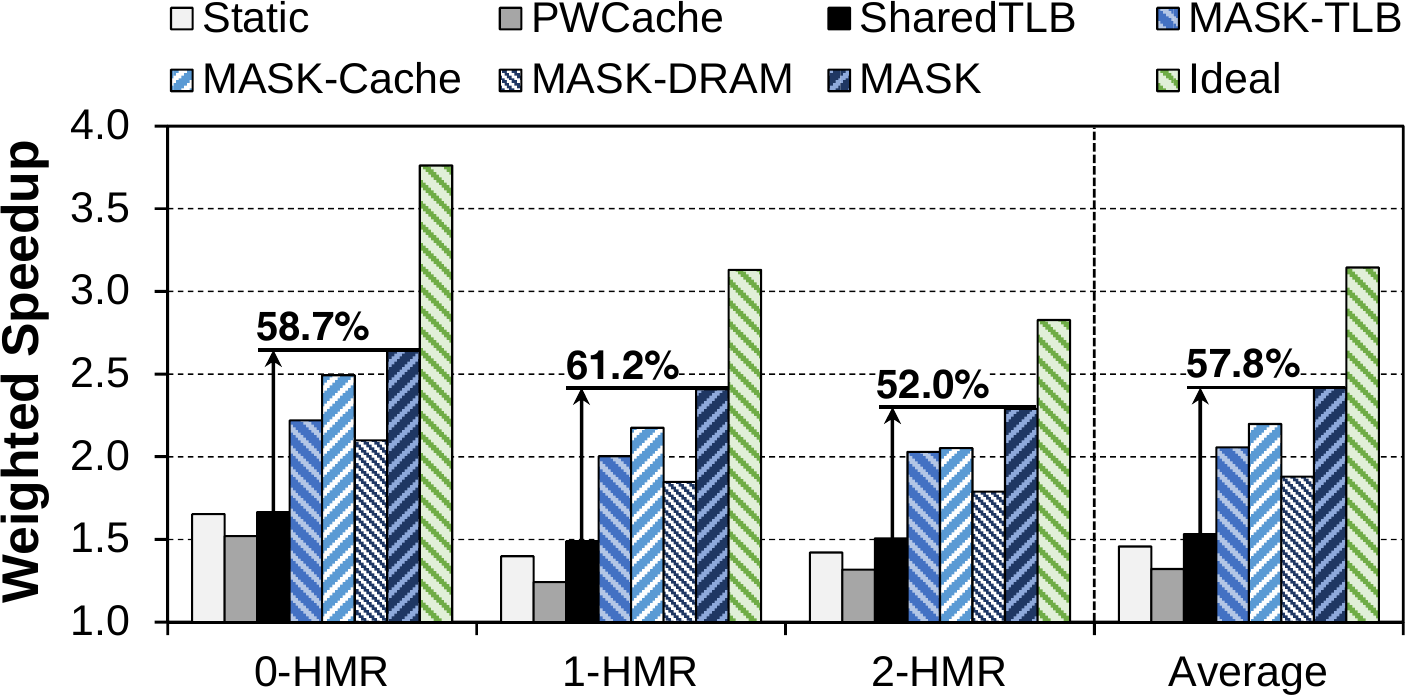}%
\caption{Multiprogrammed workload \changesIX{performance,}~\changesII{grouped
by workload category}.} 
\label{fig:multi-ws}
\end{figure}

\para{Individual Workload Performance.} \changesIV{Figures~\ref{fig:multi-ws-detail0},~\ref{fig:multi-ws-detail1}, and~\ref{fig:multi-ws-detail2} compare the weighted
speedup of \changesI{each individual} multiprogrammed workload for \titleShortMASK{}, \changesI{and the individual performance of} its three
components \changesI{(\titleShortMASK-TLB, \titleShortMASK-Cache, and \titleShortMASK-DRAM)}, 
against \emph{Static}, \emph{PWCache}, and \emph{SharedTLB} for the 0-HMR (Figure~\ref{fig:multi-ws-detail0}), 1-HMR (Figure~\ref{fig:multi-ws-detail1}), and 2-HMR (Figure~\ref{fig:multi-ws-detail2}) workload categories.
Each group of bars in Figures~\ref{fig:multi-ws-detail0}--\ref{fig:multi-ws-detail2} represents 
a pair of co-scheduled benchmarks.}
\changesIV{We make two observations from the figures.
\changesV{First, compared} to \emph{Static}, where resources are statically partitioned, 
\titleShortMASK provides better performance, because when an application stalls for
concurrent TLB misses, \changesIV{it no longer needs a large amount of} other shared resources, such as DRAM
bandwidth. During such stalls, other applications can utilize these resources.
When multiple GPGPU applications run concurrently \changesIV{using \titleShortMASK,} TLB misses from two or more
applications can be staggered, increasing the likelihood that there will be
heterogeneous and complementary \changesIV{resource demands.}
\changesV{Second,} 
\titleShortMASK provides
significant performance improvements over \changesV{\emph{both} \emph{PWCache} and \emph{SharedTLB}}
regardless of the workload type (i.e., 0-HMR to 2-HMR)}.  \changesIV{This indicates} that \titleShortMASK 
is effective at reducing the address translation overhead both when TLB contention
is high and when TLB contention is \changesIV{relatively} low.


\begin{figure*}[h!]
\centering
\includegraphics[width=\columnwidth]{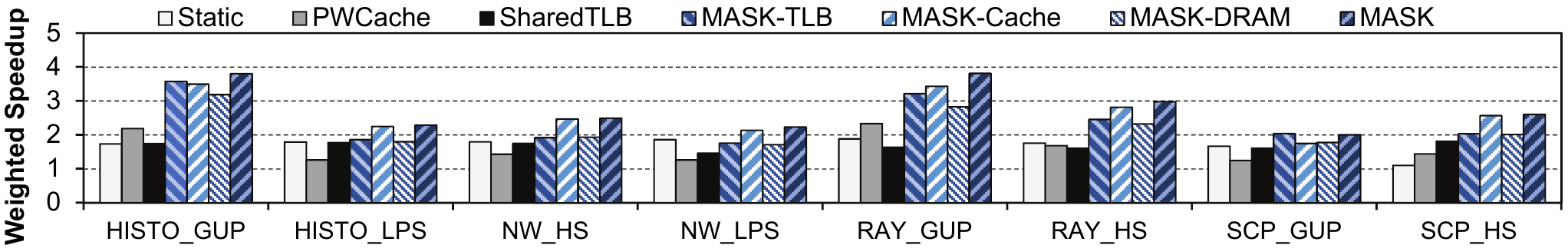}%
\caption{\changesIV{\changesVIII{Performance of} multiprogrammed workloads in the 0-HMR workload category.}}
\label{fig:multi-ws-detail0}

\centering
\includegraphics[width=\columnwidth]{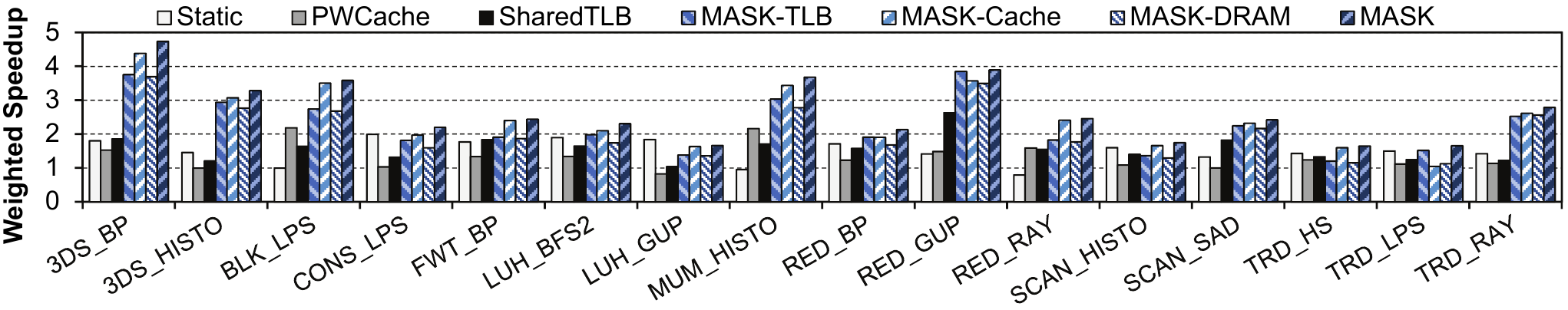}%
\caption{\changesIV{\changesVIII{Performance of} multiprogrammed workloads in the 1-HMR workload category.}}
\label{fig:multi-ws-detail1}

\centering
\includegraphics[width=\columnwidth]{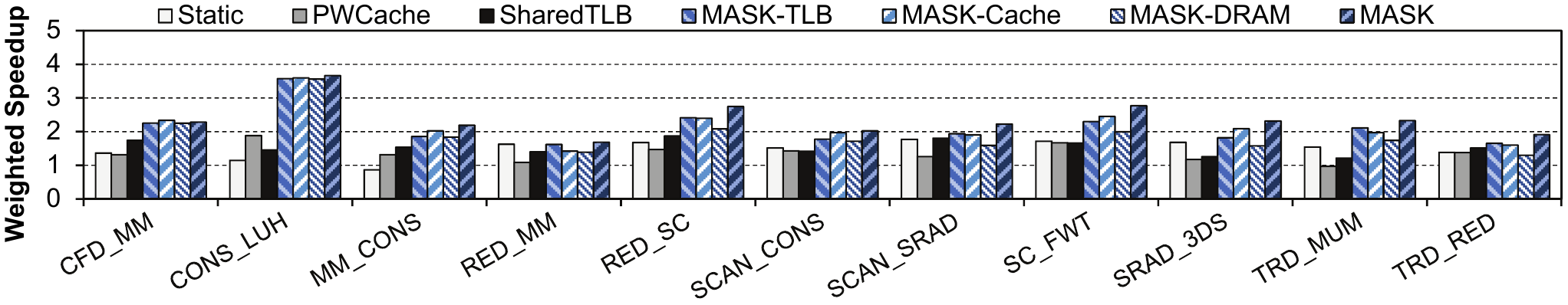}%
\caption{\changesIV{\changesVIII{Performance of} multiprogrammed workloads in the 2-HMR workload category.}}
\label{fig:multi-ws-detail2}
\end{figure*}

\changesIV{Our technical
report~\cite{mask-tech-report} provides additional analysis on the aggregate
throughput (system-wide IPC).  In the report, we show that \titleShortMASK{} provides 43.4\%
better aggregate throughput compared to \emph{SharedTLB}.}




Figure~\ref{fig:multi-unfair} compares \changesVII{the} unfairness \changesIV{of} \titleShortMASK{} to \changesIV{that of}
\changesV{\emph{Static}, \emph{PWCache}, and \emph{SharedTLB}. We make two observations. First, compared to 
statically partitioning resources (\emph{Static}), \titleShortMASK provides \changesV{better fairness 
\changesVIII{by allowing \emph{both} applications to 
access all shared resources}.} Second, compared to \emph{SharedTLB},
which is the baseline that \changesVII{provides} the best fairness, \titleShortMASK reduces
unfairness by 22.4\% on average}.  As the number of tokens for each
application changes based on the \changesIV{L2} TLB miss rate, applications that benefit more
from the shared L2 TLB are more likely to get more tokens, causing applications
that do not benefit from shared L2 TLB space to yield that shared L2 TLB space to
other applications. Our application-aware token distribution mechanism and \changesI{TLB-fill} 
bypassing mechanism work in tandem to reduce the amount of
\changesIV{shared L2 TLB thrashing} observed in
Section~\ref{sec:motiv-inter-thrashing}. 

\begin{figure}[h!]
\centering
\includegraphics[width=\columnwidth]{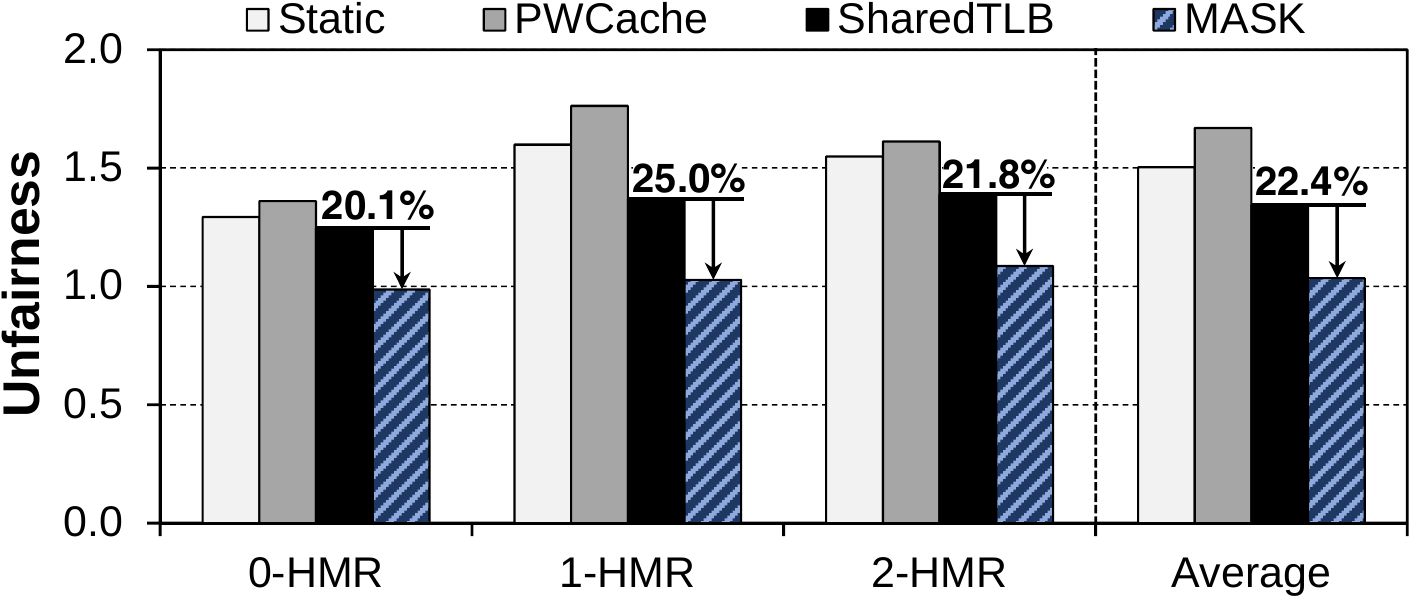}%
	\label{fig:bypass-l2-fill}
\caption{Multiprogrammed workload unfairness.} 
\label{fig:multi-unfair}
\end{figure}


\para{Individual Application Analysis.} \titleShortMASK{} provides better
throughput for \changesIV{\emph{all} individual} applications sharing the GPU due to reduced TLB miss
rates for each application \changesII{(shown in our technical report~\cite{mask-tech-report})}. The per-application L2 TLB miss rates are reduced
by over 50\% on average, which is in line with the \changesIV{reduction in} system-wide \changesIV{L2 TLB} miss rates \changesI{(see Section~\ref{sec:component-analysis})}. 
Reducing the number of TLB misses \changesIV{via}
\changesIV{our} \changesI{TLB-fill} bypassing policy (Section~\ref{sec:fill-bypassing}), and
reducing the latency of TLB misses \changesIV{via our} shared L2 bypassing
(Section~\ref{sec:data-bypassing}) and TLB- and application-aware DRAM
scheduling (Section~\ref{sec:dram-sched}) \changesVIII{policies}, enables significant
performance improvement.

In some cases, running two applications concurrently provides \changesIV{\emph{better}} \changesIV{performance
as well as lower unfairness} than running \changesI{each} application alone (e.g., \changesIV{for the 
\changesVIII{RED\_BP and RED\_RAY} workloads in Figure~\ref{fig:multi-ws-detail1}}, and the \changesVIII{SC\_FWT} workload \changesIV{in Figure~\ref{fig:multi-ws-detail2}}). We
attribute \changesI{such} cases to substantial improvements (more than 10\%) of two
factors: a lower L2 \changesV{cache} queuing latency for bypassed \changesII{address translation} requests, and a higher L1 \changesV{cache}
hit rate \changesV{of data demand requests} when applications share the L2 \changesV{cache} and main memory with other
applications.

\changesI{We conclude that \titleShortMASK is effective at \changesV{reducing the} address
translation \changesIV{overheads in modern GPUs, and thus \changesVIII{at} improving both performance and fairness,}
\changesII{by introducing \changesII{address translation} request awareness throughout the
\changesIV{GPU} memory hierarchy}.}


\subsection{Component-by-Component Analysis}
\label{sec:component-analysis}

\changesI{This section characterizes 
\titleShortMASK's underlying mechanisms (\titleShortMASK-TLB, \titleShortMASK-Cache, and
\titleShortMASK-DRAM). Figure~\ref{fig:multi-ws} shows the average performance improvement of
each individual component of \titleShortMASK compared to \emph{Static},
\changesIV{\emph{PWCache}, \emph{SharedTLB},} and \titleShortMASK. \changesIV{We
summarize our key findings here, and provide a more detailed analysis \changesVIII{in}
our technical report~\cite{mask-tech-report}.}


\para{Effectiveness of \tlbtokenname.} 
\titleShortMASK uses \tlbtokenname to reduce thrashing. 
We compare TLB hit rates for \emph{Static}, \changesIV{\emph{SharedTLB},} and \titleShortMASK-TLB.
\changesI{The hit} rates for \changesI{\emph{Static}} and \changesIV{\emph{SharedTLB}} are substantially similar.
\titleShortMASK-TLB increases \changesV{shared L2} TLB hit rates by 49.9\% on \changesIV{average over \emph{SharedTLB}~\cite{mask-tech-report}}, \changesV{because} \changesVII{the \tlbtokenname mechanism reduces} the number of warps utilizing the shared L2 TLB entries,
in turn reducing the miss rate. \changesIV{The TLB bypass cache stores}
frequently-used TLB entries that cannot be filled in the traditional TLB.
Measurement of the \changesIV{average TLB} bypass cache hit rate \changesIV{(66.5\%)} 
confirms this \changesIV{conclusion~\cite{mask-tech-report}}.\footnote{\changesI{We find that the performance of \titleShortMASK-TLB saturates when we increase the
\changesIV{TLB} bypass cache beyond 32 entries for the workloads that we evaluate.}}

\para{Effectiveness of \cachebypass.}
\titleShortMASK uses \cachebypass
with the goal of prioritizing \changesII{address translation requests}. \changesI{We measure the} average
\changesVI{L2 cache hit rate} for \changesII{address translation} requests. \changesI{We find} that 
for \changesII{address translation} requests that fill into the shared L2 cache, \cachebypass is
\changesIV{very} effective \changesV{at} selecting which blocks to cache, resulting in \changesIV{an} \changesII{address translation} request hit rate
that is higher than 99\% for all of our workloads.
At the same time, \cachebypass minimizes the \changesI{impact of long
L2 cache queuing latency~\cite{medic}}, leading to \changesIV{a 43.6\%} performance  
improvement compared to \changesIV{\emph{SharedTLB} (as shown in Figure~\ref{fig:multi-ws})}.




\para{Effectiveness of \dramsched.}
To characterize the performance impact of \titleShortMASK's DRAM scheduler,
we compare \changesIV{the} DRAM bandwidth utilization and average DRAM latency \changesIV{of (1)~address translation requests and (2)~data demand
requests} for the baseline designs and \titleShortMASK, \changesIV{and \changesVIII{make} two observations}.
\changesIV{First, we find that \titleShortMASK is effective at reducing the DRAM latency of address translation
requests, which contributes to \changesVII{the}~\changesV{22.7\% performance improvement \changesVII{of \titleShortMASK-DRAM} over \emph{SharedTLB}}, as shown in Figure~\ref{fig:multi-ws}.}
In cases where the DRAM latency is high, \changesV{our} DRAM \changesVII{scheduling} policy reduces the
latency of \changesII{address translation} requests by up to 10.6\% (\changesVIII{SCAN\_SAD}),
while increasing DRAM bandwidth utilization by up to 5.6\% (\changesVIII{SCAN\_HISTO}). 
\changesVII{Second, we find that when an application is suffering severely 
from interference due to another concurrently-executing application, the \silverQ
significantly reduces the latency of data demand requests from the suffering
application.}
For example, when \changesIV{the \silverQ} is employed, SRAD from the \changesVIII{SCAN\_SRAD} application-pair
performs 18.7\% better, while both SCAN and CONS from
\changesVIII{SCAN\_CONS} \changesVII{perform} 8.9\% and 30.2\%~\changesVII{better}, respectively.}
\changesIV{Our technical report~\cite{mask-tech-report}~\changesV{provides a} more detailed analysis 
of the impact of our \dramsched}.

\changesI{We conclude that each component of \titleShortMASK provides complementary performance
improvements by introducing \changesII{address-translation-aware} policies at different \changesV{memory hierarchy} levels.}



\subsection{Scalability and Generality}
\label{sec:750-eval}

This section evaluates the scalability of \titleShortMASK and provides evidence that
the design generalizes well across different \changesIV{architectures}. \changesIV{We summarize
our key findings here, \changesVII{and} provide a more detailed analysis in our technical
report~\cite{mask-tech-report}}.


\para{Scalability.} We compare the performance of \changesIV{\emph{SharedTLB}, which is the
best-performing state-of-the-art baseline design,} and \titleShortMASK,
normalized to \changesIV{\emph{Ideal}} performance, as the number of concurrently-running applications
\changesV{increases from one to five.} In general, as the application count increases, contention for shared
resources (e.g., shared L2 TLB, \changesIV{shared \changesV{L2} cache)} draws the performance for
both \changesIV{\emph{SharedTLB}} and \titleShortMASK further from \changesIV{the performance of \emph{Ideal}}. 
\changesI{However, \titleShortMASK maintains a consistent performance advantage relative to \changesIV{\emph{SharedTLB}},
as shown in Table~\ref{table:scaling-core}.
The performance gain \changesIV{of \titleShortMASK relative to \emph{SharedTLB}} is 
more pronounced at higher levels of \changesII{multi-application} concurrency 
because (1)~\changesIV{the shared L2 TLB} becomes 
heavily contended as the number of concurrent applications increases, and (2)~\titleShortMASK is effective
at reducing the amount of contention at the heavily-contended shared TLB.}

\begin{table}[h!]
\begin{footnotesize}
\centering
\begin{tabular}{|c||c|c|c|c|c|}
        \hline
\textbf{Number of Applications} & \textbf{1}   & \textbf{2} & \textbf{3} & \textbf{4} & \textbf{5} \\
\hhline{|=#=|=|=|=|=|}
\emph{SharedTLB} performance & 47.1\% & 48.7\% & 38.8\% & 34.2\%  & 33.1\% \\
normalized to \emph{Ideal} & & & & &  \\
        \hline
\titleShortMASK performance  & 68.5\% & 76.8\% & 62.3\% & 55.0\%  & 52.9\% \\
normalized to \emph{Ideal} & & & & & \\
        \hline
\end{tabular}%
\caption{\changesIV{Normalized performance of \emph{SharedTLB} and \titleShortMASK as the number of concurrently-executing applications increases.}}
\label{table:scaling-core}%
\end{footnotesize}%
\end{table}%

\para{Generality.} \titleShortMASK{} is an architecture-independent design: our
techniques \changesIV{are} applicable to any \changesIV{SIMT} machine~\cite{fermi,kepler,maxwell,pascal,amdr9,radeon,vivante-gpgpu,arm-mali,powervr}. 
\changesI{To demonstrate this, we} evaluate \changesIV{our two baseline variants
(\emph{PWCache} and \emph{SharedTLB}) and \titleShortMASK} on \changesIV{two additional GPU architectures: the
\changesIV{GTX480 (Fermi architecture~\cite{fermi}), 
and an integrated} GPU architecture~\cite{powers-hpca14,amd-fusion,apu,kaveri,haswell,amdzen,skylake,powervr,
arm-mali,tegra,tegrax1},} as shown in Table~\ref{fig:mask-pwcache-sweep}.
\changesIV{We make three key conclusions.}
\changesIV{First, address translation leads to significant performance overhead in both \changesVIII{\emph{PWCache} and \emph{SharedTLB}}.
Second,} \titleShortMASK{} provides \changesVIII{a}
\changesV{46.9\%}~\changesIV{average} performance improvement over \changesIV{\emph{PWCache} and \changesVIII{a} \changesV{29.1\%} average
performance improvement over \emph{SharedTLB}} on the Fermi
architecture, getting to within 22\% of the \changesIV{performance of \emph{Ideal}}. 
\changesIV{Third, on the integrated GPU configuration
used in previous work~\cite{powers-hpca14}, we find that \titleShortMASK provides \changesVIII{a} 23.8\% performance
improvement over \emph{PWCache} and \changesVIII{a} 68.8\% performance improvement
over \emph{SharedTLB}, and gets within 35.5\% of the performance of \emph{Ideal}.} 

\begin{table}[h!]
\begin{footnotesize}
  \centering
    \begin{tabular}{|c||c|c|}
\hline
\textbf{Relative Performance} & \textbf{Fermi} & \textbf{Integrated GPU~\cite{powers-hpca14}}  \\ 
\hhline{|=#=|=|}
\textbf{PWCache} & 53.1\% & 52.1\%                                                  \\ \hline
\textbf{SharedTLB} & 60.4\% & 38.2\%                                                         \\ \hline
\textbf{\titleShortMASK} & 78.0\% & 64.5\%                                            \\ \hline
    \end{tabular}%
  \caption{\changesIV{Average performance of \emph{PWCache}, \emph{SharedTLB}, and \titleShortMASK, normalized to \emph{Ideal}}.}
  \label{fig:mask-pwcache-sweep}
\end{footnotesize}%
\end{table}%

\changesIV{We conclude that \titleShortMASK{} is
effective \changesV{at}~\changesVII{(1)~reducing} the \changesV{performance overhead} of address translation, and \changesV{\changesVII{(2)~significantly improving system}
performance over \emph{both} \changesVIII{the} \emph{PWCache} and \emph{SharedTLB} \changesVIII{designs, regardless} of the GPU architecture.}}

\para{Sensitivity to L1 and L2 TLB Sizes.}
\changesI{We evaluate the benefit of MASK over many different TLB sizes in our technical \changesIV{report~\cite{mask-tech-report}.}
\changesIV{We make two observations. First, \titleShortMASK is effective at \changesV{reducing (1)~TLB} thrashing 
at the shared L2 TLB, and \changesV{(2)~the} latency of address translation requests regardless of TLB size. 
Second, as we increase the shared L2 TLB size from 64 to 8192~entries, \titleShortMASK
outperforms \emph{SharedTLB} for all TLB sizes 
except \changesV{the} 8192-entry shared L2 TLB. At 8192~entries, \titleShortMASK and \emph{SharedTLB} perform equally, 
because the working set fits completely within the 8192-entry shared L2 TLB.}



\para{Sensitivity to Memory Policies.} We study the sensitivity
of \titleShortMASK{} to (1)~main memory row policy, and (2)~memory scheduling
policies. We find that for \changesIV{all of our baselines and for \titleShortMASK{},} \changesIV{performance 
with an open-row policy~\cite{atlas} is similar (within 0.8\%) to the performance with a \changesV{closed-row} 
policy,} which is used in various \changesV{CPUs}~\cite{ivybridge,intel-sandybridge,skylake}. Aside from the FR-FCFS
scheduler~\cite{fr-fcfs,frfcfs-patent}, we \changesI{use} \titleShortMASK~\changesI{in conjunction with another}
state-of-the-art GPU memory scheduler~\cite{mafia}, and \changesI{find} that \changesII{with this scheduler,} \titleShortMASK{}
\changesI{improves performance by} 44.2\% over \changesIV{\emph{SharedTLB}}. We conclude that
\titleShortMASK is effective across different memory policies. 

\para{Sensitivity to Different Page \changesV{Sizes.}} \yellow{We evaluate the performance
of \titleShortMASK with 2MB large pages assuming an ideal page fault
latency~\cite{mask-tech-report,rachata-thesis}~\changesII{(not shown)}. \changesIV{We provide
two observations. First, even with the larger page size, \emph{SharedTLB} 
continues to experience high contention during address translation, causing its average
performance to fall 44.5\% short of \changesIV{Ideal}. Second, we find that using \titleShortMASK{} allows 
the GPU to perform within 1.8\% of Ideal.}

\subsection{Hardware Overheads}
\label{sec:overhead}

To support memory protection, each L2 TLB \changesIV{entry has an 9-bit address space identifier (ASID)},
\changesIV{which} translates to \changesV{an overhead of} 7\% of the L2 TLB size in total.} 
 
\changesII{At each core,} our \tlbtokenname mechanism uses
(1)~two 16-bit counters \changesII{to track the \changesIV{shared L2 TLB} hit rate, with one counter 
tracking the number of \changesIV{shared L2 TLB} hits, and the other counter tracking the number of \changesIV{shared L2 TLB} misses;
(2)~a 256-bit vector addressable by warp ID \changesIV{to track the number of active warps, where} each bit is set when a warp uses
the shader core for the first time, and is reset every epoch; and
(3)~an 8-bit incrementer that tracks the total number of unique warps executed
by the core (i.e., its counter value is incremented each time a bit is set in
the bit vector)}.

We augment the shared cache with \changesI{a}
32-entry fully-associative content addressable memory (CAM) for the bypass
cache, 30 15-bit token \changesI{counters}, \changesIV{and 30 1-bit direction registers to record
whether the token count increased or decreased during the previous epoch.} 
\changesIV{These structures allow the GPU} to distribute tokens \changesIV{among} up to 30 concurrent applications. 
In total, we add \changesII{706}~bytes \changesI{of storage}
(\changesII{13}~bytes per core in the L1 TLB, and 316 bytes \changesIV{total} in the
shared L2 TLB), which adds \changesII{1.6\%} to the baseline L1 TLB \changesII{size} and
3.8\% \changesIV{to} the \changesII{baseline L2 TLB size (\changesV{in \changesVII{addition to the 7\% overhead due to the} ASID bits})}.

\cachebypass uses ten 8-byte counters per core to track
\changesIV{L2} cache hits and \changesIV{L2} cache accesses per level.
The resulting 80~bytes add less than 0.1\% 
\changesII{to the baseline shared L2 cache size}. Each \changesIV{L2} cache and memory request requires an
additional 3~bits \changesV{to specify} the page walk level, as we discuss in
Section~\ref{sec:data-bypassing}.

\changesIV{For each memory channel, \changesVIII{our} \dramsched contains a 16-entry FIFO queue for the
\goldQ,
a 64-entry memory request buffer \changesVI{for the
\silverQ, and a 192-entry memory \changesV{request} buffer for the \normalQ.}
This adds an extra 6\% of storage overhead to the DRAM request queue per
memory controller.}

\para{Area and Power Consumption.}
We compare the area and power
consumption of \titleShortMASK{} to \changesIV{\emph{PWCache} and \emph{SharedTLB}} using
CACTI~\cite{cacti}. 
\changesVII{\emph{PWCache} and \emph{SharedTLB} have near-identical area and
power consumption, as we size the page walk cache and shared L2 TLB (see
Section~\ref{sec:baseline}) such that they both use the same total area.
We find that \titleShortMASK{} introduces a negligible overhead to both baselines,
consuming less than 0.1\% additional area and 0.01\% additional power in
each baseline.
We provide a detailed analysis of area and power consumption in our
technical report~\cite{mask-tech-report}.}

~\cjr{I guess we have a
smaller L1 TLB, which somewhat compensates the additional L2? Might be worth
saying that. }

\section{MASK: Conclusion}

Spatial multiplexing support, which allows multiple applications to run
concurrently, is needed to efficiently deploy GPUs in a
large-scale computing environment.
\changesII{\changesIV{Unfortunately,} due to the primitive existing support for memory virtualization,
many of the performance benefits of spatial multiplexing are lost in state-of-the-art
GPUs.  We perform a detailed analysis of state-of-the-art mechanisms for
memory virtualization, and find that current address translation mechanisms 
(1)~are highly susceptible to interference across the different
address spaces of applications \changesIV{in the shared TLB structures}, which leads to a high number of page table walks; and
(2)~undermine the fundamental latency-hiding techniques of GPUs, by often
stalling hundreds of threads at once.}
To alleviate these problems, we propose \titleShortMASK{},
a new memory hierarchy designed \changesV{carefully to support multi-application concurrency at low overhead.}
\titleShortMASK{} consists of three major components \changesII{in different parts
of the memory \changesVIII{hierarchy,} all of which incorporate \changesII{address translation request} awareness.
These three components work together to} lower inter-application 
interference during address translation, and improve L2 cache utilization \changesIV{and memory latency}
for \changesII{address} translation requests.
\changesV{\titleShortMASK} improves  performance by 57.8\%, on average across a wide range
of multiprogrammed workloads, 
over the state-of-the-art. \changesIV{We conclude that \titleShortMASK provides a promising and effective substrate for
multi-application execution on GPUs, and hope future work builds on the mechanism we provide and open source~\cite{mosaic.github}.}


\clearpage

\chapter{Reducing Inter-address-space Interference with Mosaic}
\label{sec:mosaic}

Graphics Processing Units (GPUs) are used for an ever-growing
range of application domains due to steady increases in GPU compute density and
continued improvements \changesI{in programming} 
tools~\cite{programmingguide,khronos2008opencl,huma}.
The growing adoption of GPUs has in part been due to better high-level 
language support~\cite{programmingguide, dandelion, delite, copperhead}, 
which has improved GPU programmability.
\changesI{Recent support} within GPUs for \emph{memory virtualization} features,
such as a unified virtual address 
space~\cite{fermi,huma}, demand paging~\cite{pascal}, and
preemption~\cite{pascal,firepro},
can provide \emph{fundamental} improvements \changesI{that can ease programming.}
\changesIIIII{These} features allow developers to exploit \changesI{key} benefits \changesI{that have long been} taken for granted in CPUs
(e.g., application portability, multi-application execution). 
Such familiar features can dramatically improve programmer productivity and 
further boost GPU adoption.
However, \changesI{a number of challenges have kept GPU memory
virtualization from achieving performance similar to \changesI{that in}
CPUs~\cite{abhishek-ispass16,gpu-arch-microbenchmarking}.
In this work, we focus on two fundamental challenges: (1)~the address translation
challenge, and (2)~the demand paging challenge.}

\paragraphbe{\changesI{Address Translation Challenge.}}
\changesI{Memory} virtualization relies on \emph{page tables} to store
virtual-to-physical address translations. 
\changesI{Conventionally,
systems store one translation
for every \emph{base page} (e.g., a 4KB page).}
To translate \changesI{a} virtual address on demand, a series of \emph{serialized} memory
accesses are required to traverse \changesI{(i.e., \changesIIIII{\emph{walk}})} the page table~\cite{powers-hpca14, pichai-asplos14}.
These serialized accesses clash with the \changesI{single-instruction multiple-thread (SIMT)} 
execution model\changesI{~\cite{largewarps,lindholm,dwf}} used by GPU-based systems,
which relies on \emph{high degrees of concurrency} through \changesIIIII{\emph{thread-level parallelism}} (TLP) to 
hide long memory latencies during GPU execution.  
\changesI{\emph{Translation lookaside
buffers}} (TLBs) can reduce the latency of address translation 
by caching recently-used \changesI{address} translation information.
Unfortunately, as application working sets and DRAM capacity have increased
in recent years,
state-of-the-art GPU TLB designs~\cite{powers-hpca14,pichai-asplos14} 
suffer due to \emph{inter-application} interference and stagnant TLB sizes.
Consequently, GPUs have \emph{poor TLB reach}, \changesI{i.e., the} TLB covers only a
small fraction of the physical memory \changesIII{working set of} an application.
Poor TLB reach is particularly detrimental \changesI{with the SIMT execution model}, as a \emph{single} TLB miss can stall
\changesI{\emph{hundreds}} of threads at once, undermining TLP within a GPU and
significantly reducing performance~\cite{abhishek-ispass16,gpu-arch-microbenchmarking}. 

\changesI{\emph{Large pages} (e.g., the 2MB or 1GB pages in modern CPUs~\cite{haswell,skylake})
can significantly reduce the overhead of address translation. A major constraint
for TLB reach is the small, fixed number of translations that a TLB can hold.
If we store one translation for every large page instead of one translation for
every base page, the TLB can cover a \changesIII{much} larger fraction of \changesI{the}
virtual address space using the same number of page translation entries.
Large pages have been supported by CPUs for decades~\cite{pentium-pro, alpha},
and large page support is emerging for GPUs~\cite{powers-hpca14, tianhao-hpca16,
pichai-asplos14}.
However, large pages increase the risk of \emph{internal fragmentation}, where
a portion of the large page is unallocated (or unused). Internal fragmentation occurs because it is often
difficult for an application to completely utilize large contiguous regions of memory.
This fragmentation leads to \changesI{(1)~\emph{memory bloat},
\changesI{where a much greater amount of physical memory is allocated 
than the amount of memory that the application needs;}
and \changesI{(2)~longer} memory access latencies,
due to \changesIII{a lower effective TLB reach and more} page faults~\cite{ingens}.}

\paragraphbe{\changesI{Demand Paging Challenge.}}
\changesI{For} \emph{discrete GPUs} (i.e., \changesI{GPUs that are not in the same package/die as} the CPU), 
\emph{demand paging} can incur significant overhead.  With demand paging, an 
application can request data that is not currently resident in GPU memory.
This triggers a \emph{page fault}, which requires a \changesI{long-latency data 
transfer for an entire page over the system I/O bus, \changesI{which, in today's systems, is 
also called the PCIe bus~\cite{pcie}.}}
\changesI{A single page fault can cause multiple threads to stall at once,
\changesI{as threads} often access data in the same page due to data locality.
As a result, the page fault can significantly reduce the amount of TLP that 
the GPU can exploit, and the long latency of a 
page fault harms performance~\cite{tianhao-hpca16}.}

\changesI{Unlike address translation, which benefits from \changesI{\emph{larger}} pages, demand paging
benefits from \emph{smaller pages}.  Demand paging for large pages requires a
greater amount of data to be transferred over the system I/O bus during a page fault 
than for conventional base pages.  The larger data transfer size increases the
transfer \changesI{time significantly}, due to the long latency and limited
bandwidth of the system I/O bus.  This, in turn, significantly increases the
amount of time that GPU threads stall, and can further decrease the amount of
TLP.  To make matters worse, as the size of a page increases, there is a greater 
probability that an application does not need all of the data in the page.
As a result, threads may stall for a longer time without gaining \changesI{any further
benefit} in return.}




\paragraphbe{\changesI{Page Size Trade-Off.}}
\changesI{We find that memory virtualization in state-of-the-art GPU systems
has a fundamental trade-off due to the page size choice.}
\changesI{A \changesI{\emph{larger}} page size reduces address translation stalls by increasing 
TLB reach and reducing the number of high-latency TLB misses.
\changesI{In contrast, a \emph{smaller}} 
page size reduces demand paging stalls by decreasing
the amount of unnecessary data transferred over the system I/O 
bus~\cite{powers-hpca14, tianhao-hpca16}.}
\changesI{We can relax the page size trade-off} by using \emph{multiple} page sizes
\changesI{\emph{transparently}} to the application, and, thus, to the \changesI{programmer.}
\changesI{In a system that supports multiple page sizes, several base pages
that are \emph{contiguous in both virtual and physical memory}
can be \emph{coalesced} (i.e., combined) into a single large page, and a 
large page can be \emph{splintered} (i.e., split) into multiple base \changesI{pages.}}
\changesIII{With} multiple page sizes,
and \changesIIIII{the ability to change} \changesI{virtual-to-physical} mappings dynamically, the \changesI{GPU} system
can support good TLB reach \changesI{by using large pages for address translation}, while providing better
demand paging performance \changesI{by using base pages for data transfer}.

\changesI{Application-transparent} \changesI{support for multiple page sizes}
has \changesI{proven} challenging for CPUs~\cite{ingens, superpage}.
\changesI{A key property of memory virtualization is to enforce \emph{memory
protection}, where a distinct virtual address space (i.e., a \emph{memory
protection domain}) is allocated to
an individual application or \changesI{a} virtual machine,
and memory is shared \emph{safely} (i.e., only with explicit
permissions \changesI{for accesses} across \changesI{different} address spaces).
In order to \changesI{ensure} that memory protection guarantees are not violated,
coalescing operations can combine contiguous physical base pages into a single
physical large page \emph{only if all base pages 
belong to the same virtual address space}.

\changesI{Unfortunately, in} both CPU and state-of-the-art GPU memory managers, existing memory
access patterns and allocation mechanisms make it difficult to find
regions of physical memory where base pages can be coalesced.
We show an example of this in Figure~\ref{fig:tlb-mosaic-intro-base}, which
illustrates how a state-of-the-art GPU memory manager~\cite{powers-hpca14}
allocates memory for two applications.  Within a single \emph{large page frame} 
(i.e., a contiguous piece of
physical memory that is the size of a large page and whose starting address is
page aligned), the GPU memory manager allocates base pages from both
Applications~1 and 2 (\mycirc{1} in the figure).  
As a result, the memory manager \changesI{\emph{cannot}} coalesce the
base pages into a large page (\mycirc{2}) without first migrating 
\changesI{some of the base pages, which would incur} a
high latency.}

\begin{figure}[h!]%
\centering
\subfloat[\changesI{State-of-the-art GPU memory management~\cite{powers-hpca14}}.]{{
	\includegraphics[width=0.5\columnwidth]{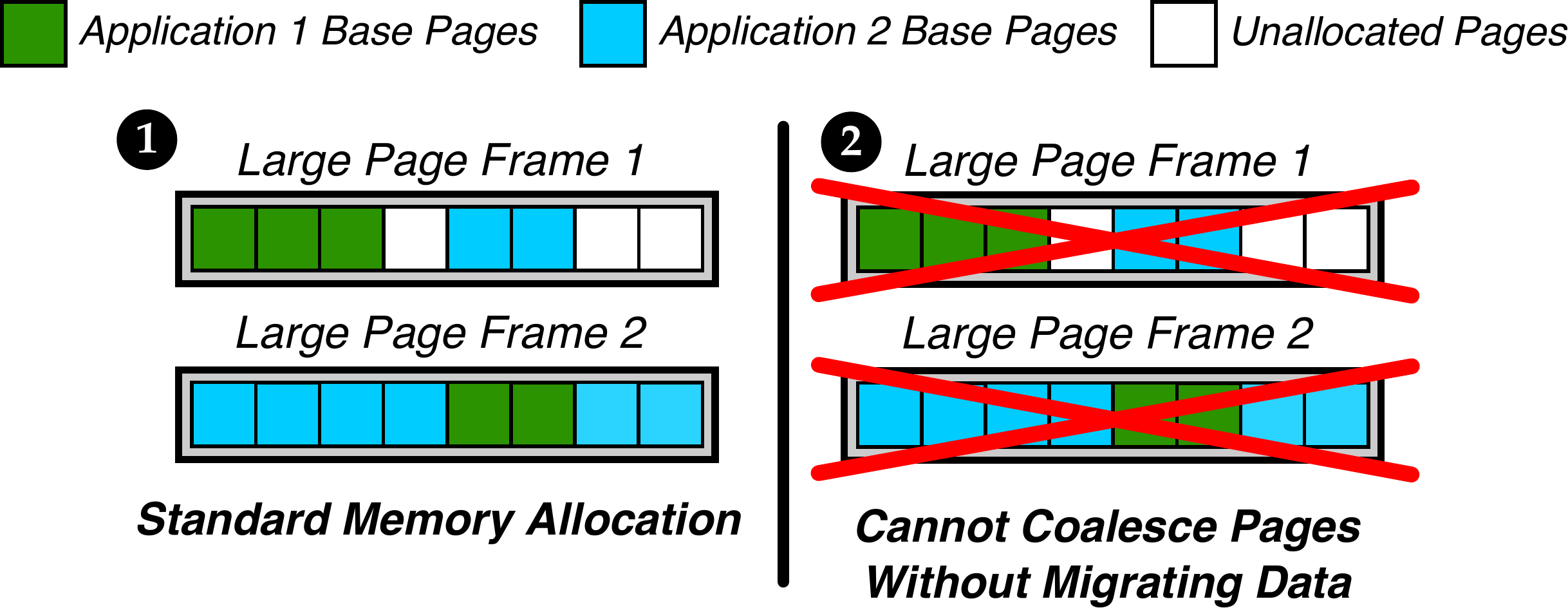}%
	\label{fig:tlb-mosaic-intro-base}
}}%
\subfloat[\changesI{Memory management with \titleShortMOSAIC.}]{{
	\includegraphics[width=0.396\columnwidth]{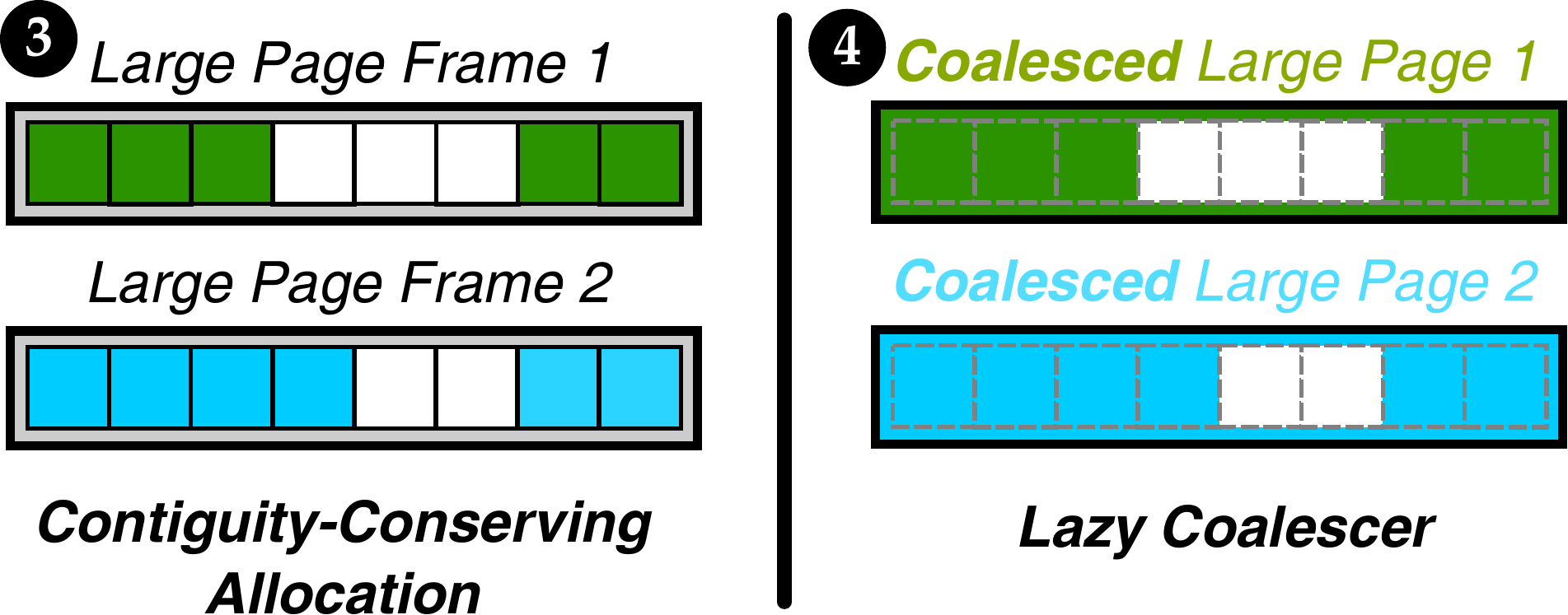}%
	\label{fig:tlb-mosaic-intro}
}}%
\caption{\changesI{Page allocation and coalescing behavior \changesI{of GPU memory managers: 
(a)~state-of-the-art~\cite{powers-hpca14}, (b)~\titleShortMOSAIC}.}}%
\end{figure}

\changesI{We make a \emph{key observation} about the memory behavior of
contemporary general-purpose GPU (GPGPU) applications.  The vast majority of
memory allocations in \changesIIIII{GPGPU applications} are performed \emph{en masse} (i.e., a large
number of pages are allocated at the same time).  The \emph{en masse} memory
allocation presents us with an opportunity: with so many pages being allocated
at once, we can rearrange how we allocate the \changesI{base pages} to ensure that
(1)~\changesI{\emph{all}} of the base pages allocated within a large page frame belong to the
\changesI{\emph{same}} virtual address space, and
(2)~base pages that are contiguous in virtual memory are allocated to a
contiguous portion of physical memory and aligned within the large page frame.
\emph{Our goal} in this work is to develop an application-transparent memory manager that
performs such memory allocation, and uses this allocation \changesIII{property} to
efficiently support multiple page sizes in order to improve TLB reach 
\changesI{\changesIIIII{and efficiently support} demand paging.}}

\changesI{To this end,} we present \titleShortMOSAIC, a \changesI{new} GPU memory manager that uses
our \changesI{key observation} to provide application-transparent support for 
multiple page sizes in GPUs while avoiding high overhead for
coalescing and splintering pages.
\changesI{The key idea of \titleShortMOSAIC is to (1)~transfer data to GPU memory 
at \changesI{the small base page (e.g., 4KB)} granularity, 
(2)~allocate physical base pages in a way that avoids the need to migrate data 
during coalescing, and 
(3)~use \changesIIIII{a} simple coalescing \changesIIIII{mechanism} to combine base pages into large pages \changesI{(e.g., 2MB)
and thus} increase TLB reach.}
Figure~\ref{fig:tlb-mosaic-intro} shows a high-level overview of 
\changesI{how \titleShortMOSAIC allocates and coalesces pages}.
\titleShortMOSAIC consists of three key design components:
(1)~\allocatorNameLong{} (\allocatorName{}), a memory allocator which provides a \emph{soft guarantee}
that \changesI{\emph{all}} of the base pages within the same large page range belong to only a single application
\changesI{(\mycirc{3} in the figure)};
(2)~\policyName{}, 
\changesI{a page size \changesIII{selection mechanism}}
\changesI{that \changesIII{merges base pages into a large page} immediately after allocation (\mycirc{4}),
and thus does \changesIII{\emph{not} need to monitor} base pages to make coalescing
decisions or migrate base pages;} and
(3)~\compactionNameLong{} (\compactionName{}), a memory compaction mechanism that transparently \changesI{migrates data} to avoid
\changesI{internal fragmentation within a large page frame},
\changesI{which frees} up large page \changesI{frames} for \allocatorName.

\paragraphbe{\changesI{Key Results.}}
We evaluate \titleShortMOSAIC using 235~workloads.
Each workload consists of multiple GPGPU applications from a wide range of
benchmark suites.  
Our evaluations show that compared to a contemporary GPU that uses only 4KB 
base pages, a GPU with \titleShortMOSAIC reduces address translation
overheads \changesI{while efficiently achieving the benefits of demand paging}, thanks to its
use of multiple page sizes.
When we compare to a \changesIII{GPU with a state-of-the-art memory manager}
\changesI{(see Section~\ref{sec:baseline-eval})}, we find that
\changesIII{a GPU with \titleShortMOSAIC} provides an average 
speedup of \changesI{55.5\% and 29.7\% for homogeneous and heterogeneous
multi-application workloads, respectively},
\changesI{and comes within 6.8\% and 15.4\% of the performance
\changesI{of \changesIII{a GPU with an ideal TLB,} where all TLB requests are hits}.}
\changesI{Thus, by alleviating the page size trade-off between address
translation and demand paging overhead, \titleShortMOSAIC improves the
efficiency and practicality of multi-application execution on the GPU.}

This chapter makes the following contributions:
\begin{itemize}

\item We analyze fundamental trade-offs \changesI{on choosing the correct
page size to optimize \changesI{both} address translation (which benefits from larger pages)
and demand paging (which benefits from smaller pages).} 
\changesI{Based on our analyses, we motivate} the need for 
application-transparent support of \changesI{\emph{multiple}} page sizes in a GPU.

\item We present \titleShortMOSAIC, \changesI{a \changesI{new} GPU memory manager 
 that \changesI{\emph{efficiently} supports} multiple page sizes.
\titleShortMOSAIC uses a novel \changesI{mechanism to allocate 
contiguous virtual pages to contiguous physical pages in the GPU memory,}
and exploits this property to 
\changesI{coalesce \changesIII{contiguously-allocated} base pages into a large page for address translation
with low overhead and no \changesIII{data} migration, while still using base pages during demand paging.}}

\item We show that \titleShortMOSAIC's application-transparent support 
for \changesI{multiple page sizes effectively} improves
TLB reach \changesI{while efficiently achieving the benefits of demand paging}.
\changesI{Overall, \titleShortMOSAIC improves the average performance of homogeneous
and heterogeneous multi-application workloads by 55.5\% and 29.7\%, respectively,
over a state-of-the-art GPU memory manager.}


\end{itemize}


\section{Background}
\label{sec:background}
\label{sec:bkgd}

We first provide necessary background on contemporary GPU architectures.
In Section~\ref{sec:bkgd:gpu}, we discuss the GPU execution model.
In Section~\ref{sec:bkgd:virt}, we discuss state-of-the-art support for GPU
memory virtualization.

\subsection{GPU Execution Model}
\label{sec:bkgd:gpu}

GPU applications use \emph{fine-grained multithreading}\changesI{~\cite{cdc6600,cdc6600-2,smith-hep,hep}}. \changesI{A GPU} application
is made up of thousands of threads.  These threads are clustered into \emph{thread
blocks} (also known as \emph{work groups}), where each thread block consists of
multiple smaller bundles of threads that execute concurrently.  Each such thread
bundle is known as a \emph{warp}, or a \emph{wavefront}.
\changesI{Each thread
within the warp executes the same instruction at the same program counter value.}
The GPU avoids stalls due to 
dependencies and long memory latencies by taking advantage of \emph{thread-level
parallelism} (TLP), where the GPU swaps out warps that have dependencies or are waiting
on memory with other warps that are ready to execute. 

A GPU consists of multiple \emph{streaming multiprocessors} (SMs), also known as
\emph{shader cores}. Each SM executes one warp at a time using the 
single-instruction, multiple-thread (SIMT) execution model\changesI{~\cite{largewarps,lindholm,dwf}}. Under SIMT, all of
the threads within a warp are executed in \emph{lockstep}.
Due to lockstep execution, a warp stalls when \emph{any one thread} within the warp \changesI{has} to
stall. \changesI{This} means that a warp is unable to proceed to the next 
instruction until the \emph{slowest} thread in the warp completes the current
instruction.

The GPU memory hierarchy typically consists of multiple levels of memory.
In contemporary GPU architectures, each SM has a private data cache,
and has access to one or more shared \changesIIIII{memory partitions} through an interconnect 
(typically a crossbar).
A \changesIIIII{\emph{memory partition}} combines a single slice of the banked L2 cache 
with \changesIII{a memory} controller that connects the GPU to off-chip
main memory (DRAM). \changesI{More detailed information about the GPU memory
hierarchy can be found in \changesIIIIII{\cite{osp-isca13,owl-asplos13,caba,zorua,
medic,toggle-hpca16, cpugpu-micro, adwait-critical-memsched,rachata-thesis,trogers-thesis,adwait-thesis}}.}

\subsection{Virtualization Support in GPUs}
\label{sec:bkgd:virt}

Hardware-supported memory virtualization relies on address translation to map
each virtual memory address to a physical address within the GPU memory. 
Address translation uses page-granularity virtual-to-physical mappings that are 
stored within a multi-level \emph{page table}.  To look up
a mapping within the page table, the GPU performs a page table \emph{walk},
where a page table walker traverses through each level of the page table in
main memory until the walker locates the \emph{page table entry} for the
requested mapping in the last level of the table.
GPUs with virtual memory support \changesIII{have} \emph{translation lookaside buffers} 
(TLBs), which cache page table entries and avoid the need to perform a page
table walk for the cached entries, thus reducing the address translation latency.

The introduction of address translation hardware into the GPU memory hierarchy 
puts TLB misses on the critical path of application execution, as a TLB miss invokes a page table walk 
that can stall multiple threads and degrade performance significantly. (We study the impact of 
TLB misses and page table walks in Section~\ref{sec:baseline-eval}.)  
A GPU uses multiple TLB levels \changesI{to reduce} the number
of TLB misses, typically including private per-SM L1 TLBs and a shared L2 
TLB~\cite{tianhao-hpca16,pichai-asplos14,powers-hpca14}.
Traditional address translation mechanisms perform memory mapping using a
\emph{base page} size of 4KB.
Prior work for \emph{integrated GPUs} (i.e., GPUs that are in the same package or die as the
CPU) has found that using a larger page size can improve address translation
performance by improving \emph{TLB reach} (i.e., the maximum \changesIII{fraction} of memory that can be
accessed using the cached TLB entries)~\cite{tianhao-hpca16, pichai-asplos14, 
powers-hpca14}. 
For a TLB that holds a fixed number of page table entries, using the \emph{large page} 
(e.g., a page with a size of 2MB or greater) as the granularity for mapping
greatly increases the TLB reach,
and thus reduces the TLB miss rate, compared to using the base page granularity.
While memory hierarchy designs for widely-used GPU architectures from
NVIDIA, AMD, and Intel are not publicly available, 
it is widely accepted that contemporary GPUs support TLB-based address translation
and, in some models, large page sizes~\cite{firepro,fermi,kepler,maxwell,gpu-arch-microbenchmarking}.
To simplify translation hardware in a GPU that uses multiple page sizes (i.e., 
both base pages and large pages), 
\changesI{we assume that} \changesIIIIII{each TLB level contains two separate 
sets of entries}\changesI{~\cite{rmm, jayneel-isca16, karakostas.hpca16, binh-colt, binh-micro15, prediction-tlb}}, where one \changesIIIIII{set of entries stores 
only} base page translations, while the other \changesIIIIII{set of entries stores
only} large page translations.

State-of-the-art GPU memory virtualization provides support for \emph{demand 
paging}~\cite{powers-hpca14,tianhao-hpca16,cc-numa-gpu-hpca15,huma,pascal}. 
In demand paging, all of the memory used by a GPU application does not need to
be transferred to \changesI{the} GPU memory at the beginning of application execution.  Instead,
during application execution, when a GPU thread issues a memory request to a
page that has not yet been allocated in \changesI{the} GPU memory, the GPU issues a \emph{page
fault}, at which point the data for that page is transferred over the off-chip system I/O bus 
(e.g., the PCIe bus~\cite{pcie} in contemporary systems)
from the CPU memory to the GPU memory. The transfer
requires a long latency due to its use of an off-chip bus.  Once the transfer
completes, the GPU runtime allocates a physical GPU memory \changesI{address} to the page,
and the thread can complete its memory request.
 

\section{A Case for Multiple Page Sizes}
\label{sec:motivation-mosaic}

Despite increases in DRAM capacity, TLB capacity (i.e., the number of
cached page table entries) has not kept pace, and \changesI{thus}
TLB reach has been declining.  As a result, address translation 
overheads have started to significantly increase the execution time of many 
large-memory workloads~\cite{direct-segment,jayneel-micro14,powers-hpca14,
pichai-asplos14,abhishek-ispass16,gpu-arch-microbenchmarking}.
In this section, we \changesI{(1)~}analyze how the address translation overhead changes if we
use large pages instead of base pages, and \changesI{(2)~}examine the 
advantages and disadvantages of both page sizes.

\subsection{Effect of Page Size on TLB Performance}
\label{sec:baseline-eval}


To quantify the performance trade-offs between base and large pages, we simulate a number
of recently-proposed TLB designs 
that support demand paging~\cite{powers-hpca14,tianhao-hpca16}
\changesIIIII{(see Section~\ref{sec:meth-mosaic} for our methodology)}. 
%
We slightly modify Power et al.'s TLB design~\cite{powers-hpca14}
to create our baseline, which we call \emph{GPU-MMU}.  
Power et al.\changesI{~\cite{powers-hpca14}} propose a GPU memory manager that has a private 128-entry L1 TLB for each SM~,
a highly-threaded page table walker, and a page walk cache\changesI{~\cite{powers-hpca14}}.
From our experiments, we find that using a shared L2 TLB instead of
a page walk cache increases the average performance \changesI{across our workloads (described in Section~\ref{sec:meth-mosaic})} by 14\% (not shown).
As a result, our GPU-MMU baseline design (shown in Figure~\ref{fig:tlb-mosaic-baseline})
omits the page walk cache in favor of a 512-entry shared L2 TLB.

\begin{figure}[h!]%
\centering
\includegraphics[width=\columnwidth]{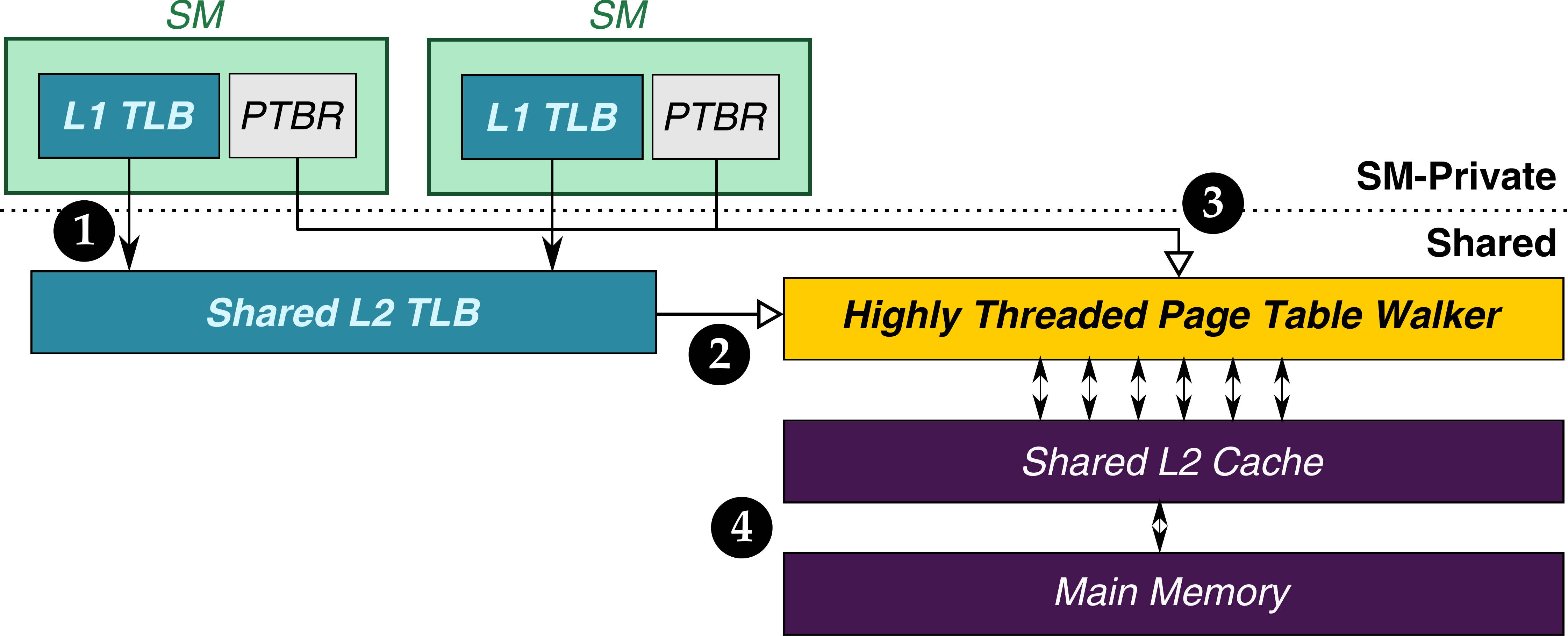}%
\caption{GPU-MMU baseline design with a two-level TLB.}%
\label{fig:tlb-mosaic-baseline}
\end{figure}


In our GPU-MMU baseline design, 
a shared L2 TLB entry is extended with address space \changesI{identifiers.
TLB} accesses from multiple threads to the same page are coalesced (i.e., combined).
On an L1 TLB miss (\mycirc{1} \changesI{in Figure~\ref{fig:tlb-mosaic-baseline}}), the shared L2 TLB is accessed. 
If the request misses in the shared L2 TLB, the page table walker begins a walk (\mycirc{2}).
The walker reads the \changesI{Page Table Base Register (PTBR)}\footnote{\changesI{CR3 in the x86 ISA~\cite{intelx86},
TTB in the ARM ISA\changesIII{~\cite{arm-cortexa}}.}} from the core that caused the TLB miss (\mycirc{3}),
which contains a pointer to the root of the page table.
The walker then accesses each level of the page table, retrieving the page table data 
from either the shared L2 cache or the GPU main memory (\mycirc{4}).




Figure~\ref{fig:mask-summary} shows the performance of two GPU-MMU designs:
(1)~a design that uses the base 4KB page size, and (2)~a design that
uses a 2MB large page size,
\changesI{where both designs have} \emph{no demand paging overhead} (i.e., \changesI{the} system I/O bus
transfer takes zero cycles \changesIII{to transfer a page}). We normalize
the performance of the two designs to \changesI{a GPU with an ideal TLB,} where \emph{all} TLB requests hit in the L1 TLB.
\changesI{We make two observations from the figure.}

\begin{figure}[h!]
\centering
  \includegraphics[width=\columnwidth]{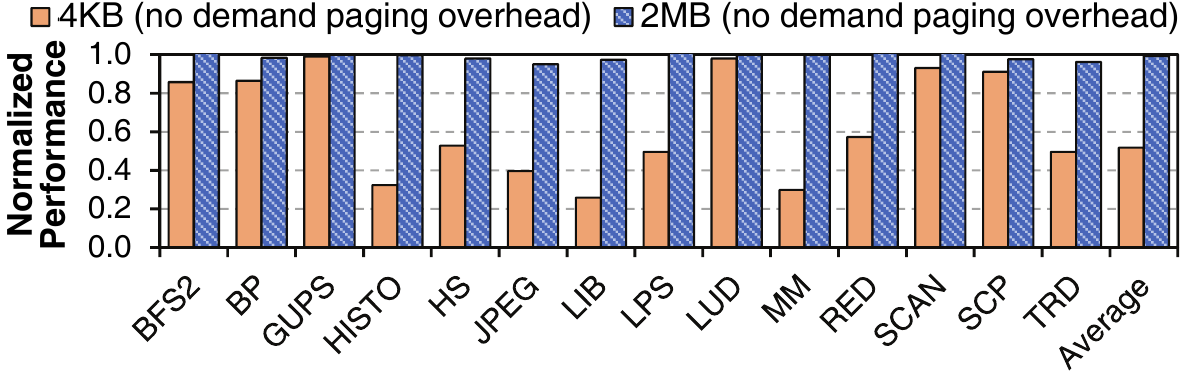}%
  \caption{\changesI{Performance of a GPU with \emph{no demand paging
   overhead}, using (1)~4KB base pages and
   (2)~2MB large pages,
   normalized to the performance of a GPU with an ideal TLB}.}%
  \label{fig:mask-summary}%
\end{figure}

\changesI{First, compared to the ideal TLB, 
the GPU-MMU with 4KB base pages experiences
an average performance \changesIII{loss} of 48.1\%. 
We observe that with 4KB base pages,
a single TLB miss often stalls \emph{many} of the warps, which undermines
the latency hiding behavior of the SIMT execution model used by GPUs.}
\changesI{Second, the figure shows that} 
using a 2MB page size with the same number of TLB entries 
as the 4KB design \changesI{allows applications to come within 2\% of the 
ideal TLB performance.  We find that with 2MB pages, the TLB has a much larger
reach, which
reduces the TLB miss rate substantially.}
Thus, there is strong incentive to use large
pages for address translation.

\subsection{Large Pages Alone Are Not the Answer}
\label{sec:motivation:largepageissues}

A natural solution to consider is to use \emph{only} large pages for GPU memory management.
Using only large pages would reduce address translation
overhead significantly, with minimal changes to the hardware or runtime.
Unfortunately, this solution is impractical because large pages (1)~\changesI{greatly} increase the \changesI{data transfer} size of
\changesI{\emph{each} demand paging request}, \changesI{causing contention on the system I/O bus, and} harming performance; and (2)~waste
memory \changesI{by causing memory} bloat \changesI{due to} internal fragmentation.


\paragraphbe{Demand Paging at a Large Page Granularity.} 
Following the nomenclature from \cite{tianhao-hpca16}, we denote GPU-side page 
faults \changesI{that} induce demand paging transfers across the system I/O bus as 
\emph{far-faults}.
Prior work 
observes that while
a 2MB large page size reduces the number of far-faults in GPU applications that
exhibit locality, the \emph{load-to-use latency} (i.e., the time between when a thread issues a
load request and when the data is returned to the thread) increases significantly
when a far-fault does occur~\cite{tianhao-hpca16}.
The impact of far-faults is particularly harmful for workloads with high locality,
as \emph{all} warps touching the 2MB \emph{large page frame} (i.e., a contiguous, page-aligned
2MB region of physical memory) must stall,
which limits the GPU's ability to overlap the system I/O bus transfer by executing other
warps.
Based on PCIe latency measurements from \changesI{a real GTX 1080 system~\cite{gtx1080}}, we determine that 
the load-to-use latency with 2MB large pages \changesI{(\SI{318}{\micro\second})} 
is six times the latency with 4KB base pages \changesI{(\SI{55}{\micro\second})}.


Figure~\ref{fig:overhead-page-size} shows how \changesIIIII{the} \changesI{GPU} performance changes when we use
different page sizes and include the effect of the demand paging overhead 
\changesI{(see Section~\ref{sec:meth-mosaic} for our methodology)}.
\changesI{We make three observations from the figure.
First, for 4KB base pages, the demand paging overhead reduces performance,
by an average of 40.0\% for our single-application workloads, and 82.3\% for
workloads with five concurrently-executing applications.}
\changesI{Second, for} our single-application workloads, we find that 
\changesI{with demand paging overhead, 2MB pages slow} down the 
execution time by an average of 92.5\% compared to \changesIII{4KB pages with demand paging}, as the GPU cores now spend
most of their time stalling on the system I/O bus transfers.
\changesI{Third, the} overhead \changesIII{\changesIIII{of} demand paging} \changesI{for larger pages} gets significantly 
worse as more applications share the GPU. With two 
applications concurrently executing on the GPU, the average performance degradation \changesIII{of demand
paging with 2MB pages instead of 4KB pages} is 98.0\%, and with five applications, the average \changesI{degradation} is 99.8\%.

 \begin{figure}[h]
   \centering
         \includegraphics[width=\columnwidth]{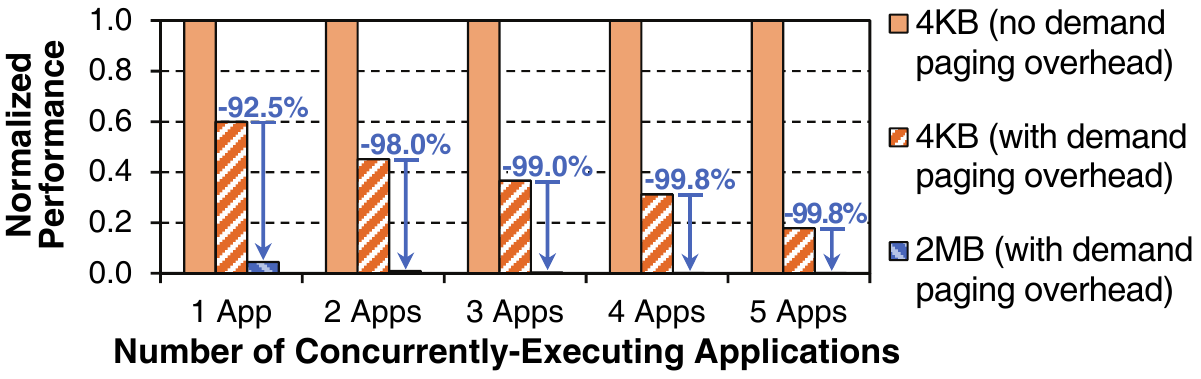}
   \caption{Performance impact of system I/O bus transfer during demand paging for base pages
   and large pages, normalized to base page performance \changesI{with no demand paging overhead}.}
   \label{fig:overhead-page-size}
 \end{figure}

\paragraphbe{\changesI{Memory} Bloat.} 
Large pages expose the system to internal fragmentation and \emph{memory
bloat}, \changesI{where a much greater amount of physical memory is allocated
than the amount of memory actually needed by an application}.
To understand the impact of memory bloat, 
we evaluate the amount of memory allocated to each application when run in
isolation, using 4KB and 2MB page sizes.
\changesI{When we use the 4KB base page size, our} applications have 
working sets ranging from 10MB to 362MB, with an average 
of 81.5MB (see Section~\ref{sec:meth-mosaic} and~\cite{mosaic-tech} for details). We
find that the amount of allocated memory inflates by 40.2\% on average, and up to
367\% in the worst case, when we use 2MB pages instead of 4KB pages (not shown). 
These numbers are likely conservative, as we expect that the 
fragmentation would worsen as an application continues to run
for longer time scales \changesI{than we can realistically simulate}. Such waste is unacceptable, particularly when there is
an increasing demand for GPU memory \changesI{due to other concurrently-running
applications}. 

\changesI{We conclude that despite the potential performance gain of 2MB large
pages (when the overhead of demand paging is ignored), the demand paging overhead
actually causes 2MB large pages to perform \changesIII{\emph{much worse}} than 4KB base pages.  As a
result, it is impractical to use \emph{only} 2MB large pages in the GPU.}
Therefore, a design that delivers the best of both page sizes is needed.

\subsection{Challenges for Multiple Page Size Support}
\label{sec:challenges}

As \changesI{Sections~\ref{sec:baseline-eval} and \ref{sec:motivation:largepageissues} demonstrate}, 
we cannot \changesIII{efficiently optimize} GPU performance by employing only a single page size.
Recent works on TLB design for integrated GPUs~\cite{pichai-asplos14,powers-hpca14}
and on GPU demand paging support~\cite{powers-hpca14,tianhao-hpca16,cc-numa-gpu-hpca15,huma,pascal} 
corroborate our own findings on the performance cost of address translation 
and the performance opportunity of \changesI{large pages}.
\changesI{\textbf{Our goal}} is to \changesIIIII{design} a \changesIII{new} memory \changesIIIII{manager}
for GPUs that \changesI{efficiently} supports \emph{multiple page sizes}, to exploit the benefits
of both \changesI{small} and large page sizes, \changesIII{while avoiding the disadvantages of each}. In order to \changesI{(1)~}\changesIII{not burden programmers} 
and \changesI{(2)~}provide performance improvements for legacy applications, 
we \changesI{would like to enable} multiple page size support \changesI{\emph{transparently to
the application}}. This \changesI{constraint} introduces several design challenges that must be taken
into account.


\paragraphbe{Page Size Selection.} 
\label{sec:mosaic-back-transparent}
While conceptually simple, multiple page size support
introduces complexity for memory management that has \changesI{traditionally} been difficult to 
\changesIII{handle}.  Despite architectural support within \changesI{CPUs~\cite{ingens, superpage}} for several decades, 
the adoption of multiple page sizes has been \changesI{quite} slow \changesI{and application-domain specific~\cite{direct-segment,jayneel-micro14}.}  
The availability of \changesI{large} pages can either be \changesI{exposed to application programmers, or} managed
\changesI{transparently} to an application.
Application-exposed management forces programmers to reason about 
physical memory and use specialized APIs~\cite{win-huge,osx-huge} for page management, which \changesI{usually sacrifices code} portability and \changesI{increases} programmer burden. 
In contrast, application-transparent support (e.g., management by the OS) requires no changes to existing programs
to use large pages, but it does require the memory manager to make \emph{predictive} decisions about whether applications 
\changesI{would} benefit from large pages. OS-level large page management remains an 
active research area~\cite{ingens,superpage},
and the optimization guidance for many
modern applications continues to advise strongly \emph{against} using large
pages~\cite{Mongodb-nothp,couchbase-nothp,dokudb-nothp,redis-nothp,nuodb-nothp,sap-nothp,splunk-nothp,voltdb-nothp},
due to \changesIII{high-latency data transfers over the system I/O bus 
and \changesIIII{memory bloat} (as \changesIIIII{described} in Section~\ref{sec:motivation:largepageissues})}. 
In order to provide effective \changesI{application-transparent} support for multiple page sizes in GPUs, we must develop
a policy for selecting page sizes that \changesIII{avoids high-latency data transfer over the system
I/O bus, and does not introduce significant memory bloat}.


\paragraphbe{Hardware Implementation.}
Application-transparent support for multiple page sizes requires (1)~primitives
that implement the transition between different page sizes, and (2)~mechanisms
\changesI{that} create and preserve contiguity in both the
virtual and physical address \changesI{spaces}.
\changesI{We must add support in the GPU to} \emph{coalesce} (i.e., combine) multiple base
pages into a single large page, and \emph{splinter} (i.e., split) a large page back into multiple
base pages. 
\changesI{While the GPU memory manager can \changesIII{\emph{migrate}} base pages in order to
create opportunities for coalescing, base page migration incurs a high latency overhead\changesIII{~\cite{seshadri2013rowclone,lisa}}.
In order to avoid the migration overhead without sacrificing coalescing opportunities, 
the GPU needs to initially \changesIII{\emph{allocate}} data in a \changesIII{\emph{coalescing-friendly manner.}}}


GPUs face additional implementation challenges 
over CPUs, as they
rely on hardware-based memory allocation mechanisms and management. 
In CPU-based \changesI{application-transparent} large page management,
coalescing and splintering are performed by the operating \changesI{system~\cite{ingens,superpage}}, which can 
(1)~use locks and inter-processor interrupts (IPIs) to implement atomic updates to page tables, 
(2)~\changesIII{stall any accesses to the} virtual addresses whose mappings are changing, and 
(3)~use background threads to perform \changesI{coalescing} and \changesI{splintering}.
GPUs currently have no mechanism to
atomically move pages or change page mappings for \changesI{coalescing or splintering}.



\section{Mosaic}
\label{sec:mechanism-mosaic}
\label{sec:mech}

In this section, we \changesI{describe} \titleShortMOSAIC, a GPU memory manager
that provides application-transparent support for multiple page sizes and solves the challenges
that we discuss in Section~\ref{sec:challenges}.  At runtime, \titleShortMOSAIC
(1)~\emph{allocates} memory in the GPU such that base pages that are contiguous
in virtual memory \changesIII{are} contiguous within a large page frame in physical memory
(which we call \emph{contiguity\changesIIII{-}\changesIII{conserving allocation}}; \changesIII{Section~\ref{sec:mech:allocation}});
(2)~\emph{coalesces} base pages into a large page frame as soon as the data is
allocated, only if all of the pages are \changesIII{\emph{i)}~}contiguous in both virtual and physical 
memory, and \changesIII{\emph{ii)}~}belong to the same application \changesIII{(Section~\ref{sec:mech:coalescing})}; and
(3)~\emph{compacts} a large page \changesIII{(i.e., moves the allocated base 
pages within the large page frame to make them contiguous)} if internal fragmentation within the page is
high after \changesIII{one of its constituent base pages} is deallocated \changesIII{(Section~\ref{sec:mech:compaction})}.

\subsection{High-Level Overview of Mosaic}
\label{sec:mech:highlevel}
\label{sec:mosaic-overall-design}

Figure~\ref{fig:mosaic-overall} shows the major components of \titleShortMOSAIC, and
how they interact with the GPU memory.  
\titleShortMOSAIC consists of three components: \allocatorNameLong (\allocatorName),
the \policyName, and \compactionNameLong (\compactionName).
These three components work together to coalesce (i.e., combine) and splinter
(i.e., split apart) base pages to/from large pages during memory management.
Memory management operations for 
\titleShortMOSAIC take place at two times:
(1)~when memory is \emph{allocated}, and
(2)~when memory is \emph{deallocated}.

\begin{figure*}[h!]
  \centering
  \includegraphics[width=\columnwidth]{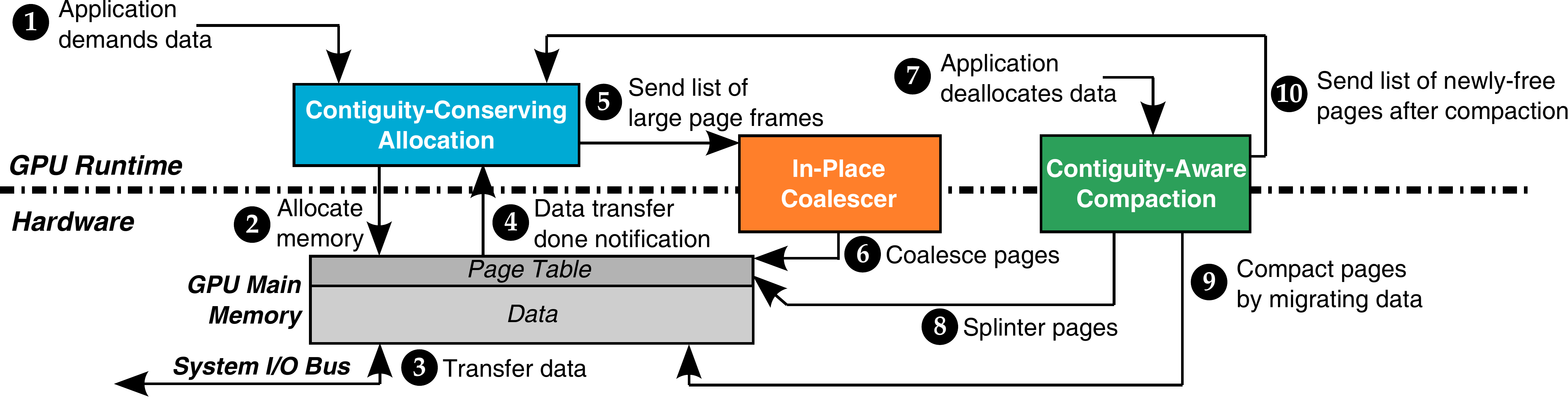}%
  \caption{High-level overview of \titleShortMOSAIC, showing how and when its three 
  components interact with the GPU memory.}
  \label{fig:mosaic-overall}
\end{figure*}

\paragraphbe{Memory Allocation.}
When a GPGPU application wants to access data that is not currently in the GPU
memory, it sends a request to the GPU runtime (e.g., OpenCL, CUDA runtimes) to
transfer the data from the CPU memory to the GPU memory (\mycirc{1} in 
Figure~\ref{fig:mosaic-overall}).
A GPGPU application typically allocates a large number of base pages at the same time.
\allocatorName allocates space within the GPU memory (\mycirc{2})
for the base pages, \changesIII{working to conserve the contiguity of base 
pages\changesIIII{,} if possible during allocation}.  Regardless of contiguity, 
\allocatorName provides a \emph{soft guarantee} that a single large 
page frame contains base pages from only a single application.
Once the base page is allocated, \allocatorName initiates the data transfer 
across the system I/O bus (\mycirc{3}).
When the data transfer is complete (\mycirc{4}), \allocatorName notifies 
the \policyName that allocation is done \changesIIII{by sending a list of the large page
frame addresses that were allocated} (\mycirc{5}).
For each \changesIIII{of these large page frames}, the \changesIII{runtime portion of the}
\policyName then checks to see whether 
(1)~\emph{all} base pages \changesIIIII{within the large page frame} have been allocated, and
(2)~\changesIII{the base pages within the large page frame are contiguous in
both virtual and physical memory}.
If both conditions are true, the \changesIII{hardware portion of the \policyName 
updates} the page table to coalesce the base pages into a large page 
\changesIII{(\mycirc{6})}.

\paragraphbe{Memory Deallocation.}
When a GPGPU application wants to deallocate memory (e.g., when an application
kernel finishes), it sends a deallocation request to the GPU runtime (\mycirc{7}).
For all deallocated base pages that are coalesced into a large page, the runtime 
invokes \compactionName for the corresponding large page.  \changesIII{The runtime portion of}
\compactionName checks to see whether the large page has a high degree of
\emph{internal fragmentation} (i.e., if the number of unallocated base pages
within the large page exceeds a predetermined threshold).
For each large page with high internal fragmentation, \changesIII{the hardware
portion of \compactionName updates} the
page table to splinter the large page back into its constituent base pages \changesIII{(\mycirc{8})}.
Next, \compactionName compacts the splintered large page frames, by migrating 
data from multiple splintered large page frames into a single large page frame
(\mycirc{9}).  Finally, \compactionName notifies \allocatorName of the large
page frames that are now free after compaction (\mycirctwo{10}), which 
\allocatorName can use for future memory allocations.

\subsection{Contiguity-Conserving Allocation}
\label{sec:mech:allocation}
\label{sec:allocator-sec}

\changesIII{Base pages can be coalesced into a large page frame} only if
(1)~all base pages within the frame are contiguous in both virtual and physical
memory, 
(2)~the data within the large page frame is page aligned with the corresponding
large page within virtual memory (i.e., the first base page within the large 
page frame is also the first base page of a virtual large page), and
(3)~all base pages within the frame come from the same virtual address space
(e.g., the same application, or the same virtual machine).
As Figure~\ref{fig:tlb-mosaic-intro-base} shows, traditional memory managers allocate base pages
without conserving contiguity or ensuring that the base pages within a large page 
frame belong to the same application.  For example, if the memory manager wants
to coalesce \changesIII{base pages of Application~1 into a large page frame
(e.g., Large Page Frame~1)}, it must first migrate 
Application~2's base pages to \changesIII{\emph{another}} large page frame, and may need to migrate
some of Application~1's \changesIII{base pages within the} large page frame to create contiguity.  Only after this
data migration, the \changesIII{base pages would be ready to be coalesced into
a large page frame}.

In \titleShortMOSAIC, we minimize the overhead of coalescing pages by designing
\allocatorName to take advantage of the memory allocation behavior of 
GPGPU applications.  Similar to many data-intensive
applications~\cite{yak,gay-pldi1998}, GPGPU
applications typically allocate memory \emph{en masse} (i.e., they allocate
a large number of pages at a time).  The \emph{en masse} allocation takes place
when an application kernel is about to be launched, and the allocation requests
are often for a large contiguous region of virtual memory.  This region is
much larger than the large page size (e.g., 2MB), and \titleShortMOSAIC allocates 
multiple page-aligned 2MB portions of contiguous virtual memory from the 
region to large page frames in physical memory, as shown in
Figure~\ref{fig:tlb-mosaic-intro}. 
With \allocatorName, the large page frames for Application~1 and Application~2
are ready to be coalesced as soon as their base pages are allocated,
\changesIII{\emph{without}} the need for any \changesIIII{data migration.}
\changesIII{For all other base pages (e.g., base pages not aligned in the
virtual address space, allocation requests that are smaller than a large page),}
\titleShortMOSAIC simply allocates these pages to any free page, \changesIII{and does
not exploit any contiguity}.

\titleShortMOSAIC provides a \emph{soft guarantee} that \emph{all} base pages within a
large page frame belong to the same application, which reduces the cost of 
performing coalescing and compaction, and ensures that these operations do not
violate memory protection.  To meet this guarantee during allocation, 
\allocatorName needs to track the application that each large page frame with
unallocated base pages is assigned to.
The allocator maintains two sets of lists to track this information:
(1)~the \emph{free frame list}, a list of free large page frames (i.e., frames 
where no base pages have been allocated) that are not yet assigned to any
application; and
(2)~\emph{free base page lists}, per-application lists of free base pages within large page 
frames where some (but not all) base pages are allocated.
When \allocatorName allocates a page-aligned 2MB region of virtual memory, 
it takes a large page frame from the free frame list and maps the virtual 
memory region to the frame.  When \allocatorName allocates \changesIII{base pages
\changesIIII{in a manner such that it cannot exploit} contiguity}, it takes a page from 
the free base page list \changesIIIII{for the application performing the memory request,}
to ensure that the soft guarantee is met.  If the free base page 
list for an application is empty, \allocatorName removes a large page frame from
the free frame list, and adds the frame's base pages to the free base page list.

Note that there may be cases where the free frame list runs out of large page
frames for allocation.  We discuss how \titleShortMOSAIC handles such situations in
Section~\ref{sec:mech:compaction}.

\subsection{In-Place Coalescer}
\label{sec:mech:coalescing}
\label{sec-mech-coalesce}

In \titleShortMOSAIC, \changesIII{due to \allocatorName (Section~\ref{sec:mech:allocation}),} 
we find that we can simplify how \changesIIII{the page size is selected for each
large page frame (i.e., \changesIIIII{decide} which pages should be coalesced),}
compared to state-of-the-art memory managers.  In state-of-the-art
memory managers, such as our GPU-MMU baseline based on Power et 
al.~\cite{powers-hpca14}, there is no guarantee that \changesIIIII{base} pages within a large page frame
belong to the \changesIII{\emph{same}} application, and memory allocators do not conserve
virtual memory contiguity in physical memory.  As a result, \changesIII{state-of-the-art memory} managers must 
perform four steps to coalesce pages, as shown under the \emph{Baseline} timeline in 
\changesIIIII{Figure~\ref{fig:coalescing-timeline}a}.
First, the manager must identify opportunities for coalescing \changesIIII{\emph{across}} multiple pages
(not shown in the \changesIIIII{timeline}, as this can be performed in the background).
This is done by a hardware memory management unit (MMU), such as the Falcon
coprocessor in recent GPU architectures~\cite{falcon}, which tallies page
utilization information from the page table entries of each base page.  The
most-utilized contiguous base pages are chosen for coalescing \changesIII{(\emph{Pages A--G} in
Figure~\ref{fig:coalescing-timeline}a)}.
Second, the manager must identify a large page frame where the \changesIII{coalesced} base pages
will reside, and then migrate the base pages to this new large page frame,
which uses DRAM channel bandwidth \changesIII{(\mycirc{1} in the figure)}.
Third, the manager must update the page table entries \changesIII{(PTEs)} to reflect the coalescing,
which again uses DRAM channel bandwidth \changesIII{(\mycirc{2})}.
Fourth, the manager invokes a TLB flush to invalidate stale virtual-to-physical 
mappings (which point to the base page locations \changesIIIII{prior to} migration), during which
the SMs stall \changesIII{(\mycirc{3})}.
\changesIII{\changesIIII{Thus, coalescing using a state-of-the-art memory manager causes} significant DRAM channel utilization and SM
\changesIIIII{stalls}, as Figure~\ref{fig:coalescing-timeline}a shows.}

\begin{figure*}[h!]
\centering
\includegraphics[width=\textwidth]{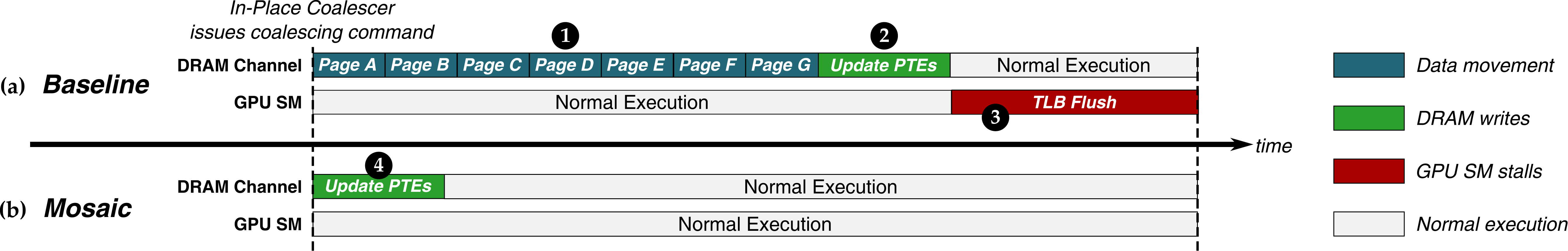}%
\caption{Coalescing timeline for \changesIII{(a)}~GPU-MMU baseline and for \changesIII{(b)}~\titleShortMOSAIC.} 
\label{fig:coalescing-timeline}
\end{figure*}

\changesIII{In contrast, \titleShortMOSAIC can perform coalescing in-place, i.e.,
base pages do \changesIIII{\emph{not}} need to be migrated in order to be coalesced into a large
page.  Hence, we call the \changesIIII{page size selection mechanism} of \titleShortMOSAIC the
\policyName.  As shown in Figure~\ref{fig:coalescing-timeline}b, the \policyName
causes much less DRAM channel utilization and no SM stalls, saving significant 
waste compared to state-of-the-art memory managers.
We} describe how the \policyName (1)~decides which pages to coalesce, and
(2)~updates the page table for pages that are coalesced.

\paragraphbe{\changesIIII{Deciding When to Coalesce.}}
Unlike existing memory managers, \titleShortMOSAIC does not need to monitor \changesIII{base} page 
utilization information to identify opportunities for coalescing.  Instead, we
design \allocatorName to ensure that the \changesIII{base} pages that we coalesce are
already allocated to the same large page frame.
Once \allocatorName has allocated data within a large page frame, it sends the
address of the frame to the \policyName.  The \policyName then checks to see
whether the base pages within the frame are contiguous in both virtual and
physical memory.\footnote{\changesIII{Coalescing decisions are made purely in
the software runtime portion of the \policyName, and thus system designers can
easily use a different coalescing policy\changesIIIIII{,} if desired.}}
As mentioned in Section~\ref{sec:mech:allocation}, \titleShortMOSAIC coalesces
\changesIII{base pages into a large page only if all of the base pages within 
the large page frame} are
allocated (i.e., the \changesIII{frame is} fully populated).  We \changesIII{empirically} find that for GPGPU 
applications, coalescing only \changesIII{contiguous base pages in} fully-populated large page frames 
achieves similar TLB reach to the coalescing performed by
existing memory managers \changesIII{(not shown)}, and avoids the need to employ an MMU or perform
page migration, which greatly reduces the overhead of \titleShortMOSAIC.

\paragraphbe{\changesIII{Coalescing in Hardware}.}
Once the \policyName selects a large page frame for coalescing, it then 
\changesIII{performs the coalescing operation in hardware}.  
\changesIII{Figure~\ref{fig:coalescing-timeline}b} shows the steps required for coalescing
with the \policyName under the \titleShortMOSAIC timeline.  Unlike coalescing in 
existing memory managers, the \policyName does \changesIIII{\emph{not}} need to perform any data
migration, as \allocatorName has already conserved contiguity within all large
page frames selected for \changesIII{coalescing: the} coalescing 
operation needs to only update the page table entries corresponding to the 
large page frame \changesIII{and the base pages (\mycirc{4} in the figure)}.

We modify the L3 and L4 page table entries (PTEs) to simplify updates during the 
coalescing operation, as shown in Figure~\ref{fig:pte}.  We add a \changesIII{\emph{large page bit}}
to each L3 PTE \changesIIII{(corresponding to a large page)}, which is initially \changesIII{set to 0} (to indicate a page 
that is \changesIII{\emph{not}} coalesced), and we add a \changesIII{\emph{disabled bit} to} each L4 PTE \changesIIII{(corresponding to a base page)},
which is initially \changesIII{set to 0} (to indicate that a page table walker should use the
base page virtual-to-physical mapping in the L4 PTE).  The \changesIII{coalescing hardware}
simply needs to locate the L3 PTE for the large page frame being
coalesced \changesIII{(\mycirc{1} in the figure)}, and set the \changesIII{\emph{large page} 
bit to 1} for the PTE \changesIII{(\mycirc{2})}.  (We discuss how page table
lookups occur below.)  \changesIII{We perform this bit setting operation atomically}, with a
single memory operation \changesIII{to minimize the amount of time before the
large page mapping can be used}.  Then, the \changesIII{coalescing hardware}
sets the \changesIII{\emph{disabled} bit to 1} for all L4 PTEs \changesIII{(\mycirc{3})}.

\begin{figure}[h!]
  \centering
  \subfloat[Page table entries.\label{fig:pte}]{\includegraphics[width=0.59\columnwidth]{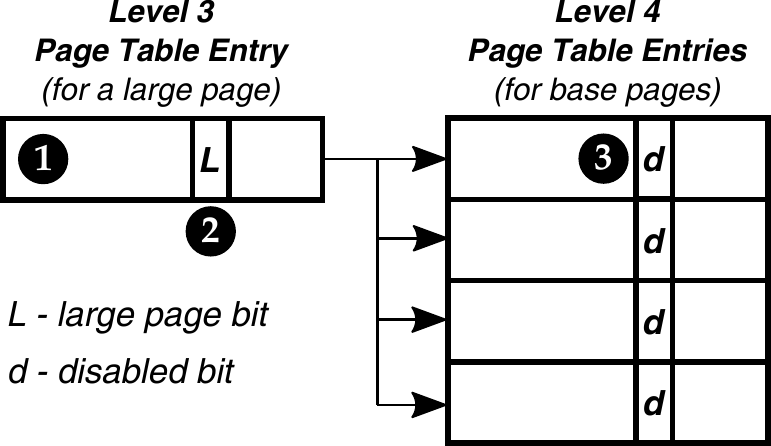}}%
  \hfill%
  \subfloat[Virtual-to-physical mappings \changesIIIII{in an L4 PTE}.\label{fig:map}]{\includegraphics[width=0.38\columnwidth]{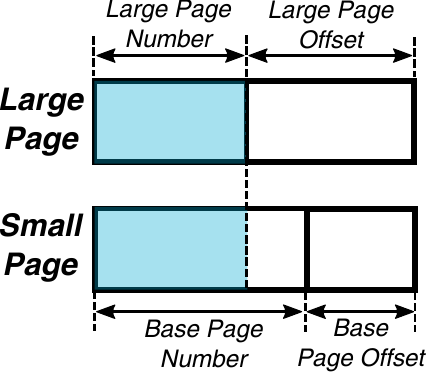}}%
  \caption{L3 and L4 page table structure in \titleShortMOSAIC.} 
  \label{fig:coalescing-sample}
\end{figure}

\changesIII{The} virtual-to-physical mapping for the large page
can be used as soon as the large page bit is set, \changesIII{without}
(1)~\changesIII{waiting for} the disabled bits in the L4 PTEs to be set, or
(2)~\changesIII{requiring a TLB flush to remove the base page mappings \changesIIII{from the TLB}}.
This is because no migration was needed to coalesce the \changesIII{base pages
into the large page}.  As a result,
the existing virtual-to-physical mappings for the coalesced base pages \changesIIIII{still
point} to the correct memory \changesIIII{locations}.  While we set the disabled bits
in the \changesIIII{PTEs} to discourage using these mappings, as the mappings consume 
\changesIIIIII{a portion of the limited number of base page entries in the TLB,}
we can continue to use the mappings \changesIII{\emph{safely}}
until they are evicted from the TLB.  
\changesIII{As shown in Figure~\ref{fig:coalescing-timeline}b, since we do not
flush the TLB, we do not need to stall the SMs.}
\changesIIIIII{\titleShortMOSAIC ensures that if the coalesced page is subsequently splintered,
the large page virtual-to-physical mapping is removed (see 
Section~\ref{sec:mech:compaction}).\footnote{\changesIIIII{As there is a chance 
that base pages within a splintered page can migrated during compaction, the
large page virtual-to-physical mapping may no longer be valid. To avoid
correctness issues when this happens, \titleShortMOSAIC flushes \changesIIIIII{the TLB large page entry} for the
mapping as soon as a coalesced page is splintered.}}}

As we can see from Figure~\ref{fig:coalescing-timeline}, the lack of data 
migration and TLB flushes in \titleShortMOSAIC greatly \changesIIIII{reduces} the time required for 
the coalescing operation in \titleShortMOSAIC, with respect to coalescing in
existing MMUs.

\paragraphbe{TLB Lookups After Coalescing.}
\changesIIIIII{As mentioned in Section~\ref{sec:bkgd:virt}, each TLB level contains two
separate sets of entries, with one set of entries for each page size.  
In order to improve TLB reach, we need to ensure that an SM does not 
fetch the \emph{base} page PTEs for coalesced base pages (even
though these are safe to use) into the TLBs, as these PTEs
contend with PTEs of uncoalesced base pages for the limited TLB space.
\changesIIIII{When a GPU with \titleShortMOSAIC needs to translate a
memory address, it first \changesIIIIII{checks if the address belongs to a
coalesced page by looking up the TLB large page entries.}}}
\changesIIIII{If} \changesIII{the SM locates} a valid large page entry for the request
\changesIIIIII{(i.e., the page is coalesced)},
\changesIII{it avoids} looking up \changesIIIIII{TLB base page entries}.

If a TLB miss occurs \changesIIIIII{in both the TLB large page and base page entries} for a coalesced page, the page walker traverses the page 
table.  At the L3 PTE \changesIII{(\mycirc{1} in Figure~\ref{fig:pte})}, 
the walker reads the large page bit \changesIII{(\mycirc{2})}. As the bit is set,
the walker \changesIIII{needs to read the virtual-to-physical mapping for the
large page.  The L3 PTE does \changesIIIII{\emph{not}} typically contain space for a 
virtual-to-physical mapping, so the walker instead reads the virtual-to-physical
mapping from the first PTE of the L4 page table} that the L3 PTE points 
to.
Figure~\ref{fig:map} shows why we can use the mapping in the L4 PTE
for the large page.  
A virtual-to-physical mapping for a large page consists of
a page number and an offset.  As the base pages within the large page were not
migrated, their mappings point to physical memory locations within the large
page frame.  As a result, if we look at only the bits of the mapping used for 
the large page number, they are identical for both the large page mapping and
the base page mapping.  When the large page bit is set, the page walker reads
\changesIII{the} large page number from the L4 PTE \changesIII{(along with other 
fields of the PTE, e.g., for access permissions)}, and returns the \changesIIII{PTE} to the
TLB.  In doing so, we do not need to allocate any extra storage for the 
virtual-to-physical mapping of the large page.  Note that for pages that are \changesIII{\emph{not}}
coalesced, the page walker behavior is \changesIII{\emph{not}} modified.

\subsection{Contiguity-Aware Compaction}
\label{sec:mech:compaction}
\label{sec:mosaic-motiv-fragmentation}

After an application kernel finishes, it can deallocate some of the base pages
that it previously allocated.  This deallocation can lead to internal 
fragmentation within a large page frame that was coalesced, as some of \changesIIIII{the frame's}
constituent base pages are no longer valid.  While the page \changesIII{could still benefit} from
coalescing \changesIII{(\changesIIIII{as this improves} TLB reach)}, the unallocated base pages within the large page frame cannot be
allocated to another virtual address as long as the page remains coalesced.
If significant memory fragmentation exists, this can cause \allocatorName to
run out of free large page frames, even though it has not allocated all of the
available base pages in GPU memory.  To avoid an out-of-memory error in the
application, \titleShortMOSAIC uses \compactionName to \changesIII{splinter and 
compact highly-fragmented large page frames, freeing up large page frames for
\allocatorName to use}.

\paragraphbe{Deciding When to \changesIIIII{Splinter and} Compact a \changesIII{Coalesced} Page.}
Whenever an application deallocates a base page within a coalesced large page frame,
\compactionName checks to see how many base pages remain allocated within the
frame.  If the number of allocated base pages falls below a predetermined 
threshold (which is configurable in the \changesIII{GPU} runtime), \compactionName decides to 
splinter the large page frame into base pages (see below).  Once the splintering
operation completes, \compactionName \changesIII{performs compaction by
migrating} the remaining base pages to
another uncoalesced large page frame that belongs to the same application.
In order to avoid occupying multiple memory channels \changesIII{while 
performing this migration}, which can hurt the 
performance of other threads that are executing concurrently, we restrict
\compactionName to migrate base pages between only large page frames that
reside within the \changesIII{\emph{same}} memory channel.
After the migration is complete, the original large page frame no longer contains
any allocated base pages, and \compactionName sends the address of the large
page frame to \allocatorName, which adds the address to its free frame list.

If the number of allocated base pages within a coalesced page is greater than 
or equal to the threshold, \compactionName does \changesIII{\emph{not}} splinter the page, but
notifies \allocatorName of the \changesIII{large page} frame address.  \allocatorName then stores the 
coalesced \changesIII{large page frame}'s address in a \emph{\changesIII{emergency} frame list}.  As a failsafe, if \allocatorName
runs out of free large pages, and \compactionName does not have any large pages
that it can compact, \allocatorName pulls a coalesced page from the \changesIII{emergency} 
frame list, asks \compactionName to splinter the page, and then uses any
unallocated base pages within the \changesIII{splintered} large page frame to allocate new virtual
base pages.

\paragraphbe{Splintering the Page in Hardware.}
Similar to the \policyName, when \compactionName selects a coalesced page for
splintering, it \changesIII{then performs the splintering operation in hardware}.
The splintering operation essentially reverses the coalescing operation.  
First, the splintering \changesIII{hardware} clears the disabled bit in the L4 PTEs of the
constituent base pages.  Then, the \changesIII{splintering hardware} clears the large page bit atomically,
which causes \changesIIII{the} subsequent page table \changesIIII{walks} to look up the virtual-to-physical
mapping for the base page.  Unlike coalescing, when the \changesIII{hardware} splinters a
\changesIIIII{coalesced} page, it must also issue a TLB flush request \changesIIIII{for the coalesced page}.  As we discuss in
Section~\ref{sec:mech:coalescing}, a large page mapping can be present in the
\changesIIIIII{TLB} only when a page is coalesced.  The flush \changesIIIIII{to the TLB removes
the large page entry for this mapping}, to ensure synchronization across all SMs with the
current state of the page table.

\paragraphbe{Optimizing Compaction with Bulk Copy Mechanisms.}
The migration of \changesIII{each base page} during compaction requires several long-latency
memory operations, where the contents of the page are copied \changesIII{to a destination location} \changesIIIII{only 64 bits} 
at a time, due to the \changesIII{narrow} width of the memory channel~\cite{seshadri2013rowclone,donghyuk-ddma,vivek-chapter}.
To optimize the performance of \compactionName, we can take advantage of 
in-DRAM bulk copy techniques such as RowClone~\cite{seshadri2013rowclone,vivek-chapter}
or LISA~\cite{lisa}, which provide \changesIII{very low-latency (e.g., 
\SI{80}{\nano\second}}\changesIIII{)} memory copy within a single \changesIII{DRAM module}.
These mechanisms use existing internal buses within DRAM to copy an entire
base page of memory \changesIII{with a single bulk} memory operation.  While \changesIII{such bulk data} copy mechanisms
are not essential \changesIII{for our proposal}, they have the potential to improve performance when a large
amount of compaction takes place.
\changesIII{Section~\ref{sec:mosaic-eval-worst-case} evaluates} the benefits of using in-DRAM
bulk copy with \compactionName.


\section{Methodology}
\label{sec:meth-mosaic}
\label{sec:mosaic-mech}

We modify the MAFIA framework~\cite{mafia}, which uses GPGPU-Sim 3.2.2~\cite{gpgpu-sim},
to evaluate \changesIIIII{\titleShortMOSAIC} on a GPU that concurrently
executes multiple applications. 
We have released our simulator modifications~\cite{mosaic.github, safari.github}.
Table~\ref{table:config} \changesIII{shows the system \changesIIII{configuration} we simulate
for our evaluations, \changesIIIII{including the configurations of the GPU core and memory partition 
(see Section~\ref{sec:bkgd:gpu})}.}


\begin{table}[h!]
\begin{footnotesize}
\centering
\begin{tabular}{ll}
         \cmidrule(rl){1-2} 
\multicolumn{2}{c}{\textbf{\changesIII{GPU Core Configuration}}} \\ 
        \cmidrule(rl){1-2}\morecmidrules\cmidrule(rl){1-2}
\textbf{Shader Core Config}           &  30 cores\changesIII{,} 1020 MHz, GTO warp scheduler~\cite{ccws}\\
        \cmidrule(rl){1-2}
\textbf{Private L1 Cache}    &  16KB, 4-way associative, LRU, L1 misses are \\ & coalesced before accessing L2, 1-cycle latency \\
        \cmidrule(rl){1-2} 
\textbf{Private L1 TLB}    &  \changesIIIII{128 base page/16 large page} entries per core,\\
 & fully associative, LRU, single port, 1-cycle latency \\ 
        \cmidrule(rl){1-2}
& \\
\cmidrule(rl){1-2}
\multicolumn{2}{c}{\textbf{\changesIII{Memory Partition Configuration}}}\\
\multicolumn{2}{c}{\changesIIIII{(6 memory partitions in total, with each partition accessible by all 30~cores)}} \\
        \cmidrule(rl){1-2}\morecmidrules\cmidrule(rl){1-2}
\textbf{Shared L2 Cache}   &  2MB total, 16-way associative, LRU, 2 cache banks and \\ & 2 ports per memory partition, 10-cycle latency \\
        \cmidrule(rl){1-2} 
\textbf{Shared L2 TLB}   &  \changesIIIII{512 base page/256 large page} entries, \changesIIII{non-inclusive,}\\ & 16-way/fully-associative (base page/large page), LRU, \\ & \changesIII{2 ports,} 10-cycle latency \\
        \cmidrule(rl){1-2} 
        \textbf{DRAM}   & 3GB GDDR5, 1674 MHz, 6 channels, 8 banks per rank, \\ & FR-FCFS scheduler~\cite{fr-fcfs,frfcfs-patent}, burst length 8\\
         \cmidrule(rl){1-2} 
    \end{tabular}%
\caption{Configuration of the simulated system.}
  \label{table:config}%
\end{footnotesize}%
\end{table}%

\paragraphbe{\changesIIIII{Simulator Modifications.}}
\changesIIIII{We modify GPGPU-Sim~\cite{gpgpu-sim} to model the behavior of Unified
Virtual Address Space~\cite{fermi}. We add a memory allocator into
cuda-sim, the CUDA simulator within GPGPU-Sim, to handle all 
virtual-to-physical address translations and to provide memory protection.
We add an accurate model of address translation to GPGPU-Sim, including TLBs,
page tables, and a page table walker.  The page table walker is shared across
all SMs, and allows up to 64~concurrent walks.  Both the L1 and
L2 TLBs have separate entries for base pages and large 
pages~\cite{rmm, jayneel-isca16, karakostas.hpca16, binh-colt, binh-micro15, prediction-tlb}.
Each TLB contains miss status holding registers \changesIIIIII{(MSHRs)~\cite{kroft-isca81}} to track in-flight 
page table walks.
Our simulation infrastructure supports demand paging, by detecting page faults
and faithfully modeling the system I/O bus (i.e., PCIe) latency
based on measurements from NVIDIA GTX 1080 cards~\cite{gtx1080}
(see Section~\ref{sec:motivation:largepageissues}).\footnote{Our 
experience with the NVIDIA GTX 1080 suggests that production \changesIIIII{GPUs} 
perform significant prefetching to reduce latencies when reference patterns 
are predictable. This feature is not modeled \changesIII{in our simulations}.}
We use a worst-case model for the performance of our compaction mechanism
(\compactionName, see Section~\ref{sec:mosaic-motiv-fragmentation}) 
conservatively, by stalling the \emph{entire} GPU \changesIII{(all SMs)} and 
flushing the pipeline.  More details about our modifications can be found in
our extended technical report~\cite{mosaic-tech}.}

\paragraphbe{Workloads.}
We evaluate the performance of \titleShortMOSAIC using both \emph{homogeneous} and 
\emph{heterogeneous} workloads.
\changesIII{We categorize each workload based on the number of 
concurrently-executing applications, which ranges from one to five for our
homogeneous workloads, and from two to five for our heterogeneous workloads.}
We form our homogeneous workloads using multiple copies of the same
application. We build 27~homogeneous workloads \changesIII{for each category} 
using GPGPU applications from the Parboil~\cite{parboil}, SHOC~\cite{shoc},
LULESH~\cite{lulesh,lulesh-origin}, Rodinia~\cite{rodinia}, and CUDA
SDK~\cite{cuda-sdk} suites. We form our heterogeneous workloads by \changesIII{randomly} selecting a
number of \changesIIII{applications} out of these 27~GPGPU applications.
\changesIII{We build 25 heterogeneous workloads per category.}
Each workload has a combined working set size that ranges from 10MB to 2GB.
The average working set size of a workload is 217MB. \changesIIIII{In total
we evaluate 235~homogeneous and heterogeneous workloads.}
\changesIIII{We provide a list of all our workloads in our extended technical report~\cite{mosaic-tech}.}


\paragraphbe{Evaluation Metrics.} 
We report workload performance using the weighted speedup metric~\cite{harmonic_speedup,ws-metric2},
\changesIIIIIII{which is a commonly-used metric to evaluate the performance of a
multi-application workload\changesIII{~\cite{sms,cpugpu-micro,tcm,atlas,parbs,stfm,mcp,aergia,Reetu-MICRO2009}}.}
Weighted speedup is calculated as:
\begin{equation}
    \text{Weighted Speedup} = \sum{\frac{IPC_{shared}}{IPC_{alone}}}
\end{equation}
where
$IPC_{alone}$ is the IPC of an application in the workload that runs on the same number of
shader cores using the baseline state-of-the-art
configuration~\cite{powers-hpca14}, but does not share GPU resources with any
other applications; and $IPC_{shared}$ is the IPC of the application when
it runs concurrently with other applications. 
We report the performance of each application within a workload using IPC.



\paragraphbe{Scheduling and Partitioning of Cores.} As scheduling is not the
focus of this work, we assume that SMs are equally partitioned
across the applications within a workload,
and use the greedy-then-oldest (GTO) warp scheduler~\cite{ccws}. 
We speculate that if we use other scheduling or partitioning policies, \titleShortMOSAIC
would still increase the TLB reach and achieve the benefits of demand paging effectively,
though we leave such studies for future work.

\section{Evaluation}
\label{s:eval}

In this section, we evaluate how \titleShortMOSAIC improves the performance of
homogeneous and heterogeneous workloads  (see Section~\ref{sec:meth-mosaic} 
for more detail).  We compare \titleShortMOSAIC to two mechanisms:
(1)~\emph{GPU-MMU}, a \changesIIII{baseline GPU with a state-of-the-art memory manager} based \changesIIII{on}
the work by Power et al.~\cite{powers-hpca14}, which we explain in detail in 
Section~\ref{sec:baseline-eval}; and
(2)~\emph{Ideal TLB}, a GPU with an ideal TLB, where every address 
translation request hits in the L1 TLB (i.e., there are no TLB misses).



\subsection{Homogeneous Workloads}
\label{s:eval-homo}

Figure~\ref{fig:homo-mosaic-eval} shows the performance of \titleShortMOSAIC \changesIIII{for the}
homogeneous workloads.  
We make two observations from the figure.
First, we observe that \titleShortMOSAIC{} is able to recover most of the performance lost \changesIIII{due to}
the overhead of address translation \changesIIII{(i.e., an ideal TLB)} in homogeneous workloads.  
Compared to the GPU-MMU baseline, \titleShortMOSAIC improves the performance by 55.5\%,
averaged across all \changesIIII{135} of our homogeneous workloads.
The performance of \titleShortMOSAIC comes within 6.8\% of the \changesIIIII{\emph{Ideal TLB}} performance, 
indicating that \titleShortMOSAIC is effective at extending the TLB reach.
\changesIIIII{Second, we observe that \titleShortMOSAIC provides good scalability.
As we increase the number of concurrently-executing applications, 
we observe that the performance of \titleShortMOSAIC remains close to
the \changesIIIII{\emph{Ideal TLB}} performance.}


\begin{figure}[h!]
\centering
  \includegraphics[width=\columnwidth]{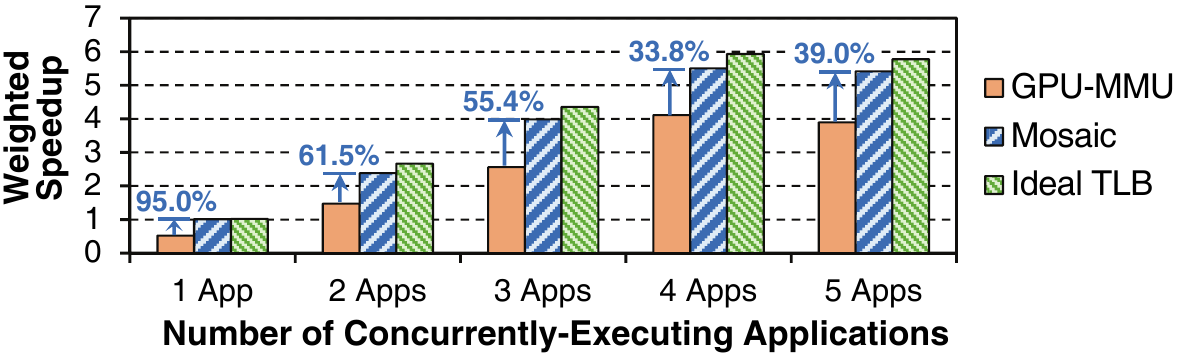}%
  \caption{Homogeneous workload performance of the GPU memory managers as
we vary the number of concurrently-executing applications in each workload.}%
  \label{fig:homo-mosaic-eval}
\end{figure}






We conclude that for homogeneous workloads, \titleShortMOSAIC effectively approaches
the performance of a GPU with \changesIIII{\changesIIIII{the \emph{Ideal TLB}}, by employing multiple page sizes to
simultaneously increase the reach of both the L1 private TLB and the shared L2 TLB.} 

\subsection{Heterogeneous Workloads}
\label{s:eval-hetero}

Figure~\ref{fig:mosaic-eval} shows the performance of \titleShortMOSAIC for
heterogeneous workloads that consist of multiple randomly-selected 
GPGPU applications. From the figure, we observe that on average across all
of the workloads, \titleShortMOSAIC{}
provides a performance improvement of 29.7\% over GPU-MMU, and
comes within 15.4\% of the \changesIIIII{\emph{Ideal TLB}} performance. We find that the
improvement comes from the significant reduction in the TLB miss 
rate with \titleShortMOSAIC, as we discuss below.

\begin{figure}[h!]
\centering
  \includegraphics[width=\columnwidth]{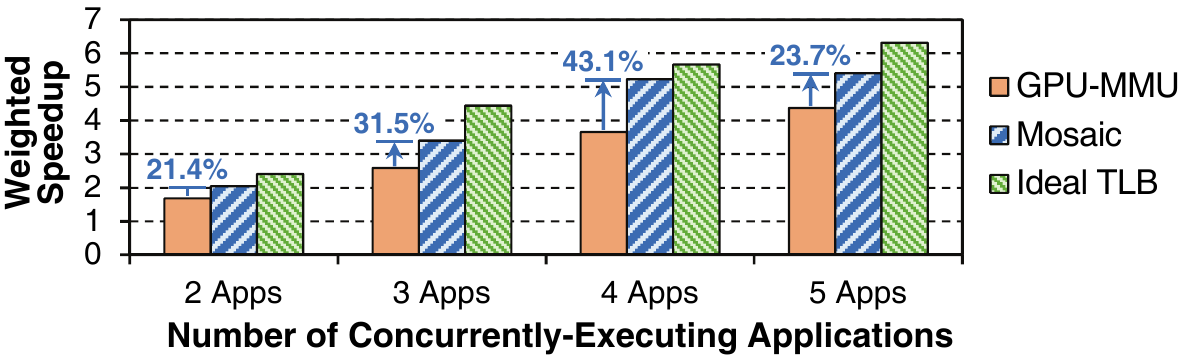}%
  \caption{Heterogeneous workload performance of the GPU memory managers.}%
  \label{fig:mosaic-eval}
\end{figure}


The performance gap between \titleShortMOSAIC and \changesIIIII{\emph{Ideal TLB}} is greater for
heterogeneous workloads than it is for homogeneous workloads.
To understand why, we examine the performance of the \changesIIIII{each workload}
in greater detail.
Figure~\ref{fig:mosaic-indiv} shows the performance improvement
of 15~\changesIIIII{randomly-}selected two-application workloads. \changesIIII{We categorize the workloads
as either \emph{TLB-friendly} or \emph{TLB-sensitive}.
The majority of the workloads are TLB-friendly,
which means that they} benefit from utilizing
large pages.  \changesIIII{The TLB hit rate increases significantly with \titleShortMOSAIC
(see \changesIIIII{Section~\ref{sec:eval:tlb}}) for TLB-friendly workloads, allowing the workload performance
to approach \changesIIIII{\emph{Ideal TLB}}.}
However, \changesIIII{for TLB-sensitive workloads, such as HS--CONS and NW--HISTO,
there is still a performance gap between \titleShortMOSAIC and the \changesIIIII{\emph{Ideal TLB}}, even
though \titleShortMOSAIC improves the TLB hit rate.}
We discover two main factors that lead to this performance gap. 
First, in these workloads, one of the applications is highly sensitive to
shared L2 TLB misses (e.g., HS in HS--CONS, HISTO in NW--HISTO),
while the other application (e.g., CONS, NW) is memory intensive.
The memory-intensive application introduces a high number of conflict misses 
on the shared L2 TLB, which harms the performance of the TLB-sensitive
application significantly, and causes the workload's performance under
\titleShortMOSAIC to drop significantly below the \changesIIIII{\emph{Ideal TLB}} performance.
Second, the \changesIIII{high latency of page walks due to} compulsory TLB
misses and \changesIIII{higher access latency to the shared L2 TLB (which increases because
TLB requests have to probe both the large page and base page TLBs)} have a high impact on the
TLB-sensitive application.  Hence, for these workloads, \changesIIIII{the \emph{Ideal TLB}} still has
significant advantages over \titleShortMOSAIC.

\begin{figure}[h!]
\centering
  \includegraphics[width=\columnwidth]{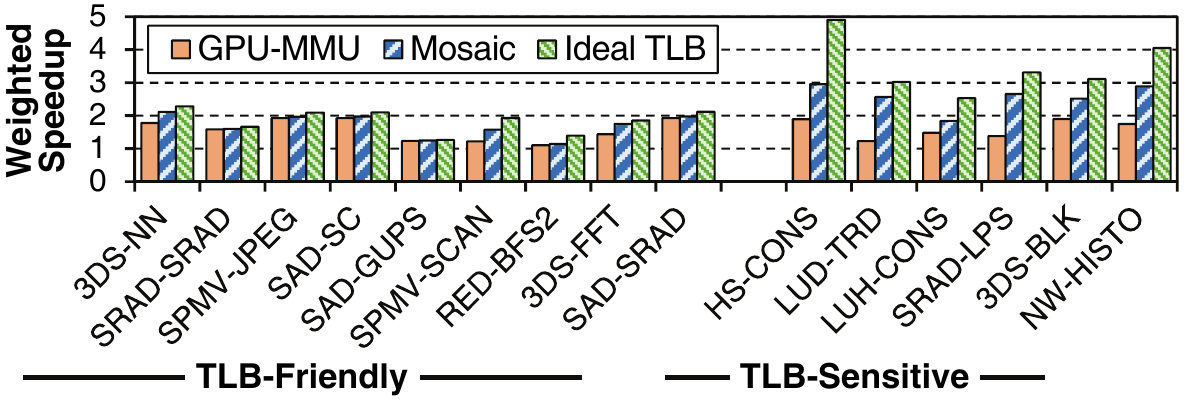}%
  \caption{Performance of selected two-application heterogeneous workloads.}%
  \label{fig:mosaic-indiv}
\end{figure}

\paragraphbe{Summary of Impact on Individual \changesIIII{Applications}.}
To determine how \titleShortMOSAIC affects \changesIIII{the individual applications within the heterogeneous workloads we evaluate,
we study the IPC of each application in all of our heterogeneous workloads.
In all, this represents a total of \changesIIII{350} individual applications.}
Figure~\ref{fig:mosaic-s-curve} shows \changesIIIII{the} per-application IPC of \titleShortMOSAIC and \changesIIIII{\emph{Ideal TLB}} normalized to the application's performance
under GPU-MMU, and sorted in ascending order. 
\changesIIIII{We show four graphs in the figure, where each graph corresponds
to individual applications from workloads with the same number of 
concurrently-executing applications.}
We make three observations from these secs/mosaic/results.
First, \titleShortMOSAIC improves performance
relative to GPU-MMU for \changesIIII{93.6\% of the \changesIIIII{350} individual applications}.
We find that the application IPC relative to the baseline
GPU-MMU for each application ranges from \changesIIII{66.3\% to 860\%, with an average of
133.0\%}. \changesIIII{Second, for the 6.4\% of the applications where \titleShortMOSAIC performs worse than GPU-MMU, we find that for each application, the other concurrently-executing
applications in the same workload experience a significant performance improvement. For example, the
worst-performing application, for which \titleShortMOSAIC hurts performance by 33.6\% compared to GPU-MMU,
is from a workload with three concurrently-executing applications. We find that the other two 
applications perform 66.3\% and 7.8\% better under \titleShortMOSAIC, compared to GPU-MMU.}
\changesIIII{Third, we find that\changesIIIII{,} on average across all heterogeneous workloads,} 48\%, 68.9\% and 82.3\% of the 
\changesIIII{applications} perform within 90\%, 80\% and 70\% of
\changesIIIII{\emph{Ideal TLB}}, respectively.

 \begin{figure}[h!]
   \centering
      \subfloat[\changesIIIIII{2 concurrent apps.}]{{
         \includegraphics[width=0.45\columnwidth]{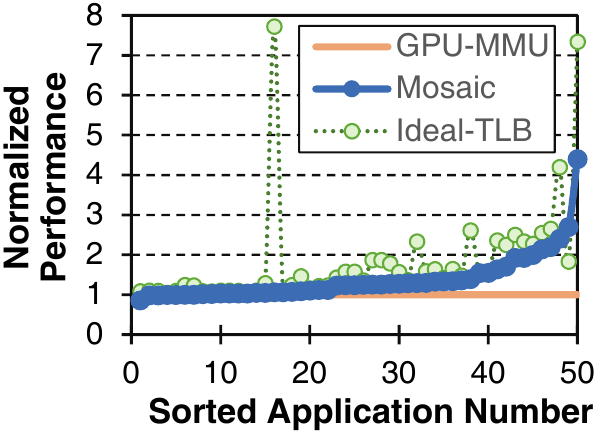}
         \label{fig:mosaic-s-curve-baseline}
     }}%
      \subfloat[\changesIIIIII{3 concurrent apps.}]{{
         \includegraphics[width=0.45\columnwidth]{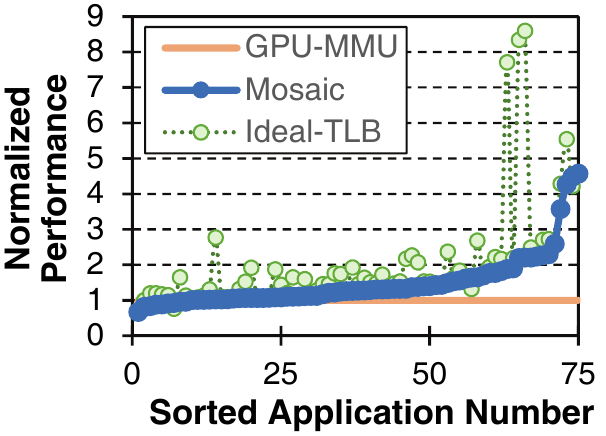}
         \label{fig:mosaic-s-curve-ideal}
     }}%
     \hfill%
      \subfloat[\changesIIIIII{4 concurrent apps.}]{{
         \includegraphics[width=0.45\columnwidth]{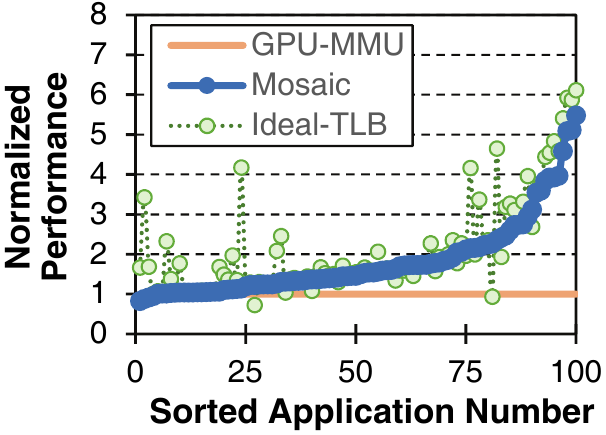}
         \label{fig:mosaic-s-curve-baseline}
     }}%
      \subfloat[\changesIIIIII{5 concurrent apps.}]{{
         \includegraphics[width=0.45\columnwidth]{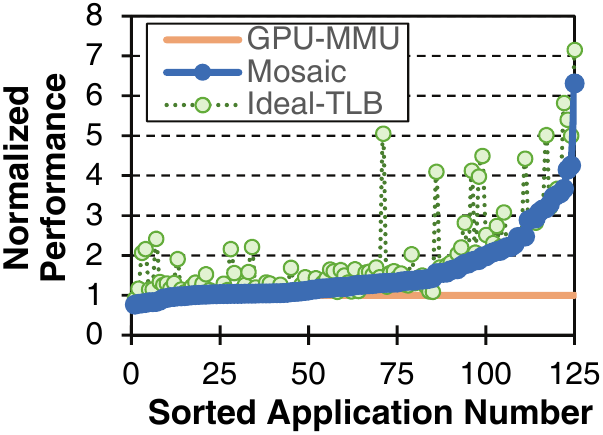}
         \label{fig:mosaic-s-curve-ideal}
     }}%
   \caption{\changesIIII{Sorted normalized per-application IPC \changesIIIIII{for applications in heterogeneous workloads, categorized by the number of applications in a workload}.}}
   \label{fig:mosaic-s-curve}
 \end{figure}

We conclude that \titleShortMOSAIC is effective at increasing the TLB reach for
heterogeneous workloads, and delivers significant performance improvements
over a state-of-the-art GPU memory manager.

\paragraphbe{\changesIIII{Impact of Demand Paging on Performance.}} 
\changesIIII{%
All of our secs/mosaic/results so far show the
performance of the GPU-MMU baseline and \titleShortMOSAIC when demand paging is \changesIIIII{\emph{enabled}}.
Figure~\ref{fig:paging} shows the normalized weighted speedup of the GPU-MMU
baseline and \titleShortMOSAIC, compared to GPU-MMU \emph{without demand paging}, 
\changesIIIII{where all data required by an application is moved to the GPU
memory \changesIIIIII{\emph{before}} the application starts executing}. 
We make two
observations from the figure. First, we find that \titleShortMOSAIC outperforms GPU-MMU without
demand paging by 58.5\% on average for homogeneous workloads and 47.5\% on average for heterogeneous workloads}. 
Second, we find that demand paging has little impact on the weighted speedup.
This is because demand paging latency occurs only when a kernel launches, at which
point the GPU retrieves data from the CPU memory.  The data transfer overhead is
required regardless of whether demand paging is enabled, and thus the GPU incurs
similar overhead with and without demand paging.}

\begin{figure}[h!]
\centering
  \includegraphics[width=\columnwidth]{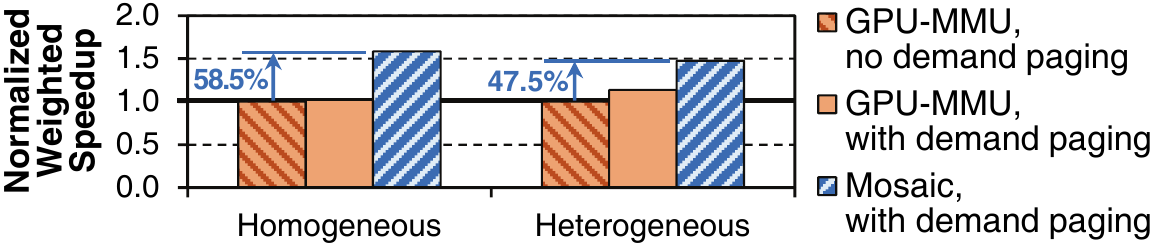}%
  \caption{Performance of GPU-MMU and \titleShortMOSAIC compared to GPU-MMU without demand paging.}%
  \label{fig:paging}
\end{figure}

\subsection{\changesIIII{Analysis of TLB Impact}}
\label{sec:eval:tlb}

\paragraphbe{\changesIIII{TLB Hit Rate.}}
Figure~\ref{fig:mosaic-eval-tlb}
compares the overall TLB \changesIIII{hit rate} of GPU-MMU to \titleShortMOSAIC{}
for \changesIIIII{214 of our 235~workloads, which} suffer from \emph{limited TLB reach}
(i.e., workloads that have \changesIIIII{an L2} TLB hit rate lower than 98\%).
We make two observations from the figure. First, we observe
\titleShortMOSAIC is very effective at increasing the TLB reach of these workloads.
\changesIIII{We find that for the GPU-MMU baseline, \emph{every} fully-mapped 
large page frame contains pages from multiple applications, as the GPU-MMU allocator
does not provide the soft guarantee of \allocatorName.  As a result, GPU-MMU does not have
any opportunities to coalesce base pages into a large page without performing
significant \changesIIIII{amounts of data} migration.
In contrast}, \titleShortMOSAIC can coalesce \changesIIIIII{a} vast majority of \changesIIIII{base pages} thanks to \allocatorName.
As a result, \titleShortMOSAIC reduces the TLB miss rate dramatically for these workloads,
with the average miss rate falling below 1\% in both the L1 and \changesIIIII{L2} TLBs.
Second, we observe an increasing amount of interference in
GPU-MMU when more \changesIIII{than three} applications are running concurrently.
This secs/mosaic/results in a lower TLB hit rate as the number of applications increases \changesIIII{from three to four applications,
and from four to five applications}.
The \changesIIIII{L2} TLB hit rate drops from 81\% in workloads with two \changesIIIII{concurrently-executing} applications
to 62\% in workloads with five \changesIIIII{concurrently-executing} applications. 
\titleShortMOSAIC experiences no such drop due to interference as we increase
the number of \changesIIIII{concurrently-executing} applications, \changesIIIII{since} it makes
much greater use of large page coalescing and enables a much larger
TLB reach.

\begin{figure}[h!]
\centering
  \includegraphics[width=\columnwidth]{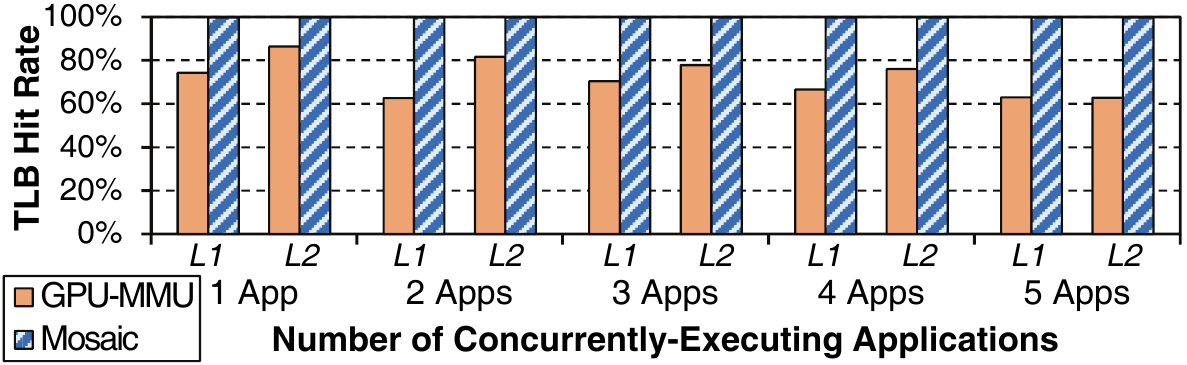}%
  \caption{\changesIIII{\changesIIIII{L1 and L2 TLB hit} rate for GPU-MMU and \titleShortMOSAIC.}}
  \label{fig:mosaic-eval-tlb}
\end{figure}

\paragraphbe{TLB Size Sensitivity.} 
A major benefit of \titleShortMOSAIC is its ability to improve TLB reach by increasing
opportunities to coalesce base pages into a large page.  
\changesIIII{After the base pages are coalesced, the GPU uses the large page TLB to
cache the virtual-to-physical mapping of the large page, which frees up base page
TLB entries so that they can be used to cache mappings for the uncoalesced 
base pages.}
We now evaluate
how sensitive the performance of \titleShortMOSAIC is to \changesIIII{the number of 
base page and large page entries in each TLB level}.

Figure~\ref{fig:mosaic-tlb-sweep} \changesIIII{shows} the performance of 
both GPU-MMU and \titleShortMOSAIC as we vary the number of 
\changesIIII{\emph{base page entries}} in the per-SM L1 TLBs 
\changesIIII{(Figure~\ref{fig:mosaic-l1-tlb})} and in the shared L2 TLB 
\changesIIII{(Figure~\ref{fig:mosaic-l2-tlb})}.
We normalize all secs/mosaic/results to the GPU-MMU performance with the baseline
\changesIIII{128-base-page-entry} L1 TLBs per SM and a 
\changesIIII{512-base-page-entry} shared L2 TLB.
From the figure, \changesIIII{we make two observations.  
First, we find that for the L1 TLB, GPU-MMU is sensitive to the
number of base page entries, while \titleShortMOSAIC is \emph{not sensitive} to the
number of base page entries.  This is because \titleShortMOSAIC successfully coalesces
most of its base pages into large pages, which significantly reduces the 
pressure on TLB base page capacity.  In fact, the number of L1 TLB base page
entries has a minimal impact on the performance of \titleShortMOSAIC until we scale \changesIIIII{it}
all the way down to 8~entries.  Even then, compared to an L1 TLB with 128~base
page entries, \titleShortMOSAIC loses only 7.6\% performance on average with 8~entries.
In contrast, we find that GPU-MMU is unable to coalesce base pages, and as a 
result, its performance scales \changesIIIII{poorly as we reduce the} number of TLB base page entries.
Second, we find that the performance of both GPU-MMU and \titleShortMOSAIC is 
sensitive to the number of L2 TLB base page entries.  This is because even
though \titleShortMOSAIC does not need many L1 TLB base page entries per SM, 
the base pages are often \textbf{shared} across multiple SMs.  The L2 TLB allows SMs
to share page table entries (PTEs) with each other, so that once an SM retrieves
a PTE from memory using a page walk, the other SMs do not need to wait on 
a page walk.  The larger the number of L2 TLB base page \changesIIIII{entries, the} more
likely it is that a TLB request can avoid the need for a page walk.} 
\changesIIIII{Since \titleShortMOSAIC does not directly have an effect on the number of page walks, it
benefits from a mechanism (e.g., a large L2 TLB) that can reduce the number of page walks
and hence is sensitive to the size of the L2 TLB.}

\begin{figure}[h!]
  \centering
    \subfloat{{
        \includegraphics[width=0.47\columnwidth]{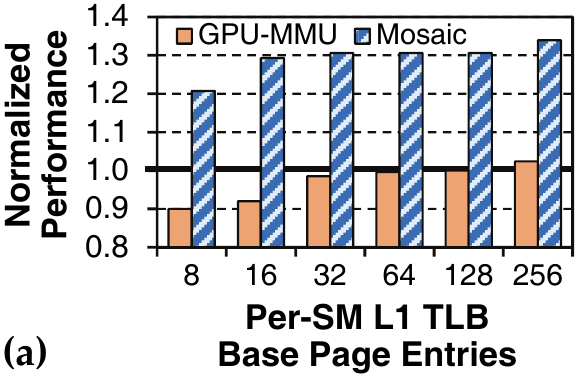}
        \label{fig:mosaic-l1-tlb}
    }}%
    \hfill%
    \subfloat{{
        \includegraphics[width=0.47\columnwidth]{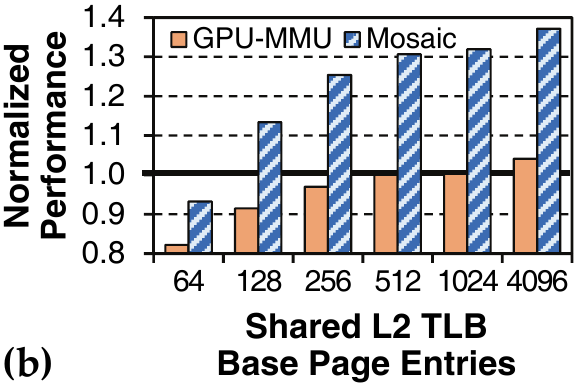}
        \label{fig:mosaic-l2-tlb}
    }}%
  \caption{Sensitivity of GPU-MMU and \titleShortMOSAIC performance to L1 and L2 TLB \changesIIIIII{base page} entries,
normalized to GPU-MMU with 128~L1 and 512~L2 \changesIIIIII{TLB base page} entries.}
  \label{fig:mosaic-tlb-sweep}
\end{figure}


\changesIIII{Figure~\ref{fig:mosaic-large-tlb-sweep} shows the performance of 
both GPU-MMU and \titleShortMOSAIC as we vary the number of \emph{large page entries}
in the per-SM L1 TLBs (Figure~\ref{fig:mosaic-l1l-tlb}) and in the shared L2 TLB 
(Figure~\ref{fig:mosaic-l2l-tlb}).
We normalize all secs/mosaic/results to the GPU-MMU performance with the baseline
16-large-page-entry L1 TLBs per SM and a 256-large-page-entry shared L2 TLB.
We make two observations from the figure.
First, for both the L1 and L2 TLBs, \titleShortMOSAIC is sensitive to the number of
large page entries.  This is because \titleShortMOSAIC successfully coalesces most of
its base pages into large pages.  We note that the sensitivity is not as high
as \titleShortMOSAIC's sensitivity to L2 TLB base page entries 
(Figure~\ref{fig:mosaic-l2-tlb}), because each large page entry covers a much
larger portion of memory, which allows a smaller number of large page entries
to still cover a majority of the total application memory.
Second, GPU-MMU is insensitive to the large page entry count.  This is because
GPU-MMU is unable to coalesce any base pages into large pages, due to its
coalescing-unfriendly allocation (see Figure~\ref{fig:tlb-mosaic-intro-base}).
As a result, GPU-MMU makes \changesIIIII{\emph{no use}} of the large page entries in the TLB.}

\begin{figure}[t]
  \centering
    \subfloat{{
        \includegraphics[width=0.47\columnwidth]{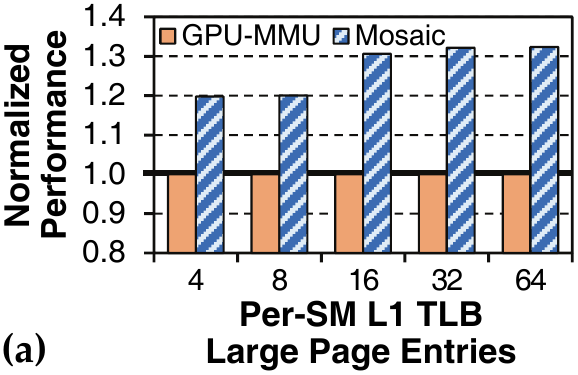}
        \label{fig:mosaic-l1l-tlb}
    }}%
    \hfill%
    \subfloat{{
        \includegraphics[width=0.47\columnwidth]{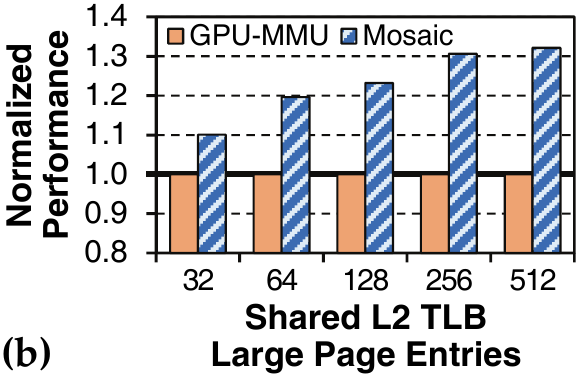}
        \label{fig:mosaic-l2l-tlb}
    }}%
  \caption{Sensitivity of GPU-MMU and \titleShortMOSAIC performance to L1 and L2 TLB \changesIIII{large page entries},
normalized to GPU-MMU with 16~L1 and 256~L2 TLB large page entries.}
  \label{fig:mosaic-large-tlb-sweep}
\end{figure}

\subsection{\changesIIII{Analysis of the Effect of Fragmentation}}
\label{sec:mosaic-eval-worst-case}

When multiple concurrently-executing GPGPU applications share the GPU, 
a series of memory allocation and deallocation requests could create significant data
fragmentation, and \changesIIIII{could} cause \allocatorName to violate its soft guarantee,
as discussed in Section~\ref{sec:allocator-sec}. 
While we do not observe this behavior in any of the workloads that we evaluate,
\titleShortMOSAIC can potentially introduce data fragmentation and memory bloat
for very long running applications.
In this section, we design stress-test experiments that induce a \changesIIII{large} amount of fragmentation
in large page \changesIIIII{frames}, to study the behavior of \allocatorName and \compactionName.

To induce \changesIIII{a large amount of} fragmentation, we allow the memory allocator to
\emph{pre-fragment} a fraction of the main memory. \changesIIIII{We randomly
place pre-fragmented data} throughout the physical memory. \changesIIIII{This data (1)~does} not conform
to \titleShortMOSAIC's soft guarantee, and \changesIIII{(2)~}\changesIIIII{cannot be coalesced
with any other base pages within the same large page frame.}
To \changesIIII{vary} the degree of large page fragmentation,
we define two metrics:
(1)~the \emph{fragmentation index}, which is the fraction of large page \changesIIIIII{frames} that
contain pre-fragmented data; and (2)~\emph{large page \changesIIIIII{frame} occupancy}, which is the \changesIIIII{fraction}
of \changesIIII{the pre-fragmented data that occupies each fragmented large page.}


\changesIIIII{We evaluate} the performance \changesIIII{of all} our workloads \changesIIIII{on}
(1)~\titleShortMOSAIC with the baseline \compactionName; and 
(2)~\titleShortMOSAIC with an optimized \compactionName that takes advantage of 
in-DRAM bulk copy mechanisms (see Section~\ref{sec:mosaic-motiv-fragmentation}), which we call \emph{{\compactionName}-BC}.
We provide a comparison against two configurations: 
(1)~\emph{Ideal \compactionName}, a compaction mechanism where data migration
incurs zero latency; and 
(2)~\emph{No \compactionName}, where \compactionName is not applied.

\changesIIII{Figure~\ref{fig:fragmentation} shows the performance of \compactionName when we vary
the fragmentation index. For these experiments, we set the large page \changesIIIIII{frame} occupancy to 50\%.}
We make \changesIIII{three observations from Figure~\ref{fig:fragmentation}.
First, we observe that there is minimal performance impact when the 
fragmentation index is less than 90\%, indicating that it is unnecessary to apply \compactionName
unless the main memory is heavily fragmented. Second, as we increase the
fragmentation index above 90\%, \compactionName provides performance improvements for
\titleShortMOSAIC, as \compactionName effectively frees up
large page frames and prevents \allocatorName from running out of frames. Third,
we observe that as the fragmentation index approaches 100\%, \compactionName becomes
less effective, due to the fact that compaction needs to be performed very frequently,}
\changesIIIII{causing a significant amount of data migration.}

 \begin{figure}[h]
   \centering
      \subfloat{{
         \includegraphics[width=0.47\columnwidth]{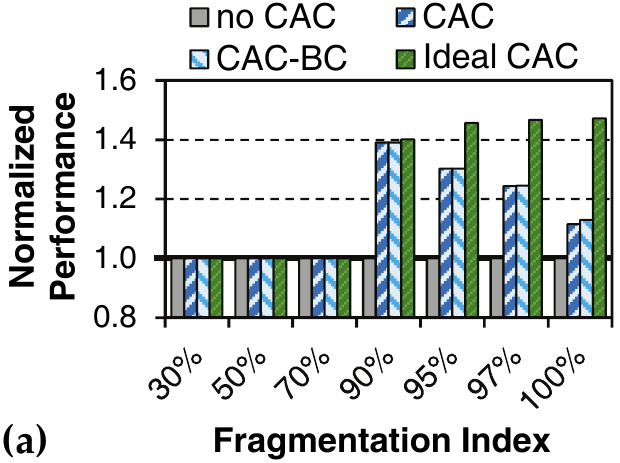}
         \label{fig:fragmentation}
     }}%
     \hfill%
      \subfloat{{
         \includegraphics[width=0.47\columnwidth]{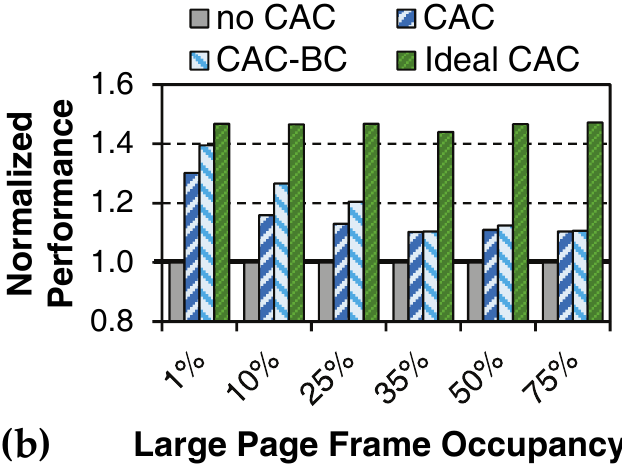}
         \label{fig:occupancy}
     }}%
   \caption{Performance of \compactionName under varying degrees of 
   \changesIIIII{(a)~fragmentation and (b)~large page \changesIIIIII{frame} occupancy}.}
   \label{fig:mosaic-eval-fragmentation}
 \end{figure}

\changesIIII{Figure~\ref{fig:occupancy} shows 
the performance of \compactionName as the large page \changesIIIII{frame} occupancy changes
\changesIIIIII{when we set the fragmentation index to} 100\% (i.e., \emph{every}
large page frame is pre-fragmented). We make two observations from the figure.
First, we observe that \compactionName-BC \changesIIIII{is effective when occupancy is
no greater than 25\%}. 
\changesIIIII{When the occupancy is low,}
in-DRAM bulk-copy operations \changesIIIII{can effectively} reduce the 
\changesIIIIII{overhead of \compactionName, as there are many opportunities to 
free up large page frames that require data migration.}
Second, we observe that as the occupancy \changesIIIII{increases beyond 35\%} 
(i.e., \changesIIIII{many} base pages are
	already allocated), the \changesIIIII{benefits} of \compactionName and \compactionName-BC decrease,
as \changesIIIII{(1)~fewer} large page frames can be freed up by compaction,
and \changesIIIIII{(2)~more} base pages need to be moved in order to free a large page frame.}

\changesIIII{Table~\ref{fig:mosaic-bloat} shows 
\changesIIIII{how \compactionName controls memory bloat for}
different large page \changesIIIIII{frame} occupancies, when we set the fragmentation index to 100\%. 
\changesIIIII{When large page \changesIIIIII{frames} are used, memory bloat can increase as a result
of high fragmentation.}
We observe that \changesIIIIII{when pages are aggressively pre-fragmented,}
\compactionName is effective at reducing the memory bloat \changesIIIIII{resulting from high levels of fragmentation}.
\changesIIIII{For example, when \changesIIIIII{the large page frame occupancy is very high (e.g., above 75\%),} \compactionName compacts the
pages effectively, reducing memory bloat to within 2.2\% of the memory \changesIIIIII{that would be}
allocated \changesIIIIII{if we were to} use only 4KB pages (i.e., \changesIIIIIII{when no large page fragmentation exists}).}
We observe negligible (<1\%) memory bloat when 
the fragmentation index is less than 100\% \changesIIIII{(not shown)}, indicating that \compactionName is 
effective at mitigating large page fragmentation.}

\begin{table}[h!]
\begin{small}
  \centering
\setlength{\tabcolsep}{4.2pt}
    \begin{tabular}{|c||c|c|c|c|c|c|}
\hline
\textbf{Large Page \changesIIIIII{Frame}}       & \multirow{2}{*}{\textbf{1\%}} & \multirow{2}{*}{\textbf{10\%}} & \multirow{2}{*}{\textbf{25\%}} & \multirow{2}{*}{\textbf{35\%}} & \multirow{2}{*}{\textbf{50\%}} & \multirow{2}{*}{\textbf{75\%}} \\
\textbf{Occupancy \changesIIIII{(\%)}}       & & & & & &  \\ \hline
\textbf{Memory Bloat} & 10.66\%     & 7.56\%      & 7.20\%      & 5.22\%      & 3.37\% & 2.22\%      \\ \hline
    \end{tabular}%
\caption{Memory bloat of \titleShortMOSAIC, compared to a GPU-MMU memory manager that uses only 4KB base pages.}
  \label{fig:mosaic-bloat}
\end{small}%
\end{table}%



We conclude that \allocatorName and \compactionName work together effectively
to preserve virtual and physical \changesIIIII{address} contiguity within a 
large page \changesIIIII{frame}, without incurring high data migration overhead \changesIIII{and memory bloat}.

\section{Mosaic: Conclusion}

\changesIII{We introduce \titleShortMOSAIC, a new GPU memory manager that provides 
application-transparent support for multiple page sizes.
The key idea of \titleShortMOSAIC is to perform demand paging using \emph{smaller} page
sizes, and then coalesce small (i.e., base) pages into a \emph{larger} page
immediately after allocation, which allows address translation to use large 
pages and thus increase TLB reach.
We have shown that \titleShortMOSAIC significantly outperforms state-of-the-art GPU 
address translation designs and achieves performance close to an ideal TLB, 
across a wide variety of workloads. We conclude that \titleShortMOSAIC effectively combines 
the benefits of large pages and demand paging in GPUs, thereby breaking the 
conventional tension that exists between these two concepts. 
We hope the ideas presented in this chapter can lead to future works that analyze 
\titleShortMOSAIC in detail and provide even lower-overhead support for synergistic 
address translation and demand paging in heterogeneous systems.}

\chapter{Common Principles and Lessons Learned} 
\label{sec:lesson}

This dissertation introduces several techniques that reduce memory interference
in GPU-based systems. 
In this chapter, we provide a list of common design principles that are used
throughout this dissertation as well as a summary of key lessons learned. 


\section{Common Design Principles}

While techniques proposed in this dissertation are applied in different parts
of the memory hierarchy, they share several key common principles. In this
section, we reiterate over these common principles.

\paragraphbe{Identification of the Benefits of Threads from Using Shared Resources.}
The first common principle in this dissertation is to give shared resources
only to threads that benefit from such shared resources. In many throughput
processors, shared resources throughout the memory hierarchy are heavily contended due
to the parallelism of these throughput processors. As a result, allowing all
threads to freely use these shared resources usually leads to memory
interference 
as we have analyzed thoroughly in
Chapters~\ref{sec:medic},~\ref{sec:sms},~\ref{sec:mask} and~\ref{sec:mosaic}.
We observed that intelligently limiting the number of threads that use these shared resources 
often leads to significant performance improvement of GPU-based systems.

To this end, all mechanisms proposed in this dissertation modify shared
resources such that they 1) always prioritize threads that benefit from
utilizing shared resources and 2) deprioritize threads that do not benefit from
utilizing shared resources to avoid memory interference. 

\paragraphbe{Division of Key Tasks of a Monolithic Structure into Simpler Structures.}
Another common principle that is utilized is the decoupling of key tasks on
monolithic structures throughout the memory hierarchy. In MeDiC and MASK (See
Chapters~\ref{sec:medic} and~\ref{sec:mask}), we provide a mechanism that
decouples the monolithic memory request buffer commonly used in modern systems
into multiple queues, where different queues deal with different types of GPU
memory requests. We found that the division of the monolithic request buffer
simplifies the design of the memory scheduler. Specifically, it simplifies
memory scheduler logic as the logic can now apply the same scheduling policy
on each queue. A similar technique applies to SMS (See
Chapter~\ref{sec:sms}), which is a memory controller design for heterogeneous
CPU-GPU systems.

\section{Lessons Learned}

This dissertation provides several techniques that together attempt to
mitigate the performance impact of memory interference. While our analysis and
evaluation have shown that our proposed techniques are effective in reducing
the memory interference on various types of GPU-based systems, this
dissertation also provides two important lessons. In this section, we
summarize these two major lessons learned from our analysis.

\paragraphbe{Memory Latency is Important for the Performance of Throughput
Processors.} Typically, limited off-chip memory bandwidth is the major
performance bottleneck of throughput processors. 
In this work, we show that the latency of
memory requests also plays an important role in increasing the performance of
throughput processors. First,
we show that it is possible to reduce the number of cycles many warps are stalled
by prioritizing the slowest thread within each warp. Our techniques allow these slow threads
to benefit from the lower latency of the shared cache. 

Second, we show that the memory latency of the page-walk-related requests is
very important to the performance of GPU-based systems. In Chapters~\ref{sec:mask}
and~\ref{sec:mosaic}, we show that page walks can significantly reduces the
memory hiding capability of GPU-based systems. As a result, it is crucial to
reduce the latency of these page-walk-related memory requests. 

\paragraphbe{How to Design the GPU Memory Hierarchy to Avoid Memory Interference?}
This dissertation introduces several techniques across the main memory
hierarchy of GPU-based systems. 
In this section, we provide recommended modifications for the memory hierarchy
for both discrete GPUs as well as heterogeneous CPU-GPU systems. 

The memory hierarchy of a discrete GPU should be designed to provide high
throughput on both single-application and multi-application setups. As a
result, the shared L2 data cache, the off-chip main memory and the shared TLB should
be designed to minimize memory interference. 
To this end, MeDiC, MASK, and Mosaic (See Chapters~\ref{sec:medic},~\ref{sec:mask} and ~\ref{sec:mosaic} for the detailed
designs and analyses of these mechanisms) can be combined together to improve the
efficiency of shared resources (the shared L2 cache, the shared
TLB and the main memory). Specifically, we recommend system
designers to modify the shared cache to 1) prioritize to threads that
benefit from the shared L2 cache (e.g., threads from the \emph{mostly-hit} and \emph{all-hit}
warp types), 2) deprioritize threads that are less likely to benefit from the
shared L2 cache (e.g., threads from the \emph{mostly-miss} and \emph{all-miss}
warp types), and 3) only cache page-walk-related data that would only benefit
from using the shared data cache. Additionally, we recommend system
designers to decouple the memory controller to perform two tasks
hierarchically. The first task is to divide GPU memory request buffer into three
different queues (\emph{Golden}, \emph{Silver} and \emph{Normal} queues) similar to the design of MASK (See Section~\ref{sec:mask}). To
combine MASK with MeDiC, requests from the \emph{mostly-hit} and \emph{all-hit} warp types
should be inserted into the \emph{Silver Queue} to ensure that these requests
have more priority than other data requests. Lastly, system designers should
modify the GPU memory allocator to enforce the \emph{soft guarantee} as defined
in Section~\ref{sec:mosaic-mech}, which enables the GPU to provide low-overhead
multi-page-size support.

To integrate our techniques into a CPU-GPU heterogeneous system, additional
per-application FIFO queues can be integrated into the memory hierarchy as
described in Section~\ref{sec:mechanism-sms}. This results in a memory
hierarchy design that minimizes all types of memory interference that occur in
GPU-based systems.

\rachata{Might be a good idea to put a figure here showing the overall recommended design}

\chapter{Conclusions and Future Directions} \label{sec:conclusion}

In summary, the goal of this dissertation is to develop shared resource
management mechanisms that can reduce memory interference in current and future
throughput processors. To this end, we analyze memory interference that occurs in Graphics Processing Units,
which are the prime example of throughput processors. Based on our analysis of
GPU characteristics and the source of memory interference, we categorize memory
interference into three different types: intra-application interference, inter-application
interference and inter-address-space interference. We propose changes
to the cache management and memory scheduling mechanisms to mitigate
intra-application interference in GPGPU applications. We propose changes to the
memory controller design and its scheduling policy to mitigate
inter-application interference in heterogeneous CPU-GPU systems. 
We redesign the memory management unit and the memory hierarchy in GPUs to be
aware of TLB-related data in order to mitigate the inter-address-space interference that
originates from the address translation process. We introduce a hardware-software
cooperative technique that modifies the memory allocation policy to enable
large page support in order to further reduce the inter-address-space
interference at the shared TLB. Our evaluations show that the GPU-aware cache and 
memory management techniques proposed in this dissertation are effective at
mitigating the interference caused by GPUs on current and future GPU-based
systems.

\section{Future Research Directions}

While this dissertation focuses on methods to mitigate memory interference in
various GPU-based systems, this dissertation also uncovers new research topics.
In this section, we describe potential research directions to further increase
the performance of GPU-based systems.

\subsection{Improving the Performance of the Memory Hierarchy in GPU-based Systems}

\para{Ways to Exploit Emerging High-Bandwidth Memory Technologies.} 3D-stacked
DRAM~\cite{hbm2,hmc1,hmc2,stackjdec,loh-stack,donghyuk-stack} is an emerging main
memory design that provides high bandwidth and high energy efficiency.
We believe that analyzing how this new type of DRAM operates can
expose techniques that might benefit modern GPU-based systems. 

Aside from 3D-stacked memory, recent proposals provide methods to reduce DRAM
latency~\cite{ava-dram,dsarp,al-dram,tl-dram,salp}, a method to utilize
multi-ported DRAM~\cite{donghyuk-ddma}, or methods to perform some computations
within DRAM in order to reduce the amount of DRAM
bandwidth~\cite{seshadri-cal2015,seshadri2013rowclone,lisa,tom-isca16}. We think that these
techniques, combined with observations on GPU applications' characteristics
provided in this dissertation, can be applied to GPUs and should provide
significant performance improvement for GPU-based systems.

\para{Other Methods to Exploit Warp-type Heterogeneity and TLB-related Data in
GPU-based system.} In this dissertation, we show in Chapter~\ref{sec:medic} how
GPU-based systems exploit warp-type heterogeneity to reduce
\emph{intra-application} interference and improve the effectiveness of the
cache and the main memory. We also show in Chapter~\ref{sec:mask} how to design
a GPU memory hierarchy that is aware of TLB data to minimize \emph{inter-address-space}
interference. We believe that it is beneficial to integrate these warp-type and
TLB-awareness characteristics to the memory hierarchy in GPU-based systems to
further improve system performance. 


\para{Potential Denial-of-service in Software Managed Shared Memory.} Allowing
GPU-based systems to be shared across multiple GPGPU applications potentially introduces
new performance bottlenecks. Concurrently
running multiple GPU applications creates a unique resource contention at GPU's
software-managed Shared Scratchpad Memory. 
Because this particular resource is managed by the GPGPU applications (in
software), GPGPU applications that share the GPU all contend for this resource.
The lack of communication between each GPGPU application
prevents one application to inform its demand
for the Shared Scratchpad Memory to other applications. As a result, one application can
completely block other applications by using \emph{all} Shared Scratchpad Memory.

It is possible to solve this unique problem through
modifications in the hypervisor. For example, additional kernel scheduling
techniques can be applied to 1) probe how much Shared Scratchpad Memory is needed by each
application and 2) enforce a proper policy that only grants each
application a portion of the Shared Scratchpad Memory.



\para{Interference Management in GPUs for Emerging Applications.} The emergence
of embedded applications introduces a new requirement: real-time deadlines.
Traditionally, these applications run on an embedded device which contains
multiple application-specific integrated circuits (ASICs) to handle most of the
computations. However, the rise of integrated GPUs in modern System-on-Chips
(e.g.,~\cite{arm-mali,amdzen,tegra,tegrax1}) as well as better GPU support in
several cloud infrastructures (e.g.,~\cite{ec2,amazonec2,vmCUDA,gVirt})
allow these applications to perform these computations on the GPUs. While
the GPUs can provide good IPC throughput due to their parallelism, the GPUs and
the GPUs' memory hierarchy, also need to provide a low response
time, or in many cases enforce \emph{hard} performance guarantees (i.e., an
application must finish its execution within a certain time limit). 

Even though mechanisms proposed in this dissertation aim to minimize the
slowdown caused by interference, these mechanisms do not provide actual
performance guarantees. However, we believe it is possible to use observations
in this dissertation to aid in designing mechanisms to provide a
\emph{hard} performance guarantee and limit the amount of memory interference
when multiple of these new embedded applications are concurrently sharing
GPU-based systems.


\subsection{Low-overhead Virtualization Support in GPU-based Systems}
%


While this dissertation proposes mechanisms to minimize inter-address-space
interference in GPU-based systems, there are several open-ended research
questions on how to efficiently virtualize GPU-based systems and how to
efficiently shared other non-memory resources across multiple applications.



\para{Maintaining Virtual Address Space Contiguity.}
While Chapter~\ref{sec:mosaic} provides a mechanism that maintains contiguous
physical address, \titleShortMOSAIC does not perform compaction in the
virtual address space as this dissertation does not observe virtual address space
fragmentation in current GPGPU applications. However, it might be possible that
a long chain of small size memory allocations and deallocations can break
contiguity within the virtual address space. In this case, the virtual address
space has to be remapped in order to create a contiguous chunk of unallocated
virtual memory. This can lower the performance of GPU-based systems.


\para{Utilizing High-bandwidth Interconnects to Transfer Data between CPU
Memory and GPU Memory.} As shown in Chapter~\ref{sec:mosaic}, demand paging can
be costly, especially when a large amount of data has to be transferred to the
GPU. The long latency of demand paging can lead to significant stall time for
GPU cores. 
Methods to improve the performance of demand paging remain a potential research problem. Emerging technologies such as
NVIDIA's NVLink~\cite{nvlink} and AMD's Infinity~\cite{amdzen} 
can improve the data transfer rate between the CPUs and the GPUs. However,
there is a lack of details on how to integrate these high-bandwidth
interconnects to existing GPU hardware. Analyzing how these technologies
operate, and providing a detailed study of their potential benefits and
limitations is crucial for the integration of these new technologies in
GPU-based systems. 

Aside from techniques that utilize new technologies, architectural techniques
can also mitigate the long data transfer latency between CPU memory and GPU
memory. We believe that methods such as preemptively fetching the data of
potential pages or proactively evicting potentially unused data in GPU memory
can be effective in reducing the performance impact of demand paging.

\subsection{Providing an Optimal Method to Concurrently Execute GPGPU Applications}

While this dissertation allows applications to share the GPUs more efficiently
by limiting the memory interference, how to schedule kernels and how to map these
kernels to GPU cores remain an open research problem. In this work, we assume
1) an equal partitioning of GPU cores for each GPGPU application, and 2) every
application is scheduled to start at the same time. Because applications have a
different amount of parallelism as well as bandwidth demand, the optimal number
of GPU cores that should be assigned to each application varies not only across
different applications, but also across different workload setups.


As a result, providing an optimal method to manage the execution of GPGPU
applications on GPU-based systems is a very complex problem. However, we
believe that using the knowledge of the resource demand of each application
between system software and the GPU hardware can significantly reduce the
complexity of the scheduler. Information such as the amount of thread-level
parallelism, the expected amount of data parallelism, the expected memory usage, cache
locality, memory locality, etc. can be used as hints to assist in providing
desirable application-to-GPU-core mappings and kernel scheduling decisions. In
this dissertation, we provide several observations regarding GPGPU
applications' characteristics that might be useful for assisting the system
software to provide better mapping and scheduling decisions (e.g., memory
allocation behavior, warp characteristics).



\section{Final Summary}

We conclude and hope that this dissertation, with the analyses of memory
interference and mechanisms to mitigate this memory interference, enables many
new research directions that further improve the capability of GPU-based
systems.



\chapter*{Other Contributions by the Author}

During my Ph.D., I had opportunities to be involved in many other research
projects. While these projects do not fit into the theme of this dissertation,
they have helped me tremendously in learning an in-depth knowledge about the
memory hierarchy as well as the GPU architecture. I would like to acknowledge these
projects as well as my early works on Network-on-Chip (NoCs) that kicked start
my Ph.D.

My interest in studying memory interference in the memory hierarchy starts from the
interests in Network-on-Chip. I have an opportunity in collaborating with Kevin
Chang and Chris Fallin on two power-efficient network-on-chip designs that
focus on bufferless network-on-chip: HAT~\cite{HAT-SBAC_PAD-2012} and MinBD~\cite{minbd}. In
addition, I have authored another work on a hierarchical bufferless
network-on-chip design called
HiRD~\cite{hird,hird-journal} and have released NOCulator, which is the
simulation infrastructure for both MinBD and HiRD~\cite{noculator}. All these works focus on
mechanisms to improve power efficiency and simplifying the design of NoCs
without sacrificing system performance. I also have an opportunity
collaborating with Reetuparna Das on another work called A2C~\cite{a2c}, which
studies the placement of applications to cores in NoCs. A2C allows operating
systems to be able to place applications to cores in a way that minimize
interference, which is also the main theme in this thesis. I worked with Mohammad Fattah
on a low-overhead fault-tolerant routing mechanism for network-on-chip~\cite{maze-routing}.
I worked with Besta Maciej on a scalable and energy efficient topology for network-on-chip~\cite{slimnoc}.

In collaboration with Amirali Boroumand, I have worked on using
processing-in-memory to improve the energy efficiency of mobile
workloads~\cite{googlePIM}.

In collaboration with Vivek Seshadri, I have worked on techniques to allow
in-DRAM bulk copy called RowClone~\cite{seshadri2013rowclone}. 

In collaboration with Donghyuk Lee, I have worked on a study that characterizes
latency variation in DRAM cells and provides techniques to improve the
performance of DRAM by incorporating latency variation~\cite{ava-dram}

In collaboration with Justin Meza and Hanbin Yoon, I have worked on techniques
to manage resources for hybrid memory that consists of DRAM and Phased
changed memory (PCM)~\cite{yoon-iccd2012}. 

In collaboration with Nandita Vijaykumar, I have worked on a technique that
allows better utilization of GPU cores called CABA~\cite{caba}. CABA uses a
technique similar to helper threads in order to improve the utilization of
GPUs.

In collaboration with Mohammad Sadrosadati, I have worked on a technique to
improve the GPU register file performance~\cite{gpu-regfile-mohammad}.

In collaboration with Onur Kayiran and Gabriel H. Loh, I have worked on a
technique that manages GPU concurrency in a heterogeneous architecture in 
order to reduce interference~\cite{cpugpu-micro}. In addition, I also worked
on a GPU power management technique that turns down datapath components that
are not in the bottleneck~\cite{ucstate}.

In additional to these works, I have co-authored three book chapters on the topics of
GPUs~\cite{nandita-chapter}, processing-in-memory~\cite{pim-chapter} and bufferless network-on-chip~\cite{minbd-book}.

\small
\singlespacing
\bibliography{references}

\begin{thebibliography}{100}

\bibitem{noculator}
{NOCulator}.
\newblock \url{https://github.com/CMU-SAFARI/NOCulator}, 2014.

\bibitem{rotary-router}
P.~Abad et~al.
\newblock {Rotary router}: an efficient architecture for {CMP} interconnection
  networks.
\newblock {\em ISCA}, 2007.

\bibitem{tensorflow}
M.~Abadi, A.~Agarwal, P.~Barham, E.~Brevdo, Z.~Chen, C.~Citro, G.~Corrado,
  A.~Davis, J.~Dean, M.~Devin, S.~Ghemawat, I.~Goodfellow, A.~Harp, G.~Irving,
  M.~Isard, Y.~Jia, R.~Jozefowicz, L.~Kaiser, M.~Kudlur, J.~Levenberg,
  D.~Mané, R.~Monga, S.~Moore, D.~Murray, C.~Olah, M.~Schuster, J.~Shlens,
  B.~Steiner, I.~Sutskever, K.~Talwar, P.~Tucker, V.~Vanhoucke, V.~Vasudevan,
  F.~Viégas, O.~Vinyals, P.~Warden, M.~Wattenberg, M.~Wicke, Y.~Yu, and
  X.~Zheng.
\newblock {TensorFlow: Large-Scale Machine Learning on Heterogeneous
  Distributed Systems}, 2015.

\bibitem{gpu-multitasking}
J.~Adriaens, K.~Compton, N.~S. Kim, and M.~Schulte.
\newblock {The Case for GPGPU Spatial Multitasking}.
\newblock In {\em HPCA}, 2012.

\bibitem{amd-fusion}
{Advanced Micro Devices}.
\newblock {AMD Accelerated Processing Units}.

\bibitem{amd-io-virt}
{Advanced Micro Devices}.
\newblock {AMD I/O Virtualization Technology (IOMMU) Specification}.

\bibitem{amdr9}
{Advanced Micro Devices}.
\newblock {AMD Radeon R9 290X}.
\newblock
  \url{http://www.amd.com/us/press-releases/Pages/amd-radeon-r9-290x-2013oct24.aspx}.

\bibitem{radeon}
{Advanced Micro Devices}.
\newblock {ATI Radeon GPGPUs}.
\newblock {\em
  http://www.amd.com/us/products/desktop/graphics/amd-radeon-hd-6000/Pages/amd-radeon-hd-6000.aspx}.

\bibitem{firepro}
{Advanced Micro Devices}.
\newblock {OpenCL™: The Future of Accelerated Application Performance Is
  Now}.

\bibitem{npt}
{Advanced Micro Devices}.
\newblock {\em {AMD-V Nested Paging}}, 2010.
\newblock
  \url{http://developer.amd.com/wordpress/media/2012/10/NPT-WP-1%201-final-TM.pdf}.

\bibitem{ati-wavefront}
{Advanced Micro Devices}.
\newblock {AMD Graphics Cores Next (GCN) Architecture}.
\newblock \url{http://www.amd.com/Documents/GCN_Architecture_whitepaper.pdf},
  2012.

\bibitem{huma}
{Advanced Micro Devices}.
\newblock {Heterogeneous System Architecture: A Technical Review}.
\newblock
  \url{http://amd-dev.wpengine.netdna-cdn.com/wordpress/media/2012/10/hsa10.pdf},
  2012.

\bibitem{april-fmt}
A.~Agarwal, B.~H. Lim, D.~Kranz, and J.~Kubiatowicz.
\newblock {APRIL: A Processor Architecture for Multiprocessing}.
\newblock Technical report, Cambridge, MA, USA, 1991.

\bibitem{cc-numa-gpu-hpca15}
N.~Agarwal, D.~Nellans, M.~O'Connor, S.~W. Keckler, and T.~F. Wenisch.
\newblock {Unlocking Bandwidth for GPUs in CC-NUMA Systems}.
\newblock In {\em HPCA}, 2015.

\bibitem{agrawal-hpca2014}
A.~Agrawal, A.~Ansari, and J.~Torrellas.
\newblock {Mosaic: Exploiting the Spatial Locality of Process Variation to
  Reduce Refresh Energy in On-chip eDRAM Modules}.
\newblock In {\em HPCA}, 2014.

\bibitem{agrawal-memsys2016}
A.~Agrawal, M.~O'Connor, E.~Bolotin, N.~Chatterjee, J.~Emer, and S.~Keckler.
\newblock {CLARA: Circular Linked-List Auto and Self Refresh Architecture}.
\newblock In {\em MEMSYS}, 2016.

\bibitem{ahn-isca2015}
J.~Ahn, S.~Hong, S.~Yoo, O.~Mutlu, and K.~Choi.
\newblock {A Scalable Processing-in-memory Accelerator for Parallel Graph
  Processing}.
\newblock In {\em ISCA}, 2015.

\bibitem{ahn-isca12}
J.~Ahn, S.~Jin, and J.~Huh.
\newblock {Revisiting Hardware-Assisted Page Walks for Virtualized Systems}.
\newblock In {\em ISCA}, 2012.

\bibitem{ahn-tocs15}
J.~Ahn, S.~Jin, and J.~Huh.
\newblock {Fast Two-Level Address Translation for Virtualized Systems}.
\newblock In {\em IEEE TC}, 2015.

\bibitem{ahn-isca2015-2}
J.~Ahn, S.~Yoo, O.~Mutlu, and K.~Choi.
\newblock {PIM-enabled Instructions: A Low-overhead, Locality-aware
  Processing-in-memory Architecture}.
\newblock In {\em ISCA}, 2015.

\bibitem{ahn-taco2012}
J.~H. Ahn, N.~P. Jouppi, C.~Kozyrakis, J.~Leverich, and R.~S. Schreiber.
\newblock {Improving System Energy Efficiency with Memory Rank Subsetting}.
\newblock {\em ACM TACO}, 9(1):4:1--4:28, 2012.

\bibitem{ahn-cal2009}
J.~H. Ahn, J.~Leverich, R.~Schreiber, and N.~P. Jouppi.
\newblock {Multicore DIMM: an Energy Efficient Memory Module with Independently
  Controlled DRAMs}.
\newblock {\em IEEE CAL}, 2009.

\bibitem{akin-isca2015}
B.~Akin, F.~Franchetti, and J.~C. Hoe.
\newblock {Data Reorganization in Memory Using 3D-stacked DRAM}.
\newblock In {\em ISCA}, 2015.

\bibitem{alameldeen-hpca2007}
A.~R. Alameldeen and D.~A. Wood.
\newblock {Interactions Between Compression and Prefetching in Chip
  Multiprocessors}.
\newblock In {\em HPCA}, 2007.

\bibitem{netflix}
J.~B. Alex~Chen and X.~Amatriain.
\newblock {Distributed Neural Networks with GPUs in the AWS cloud}.
\newblock 2014.

\bibitem{tera-mta}
R.~Alverson, D.~Callahan, D.~Cummings, B.~Koblenz, A.~Porterfield, and
  B.~Smith.
\newblock {The Tera Computer System}.
\newblock In {\em ICS}, 1990.

\bibitem{amazonec2}
Amazon.
\newblock {Amazon EC2 GPU Instance}.
\newblock {\em
  http://aws.amazon.com/about-aws/whats-new/2013/11/04/announcing-new-amazon-ec2-gpu-instance-type/}.

\bibitem{ec2}
{Amazon}.
\newblock {An Introduction to High Performance Computing on AWS}.
\newblock \url{https://d0.awsstatic.com/whitepapers/Intro_to_HPC_on_AWS.pdf},
  2015.

\bibitem{osx-huge}
Apple Inc.
\newblock {\em {Huge Page Support in Mac OS X}}.
\newblock [Accessed April-2017].

\bibitem{arm-cortexa}
{ARM Holdings}.
\newblock {ARM Cortex-A Series}.
\newblock
  \url{http://infocenter.arm.com/help/topic/com.arm.doc.den0024a/DEN0024A_v8_architecture_PG.pdf},
  2015.

\bibitem{arunkumar-isca17}
A.~Arunkumar, E.~Bolotin, B.~Cho, U.~Milic, E.~Ebrahimi, O.~Villa, A.~Jaleel,
  and C.-J. Wu.
\newblock {MCM-GPU: Multi-Chip-Module GPUs for Continued Performance
  Scalability}.
\newblock In {\em ISCA}, 2017.

\bibitem{rachata-thesis}
R.~Ausavarungnirun.
\newblock {\em {Techniques for Shared Resource Management in Systems with
  Throughput Processors}}.
\newblock PhD thesis, Carnegie Mellon Univ., 2017.

\bibitem{sms}
R.~Ausavarungnirun, K.~Chang, L.~Subramanian, G.~Loh, and O.~Mutlu.
\newblock {Staged Memory Scheduling: Achieving High Performance and Scalability
  in Heterogeneous Systems}.
\newblock In {\em ISCA}, 2012.

\bibitem{hird}
R.~Ausavarungnirun, C.~Fallin, X.~Yu, K.~Chang, G.~Nazario, R.~Das, G.~H. Loh,
  and O.~Mutlu.
\newblock {Design and Evaluation of Hierarchical Rings with Deflection
  Routing}.
\newblock In {\em SBAC-PAD}, 2014.

\bibitem{hird-journal}
R.~Ausavarungnirun, C.~Fallin, X.~Yu, K.~Chang, G.~Nazario, R.~Das, G.~H. Loh,
  and O.~Mutlu.
\newblock {A Case for Hierarchical Rings with Deflection Routing}.
\newblock {\em PARCO}, 54(C):29--45, May 2016.

\bibitem{medic}
R.~Ausavarungnirun, S.~Ghose, O.~Kayıran, G.~H. Loh, C.~R. Das, M.~T.
  Kandemir, and O.~Mutlu.
\newblock {Exploiting Inter-Warp Heterogeneity to Improve GPGPU Performance}.
\newblock In {\em PACT}, 2015.

\bibitem{mosaic}
R.~Ausavarungnirun, J.~Landgraf, V.~Miller, S.~Ghose, J.~Gandhi, C.~J.
  Rossbach, and O.~Mutlu.
\newblock {Mosaic: A GPU Memory Manager with Application-Transparent Support
  for Multiple Page Sizes}.
\newblock In {\em MICRO}, 2017.

\bibitem{mosaic-tech}
R.~Ausavarungnirun, J.~Landgraf, V.~Miller, S.~Ghose, J.~Gandhi, C.~J.
  Rossbach, and O.~Mutlu.
\newblock {Mosaic: A GPU Memory Manager with Application-Transparent Support
  for Multiple Page Sizes}.
\newblock Technical Report TR-2017-003, Carnegie Mellon Univ., SAFARI Research
  Group, 2017.

\bibitem{mask}
R.~Ausavarungnirun, V.~Miller, J.~Landgraf, S.~Ghose, J.~Gandhi, A.~Jog,
  C.~Rossbach, and O.~Mutlu.
\newblock {MASK: Redesigning the GPU Memory Hierarchy to Support
  Multi-Application Concurrency}.
\newblock In {\em ASPLOS}, 2018.

\bibitem{mask-tech-report}
R.~Ausavarungnirun, V.~Miller, J.~Landgraf, S.~Ghose, J.~Gandhi, A.~Jog, C.~J.
  Rossbach, and O.~Mutlu.
\newblock {Spatial Multiplexing Support for Multi-Application Concurrency in
  GPUs}.
\newblock Technical Report TR-2018-002, Carnegie Mellon Univ., SAFARI Research
  Group, 2018.

\bibitem{mask-arxiv}
R.~Ausavarungnirun, C.~J. Rossbach, V.~Miller, J.~Landgraf, S.~Ghose,
  J.~Gandhi, A.~Jog, and O.~Mutlu.
\newblock {Improving Multi-Application Concurrency Support Within the GPU
  Memory System}.
\newblock arXiv:1708.04911 [cs.AR], 2017.

\bibitem{rajeev-pact10}
M.~Awasthi, D.~W. Nellans, K.~Sudan, R.~Balasubramonian, and A.~Davis.
\newblock {Handling the Problems and Opportunities Posed by Multiple On-chip
  Memory Controllers}.
\newblock In {\em PACT}, 2010.

\bibitem{babarinsa-2015}
O.~O. Babarinsa and S.~Idreos.
\newblock {JAFAR: Near-Data Processing for Databases}.
\newblock In {\em SIGMOD}, 2015.

\bibitem{baek-tc2014}
S.~Baek, S.~Cho, and R.~Melhem.
\newblock {Refresh Now and Then}.
\newblock {\em IEEE TC}, 63(12):3114--3126, 2014.

\bibitem{baer-1995}
J.-L. Baer and T.-F. Chen.
\newblock {Effective Hardware-Based Data Prefetching for High-Performance
  Processors}.
\newblock {\em IEEE TC}, 44(5):609--623, 1995.

\bibitem{gpgpu-sim}
A.~Bakhoda, G.~Yuan, W.~Fung, H.~Wong, and T.~Aamodt.
\newblock {Analyzing CUDA Workloads Using a Detailed GPU Simulator}.
\newblock In {\em ISPASS}, 2009.

\bibitem{hotpotato}
P.~Baran.
\newblock {On Distributed Communications Networks}.
\newblock 1964.

\bibitem{illiac}
G.~H. Barnes, R.~M. Brown, M.~Kato, D.~J. Kuck, D.~L. Slotnick, and R.~A.
  Stokes.
\newblock {The Illiac IV Computer}.
\newblock {\em IEEE TC}, 100(8):746--757, 1968.

\bibitem{barr-isca10}
T.~W. Barr, A.~L. Cox, and S.~Rixner.
\newblock {Translation Caching: Skip, Don'T Walk (the Page Table)}.
\newblock In {\em ISCA}, 2010.

\bibitem{spectlb}
T.~W. Barr, A.~L. Cox, and S.~Rixner.
\newblock {SpecTLB: A Mechanism for Speculative Address Translation}.
\newblock In {\em ISCA}, 2011.

\bibitem{direct-segment}
A.~Basu, J.~Gandhi, J.~Chang, M.~D. Hill, and M.~M. Swift.
\newblock {Efficient Virtual Memory for Big Memory Servers}.
\newblock In {\em ISCA}, 2013.

\bibitem{slimnoc}
M.~Besta, S.~M. Hassan, S.~Yalamanchili, R.~Ausavarungnirun, O.~Mutlu, and
  T.~Hoefler.
\newblock {Slim NoC: A Low-Diameter On-Chip Network Topology for High Energy
  Efficiency and Scalability}.
\newblock In {\em ASPLOS}, 2018.

\bibitem{bhati-isca2015}
I.~Bhati, Z.~Chishti, S.-L. Lu, and B.~Jacob.
\newblock {Flexible Auto-refresh: Enabling Scalable and Energy-efficient DRAM
  Refresh Reductions}.
\newblock In {\em ISCA}, 2015.

\bibitem{large-reach}
A.~Bhattacharjee.
\newblock {Large-reach Memory Management Unit Caches}.
\newblock In {\em MICRO}, 2013.

\bibitem{bhattacharjee-hpca11}
A.~Bhattacharjee, D.~Lustig, and M.~Martonosi.
\newblock {Shared Last-level TLBs for Chip Multiprocessors}.
\newblock In {\em HPCA}, 2011.

\bibitem{bhattacharjee-pact09}
A.~Bhattacharjee and M.~Martonosi.
\newblock {Characterizing the TLB Behavior of Emerging Parallel Workloads on
  Chip Multiprocessors}.
\newblock In {\em PACT}, 2009.

\bibitem{inter-core-tlb}
A.~Bhattacharjee and M.~Martonosi.
\newblock {Inter-core Cooperative TLB for Chip Multiprocessors}.
\newblock In {\em ASPLOS}, 2010.

\bibitem{tlb-consistency}
D.~L. Black, R.~F. Rashid, D.~B. Golub, and C.~R. Hill.
\newblock {Translation Lookaside Buffer Consistency: A Software Approach}.
\newblock In {\em ASPLOS}, 1989.

\bibitem{googlePIM}
A.~Boroumand, S.~Ghose, Y.~Kim, R.~Ausavarungnirun, E.~Shiu, R.~Thakur, D.~Kim,
  A.~Kuusela, A.~Knies, P.~Ranganathan, and O.~Mutlu.
\newblock {Google Workloads for Consumer Devices: Mitigating Data Movement
  Bottlenecks}.
\newblock In {\em ASPLOS}, 2018.

\bibitem{amirali-cal2016}
A.~Boroumand, S.~Ghose, B.~Lucia, K.~Hsieh, K.~Malladi, H.~Zheng, and O.~Mutlu.
\newblock {LazyPIM: An Efficient Cache Coherence Mechanism for
  Processing-in-Memory}.
\newblock {\em IEEE CAL}, 2016.

\bibitem{kaveri}
D.~Bouvier and B.~Sander.
\newblock {Applying AMD's "Kaveri" APU for Heterogeneous Computing}.
\newblock 2014.

\bibitem{bobcat}
B.~Burgess, B.~Cohen, J.~Dundas, J.~Rupley, D.~Kaplan, and M.~Denman.
\newblock {Bobcat: AMD's Low-Power x86 Processor}.
\newblock {\em IEEE Micro}, 2011.

\bibitem{lonestar}
M.~Burtscher, R.~Nasre, and K.~Pingali.
\newblock {A Quantitative Study of Irregular Programs on {GPUs}}.
\newblock In {\em IISWC}, 2012.

\bibitem{cao-sigmetrics1995}
P.~Cao, E.~W. Felten, A.~R. Karlin, and K.~Li.
\newblock {A Study of Integrated Prefetching and Caching Strategies}.
\newblock In {\em SIGMETRICS}, 1995.

\bibitem{carter-hpca1999}
J.~Carter, W.~Hsieh, L.~Stoller, M.~Swanson, L.~Zhang, E.~Brunvand, A.~Davis,
  C.-C. Kuo, R.~Kuramkote, M.~Parker, L.~Schaelicke, and T.~Tateyama.
\newblock {Impulse: Building a Smarter Memory Controller}.
\newblock In {\em HPCA}, 1999.

\bibitem{copperhead}
B.~Catanzaro, M.~Garland, and K.~Keutzer.
\newblock {Copperhead: Compiling an Embedded Data Parallel Language}.
\newblock In {\em SIGPLAN}, 2011.

\bibitem{chandrasekar-date2014}
K.~Chandrasekar, S.~Goossens, C.~Weis, M.~Koedam, B.~Akesson, N.~Wehn, and
  K.~Goossens.
\newblock {Exploiting Expendable Process-Margins in DRAMs for Run-Time
  Performance Optimization}.
\newblock In {\em DATE}, 2014.

\bibitem{HAT-SBAC_PAD-2012}
K.~Chang, R.~Ausavarungnirun, C.~Fallin, and O.~Mutlu.
\newblock {HAT: Heterogeneous Adaptive Throttling for On-Chip Networks}.
\newblock In {\em SBAC-PAD}, 2012.

\bibitem{chang-sigmetric16}
K.~Chang, A.~Kashyap, H.~Hassan, S.~Ghose, K.~Hsieh, D.~Lee, T.~Li,
  G.~Pekhimenko, S.~Khan, and O.~Mutlu.
\newblock {Understanding Latency Variation in Modern DRAM Chips: Experimental
  Characterization, Analysis, and Optimization}.
\newblock In {\em SIGMETRICS}, 2016.

\bibitem{dsarp}
K.~Chang, D.~Lee, Z.~Chishti, A.~Alameldeen, C.~Wilkerson, Y.~Kim, and
  O.~Mutlu.
\newblock {Improving DRAM Performance by Parallelizing Refreshes with Accesses
  }.
\newblock In {\em HPCA}, 2014.

\bibitem{lisa}
K.~Chang, P.~J. Nair, D.~Lee, S.~Ghose, M.~K. Qureshi, and O.~Mutlu.
\newblock {Low-cost Inter-linked Subarrays (LISA): Enabling Fast Inter-subarray
  Data Movement in DRAM}.
\newblock In {\em HPCA}, 2016.

\bibitem{chang-sigmetric17}
K.~K. Chang, A.~G. Yaglikci, S.~Ghose, A.~Agrawal, N.~Chatterjee, A.~Kashyap,
  D.~Lee, M.~O'Connor, H.~Hassan, and O.~Mutlu.
\newblock {Understanding Reduced-Voltage Operation in Modern DRAM Devices:
  Experimental Characterization, Analysis, and Mechanisms}.
\newblock In {\em SIGMETRIC}, 2017.

\bibitem{nvidia-hpca17}
N.~Chatterjee, M.~O'Connor, D.~Lee, D.~R. Johnson, S.~W. Keckler, M.~Rhu, and
  W.~J. Dally.
\newblock {Architecting an Energy-Efficient DRAM System for GPUs}.
\newblock In {\em HPCA}, 2017.

\bibitem{chatterjee-sc14}
N.~Chatterjee, M.~O'Connor, G.~H. Loh, N.~Jayasena, and R.~Balasubramonian.
\newblock {Managing DRAM Latency Divergence in Irregular GPGPU Applications}.
\newblock In {\em SC}, 2014.

\bibitem{chatterjee-micro2012}
N.~Chatterjee, M.~Shevgoor, R.~Balasubramonian, A.~Davis, Z.~Fang, R.~Illikkal,
  and R.~Iyer.
\newblock {Leveraging Heterogeneity in DRAM Main Memories to Accelerate
  Critical Word Access}.
\newblock In {\em MICRO}, 2012.

\bibitem{chaudhuri-pact12}
M.~Chaudhuri, J.~Gaur, N.~Bashyam, S.~Subramoney, and J.~Nuzman.
\newblock {Introducing Hierarchy-awareness in Replacement and Bypass Algorithms
  for Last-level Caches}.
\newblock In {\em PACT}, 2012.

\bibitem{rodinia}
S.~Che, M.~Boyer, J.~Meng, D.~Tarjan, J.~Sheaffer, S.-H. Lee, and K.~Skadron.
\newblock {Rodinia: A Benchmark Suite for Heterogeneous Computing}.
\newblock In {\em IISWC}, 2009.

\bibitem{chen-micro47}
X.~Chen, L.-W. Chang, C.~I. Rodrigues, J.~Lv, Z.~Wang, and W.~W. Hwu.
\newblock {Adaptive Cache Management for Energy-Efficient {GPU} Computing}.
\newblock In {\em MICRO}, 2014.

\bibitem{chen-mes14}
X.~Chen, S.~Wu, L.-W. Chang, W.-S. Huang, C.~Pearson, Z.~Wang, and W.~W. Hwu.
\newblock {Adaptive Cache Bypass and Insertion for Many-Core Accelerators}.
\newblock In {\em MES}, 2014.

\bibitem{amdzen}
M.~Clark.
\newblock {A New X86 Core Architecture for the Next Generation of Computing}.
\newblock In {\em HotChips}, 2016.

\bibitem{safari.github}
{CMU SAFARI Research Group}.
\newblock \url{https://github.com/CMU-SAFARI}.

\bibitem{mct}
J.~D. Collins and D.~M. Tullsen.
\newblock {Hardware Identification of Cache Conflict Misses}.
\newblock In {\em MICRO}, 1999.

\bibitem{cong-hpca17}
J.~Cong, Z.~Fang, Y.~Hao, and G.~Reinmana.
\newblock {Supporting Address Translation for Accelerator-Centric
  Architectures}.
\newblock In {\em HPCA}, 2017.

\bibitem{cdc7600}
{Control Data Corporation}.
\newblock {Control Data 7600 Computer Systems Reference Manual}, 1972.

\bibitem{cooksey-asplos2002}
R.~Cooksey, S.~Jourdan, and D.~Grunwald.
\newblock {A Stateless, Content-directed Data Prefetching Mechanism}.
\newblock In {\em ASPLOS}, 2002.

\bibitem{couchbase-nothp}
{Couchbase Inc.}
\newblock {{Often Overlooked Linux OS Tweaks}}.
\newblock [Accessed March, 2014].

\bibitem{crane-ec65}
B.~A. Crane and J.~A. Githens.
\newblock {Bulk Processing in Distributed Logic Memory}.
\newblock {\em IEEE EC}, 14(2):186--196, April 1965.

\bibitem{dahlgren-1995}
F.~Dahlgren, M.~Dubois, and P.~Stenstr\"{o}m.
\newblock {Sequential Hardware Prefetching in Shared-Memory Multiprocessors}.
\newblock {\em IEEE TPDS}, 6(7):733--746, 1995.

\bibitem{dai-dac16}
H.~Dai, C.~Li, H.~Zhou, S.~Gupta, C.~Kartsaklis, and M.~Mantor.
\newblock {A Model-driven Approach to Warp/thread-block Level GPU Cache
  Bypassing}.
\newblock In {\em DAC}, 2016.

\bibitem{shoc}
A.~Danalis, G.~Marin, C.~McCurdy, J.~S. Meredith, P.~C. Roth, K.~Spafford,
  V.~Tipparaju, and J.~S. Vetter.
\newblock {The Scalable Heterogeneous Computing (SHOC) benchmark suite}.
\newblock In {\em GPGPU}, 2010.

\bibitem{a2c}
R.~Das, R.~Ausavarungnirun, O.~Mutlu, A.~Kumar, and M.~Azimi.
\newblock {Application-to-core Mapping Policies to Reduce Memory System
  Interference in Multi-core Systems}.
\newblock In {\em HPCA}, 2013.

\bibitem{das09}
R.~Das, S.~Eachempati, A.~K. Mishra, V.~Narayanan, and C.~R. Das.
\newblock {Design and Evaluation of Hierarchical On-Chip Network Topologies for
  Next Generation CMPs}.
\newblock {\em HPCA}, 2009.

\bibitem{Reetu-MICRO2009}
R.~Das, O.~Mutlu, T.~Moscibroda, and C.~R. Das.
\newblock {Application-aware Prioritization Mechanisms for On-chip Networks}.
\newblock In {\em MICRO}, 2009.

\bibitem{aergia}
R.~Das, O.~Mutlu, T.~Moscibroda, and C.~R. Das.
\newblock {A\'ergia: Exploiting Packet Latency Slack in On-chip Networks}.
\newblock In {\em ISCA}, 2010.

\bibitem{draper-ics2002}
J.~Draper, J.~Chame, M.~Hall, C.~Steele, T.~Barrett, J.~LaCoss, J.~Granacki,
  J.~Shin, C.~Chen, C.~W. Kang, I.~Kim, and G.~Daglikoca.
\newblock {The Architecture of the DIVA Processing-in-memory Chip}.
\newblock In {\em ICS}, 2002.

\bibitem{gtsm}
Y.~Du, M.~Zhou, B.~Childers, D.~Mosse, and R.~Melhem.
\newblock {Supporting Superpages in Non-contiguous Physical Memory}.
\newblock In {\em HPCA}, 2015.

\bibitem{rcuda}
J.~Duato, A.~Pena, F.~Silla, R.~Mayo, and E.~Quintana-Orti.
\newblock {rCUDA: Reducing the Number of GPU-based Accelerators in High
  Performance Clusters}.
\newblock In {\em HPCS}, 2010.

\bibitem{ksr}
T.~H. Dunigan.
\newblock {Kendall Square Multiprocessor: Early Experiences and Performance}.
\newblock In {\em of the Intel Paragon, ORNL/TM-12194}, 1994.

\bibitem{doung-micro12}
N.~Duong, D.~Zhao, T.~Kim, R.~Cammarota, M.~Valero, and A.~V. Veidenbaum.
\newblock {Improving Cache Management Policies Using Dynamic Reuse Distances}.
\newblock In {\em MICRO}, 2012.

\bibitem{fst}
E.~Ebrahimi, C.~J. Lee, O.~Mutlu, and Y.~N. Patt.
\newblock {Fairness via Source Throttling: A Configurable and High-performance
  Fairness Substrate for Multi-core Memory Systems}.
\newblock In {\em ASPLOS}, 2010.

\bibitem{ebrahimi-isca2011}
E.~Ebrahimi, C.~J. Lee, O.~Mutlu, and Y.~N. Patt.
\newblock {Prefetch-aware Shared Resource Management for Multi-core Systems}.
\newblock In {\em ISCA}, 2011.

\bibitem{fst-tocs}
E.~Ebrahimi, C.~J. Lee, O.~Mutlu, and Y.~N. Patt.
\newblock {Fairness via Source Throttling: A Configurable and High-Performance
  Fairness Substrate for Multi-Core Memory Systems}.
\newblock {\em ACM TOCS}, 30(7), 2012.

\bibitem{pam}
E.~Ebrahimi, R.~Miftakhutdinov, C.~Fallin, C.~J. Lee, J.~A. Joao, O.~Mutlu, and
  Y.~N. Patt.
\newblock {Parallel Application Memory Scheduling}.
\newblock In {\em MICRO}, 2011.

\bibitem{ebrahimi-micro09}
E.~Ebrahimi, O.~Mutlu, C.~J. Lee, and Y.~N. Patt.
\newblock {Coordinated Control of Multiple Prefetchers in Multi-core Systems}.
\newblock In {\em MICRO}, 2009.

\bibitem{ebrahimi-hpca}
E.~Ebrahimi, O.~Mutlu, and Y.~N. Patt.
\newblock {Techniques for Bandwidth-efficient Prefetching of Linked Data
  Structures in Hybrid Prefetching Systems}.
\newblock In {\em HPCA}, 2009.

\bibitem{etsion-tc}
Y.~Etsion and D.~G. Feitelson.
\newblock {Exploiting Core Working Sets to Filter the L1 Cache with Random
  Sampling}.
\newblock {\em IEEE TC}, 61(11):1535--1550, 2012.

\bibitem{harmonic_speedup}
S.~Eyerman and L.~Eeckhout.
\newblock {System-Level Performance Metrics for Multiprogram Workloads}.
\newblock {\em IEEE Micro}, 28(3), 2008.

\bibitem{ws-metric2}
S.~Eyerman and L.~Eeckhout.
\newblock {Restating the Case for Weighted-IPC Metrics to Evaluate Multiprogram
  Workload Performance}.
\newblock {\em IEEE CAL}, 2014.

\bibitem{chipper}
C.~Fallin, C.~Craik, and O.~Mutlu.
\newblock {CHIPPER: A Low-complexity bufferless deflection router}.
\newblock In {\em {HPCA}}, 2011.

\bibitem{minbd}
C.~Fallin, G.~Nazario, X.~Yu, K.~Chang, R.~Ausavarungnirun, and O.~Mutlu.
\newblock {MinBD: Minimally-Buffered Deflection Routing for Energy-Efficient
  Interconnect}.
\newblock In {\em NoCs}, 2012.

\bibitem{minbd-book}
C.~Fallin, G.~Nazario, X.~Yu, K.~Chang, R.~Ausavarungnirun, and O.~Mutlu.
\newblock {\em {Bufferless and Minimally-Buffered Deflection Routing, in
  Routing Algorithms in Networks-on-Chip}}, pages 241--275.
\newblock Springer New York, New York, NY, 2014.

\bibitem{7056040}
A.~Farmahini-Farahani, J.~H. Ahn, K.~Morrow, and N.~S. Kim.
\newblock {NDA: Near-DRAM Acceleration Architecture Leveraging Commodity DRAM
  Devices and Standard Memory Modules}.
\newblock In {\em HPCA}, 2015.

\bibitem{maze-routing}
M.~Fattah et~al.
\newblock {A Low-Overhead, Fully-Distributed, Guaranteed-Delivery Routing
  Algorithm for Faulty Network-on-Chips}.
\newblock In {\em NOCS}, 2015.

\bibitem{feng-interact12}
M.~Feng, C.~Tian, and R.~Gupta.
\newblock {Enhancing LRU Replacement via Phantom Associativity}.
\newblock In {\em INTERACT}, Feb 2012.

\bibitem{vliw}
J.~A. Fisher.
\newblock {Very Long Instruction Word Architectures and the ELI-512}.
\newblock In {\em ISCA}, 1983.

\bibitem{flynn}
M.~Flynn.
\newblock {Very High-Speed Computing Systems}.
\newblock {\em Proc.\ of the IEEE}, 54(2), 1966.

\bibitem{microarch}
A.~Fog.
\newblock {The Microarchitecture of Intel, AMD and VIA CPUs}.

\bibitem{nvlink}
D.~Foley.
\newblock {Ultra-Performance Pascal GPU and NVLink Interconnect}.
\newblock In {\em HotChips}.

\bibitem{fraguela-2003}
B.~B. Fraguela, J.~Renau, P.~Feautrier, D.~Padua, and J.~Torrellas.
\newblock {Programming the FlexRAM Parallel Intelligent Memory System}.
\newblock In {\em PPoPP}, 2003.

\bibitem{dwf}
W.~Fung, I.~Sham, G.~Yuan, and T.~Aamodt.
\newblock {Dynamic Warp Formation and Scheduling for Efficient GPU Control
  Flow}.
\newblock In {\em MICRO}, 2007.

\bibitem{tbc}
W.~W.~L. Fung and T.~M. Aamodt.
\newblock {Thread Block Compaction for Efficient SIMT Control Flow}.
\newblock In {\em HPCA}, 2011.

\bibitem{jayneel-isca16}
J.~Gandhi, , M.~D. Hill, and M.~M. Swift.
\newblock {Exceeding the Best of Nested and Shadow Paging}.
\newblock In {\em ISCA}, 2016.

\bibitem{jayneel-micro14}
J.~Gandhi, A.~Basu, M.~D. Hill, and M.~M. Swift.
\newblock {Efficient Memory Virtualization}.
\newblock In {\em MICRO}, 2014.

\bibitem{gao2010dueling}
H.~Gao and C.~Wilkerson.
\newblock {A Dueling Segmented LRU Replacement Algorithm with Adaptive
  Bypassing}.
\newblock In {\em JWAC}, 2010.

\bibitem{7429299}
M.~Gao, G.~Ayers, and C.~Kozyrakis.
\newblock {Practical Near-Data Processing for In-Memory Analytics Frameworks}.
\newblock In {\em PACT}, 2015.

\bibitem{7446059}
M.~Gao and C.~Kozyrakis.
\newblock {HRL: Efficient and Flexible Reconfigurable Logic for Near-data
  Processing}.
\newblock In {\em HPCA}, 2016.

\bibitem{gaur-isca11}
J.~Gaur, M.~Chaudhuri, and S.~Subramoney.
\newblock {Bypass and Insertion Algorithms for Exclusive Last-Level Caches}.
\newblock In {\em ISCA}, 2011.

\bibitem{gay-pldi1998}
D.~Gay and A.~Aiken.
\newblock {Memory Management with Explicit Regions}.
\newblock In {\em PLDI}, 1998.

\bibitem{gebhart}
M.~Gebhart, D.~R. Johnson, D.~Tarjan, S.~W. Keckler, W.~J. Dally, E.~Lindholm,
  and K.~Skadron.
\newblock {Energy-Efficient Mechanisms for Managing Thread Context in
  Throughput Processors}.
\newblock In {\em ISCA}, 2011.

\bibitem{pim-chapter}
S.~Ghose, K.~Hsieh, A.~Boroumand, R.~Ausavarungnirun, and O.~Mutlu.
\newblock {The Processing-in-Memory Paradigm: Mechanisms to Enable Adoption}.
\newblock In {\em {{Beyond-CMOS Technologies for Next Generation Computer
  Design}}}. 2018.

\bibitem{ghose2013}
S.~Ghose, H.~Lee, and J.~F. Mart{\'\i}nez.
\newblock {Improving Memory Scheduling via Processor-side Load Criticality
  Information}.
\newblock In {\em ISCA}, 2013.

\bibitem{375174}
M.~Gokhale, B.~Holmes, and K.~Iobst.
\newblock {Processing in Memory: the Terasys Massively Parallel PIM Array}.
\newblock {\em Computer}, 28(4):23--31, 1995.

\bibitem{gomez08}
C.~G{\'o}mez, M.~G{\'o}mez, P.~L{\'o}pez, and J.~Duato.
\newblock {Reducing Packet Dropping in a Bufferless NoC}.
\newblock {\em EuroPar}, 2008.

\bibitem{gorman-ismm08}
M.~Gorman and P.~Healy.
\newblock {Supporting Superpage Allocation Without Additional Hardware
  Support}.
\newblock In {\em ISMM}, 2008.

\bibitem{gorman-wiosca10}
M.~Gorman and P.~Healy.
\newblock {Performance Characteristics of Explicit Superpage Support}.
\newblock In {\em WIOSCA}, 2010.

\bibitem{govindaraju-sc06}
N.~Govindaraju, S.~Larsen, J.~Gray, and D.~Manocha.
\newblock {A Memory Model for Scientific Algorithms on Graphics Processors}.
\newblock In {\em SC}, 2006.

\bibitem{grimes1989intel}
J.~D. Grimes, L.~Kohn, and R.~Bharadhwaj.
\newblock {The Intel i860 64-bit Processor: A General-purpose CPU with 3D
  Graphics Capabilities}.
\newblock {\em IEEE CGA}, 9(4):85--94, 1989.

\bibitem{numachine}
R.~Grindley, T.~Abdelrahman, S.~Brown, S.~Caranci, D.~DeVries, B.~Gamsa,
  A.~Grbic, M.~Gusat, R.~Ho, O.~Krieger, et~al.
\newblock {The NUMAchine Multiprocessor}.
\newblock In {\em ICPP}, 2000.

\bibitem{grot-hpca2009}
B.~Grot, J.~Hestness, S.~W. Keckler, and O.~Mutlu.
\newblock {{Express Cube Topologies for On-Chip Interconnects}}.
\newblock In {\em HPCA}, 2009.

\bibitem{grot-isca2011}
B.~Grot, J.~Hestness, S.~W. Keckler, and O.~Mutlu.
\newblock {Kilo-NOC: A Heterogeneous Network-on-chip Architecture for
  Scalability and Service Guarantees}.
\newblock In {\em ISCA}, 2011.

\bibitem{grot2010}
B.~Grot, S.~Keckler, and O.~Mutlu.
\newblock {Topology-aware Quality-of-service Support in Highly Integrated Chip
  Multiprocessors}.
\newblock In {\em WIOSCA}, 2010.

\bibitem{pvc}
B.~Grot, S.~W. Keckler, and O.~Mutlu.
\newblock {Preemptive Virtual Clock: A Flexible, Efficient, and Cost-effective
  QOS Scheme for Networks-on-Chip}.
\newblock In {\em MICRO}, 2009.

\bibitem{gschwind-cf2006}
M.~Gschwind.
\newblock {Chip Multiprocessing and the Cell Broadband Engine}.
\newblock In {\em CF}, 2006.

\bibitem{gummaraju-pact2007}
J.~Gummaraju, M.~Erez, J.~Coburn, M.~Rosenblum, and W.~J. Dally.
\newblock {Architectural Support for the Stream Execution Model on
  General-Purpose Processors}.
\newblock In {\em PACT}, 2007.

\bibitem{guo-wondp14}
Q.~Guo, N.~Alachiotis, B.~Akin, F.~Sadi, G.~Xu, T.-M. Low, L.~Pileggi, J.~C.
  Hoe, and F.~Franchetti.
\newblock {3D-Stacked Memory-Side Acceleration: Accelerator and System Design}.
\newblock In {\em WONDP}, 2014.

\bibitem{gupta-ipdps13}
S.~Gupta, H.~Gao, and H.~Zhou.
\newblock {Adaptive Cache Bypassing for Inclusive Last Level Caches}.
\newblock In {\em IPDPS}, 2013.

\bibitem{sci}
D.~Gustavson.
\newblock {The Scalable Coherent Interface and Related Standards Projects}.
\newblock {\em {IEEE} Micro}, 1992.

\bibitem{masa-fmt}
R.~H. Halstead and T.~Fujita.
\newblock {MASA: A Multithreaded Processor Architecture for Parallel Symbolic
  Computing}.
\newblock In {\em ISCA}, 1988.

\bibitem{hr-model}
V.~C. Hamacher and H.~Jiang.
\newblock {Hierarchical Ring Network Configuration and Performance Modeling}.
\newblock {\em IEEE TC}, 2001.

\bibitem{han-reducing-div}
T.~D. Han and T.~S. Abdelrahman.
\newblock {Reducing Branch Divergence in {GPU} Programs}.
\newblock In {\em GPGPU}, 2011.

\bibitem{hart-compcon1994}
C.~A. Hart.
\newblock {CDRAM in a Unified Memory Architecture}.
\newblock In {\em Intl.~Computer Conference}, 1994.

\bibitem{hashemi-isca2016}
M.~Hashemi, Khubaib, E.~Ebrahimi, O.~Mutlu, and Y.~N. Patt.
\newblock {Accelerating Dependent Cache Misses with an Enhanced Memory
  Controller}.
\newblock In {\em ISCA}, 2016.

\bibitem{hashemi-micro2016}
M.~Hashemi, O.~Mutlu, and Y.~N. Patt.
\newblock {Continuous Runahead: Transparent Hardware Acceleration for Memory
  Intensive Workloads}.
\newblock In {\em MICRO}, 2016.

\bibitem{chargecache}
H.~Hassan, G.~Pekhimenko, N.~Vijaykumar, V.~Seshadri, D.~Lee, O.~Ergin, and
  O.~Mutlu.
\newblock {ChargeCache: Reducing DRAM Latency by Exploiting Row Access
  Locality}.
\newblock In {\em HPCA}, 2016.

\bibitem{softmc}
H.~Hassan, N.~Vijaykumar, S.~Khan, S.~Ghose, K.~Chang, G.~Pekhimenko, D.~Lee,
  O.~Ergin, and O.~Mutlu.
\newblock {SoftMC: A Flexible and Practical Open-source Infrastructure for
  Enabling Experimental DRAM Studies}.
\newblock In {\em HPCA}, 2017.

\bibitem{scarab}
M.~Hayenga, N.~E. Jerger, and M.~Lipasti.
\newblock {SCARAB: A Single Cycle Adaptive Routing and Bufferless Network}.
\newblock In {\em MICRO}, 2009.

\bibitem{mars}
B.~He, W.~Fang, Q.~Luo, N.~K. Govindaraju, and T.~Wang.
\newblock {Mars: A MapReduce Framework on Graphics Processors}.
\newblock In {\em PACT}, 2008.

\bibitem{hellerman-ec66}
H.~Hellerman.
\newblock {Parallel Processing of Algebraic Expressions}.
\newblock {\em IEEE Transactions on Electronic Computers}, EC-15(1):82--91, Feb
  1966.

\bibitem{grid}
A.~Herrera.
\newblock {NVIDIA GRID: Graphics Accelerated VDI with the Visual Performance of
  a Workstation}.
\newblock May 2014.

\bibitem{hidaka-ieeemicro90}
H.~Hidaka, Y.~Matsuda, M.~Asakura, and K.~Fujishima.
\newblock {The Cache DRAM Architecture}.
\newblock {\em IEEE Micro}, 1990.

\bibitem{cm}
W.~Hillis.
\newblock {\em The Connection Machine}.
\newblock MIT Press, 1989.

\bibitem{hong-isca09}
S.~Hong and H.~Kim.
\newblock {An Analytical Model for a GPU Architecture with Memory-Level and
  Thread-Level Parallelism Awareness}.
\newblock In {\em ISCA}, 2009.

\bibitem{tom-isca16}
K.~Hsieh, E.~Ebrahimi, G.~Kim, N.~Chatterjee, M.~O'Connor, N.~Vijaykumar,
  O.~Mutlu, and S.~W. Keckler.
\newblock {Transparent Offloading and Mapping (TOM): Enabling
  Programmer-transparent Near-data Processing in GPU Systems}.
\newblock In {\em ISCA}, 2016.

\bibitem{hsieh-iccd2016}
K.~Hsieh, S.~Khan, N.~Vijaykumar, K.~K. Chang, A.~Boroumand, S.~Ghose, and
  O.~Mutlu.
\newblock {Accelerating Pointer Chasing in 3D-stacked Memory: Challenges,
  Mechanisms, Evaluation}.
\newblock In {\em ICCD}, 2016.

\bibitem{hsu-isca1993}
W.-C. Hsu and J.~E. Smith.
\newblock {Performance of Cached {DRAM} Organizations in Vector
  Supercomputers}.
\newblock In {\em ISCA}, 1993.

\bibitem{hur-micro2006}
I.~Hur and C.~Lin.
\newblock {Memory Prefetching Using Adaptive Stream Detection}.
\newblock In {\em MICRO}, 2006.

\bibitem{hbm2}
{Hybrid Memoty Cube Consortium}.
\newblock {High-Bandwidth Memory White Paper}.

\bibitem{hmc1}
{Hybrid Memoty Cube Consortium}.
\newblock {HMC Specification 1.1}, 2013.

\bibitem{hmc2}
{Hybrid Memoty Cube Consortium}.
\newblock {HMC Specification 2.0}, 2014.

\bibitem{gddr5}
{Hynix.}
\newblock {Hynix GDDR5 SGRAM Part H5GQ1H24AFR Revision 1.0}.

\bibitem{ikeda-icpads13}
T.~Ikeda and K.~Kise.
\newblock {Application Aware DRAM Bank Partitioning in CMP}.
\newblock In {\em ICPADS}, 2013.

\bibitem{intelioat}
{Intel Corp.}
\newblock {Intel\textregistered I/O Acceleration Technology}.
\newblock
  \url{http://www.intel.com/content/www/us/en/wireless-network/accel-technology.html}.

\bibitem{intelx86}
{Intel Corp.}
\newblock {Intel\textregistered\ 64 and IA-32 Architectures Optimization
  Reference Manual}, 2016.

\bibitem{intel-io-virt}
{Intel Corporation}.
\newblock Intel virtualization technology for directed i/o.

\bibitem{intel-sandybridge}
{Intel Corporation}.
\newblock {Intel(R) Microarchitecture Codename Sandy Bridge}.
\newblock http://www.intel.com/technology/architecture-silicon/2ndgen/.

\bibitem{sandybridge}
{{Intel Corporation}}.
\newblock {{Sandy Bridge Intel Processor Graphics Performance Developer's
  Guide}}.

\bibitem{mmx}
{Intel Corporation}.
\newblock Intel architecture mmx™ technology in business applications.
\newblock 1997.
\newblock \url{http://download.intel.com/design/PentiumII/papers/24336702.PDF}.

\bibitem{ivybridge}
{Intel Corporation}.
\newblock {Products (Formerly Ivy Bridge)}, {2012}.

\bibitem{haswell}
{Intel Corporation}.
\newblock Introduction to intel® architecture.
\newblock 2014.
\newblock
  \url{http://www.intel.com/content/dam/www/public/us/en/documents/white-papers/ia-introduction-basics-paper.pdf}.

\bibitem{ept}
{Intel Corporation}.
\newblock Intel 64 and ia-32 architectures software developer’s manual.
\newblock 2016.
\newblock
  \url{https://www-ssl.intel.com/content/dam/www/public/us/en/documents/manuals/64-ia-32-architectures-software-developer-manual-325462.pdf}.

\bibitem{skylake}
{Intel Corporation}.
\newblock 6th generation intel® core™ processor family datasheet, vol. 1.
\newblock 2017.
\newblock
  \url{http://www.intel.com/content/dam/www/public/us/en/documents/datasheets/desktop-6th-gen-core-family-datasheet-vol-1.pdf}.

\bibitem{vm-contemporary}
B.~Jacob and T.~Mudge.
\newblock {Virtual Memory in Contemporary Microprocessors}.
\newblock In {\em IEEE Micro}, 1998.

\bibitem{dip}
A.~Jaleel, W.~Hasenplaugh, M.~Qureshi, J.~Sebot, S.~Steely, Jr., and J.~Emer.
\newblock {Adaptive Insertion Policies for Managing Shared Caches}.
\newblock In {\em PACT}, 2008.

\bibitem{rrip}
A.~Jaleel, K.~B. Theobald, S.~C. Steely, Jr., and J.~Emer.
\newblock {High Performance Cache Replacement Using Re-reference Interval
  Prediction ({RRIP})}.
\newblock In {\em ISCA}, 2010.

\bibitem{jalminger-iccp03}
J.~Jalminger and P.~Stenstrom.
\newblock {A Novel Approach to Cache Block Reuse Predictions}.
\newblock In {\em ICPP}, 2003.

\bibitem{stackjdec}
{JEDEC}.
\newblock {High Bandwidth Memory (HBM)}, 2013.

\bibitem{jeong2012qos}
M.~K. Jeong, M.~Erez, C.~Sudanthi, and N.~Paver.
\newblock {A QoS-Aware Memory Controller for Dynamically Balancing GPU and CPU
  Bandwidth Use in an MPSoC}.
\newblock In {\em DAC}, 2012.

\bibitem{mrpb}
W.~Jia, K.~A. Shaw, and M.~Martonosi.
\newblock {{MRPB}: Memory Request Prioritization for Massively Parallel
  Processors}.
\newblock In {\em HPCA}, 2014.

\bibitem{jiang-pact2009}
X.~Jiang, Y.~Solihin, L.~Zhao, and R.~Iyer.
\newblock {Architecture Support for Improving Bulk Memory Copying and
  Initialization Performance}.
\newblock In {\em PACT}, 2009.

\bibitem{adwait-thesis}
A.~Jog.
\newblock {\em {Design and Analysis of Scheduling Techniques for Throughput
  Processors}}.
\newblock PhD thesis, Pennsylvania State Univ., 2015.

\bibitem{mafia}
A.~Jog, O.~Kayiran, T.~Kesten, A.~Pattnaik, E.~Bolotin, N.~Chatterjee, S.~W.
  Keckler, M.~T. Kandemir, and C.~R. Das.
\newblock {Anatomy of GPU Memory System for Multi-Application Execution}.
\newblock In {\em MEMSYS}, 2015.

\bibitem{osp-isca13}
A.~Jog, O.~Kay{\i}ran, A.~K. Mishra, M.~T. Kandemir, O.~Mutlu, R.~Iyer, and
  C.~R. Das.
\newblock {Orchestrated Scheduling and Prefetching for GPGPUs}.
\newblock In {\em ISCA}, 2013.

\bibitem{owl-asplos13}
A.~Jog, O.~Kay{\i}ran, N.~C. Nachiappan, A.~K. Mishra, M.~T. Kandemir,
  O.~Mutlu, R.~Iyer, and C.~R. Das.
\newblock {OWL: Cooperative Thread Array Aware Scheduling Techniques for
  Improving GPGPU Performance}.
\newblock In {\em ASPLOS}, 2013.

\bibitem{adwait-critical-memsched}
A.~Jog, O.~Kayiran, A.~Pattnaik, M.~T. Kandemir, O.~Mutlu, R.~Iyer, and C.~R.
  Das.
\newblock {Exploiting Core Criticality for Enhanced GPU Performance}.
\newblock In {\em SIGMETRICS}, 2016.

\bibitem{annex-cache}
L.~K. John and A.~Subramanian.
\newblock {Design and Performance Evaluation of A Cache Assist to Implement
  Selective Caching}.
\newblock In {\em ICCD}, 1997.

\bibitem{joseph-isca1997}
D.~Joseph and D.~Grunwald.
\newblock {Prefetching Using Markov Predictors}.
\newblock In {\em ISCA}, 1997.

\bibitem{jouppi-isca90}
N.~P. Jouppi.
\newblock {Improving Direct-Mapped Cache Performance by the Addition of a Small
  Fully-Associative Cache and Prefetch Buffers}.
\newblock In {\em ISCA}, 1990.

\bibitem{kahle-ibmjrd2005}
J.~A. Kahle, M.~N. Day, H.~P. Hofstee, C.~R. Johns, T.~R. Maeurer, and
  D.~Shippy.
\newblock {Introduction to the Cell Multiprocessor}.
\newblock {\em IBM JRD}, 2005.

\bibitem{kandiraju-isca02}
G.~B. Kandiraju and A.~Sivasubramaniam.
\newblock {Going the Distance for TLB Prefetching: An Application-driven
  Study}.
\newblock In {\em ISCA}, 2002.

\bibitem{808425}
Y.~Kang, W.~Huang, S.-M. Yoo, D.~Keen, Z.~Ge, V.~Lam, P.~Pattnaik, and
  J.~Torrellas.
\newblock {FlexRAM: Toward an Advanced Intelligent Memory System}.
\newblock In {\em ICCD}, 1999.

\bibitem{rmm}
V.~Karakostas, J.~Gandhi, F.~Ayar, A.~Cristal, M.~D. Hill, K.~S. McKinley,
  M.~Nemirovsky, M.~M. Swift, and O.~\"{U}nsal.
\newblock {Redundant Memory Mappings for Fast Access to Large Memories}.
\newblock In {\em ISCA}, 2015.

\bibitem{karakostas.hpca16}
V.~Karakostas, J.~Gandhi, A.~Cristal, M.~D. Hill, K.~S. McKinley,
  M.~Nemirovsky, M.~M. Swift, and O.~Unsal.
\newblock {Energy-Efficient Address Translation}.
\newblock In {\em HPCA}, 2016.

\bibitem{lulesh}
I.~Karlin, A.~Bhatele, J.~Keasler, B.~Chamberlain, J.~Cohen, Z.~DeVito,
  R.~Haque, D.~Laney, E.~Luke, F.~Wang, D.~Richards, M.~Schulz, and C.~Still.
\newblock {Exploring Traditional and Emerging Parallel Programming Models using
  a Proxy Application}.
\newblock In {\em IPDPS}, 2013.

\bibitem{lulesh-origin}
I.~Karlin, J.~Keasler, and R.~Neely.
\newblock {Lulesh 2.0 Updates and Changes}.
\newblock 2013.

\bibitem{minimalist}
D.~Kaseridis, J.~Stuecheli, and L.~K. John.
\newblock {Minimalist Open-page: A DRAM Page-mode Scheduling Policy for the
  Many-core Era}.
\newblock In {\em MICRO}, 2011.

\bibitem{gdev}
S.~Kato, M.~McThrow, C.~Maltzahn, and S.~Brandt.
\newblock {Gdev: First-Class GPU Resource Management in the Operating System}.
\newblock In {\em USENIX ATC}, 2012.

\bibitem{nmnl-pact13}
O.~Kay{\i}ran, A.~Jog, M.~T. Kandemir, and C.~R. Das.
\newblock {Neither More Nor Less: Optimizing Thread-Level Parallelism for
  GPGPUs}.
\newblock In {\em PACT}, 2013.

\bibitem{ucstate}
O.~Kayiran, A.~Jog, A.~Pattnaik, R.~Ausavarungnirun, X.~Tang, M.~T. Kandemir,
  G.~H. Loh, O.~Mutlu, and C.~R. Das.
\newblock {uC-States: Fine-grained GPU Datapath Power Management}.
\newblock In {\em PACT}, 2016.

\bibitem{cpugpu-micro}
O.~Kay{\i}ran, N.~C. Nachiappan, A.~Jog, R.~Ausavarungnirun, M.~T. Kandemir,
  G.~H. Loh, O.~Mutlu, and C.~R. Das.
\newblock {Managing GPU Concurrency in Heterogeneous Architectures}.
\newblock In {\em MICRO}, 2014.

\bibitem{kedem-1997}
G.~Kedem and R.~P. Koganti.
\newblock {{WCDRAM}: A Fully Associative Integrated Cached-{DRAM} with Wide
  Cache Lines}.
\newblock {\em CS-1997-03, Duke}, 1997.

\bibitem{khan-dsn2016}
S.~Khan et~al.
\newblock {PARBOR: An Efficient System-Level Technique to Detect Data Dependent
  Failures in DRAM}.
\newblock In {\em DSN}, 2016.

\bibitem{khan-sigmetrics2014}
S.~Khan, D.~Lee, Y.~Kim, A.~R. Alameldeen, C.~Wilkerson, and O.~Mutlu.
\newblock {The Efficacy of Error Mitigation Techniques for DRAM Retention
  Failures: A Comparative Experimental Study}.
\newblock In {\em SIGMETRICS}, 2014.

\bibitem{khan-cal2017}
S.~Khan, C.~Wilkerson, D.~Lee, A.~R. Alameldeen, and O.~Mutlu.
\newblock {A Case for Memory Content-Based Detection and Mitigation of
  Data-Dependent Failures in DRAM}.
\newblock In {\em IEEE CAL}, 2016.

\bibitem{khan-micro2017}
S.~Khan, C.~Wilkerson, Z.~Wang, A.~R. Alameldeen, D.~Lee, and O.~Mutlu.
\newblock {Detecting and Mitigating Data-dependent DRAM Failures by Exploiting
  Current Memory Content}.
\newblock In {\em MICRO}, 2017.

\bibitem{kharbutli-ieeetran}
M.~Kharbutli and Y.~Solihin.
\newblock {Counter-Based Cache Replacement and Bypassing Algorithms}.
\newblock {\em IEEE TC}, 57(4):433--447, Apr. 2008.

\bibitem{khronos2008opencl}
{Khronos OpenCL Working Group}.
\newblock {The OpenCL Specification}.
\newblock \url{http://www.khronos.org/registry/cl/specs/opencl-1.0.29.pdf},
  2008.

\bibitem{kim-asic2001}
J.~Kim and M.~C. Papaefthymiou.
\newblock {Block-based Multi-period Refresh for Energy Efficient Dynamic
  Memory}.
\newblock In {\em ASIC}, 2001.

\bibitem{kim.bmc18}
J.~S. Kim, D.~Senol, H.~Xin, D.~Lee, S.~Ghose, M.~Alser, H.~Hassan, O.~Ergin,
  C.~Alkan, and O.~Mutlu.
\newblock {GRIM-Filter: Fast Seed Location Filtering in DNA Read Mapping Using
  Processing-in-Memory Technologies}.
\newblock {\em BMC Genomics}, 2018.

\bibitem{kim09}
K.~Kim and J.~Lee.
\newblock {A New Investigation of Data Retention Time in Truly Nanoscaled
  DRAMs}.
\newblock In {\em EDL}, 2009.

\bibitem{atlas}
Y.~Kim, D.~Han, O.~Mutlu, and M.~Harchol-Balter.
\newblock {ATLAS: A Scalable and High-Performance Scheduling Algorithm for
  Multiple Memory Controllers}.
\newblock In {\em HPCA}, 2010.

\bibitem{tcm}
Y.~Kim, M.~Papamichael, O.~Mutlu, and M.~Harchol-Balter.
\newblock {Thread Cluster Memory Scheduling: Exploiting Differences in Memory
  Access Behavior}.
\newblock In {\em MICRO}, 2010.

\bibitem{salp}
Y.~Kim, V.~Seshadri, D.~Lee, J.~Liu, and O.~Mutlu.
\newblock {A Case for Exploiting Subarray-Level Parallelism (SALP) in DRAM}.
\newblock In {\em ISCA}, 2012.

\bibitem{Kodi08}
A.~K. Kodi, A.~Sarathy, and A.~Louri.
\newblock {iDEAL: Inter-router Dual-function Energy and Area-efficient Links
  for Network-on-chip (NoC) Architectures}.
\newblock In {\em ISCA}, 2008.

\bibitem{4115697}
P.~M. Kogge.
\newblock {EXECUBE-A New Architecture for Scaleable MPPs}.
\newblock In {\em ICPP}, 1994.

\bibitem{chaosrouter}
S.~Konstantinidou and L.~Snyder.
\newblock {Chaos Router: Architecture and Performance}.
\newblock In {\em ISCA}, 1991.

\bibitem{kroft-isca81}
D.~Kroft.
\newblock {Lockup-Free Instruction Fetch/Prefetch Cache Organization}.
\newblock In {\em ISCA}, 1981.

\bibitem{ku-ispass2013}
E.~Kultursay, M.~Kandemir, A.~Sivasubramaniam, and O.~Mutlu.
\newblock {Evaluating STT-RAM as an energy-efficient main memory alternative}.
\newblock In {\em ISPASS}, 2013.

\bibitem{ingens}
Y.~Kwon, H.~Yu, S.~Peter, C.~J. Rossbach, and E.~Witchel.
\newblock {Coordinated and Efficient Huge Page Management with Ingens}.
\newblock In {\em OSDI}, 2016.

\bibitem{lai-isca2001}
A.-C. Lai, C.~Fide, and B.~Falsafi.
\newblock {Dead-block Prediction \& Dead-block Correlating Prefetchers}.
\newblock In {\em ISCA}, 2001.

\bibitem{bowtie2}
B.~Langmead and S.~L. Salzberg.
\newblock {Fast Gapped-Read Alignment with Bowtie 2}.
\newblock {\em Nature Methods}, 2012.

\bibitem{lee-isca2009}
B.~C. Lee, E.~Ipek, O.~Mutlu, and D.~Burger.
\newblock {Architecting Phase Change Memory as a Scalable DRAM Alternative}.
\newblock In {\em ISCA}, 2009.

\bibitem{lee-cacm2010}
B.~C. Lee, E.~Ipek, O.~Mutlu, and D.~Burger.
\newblock {Phase Change Memory Architecture and the Quest for Scalability}.
\newblock {\em CACM}, 53(7):99--106, 2010.

\bibitem{lee-ieeemicro2010}
B.~C. Lee, P.~Zhou, J.~Yang, Y.~Zhang, B.~Zhao, E.~Ipek, O.~Mutlu, and
  D.~Burger.
\newblock {Phase-Change Technology and the Future of Main Memory}.
\newblock {\em {IEEE Micro}}, 30(1):143--143, 2010.

\bibitem{pa-micro08}
C.~J. Lee, O.~Mutlu, V.~Narasiman, and Y.~N. Patt.
\newblock {Prefetch-aware DRAM Controllers}.
\newblock In {\em MICRO}, 2008.

\bibitem{lee-tc2011}
C.~J. Lee, O.~Mutlu, V.~Narasiman, and Y.~N. Patt.
\newblock {Prefetch-aware Memory Controllers}.
\newblock {\em IEEE TC}, 60(10):1406--1430, 2011.

\bibitem{lee2010dram}
C.~J. Lee, V.~Narasiman, E.~Ebrahimi, O.~Mutlu, and Y.~N. Patt.
\newblock {DRAM-aware Last-level Cache Writeback: Reducing Write-caused
  Interference in Memory Systems}.
\newblock In {\em TR-HPS-2010-002}, April, 2010.

\bibitem{cjlee-micro09}
C.~J. Lee, V.~Narasiman, O.~Mutlu, and Y.~N. Patt.
\newblock {Improving Memory Bank-Level Parallelism in the Presence of
  Prefetching}.
\newblock In {\em MICRO}, 2009.

\bibitem{donghyuk-stack}
D.~Lee, S.~Ghose, G.~Pekhimenko, S.~Khan, and O.~Mutlu.
\newblock {Simultaneous Multi-layer Access: Improving 3D-stacked Memory
  Bandwidth at Low Cost}.
\newblock {\em ACM TACO}, 12(4):63, 2016.

\bibitem{ava-dram}
D.~Lee, S.~Khan, L.~Subramanian, S.~Ghose, R.~Ausavarungnirun, G.~Pekhimenko,
  V.~Seshadri, and O.~Mutlu.
\newblock {Design-Induced Latency Variation in Modern DRAM Chips:
  Characterization, Analysis, and Latency Reduction Mechanisms}.
\newblock In {\em SIGMETRICS}, 2017.

\bibitem{al-dram}
D.~Lee, Y.~Kim, G.~Pekhimenko, S.~Khan, V.~Seshadri, K.~Chang, and O.~Mutlu.
\newblock {Adaptive-latency DRAM: Optimizing DRAM Timing for the Common-case}.
\newblock In {\em HPCA}, 2015.

\bibitem{tl-dram}
D.~Lee, Y.~Kim, V.~Seshadri, J.~Liu, L.~Subramanian, and O.~Mutlu.
\newblock {Tiered-latency DRAM: A Low Latency and Low Cost DRAM Architecture}.
\newblock In {\em HPCA}, 2013.

\bibitem{donghyuk-ddma}
D.~Lee, L.~Subramanian, R.~Ausavarungnirun, J.~Choi, and O.~Mutlu.
\newblock {Decoupled Direct Memory Access: Isolating CPU and IO Traffic by
  Leveraging a Dual-Data-Port DRAM}.
\newblock In {\em PACT}, 2015.

\bibitem{carole-wu}
S.-Y. Lee and C.-J. Wu.
\newblock {Characterizing GPU Latency Hiding Ability}.
\newblock In {\em ISPASS}, 2014.

\bibitem{gpuwattch}
J.~Leng, T.~Hetherington, A.~ElTantawy, S.~Gilani, N.~S. Kim, T.~M. Aamodt, and
  V.~J. Reddi.
\newblock {GPUWattch: Enabling Energy Optimizations in GPGPUs}.
\newblock In {\em ISCA}, 2013.

\bibitem{li-sc15}
A.~Li, G.-J. van~den Braak, A.~Kumar, and H.~Corporaal.
\newblock {Adaptive and Transparent Cache Bypassing for {GPUs}}.
\newblock In {\em SC}, 2015.

\bibitem{li-ics15}
C.~Li, S.~L. Song, H.~Dai, A.~Sidelnik, S.~K.~S. Hari, and H.~Zhou.
\newblock {Locality-Driven Dynamic {GPU} Cache Bypassing}.
\newblock In {\em ICS}, 2015.

\bibitem{donglihpca15}
D.~Li, M.~Rhu, D.~Johnson, M.~O'Connor, M.~Erez, D.~Burger, D.~Fussell, and
  S.~Redder.
\newblock {Priority-Based Cache Allocation in Throughput Processors}.
\newblock In {\em HPCA}, 2015.

\bibitem{li2014symbiotic}
T.~Li, V.~K. Narayana, and T.~El-Ghazawi.
\newblock {Symbiotic Scheduling of Concurrent GPU Kernels for Performance and
  Energy Optimizations}.
\newblock In {\em CF}, 2014.

\bibitem{libhugetlbfs}
{Huge Pages Part 2 (Interfaces)}.
\newblock \url{https://lwn.net/Articles/375096/}.
\newblock [February, 2010].

\bibitem{lin-iccd2012}
C.~H. Lin, D.~Y. Shen, Y.~J. Chen, C.~L. Yang, and M.~Wang.
\newblock {SECRET: Selective Error Correction for Refresh Energy Reduction in
  DRAMs}.
\newblock In {\em ICCD}, 2012.

\bibitem{lindholm}
E.~Lindholm, J.~Nickolls, S.~Oberman, and J.~Montrym.
\newblock {NVIDIA Tesla: A Unified Graphics and Computing Architecture}.
\newblock {\em IEEE Micro}, 28(2), 2008.

\bibitem{liu-micro08}
H.~Liu, M.~Ferdman, J.~Huh, and D.~Burger.
\newblock {Cache Bursts: A New Approach for Eliminating Dead Blocks and
  Increasing Cache Efficiency}.
\newblock In {\em MICRO}, 2008.

\bibitem{liu-isca2013}
J.~Liu, B.~Jaiyen, Y.~Kim, C.~Wilkerson, and O.~Mutlu.
\newblock {An Experimental Study of Data Retention Behavior in Modern {DRAM}
  Devices: Implications for Retention Time Profiling Mechanisms}.
\newblock In {\em ISCA}, 2013.

\bibitem{raidr}
J.~Liu, B.~Jaiyen, R.~Veras, and O.~Mutlu.
\newblock {RAIDR: Retention-aware Intelligent DRAM Refresh}.
\newblock In {\em ISCA}, 2012.

\bibitem{liu-pact12}
L.~Liu, Z.~Cui, M.~Xing, Y.~Bao, M.~Chen, and C.~Wu.
\newblock {A Software Memory Partition Approach for Eliminating Bank-level
  Interference in Multicore Systems}.
\newblock In {\em PACT}, 2012.

\bibitem{liu-ipccc16}
W.~Liu, P.~Huang, T.~Kun, T.~Lu, K.~Zhou, C.~Li, and X.~He.
\newblock {LAMS: A Latency-aware Memory Scheduling Policy for Modern DRAM
  Systems}.
\newblock In {\em IPCCC}, 2016.

\bibitem{liu-icpp07}
W.~Liu, W.~Muller-Wittig, and B.~Schmidt.
\newblock {Performance Predictions for General-Purpose Computation on GPUs}.
\newblock In {\em ICPP}, 2007.

\bibitem{molecular-gpgpu}
W.~Liu, B.~Schmidt, G.~Voss, and W.~Muller-Wittig.
\newblock {Accelerating Molecular Dynamics Simulations using Graphics
  Processing Units with CUDA}.
\newblock {\em Computer Physics Communications}, 179(9):634--641, 2008.

\bibitem{loh-stack}
G.~H. Loh.
\newblock {3D-stacked Memory Architectures for Multi-core Processors}.
\newblock In {\em ISCA}, 2008.

\bibitem{lu-micro2015}
S.-L. Lu, Y.-C. Lin, and C.-L. Yang.
\newblock {Improving DRAM Latency with Dynamic Asymmetric Subarray}.
\newblock In {\em MICRO}, 2015.

\bibitem{pin}
C.-K. Luk, R.~Cohn, R.~Muth, H.~Patil, A.~Klauser, G.~Lowney, S.~Wallace, V.~J.
  Reddi, and K.~Hazelwood.
\newblock {Pin: Building Customized Program Analysis Tools with Dynamic
  Instrumentation}.
\newblock In {\em PLDI}, 2005.

\bibitem{luo-dsn2014}
Y.~Luo, S.~Govindan, B.~Sharma, M.~Santaniello, J.~Meza, A.~Kansal, J.~Liu,
  B.~Khessib, K.~Vaid, and O.~Mutlu.
\newblock {Characterizing Application Memory Error Vulnerability to Optimize
  Datacenter Cost via Heterogeneous-Reliability Memory}.
\newblock In {\em DSN}, 2014.

\bibitem{lustig-13}
D.~Lustig, A.~Bhattacharjee, and M.~Martonosi.
\newblock {TLB Improvements for Chip Multiprocessors: Inter-Core Cooperative
  Prefetchers and Shared Last-Level TLBs}.
\newblock {\em ACM TACO}, 2013.

\bibitem{ma-asap12}
L.~Ma and R.~Chamberlain.
\newblock {A Performance Model for Memory Bandwidth Constrained Applications on
  Graphics Engines}.
\newblock In {\em ASAP}, 2012.

\bibitem{mai-isca2000}
K.~Mai, T.~Paaske, N.~Jayasena, R.~Ho, W.~J. Dally, and M.~Horowitz.
\newblock {Smart Memories: A Modular Reconfigurable Architecture}.
\newblock In {\em ISCA}, 2000.

\bibitem{mao-temp}
M.~Mao, W.~Wen, X.~Liu, J.~Hu, D.~Wang, Y.~Chen, and H.~Li.
\newblock {TEMP: Thread Batch Enabled Memory Partitioning for GPU}.
\newblock In {\em DAC}, 2016.

\bibitem{mashimo2013molecular}
T.~Mashimo, Y.~Fukunishi, N.~Kamiya, Y.~Takano, I.~Fukuda, and H.~Nakamura.
\newblock {Molecular Dynamics Simulations Accelerated by GPU for Biological
  Macromolecules with a Non-Ewald Scheme for Electrostatic Interactions}.
\newblock {\em Journal of Chemical Theory and Computation}, 2013.

\bibitem{itanium}
C.~McNairy and D.~Soltis.
\newblock {Itanium 2 Processor Microarchitecture}.
\newblock {\em IEEE Micro}, 23(2):44--55, 2003.

\bibitem{gpu-arch-microbenchmarking}
X.~Mei and X.~Chu.
\newblock {Dissecting GPU Memory Hierarchy Through Microbenchmarking}.
\newblock {\em IEEE TPDS}, 28(1):72--86, Jan 2017.

\bibitem{mekkat-pact13}
V.~Mekkat, A.~Holey, P.-C. Yew, and A.~Zhai.
\newblock {Managing Shared Last-Level Cache in a Heterogeneous Multicore
  Processor}.
\newblock In {\em PACT}, 2013.

\bibitem{warpsub}
J.~Meng, D.~Tarjan, and K.~Skadron.
\newblock {Dynamic Warp Subdivision for Integrated Branch and Memory Divergence
  Tolerance}.
\newblock In {\em ISCA}, 2010.

\bibitem{igpu}
J.~Menon, M.~de~Kruijf, and K.~Sankaralingam.
\newblock {iGPU: Exception Support and Speculative Execution on GPUs}.
\newblock In {\em ISCA}, 2012.

\bibitem{meza-cal2012}
J.~Meza, J.~Chang, H.~Yoon, O.~Mutlu, and P.~Ranganathan.
\newblock {Enabling Efficient and Scalable Hybrid Memories Using
  Fine-Granularity DRAM Cache Management}.
\newblock {\em IEEE CAL}, 2012.

\bibitem{meza-iccd2012}
J.~Meza, J.~Li, and O.~Mutlu.
\newblock {A Case for Small Row Buffers in Non-Volatile Main Memories}.
\newblock In {\em ICCD}, 2012.

\bibitem{meza-weed2013}
J.~Meza, Y.~Luo, S.~Khan, J.~Zhao, Y.~Xie, and O.~Mutlu.
\newblock {A Case for Efficient Hardware/Software Cooperative Management of
  Storage and Memory}.
\newblock In {\em WEED}, 2013.

\bibitem{micron-rldram3}
{Micron Technology, Inc.}
\newblock {576Mb: x18, x36 RLDRAM3}, 2011.

\bibitem{win-huge}
{Microsoft Corporation}.
\newblock {\em {Large-Page Support in Windows}}.
\newblock [Accessed April-2017].

\bibitem{arm-mali}
R.~Mijat.
\newblock {Take GPU Processing Power Beyond Graphics with Mali GPU Computing},
  2012.

\bibitem{het-nocs-dac13}
A.~K. Mishra, O.~Mutlu, and C.~R. Das.
\newblock {A Heterogeneous Multiple Network-on-chip Design: An
  Application-aware Approach}.
\newblock In {\em DAC}, 2013.

\bibitem{Mongodb-nothp}
{MongoDB Inc.}
\newblock {{Disable Transparent Huge Pages (THP)}}.
\newblock [Accessed April, 2016].

\bibitem{memattack}
T.~Moscibroda and O.~Mutlu.
\newblock {Memory Performance Attacks: Denial of Memory Service in Multi-core
  Systems}.
\newblock In {\em USENIX Security}, 2007.

\bibitem{mutlu-podc08}
T.~Moscibroda and O.~Mutlu.
\newblock {Distributed Order Scheduling and Its Application to Multi-core DRAM
  Controllers}.
\newblock In {\em PODC}, 2008.

\bibitem{casebufferless}
T.~Moscibroda and O.~Mutlu.
\newblock {A Case for Bufferless Routing in On-Chip Networks}.
\newblock In {\em {ISCA}}, 2009.

\bibitem{cudaprotien2014}
D.~Mrozek, M.~Brozek, and B.~Malysiak-Mrozek.
\newblock {Parallel Implementation of 3D Protein Structure Similarity Searches
  Using a GPU and the CUDA}.
\newblock {\em Journal of Molecular Modeling}, 2014.

\bibitem{morse-hpca12}
J.~Mukundan and J.~F. Martinez.
\newblock {MORSE: Multi-objective Reconfigurable Self-optimizing Memory
  Scheduler}.
\newblock In {\em HPCA}, 2012.

\bibitem{mullins04}
R.~Mullins, A.~West, and S.~Moore.
\newblock {Low-latency Virtual-channel Routers for On-chip Networks}.
\newblock In {\em ISCA}, 2004.

\bibitem{mcp}
S.~P. Muralidhara, L.~Subramanian, O.~Mutlu, M.~Kandemir, and T.~Moscibroda.
\newblock {Reducing Memory Interference in Multicore Systems via
  Application-Aware Memory Channel Partitioning}.
\newblock In {\em MICRO}, 2011.

\bibitem{cacti}
N.~Muralimanohar, R.~Balasubramonian, and N.~Jouppi.
\newblock {Optimizing NUCA Organizations and Wiring Alternatives for Large
  Caches with CACTI 6.0}.
\newblock In {\em MICRO}, 2007.

\bibitem{imw2013}
O.~Mutlu.
\newblock {Memory Scaling: A Systems Architecture Perspective}.
\newblock In {\em IMW}, 2013.

\bibitem{mutlu-micro2005}
O.~Mutlu, H.~Kim, and Y.~N. Patt.
\newblock {Address-value Delta (AVD) Prediction: Increasing the Effectiveness
  of Runahead Execution by Exploiting Regular Memory Allocation Patterns}.
\newblock In {\em MICRO}, 2005.

\bibitem{mutlu-isca2005}
O.~Mutlu, H.~Kim, and Y.~N. Patt.
\newblock {Techniques for Efficient Processing in Runahead Execution Engines}.
\newblock In {\em ISCA}, 2005.

\bibitem{stfm}
O.~Mutlu and T.~Moscibroda.
\newblock {Stall-Time Fair Memory Access Scheduling for Chip Multiprocessors}.
\newblock In {\em {MICRO}}, 2007.

\bibitem{parbs}
O.~Mutlu and T.~Moscibroda.
\newblock {Parallelism-Aware Batch Scheduling: Enhancing Both Performance and
  Fairness of Shared DRAM Systems}.
\newblock In {\em ISCA}, 2008.

\bibitem{mutlu-hpca2003}
O.~Mutlu, J.~Stark, C.~Wilkerson, and Y.~N. Patt.
\newblock {Runahead Execution: An Alternative to Very Large Instruction Windows
  for Out-of-Order Processors}.
\newblock In {\em HPCA}, 2003.

\bibitem{superfri}
O.~Mutlu and L.~Subramanian.
\newblock {Research Problems and Opportunities in Memory Systems}.
\newblock {\em SUPERFRI}, 2015.

\bibitem{nair-isca2013}
P.~J. Nair, D.-H. Kim, and M.~K. Qureshi.
\newblock {ArchShield: Architectural Framework for Assisting DRAM Scaling by
  Tolerating High Error Rates}.
\newblock In {\em ISCA}, 2013.

\bibitem{largewarps}
V.~Narasiman, M.~Shebanow, C.~J. Lee, R.~Miftakhutdinov, O.~Mutlu, and Y.~N.
  Patt.
\newblock {Improving GPU Performance via Large Warps and Two-Level Warp
  Scheduling}.
\newblock In {\em MICRO}, 2011.

\bibitem{superpage}
J.~Navarro, S.~Iyer, P.~Druschel, and A.~Cox.
\newblock {Practical, Transparent Operating System Support for Superpages}.
\newblock In {\em OSDI}, 2002.

\bibitem{nesbit-pact2004}
K.~J. Nesbit, A.~S. Dhodapkar, and J.~E. Smith.
\newblock {AC/DC: An Adaptive Data Cache Prefetcher}.
\newblock In {\em PACT}, 2004.

\bibitem{yak}
K.~Nguyen, L.~Fang, G.~Xu, B.~Demsky, S.~Lu, S.~Alamian, and O.~Mutlu.
\newblock {Yak: A High-Performance Big-Data-Friendly Garbage Collector}.
\newblock In {\em OSDI}, 2016.

\bibitem{nobile2016graphics}
M.~S. Nobile, P.~Cazzaniga, A.~Tangherloni, and D.~Besozzi.
\newblock {Graphics Processing Units in Bioinformatics, Computational Biology
  and Systems Biology}.
\newblock {\em Briefings in Bioinformatics}, 2016.

\bibitem{nuodb-nothp}
{NuoDB Inc.}
\newblock {{Linux Transparent Huge Pages, JEMalloc and NuoDB}}.
\newblock [Accessed May, 2014].

\bibitem{k40}
{NVIDIA Corp.}
\newblock {Tesla K40 GPU Active Accelerator}.
\newblock
  \url{https://www.nvidia.com/content/PDF/kepler/Tesla-K40-Active-Board-Spec-BD-06949-001_v03.pdf},
  2013.

\bibitem{gtx1080}
{NVIDIA Corp.}
\newblock {NVIDIA GeForce GTX 1080}.
\newblock
  \url{https://international.download.nvidia.com/geforce-com/international/pdfs/GeForce_GTX_1080_Whitepaper_FINAL.pdf},
  2016.

\bibitem{falcon}
{NVIDIA Corp.}
\newblock {NVIDIA RISC-V Story}.
\newblock
  \url{https://riscv.org/wp-content/uploads/2016/07/Tue1100_Nvidia_RISCV_Story_V2.pdf},
  2016.

\bibitem{cudastream}
{NVIDIA Corp.}
\newblock {CUDA Toolkit Documentation}.
\newblock
  \url{http://docs.nvidia.com/cuda/cuda-runtime-api/stream-sync-behavior.html},
  2017.

\bibitem{tegra}
{NVIDIA Corporation}.
\newblock {NVIDIA Tegra K1}.

\bibitem{tegrax1}
{NVIDIA Corporation}.
\newblock {NVIDIA® Tegra® X1}.

\bibitem{cuda-sdk}
{NVIDIA Corporation}.
\newblock {CUDA C/C++ SDK Code Samples}.
\newblock \url{http://developer.nvidia.com/cuda-cc-sdk-code-samples}, 2011.

\bibitem{fermi}
{NVIDIA Corporation}.
\newblock {NVIDIA's Next Generation CUDA Compute Architecture: Fermi}.
\newblock
  \url{http://www.nvidia.com/content/pdf/fermi_white_papers/nvidia_fermi_compute_architecture_whitepaper.pdf},
  2011.

\bibitem{kepler}
{NVIDIA Corporation}.
\newblock {NVIDIA's Next Generation CUDA Compute Architecture: Kepler GK110}.
\newblock
  \url{http://www.nvidia.com/content/PDF/kepler/NVIDIA-Kepler-GK110-Architecture-Whitepaper.pdf},
  2012.

\bibitem{maxwell}
{NVIDIA Corporation}.
\newblock {NVIDIA GeForce GTX 750 Ti}.
\newblock 2014.

\bibitem{programmingguide}
{NVIDIA Corporation}.
\newblock {CUDA C Programming Guide}.
\newblock
  \url{http://docs.nvidia.com/cuda/cuda-c-programming-guide/index.html}, 2015.

\bibitem{mps}
{NVIDIA Corporation}.
\newblock {Multi-Process Service}.
\newblock
  \url{https://docs.nvidia.com/deploy/pdf/CUDA_Multi_Process_Service_Overview.pdf},
  2015.

\bibitem{pascal}
{NVIDIA Corporation}.
\newblock {NVIDIA Tesla P100}.
\newblock
  \url{https://images.nvidia.com/content/pdf/tesla/whitepaper/pascal-architecture-whitepaper.pdf},
  2016.

\bibitem{nvidia-bypassing}
{NVIDIA Corporation}.
\newblock {Parallel Thread Execution ISA Version 5.0}.
\newblock 2017.

\bibitem{maxwell-guide}
{NVIDIA Corporation}.
\newblock {Tuning CUDA Applications for Maxwell}.
\newblock 2017.

\bibitem{hotnets2010}
G.~Nychis, C.~Fallin, T.~Moscibroda, and O.~Mutlu.
\newblock {Next Generation On-Chip Networks: What Kind of Congestion Control Do
  We Need?}
\newblock In {\em Hotnets}, 2010.

\bibitem{sigcomm12}
G.~Nychis, C.~Fallin, T.~Moscibroda, O.~Mutlu, and S.~Seshan.
\newblock {On-chip Networks from a Networking Perspective: Congestion and
  Scalability in Many-core Interconnects}.
\newblock In {\em SIGCOMM}, 2012.

\bibitem{o-isca2014}
S.~O, Y.~H. Son, N.~S. Kim, and J.~H. Ahn.
\newblock {Row-Buffer Decoupling: A Case for Low-Latency DRAM
  Microarchitecture}.
\newblock In {\em ISCA}, 2014.

\bibitem{ohsawa-islped1998}
T.~Ohsawa, K.~Kai, and K.~Murakami.
\newblock {Optimizing the DRAM Refresh Count for Merged DRAM/Logic LSIs}.
\newblock In {\em ISLPED}, 1998.

\bibitem{694774}
M.~Oskin, F.~T. Chong, and T.~Sherwood.
\newblock {Active Pages: A Computation Model for Intelligent Memory}.
\newblock In {\em ISCA}, 1998.

\bibitem{asplos-sree}
S.~Pai, M.~J. Thazhuthaveetil, and R.~Govindarajan.
\newblock {Improving GPGPU Concurrency with Elastic Kernels}.
\newblock In {\em ASPLOS}, 2013.

\bibitem{palacharla}
S.~Palacharla, N.~P. Jouppi, and J.~E. Smith.
\newblock {Complexity-effective Superscalar Processors}.
\newblock In {\em ISCA}, 1997.

\bibitem{prediction-tlb}
M.-M. Papadopoulou, X.~Tong, A.~Seznec, and A.~Moshovos.
\newblock {Prediction-based Superpage-friendly TLB Designs}.
\newblock In {\em HPCA}, 2015.

\bibitem{park-sc13}
J.~Park, R.~M. Yoo, D.~S. Khudia, C.~J. Hughes, and D.~Kim.
\newblock {Location-aware Cache Management for Many-core Processors with Deep
  Cache Hierarchy}.
\newblock In {\em SC}, 2013.

\bibitem{patel-isca2017}
M.~Patel, J.~Kim, and O.~Mutlu.
\newblock {The Reach Profiler (REAPER): Enabling the Mitigation of DRAM
  Retention Failures via Profiling at Aggressive Conditions}.
\newblock In {\em ISCA}, 2017.

\bibitem{pinpoint}
H.~Patil, R.~Cohn, M.~Charney, R.~Kapoor, A.~Sun, and A.~Karunanidhi.
\newblock {Pinpointing Representative Portions of Large Intel Itanium Programs
  with Dynamic Instrumentation}.
\newblock In {\em MICRO}, 2004.

\bibitem{592312}
D.~Patterson, T.~Anderson, N.~Cardwell, R.~Fromm, K.~Keeton, C.~Kozyrakis,
  R.~Thomas, and K.~Yelick.
\newblock {A Case for Intelligent RAM}.
\newblock {\em IEEE Micro}, 17(2):34--44, 1997.

\bibitem{pattnaik-pact2016}
A.~Pattnaik, X.~Tang, A.~Jog, O.~Kayiran, A.~K. Mishra, M.~T. Kandemir,
  O.~Mutlu, and C.~R. Das.
\newblock {Scheduling Techniques for GPU Architectures with
  Processing-In-Memory Capabilities}.
\newblock In {\em PACT}, 2016.

\bibitem{pcie}
{PCI-SIG}.
\newblock {PCI Express Base Specification Revision 3.1a}, 2015.

\bibitem{toggle-hpca16}
G.~Pekhimenko, E.~Bolotin, N.~Vijaykumar, O.~Mutlu, T.~C. Mowry, and S.~W.
  Keckler.
\newblock {A Case for Toggle-aware Compression for GPU Systems}.
\newblock In {\em HPCA}, 2016.

\bibitem{compress-reuse-hpca15}
G.~Pekhimenko, T.~Huberty, R.~Cai, O.~Mutlu, P.~B. Gibbons, M.~A. Kozuch, and
  T.~C. Mowry.
\newblock {Exploiting Compressed Block Size as an Indicator of Future Reuse}.
\newblock In {\em HPCA}, 2015.

\bibitem{lcp-micro13}
G.~Pekhimenko, V.~Seshadri, Y.~Kim, H.~Xin, O.~Mutlu, M.~A. Kozuch, P.~B.
  Gibbons, and T.~C. Mowry.
\newblock {Linearly Compressed Pages: A Main Memory Compression Framework with
  Low Complexity and Low Latency}.
\newblock In {\em MICRO}, 2013.

\bibitem{bdi-pact12}
G.~Pekhimenko, V.~Seshadri, O.~Mutlu, P.~B. Gibbons, M.~A. Kozuch, and T.~C.
  Mowry.
\newblock {Base-delta-immediate Compression: Practical Data Compression for
  On-chip Caches}.
\newblock In {\em PACT}, 2012.

\bibitem{peleg-ieeemicro96}
A.~Peleg and U.~Weiser.
\newblock {MMX Technology Extension to the Intel Architecture}.
\newblock {\em IEEE Micro}, 16(4):51--59, August 1996.

\bibitem{dokudb-nothp}
{Percona}.
\newblock {{Why TokuDB Hates Transparent HugePages}}.
\newblock [Accessed July, 2014].

\bibitem{phadke-date2011}
S.~Phadke and S.~Narayanasamy.
\newblock {MLP Aware Heterogeneous Memory System}.
\newblock In {\em DATE}, 2011.

\bibitem{binh-hpca14}
B.~Pham, A.~Bhattacharjee, Y.~Eckert, and G.~H. Loh.
\newblock {Increasing TLB Reach by Exploiting Clustering in Page Translations}.
\newblock In {\em HPCA}, 2014.

\bibitem{binh-colt}
B.~Pham, V.~Vaidyanathan, A.~Jaleel, and A.~Bhattacharjee.
\newblock {CoLT: Coalesced Large-Reach TLBs}.
\newblock In {\em MICRO}, 2012.

\bibitem{binh-micro15}
B.~Pham, J.~Vesely, G.~Loh, and A.~Bhattacharjee.
\newblock {Large Pages and Lightweight Memory Management in Virtualized
  Systems: Can You Have it Both Ways?}
\newblock In {\em MICRO}, 2015.

\bibitem{pichai-asplos14}
B.~Pichai, L.~Hsu, and A.~Bhattacharjee.
\newblock {Architectural Support for Address Translation on GPUs: Designing
  Memory Management Units for CPU/GPUs with Unified Address Spaces}.
\newblock In {\em ASPLOS}, 2014.

\bibitem{powers-hpca14}
J.~Power, M.~D. Hill, and D.~A. Wood.
\newblock {Supporting x86-64 Address Translation for 100s of GPU Lanes}.
\newblock In {\em HPCA}, 2014.

\bibitem{powervr}
{PowerVR}.
\newblock {PowerVR Hardware Architecture Overview for Developers}.
\newblock 2016.
\newblock
  \url{http://cdn.imgtec.com/sdk-documentation/PowerVR+Hardware.Architecture+Overview+for+Developers.pdf}.

\bibitem{tobias-stock}
T.~Preis, P.~Virnau, W.~Paul, and J.~J. Schneider.
\newblock {Accelerated Fluctuation Analysis by Graphic Cards and Complex
  Pattern Formation in Financial Markets}.
\newblock {\em New Journal of Physics}, 11, 2009.

\bibitem{6844483}
S.~H. Pugsley, J.~Jestes, H.~Zhang, R.~Balasubramonian, V.~Srinivasan,
  A.~Buyuktosunoglu, A.~Davis, and F.~Li.
\newblock {NDC: Analyzing the Impact of 3D-stacked Memory+logic Devices on
  MapReduce Workloads}.
\newblock In {\em ISPASS}, 2014.

\bibitem{bip}
M.~K. Qureshi, A.~Jaleel, Y.~N. Patt, S.~C. Steely, and J.~Emer.
\newblock {Adaptive Insertion Policies for High Performance Caching}.
\newblock In {\em ISCA}, 2007.

\bibitem{qureshi-micro2009}
M.~K. Qureshi, J.~Karidis, M.~Franceschini, V.~Srinivasan, L.~Lastras, and
  B.~Abali.
\newblock {Enhancing Lifetime and Security of PCM-based Main Memory with
  Start-gap Wear Leveling}.
\newblock In {\em MICRO}, 2009.

\bibitem{qureshi-dsn2015}
M.~K. Qureshi, D.~H. Kim, S.~Khan, P.~J. Nair, and O.~Mutlu.
\newblock {AVATAR: A Variable-Retention-Time (VRT) Aware Refresh for DRAM
  Systems}.
\newblock In {\em DSN}, 2015.

\bibitem{ucp}
M.~K. Qureshi and Y.~N. Patt.
\newblock {Utility-based Cache Partitioning: A Low-overhead, High-performance,
  Runtime Mechanism to Partition Shared Caches}.
\newblock In {\em MICRO}, 2006.

\bibitem{qureshi-isca2009}
M.~K. Qureshi, V.~Srinivasan, and J.~A. Rivers.
\newblock {Scalable High Performance Main Memory System Using Phase-change
  Memory Technology}.
\newblock In {\em ISCA}, 2009.

\bibitem{rau-isca91}
B.~R. Rau.
\newblock {Pseudo-randomly Interleaved Memory}.
\newblock In {\em ISCA}, 1991.

\bibitem{ravindran97}
G.~Ravindran and M.~Stumm.
\newblock {A Performance Comparison of Hierarchical Ring- and Mesh-connected
  Multiprocessor Networks}.
\newblock In {\em HPCA}, 1997.

\bibitem{ravindran98}
G.~Ravindran and M.~Stumm.
\newblock {On Topology and Bisection Bandwidth for Hierarchical-ring Networks
  for Shared Memory Multiprocessors}.
\newblock In {\em HPCA}, 1998.

\bibitem{reddi2004pin}
V.~J. Reddi, A.~Settle, D.~A. Connors, and R.~S. Cohn.
\newblock {PIN: A Binary Instrumentation Tool for Computer Architecture
  Research and Education}.
\newblock In {\em WCAE}, 2004.

\bibitem{redis-nothp}
{Redis Labs}.
\newblock {{Redis Latency Problems Troubleshooting}}.
\newblock [Accessed April, 2016].

\bibitem{fr-fcfs}
S.~Rixner, W.~J. Dally, U.~J. Kapasi, P.~Mattson, and J.~D. Owens.
\newblock {Memory Access Scheduling}.
\newblock In {\em ISCA}, 2000.

\bibitem{trogers-thesis}
T.~G. Rogers.
\newblock {\em {Locality and Scheduling in the Massively Multithreaded Era}}.
\newblock PhD thesis, Univ. of British Columbia, 2015.

\bibitem{ccws}
T.~G. Rogers, M.~O'Connor, and T.~M. Aamodt.
\newblock {Cache-Conscious Wavefront Scheduling}.
\newblock In {\em MICRO}, 2012.

\bibitem{tor-micro13}
T.~G. Rogers, M.~O'Connor, and T.~M. Aamodt.
\newblock {Divergence-Aware Warp Scheduling}.
\newblock In {\em MICRO}, 2013.

\bibitem{unitd}
B.~F. Romanescu, A.~R. Lebeck, D.~J. Sorin, and A.~Bracy.
\newblock {UNified Instruction/Translation/Data (UNITD) Coherence: One Protocol
  to Rule them All}.
\newblock In {\em HPCA}, 2010.

\bibitem{ptask}
C.~J. Rossbach, J.~Currey, M.~Silberstein, B.~Ray, and E.~Witchel.
\newblock {PTask: Operating System Abstractions to Manage GPUs as Compute
  Devices}.
\newblock In {\em SOSP}, 2011.

\bibitem{dandelion}
C.~J. Rossbach, Y.~Yu, J.~Currey, J.-P. Martin, and D.~Fetterly.
\newblock {Dandelion: A Compiler and Runtime for Heterogeneous Systems}.
\newblock In {\em SIGOPS}, 2013.

\bibitem{cray1}
R.~M. Russell.
\newblock {The CRAY-1 Computer System}.
\newblock {\em CACM}, 21(1):63--72, 1978.

\bibitem{gpu-regfile-mohammad}
M.~Sadrosadati, A.~Mirhosseini, B.~Ehsani, H.~Sarbazi-Azad, M.~P. Drumond,
  B.~Falsafi, R.~Ausavarungnirun, and O.~Mutlu.
\newblock {LTRF: A Latency Tolerant Register File Architecture for GPUs}.
\newblock In {\em ASPLOS}, 2018.

\bibitem{mosaic.github}
{SAFARI Research Group}.
\newblock {Mosaic -- GitHub Repository}.
\newblock \url{https://github.com/CMU-SAFARI/Mosaic/}.

\bibitem{sato-vlsic1998}
Y.~Sato, T.~Suzuki, T.~Aikawa, S.~Fujioka, W.~Fujieda, H.~Kobayashi, H.~Ikeda,
  T.~Nagasawa, A.~Funyu, Y.~Fuji, K.~Kawasaki, M.~Yamazaki, and M.~Taguchi.
\newblock {Fast cycle RAM (FCRAM): A 20-ns Random Row Access, Pipe-Lined
  Operating DRAM}.
\newblock In {\em VLSIC}, 1998.

\bibitem{saulsbury-isca00}
A.~Saulsbury, F.~Dahlgren, and P.~Stenstr\"{o}m.
\newblock {Recency-based TLB Preloading}.
\newblock In {\em ISCA}, 2000.

\bibitem{cdcstar}
P.~B. Schneck.
\newblock {\em The CDC STAR-100}, pages 99--117.
\newblock Springer US, Boston, MA, 1987.

\bibitem{senzig-afips65}
D.~N. Senzig and R.~V. Smith.
\newblock {Computer Organization for Array Processing}.
\newblock In {\em AFIPS}, 1965.

\bibitem{seo-patent}
S.-Y. Seo.
\newblock {Methods of Copying a Page in a Memory Device and Methods of Managing
  Pages in a Memory System}.
\newblock U.S. Patent Application 20140185395, 2014.

\bibitem{dbi}
V.~Seshadri, A.~Bhowmick, O.~Mutlu, P.~B. Gibbons, M.~A. Kozuch, and T.~C.
  Mowry.
\newblock {The Dirty-Block Index}.
\newblock In {\em ISCA}, 2014.

\bibitem{seshadri-cal2015}
V.~Seshadri, K.~Hsieh, A.~Boroumand, D.~Lee, M.~Kozuch, O.~Mutlu, P.~Gibbons,
  and T.~Mowry.
\newblock {Fast Bulk Bitwise AND and OR in DRAM}.
\newblock {\em IEEE CAL}, 2015.

\bibitem{seshadri2013rowclone}
V.~Seshadri, Y.~Kim, C.~Fallin, D.~Lee, R.~Ausavarungnirun, G.~Pekhimenko,
  Y.~Luo, O.~Mutlu, P.~B. Gibbons, M.~A. Kozuch, et~al.
\newblock {RowClone: Fast and Energy-efficient in-DRAM Bulk Data Copy and
  Initialization}.
\newblock In {\em ISCA}, 2013.

\bibitem{seshadri-arxiv2016}
V.~Seshadri, D.~Lee, T.~Mullins, H.~Hassan, A.~Boroumand, J.~Kim, M.~A. Kozuch,
  O.~Mutlu, P.~B. Gibbons, and T.~C. Mowry.
\newblock {Buddy-RAM: Improving the Performance and Efficiency of Bulk Bitwise
  Operations Using DRAM}.
\newblock In {\em arXiv CoRR}, 2016.

\bibitem{ambit}
V.~Seshadri, D.~Lee, T.~Mullins, H.~Hassan, A.~Boroumand, J.~Kim, M.~A. Kozuch,
  O.~Mutlu, P.~B. Gibbons, and T.~C. Mowry.
\newblock {Ambit: In-Memory Accelerator for Bulk Bitwise Operations Using
  Commodity DRAM Technology}.
\newblock In {\em MICRO}, 2017.

\bibitem{seshadri-micro2015}
V.~Seshadri, T.~Mullins, A.~Boroumand, O.~Mutlu, P.~B. Gibbons, M.~A. Kozuch,
  and T.~C. Mowry.
\newblock {Gather-Scatter DRAM: In-DRAM Address Translation to Improve the
  Spatial Locality of Non-Unit Strided Accesses}.
\newblock In {\em MICRO}, 2015.

\bibitem{vivek-chapter}
V.~Seshadri and O.~Mutlu.
\newblock {{Simple Operations in Memory to Reduce Data Movement}}.
\newblock In {\em {{Advances in Computers}}}. 2017.

\bibitem{eaf-vivek}
V.~Seshadri, O.~Mutlu, M.~A. Kozuch, and T.~C. Mowry.
\newblock {The Evicted-Address Filter: A Unified Mechanism to Address Both
  Cache Pollution and Thrashing}.
\newblock In {\em PACT}, 2012.

\bibitem{seshadri-taco2015}
V.~Seshadri, S.~Yedkar, H.~Xin, O.~Mutlu, P.~B. Gibbons, M.~A. Kozuch, and
  T.~C. Mowry.
\newblock {Mitigating Prefetcher-Caused Pollution Using Informed Caching
  Policies for Prefetched Blocks}.
\newblock {\em ACM TACO}, 11(4):51:1--51:22, 2015.

\bibitem{pentium-pro}
T.~Shanley.
\newblock {\em {Pentium Pro Processor System Architecture}}.
\newblock Addison-Wesley Longman Publishing Co., Inc., Boston, MA, USA, 1st
  edition, 1996.

\bibitem{shin-hpca2014}
W.~Shin, J.~Yang, J.~Choi, and L.-S. Kim.
\newblock {NUAT: A Non-Uniform Access Time Memory Controller}.
\newblock In {\em HPCA}, 2014.

\bibitem{perf-model}
J.~Sim, A.~Dasgupta, H.~Kim, and R.~Vuduc.
\newblock {A Performance Analysis Framework for Identifying Potential Benefits
  in {GPGPU} Applications}.
\newblock In {\em PPoPP}, 2012.

\bibitem{jaewoong-micro12}
J.~Sim, G.~H. Loh, H.~Kim, M.~O'Connor, and M.~Thottethodi.
\newblock {A Mostly-Clean {DRAM} Cache for Effective Hit Speculation and
  Self-Balancing Dispatch}.
\newblock In {\em MICRO}, 2012.

\bibitem{coherence-hpca13}
I.~Singh, A.~Shriraman, W.~W.~L. Fung, M.~O'Connor, and T.~M. Aamodt.
\newblock {Cache Coherence for {GPU} Architectures}.
\newblock In {\em HPCA}, 2013.

\bibitem{sisoftware}
SiSoftware.
\newblock {Benchmarks : Measuring GP (GPU/APU) Cache and Memory Latencies}.
\newblock \url{http://www.sisoftware.net}, 2014.

\bibitem{alpha}
R.~L. Sites and R.~T. Witek.
\newblock {\em {ALPHA Architecture Reference Manual}}.
\newblock Digital Press, Boston, Oxford, Melbourne, 1998.

\bibitem{solomon62}
D.~L. Slotnick, W.~C. Borck, and R.~C. McReynolds.
\newblock {The Solomon Computer -- A Preliminary Report}.
\newblock In {\em Workshop on Computer Organization}, 1962.

\bibitem{hep}
B.~Smith.
\newblock {Architecture and Applications of the HEP Multiprocessor Computer
  System}.
\newblock {\em SPIE}, 1981.

\bibitem{smith-hep}
B.~J. Smith.
\newblock {A Pipelined, Shared Resource {MIMD} Computer}.
\newblock In {\em ICPP}, 1978.

\bibitem{son-isca2013}
Y.~H. Son, S.~O, Y.~Ro, J.~W. Lee, and J.~H. Ahn.
\newblock {Reducing Memory Access Latency with Asymmetric DRAM Bank
  Organizations}.
\newblock In {\em ISCA}, 2013.

\bibitem{splunk-nothp}
{Splunk Inc.}
\newblock {{Transparent Huge Memory Pages and Splunk Performance}}.
\newblock [Accessed December, 2013].

\bibitem{srikantaiah-micro10}
S.~Srikantaiah and M.~Kandemir.
\newblock {Synergistic TLBs for High Performance Address Translation in Chip
  Multiprocessors}.
\newblock In {\em MICRO}, 2010.

\bibitem{srinath-hpca2007}
S.~Srinath, O.~Mutlu, H.~Kim, and Y.~N. Patt.
\newblock {Feedback Directed Prefetching: Improving the Performance and
  Bandwidth-Efficiency of Hardware Prefetchers}.
\newblock In {\em HPCA}, 2007.

\bibitem{stone-1970}
H.~S. Stone.
\newblock {A Logic-in-Memory Computer}.
\newblock {\em IEEE TC}, C-19(1):73--78, 1970.

\bibitem{parboil}
J.~A. Stratton, C.~Rodrigrues, I.-J. Sung, N.~Obeid, L.~Chang, G.~Liu, and
  W.-M.~W. Hwu.
\newblock {Parboil: A Revised Benchmark Suite for Scientific and Commercial
  Throughput Computing}.
\newblock Technical Report IMPACT-12-01, University of Illinois at
  Urbana-Champaign, Urbana, Mar. 2012.

\bibitem{vwq-isca10}
J.~Stuecheli, D.~Kaseridis, D.~Daly, H.~C. Hunter, and L.~K. John.
\newblock {The Virtual Write Queue: Coordinating DRAM and Last-level Cache
  Policies}.
\newblock In {\em ISCA}, 2010.

\bibitem{bliss}
L.~Subramanian, D.~Lee, V.~Seshadri, H.~Rastogi, and O.~Mutlu.
\newblock {The Blacklisting Memory Scheduler: Achieving high performance and
  fairness at low cost}.
\newblock In {\em ICCD}, 2014.

\bibitem{bliss-arxiv}
L.~Subramanian, D.~Lee, V.~Seshadri, H.~Rastogi, and O.~Mutlu.
\newblock {The Blacklisting Memory Scheduler: Balancing Performance, Fairness
  and Complexity}.
\newblock {\em arXiv CoRR}, 2015.

\bibitem{bliss-tpds}
L.~Subramanian, D.~Lee, V.~Seshadri, H.~Rastogi, and O.~Mutlu.
\newblock {BLISS: Balancing Performance, Fairness and Complexity in Memory
  Access Scheduling}.
\newblock In {\em IEEE TPDS}, 2016.

\bibitem{lavanya-asm}
L.~Subramanian, V.~Seshadri, A.~Ghosh, S.~Khan, and O.~Mutlu.
\newblock {The Application Slowdown Model: Quantifying and Controlling the
  Impact of Inter-application Interference at Shared Caches and Main Memory}.
\newblock In {\em MICRO}, 2015.

\bibitem{mise}
L.~Subramanian, V.~Seshadri, Y.~Kim, B.~Jaiyen, and O.~Mutlu.
\newblock {MISE: Providing Performance Predictability and Improving Fairness in
  Shared Main Memory Systems}.
\newblock In {\em HPCA}, 2013.

\bibitem{delite}
A.~K. Sujeeth, K.~J. Brown, H.~Lee, T.~Rompf, H.~Chafi, M.~Odersky, and
  K.~Olukotun.
\newblock {Delite: A Compiler Architecture for Performance-oriented Embedded
  Domain-specific Languages}.
\newblock In {\em TECS}, 2014.

\bibitem{sura-2015}
Z.~Sura, A.~Jacob, T.~Chen, B.~Rosenburg, O.~Sallenave, C.~Bertolli, S.~Antao,
  J.~Brunheroto, Y.~Park, K.~O'Brien, and R.~Nair.
\newblock {Data Access Optimization in a Processing-in-memory System}.
\newblock In {\em CF}, 2015.

\bibitem{reinforcement-learning}
R.~S. Sutton and A.~G. Barto.
\newblock {\em {Reinforcement Learning: An Introduction}}, volume~1.

\bibitem{gpuvm}
Y.~Suzuki, S.~Kato, H.~Yamada, and K.~Kono.
\newblock {GPUvm: Why Not Virtualizing GPUs at the Hypervisor?}
\newblock In {\em USENIX ATC}, 2014.

\bibitem{sap-nothp}
{Sybase Inc.}
\newblock {{SAP IQ and Linux Hugepages/Transparent Hugepages}}.
\newblock [Accessed May, 2014].

\bibitem{talluri-asplos94}
M.~Talluri and M.~D. Hill.
\newblock {Surpassing the TLB Performance of Superpages with Less Operating
  System Support}.
\newblock In {\em ASPLOS}, 1994.

\bibitem{isca-2014-preemptive}
I.~Tanasic, I.~Gelado, J.~Cabezas, A.~Ramirez, N.~Navarro, and M.~Valero.
\newblock {Enabling Preemptive Multiprogramming on GPUs}.
\newblock In {\em ISCA}, 2014.

\bibitem{cdc6600}
J.~E. Thornton.
\newblock {Parallel Operation in the Control Data 6600}.
\newblock {\em AFIPS FJCC}, 1964.

\bibitem{cdc6600-2}
J.~E. Thornton.
\newblock {Design of a Computer—the Control Data 6600}.
\newblock 1970.

\bibitem{thp}
{Transparent Hugepages}.
\newblock \url{https://lwn.net/Articles/359158/}.
\newblock [October, 2009].

\bibitem{gVirt}
K.~Tian, Y.~Dong, and D.~Cowperthwaite.
\newblock {A Full GPU Virtualization Solution with Mediated Pass-Through}.
\newblock In {\em USENIX ATC}, 2014.

\bibitem{tyson-micro95}
G.~Tyson, M.~Farrens, J.~Matthews, and A.~R. Pleszkun.
\newblock {A Modified Approach to Data Cache Management}.
\newblock In {\em MICRO}, 1995.

\bibitem{gtx480-config}
{Univ.\ of British Columbia}.
\newblock {GPGPU-Sim GTX 480 Configuration}.
\newblock
  \url{http://dev.ece.ubc.ca/projects/gpgpu-sim/browser/v3.x/configs/GTX480}.

\bibitem{usui-squash}
H.~Usui, L.~Subramanian, K.~Chang, and O.~Mutlu.
\newblock {SQUASH: Simple qos-aware high-performance memory scheduler for
  heterogeneous systems with hardware accelerators}.
\newblock {\em arXiv CoRR}, 2015.

\bibitem{usui-dash}
H.~Usui, L.~Subramanian, K.~Chang, and O.~Mutlu.
\newblock {DASH: Deadline-Aware High-Performance Memory Scheduler for
  Heterogeneous Systems with Hardware Accelerators}.
\newblock {\em ACM TACO}, 12(4), Jan. 2016.

\bibitem{vandierendonck}
H.~Vandierendonck and A.~Seznec.
\newblock {Fairness Metrics for Multi-threaded Processors}.
\newblock {\em IEEE CAL}, Feb 2011.

\bibitem{venkatesan-hpca2006}
R.~Venkatesan, S.~Herr, and E.~Rotenberg.
\newblock {Retention-aware Placement in DRAM (RAPID): Software Methods for
  Quasi-non-volatile DRAM}.
\newblock In {\em HPCA}, 2006.

\bibitem{abhishek-ispass16}
J.~Vesely, A.~Basu, M.~Oskin, G.~H. Loh, and A.~Bhattacharjee.
\newblock {Observations and Opportunities in Architecting Shared Virtual Memory
  for Heterogeneous Systems}.
\newblock In {\em ISPASS}, 2016.

\bibitem{vijay-hpca17}
T.~Vijayaraghavany, Y.~Eckert, G.~H. Loh, M.~J. Schulte, M.~Ignatowski, B.~M.
  Beckmann, W.~C. Brantley, J.~L. Greathouse, W.~Huang, A.~Karunanithi,
  O.~Kayiran, M.~Meswani, I.~Paul, M.~Poremba, S.~Raasch, S.~K. Reinhardt,
  G.~Sadowski, and V.~Sridharan.
\newblock {Design and Analysis of an APU for Exascale Computing}.
\newblock In {\em HPCA}, 2017.

\bibitem{vijaykumar-isca18}
N.~Vijaykumar, E.~Ebrahimi, K.~Hsieh, P.~B. Gibbons, and O.~Mutlu.
\newblock {The Locality Descriptor: A Holistic Abstraction to Exploit Data
  Locality in GPUs}.
\newblock In {\em ISCA}, 2018.

\bibitem{zorua}
N.~Vijaykumar, K.~Hsieh, G.~Pekhimenko, S.~Khan, A.~Shrestha, S.~Ghose, A.~Jog,
  P.~B. Gibbons, and O.~Mutlu.
\newblock {Zorua: A Holistic Approach to Resource Virtualization in GPUs}.
\newblock In {\em MICRO}, 2016.

\bibitem{vijaykumar-xmem-isca18}
N.~Vijaykumar, A.~Jain, D.~Majumdar, K.~Hsieh, G.~Pekhimenko, E.~Ebrahimi,
  N.~Hajinazar, P.~B. Gibbons, and O.~Mutlu.
\newblock {A Case for Richer Cross-layer Abstractions: Bridging the Semantic
  Gap to Enhance Memory Optimization}.
\newblock In {\em ISCA}, 2018.

\bibitem{caba}
N.~Vijaykumar, G.~Pekhimenko, A.~Jog, A.~Bhowmick, R.~Ausavarungnirun, C.~Das,
  M.~Kandemir, T.~C. Mowry, and O.~Mutlu.
\newblock {A Case for Core-Assisted Bottleneck Acceleration in {GPUs}: Enabling
  Flexible Data Compression with Assist Warps}.
\newblock In {\em ISCA}, 2015.

\bibitem{nandita-chapter}
N.~Vijaykumar, G.~Pekhimenko, A.~Jog, A.~Bhowmick, R.~Ausavarungnirun, C.~Das,
  M.~Kandemir, T.~C. Mowry, and O.~Mutlu.
\newblock {A Framework for Accelerating Bottlenecks in GPU Execution with
  Assist Warps}.
\newblock In {\em {{Advances in GPU Research and Practice}}}. 2016.

\bibitem{vivante-gpgpu}
{Vivante}.
\newblock {Vivante Vega GPGPU Technology}.
\newblock 2016.
\newblock \url{http://www.vivantecorp.com/index.php/en/technology/gpgpu.html}.

\bibitem{voltdb-nothp}
{VoltDB Inc.}
\newblock {{VoltDB Documentation}}.
\newblock [Accessed April, 2016].

\bibitem{vmCUDA}
L.~Vu, H.~Sivaraman, and R.~Bidarkar.
\newblock {GPU Virtualization for High Performance General Purpose Computing on
  the ESX Hypervisor}.
\newblock In {\em HPC}, 2014.

\bibitem{wang-hpca16}
Z.~Wang, J.~Yang, R.~Melhem, B.~R. Childers, Y.~Zhang, and M.~Guo.
\newblock {Simultaneous Multikernel GPU: Multi-tasking Throughput Processors
  via Fine-Grained Sharing}.
\newblock In {\em HPCA}, 2016.

\bibitem{ware-iccd2006}
F.~A. Ware and C.~Hampel.
\newblock {Improving Power and Data Efficiency with Threaded Memory Modules}.
\newblock In {\em ICCD}, 2006.

\bibitem{apu}
S.~Wasson.
\newblock {AMD's A8-3800 Fusion APU.}, Oct. 2011.

\bibitem{demystify}
H.~Wong, M.-M. Papadopoulou, M.~Sadooghi-Alvandi, and A.~Moshovos.
\newblock {Demystifying GPU Microarchitecture Through Microbenchmarking}.
\newblock In {\em ISPASS}, 2010.

\bibitem{youfeng-micro02}
Y.~Wu, R.~Rakvic, L.-L. Chen, C.-C. Miao, G.~Chrysos, and J.~Fang.
\newblock {Compiler Managed Micro-cache Bypassing for High Performance EPIC
  Processors}.
\newblock In {\em MICRO}, 2002.

\bibitem{xiang-ics09}
L.~Xiang, T.~Chen, Q.~Shi, and W.~Hu.
\newblock {Less Reused Filter: Improving L2 Cache Performance via Filtering
  Less Reused Lines}.
\newblock In {\em ICS}, 2009.

\bibitem{warp-level-div}
P.~Xiang, Y.~Yang, and H.~Zhou.
\newblock {Warp-Level Divergence in {GPUs}: Characterization, Impact, and
  Mitigation}.
\newblock In {\em HPCA}, 2014.

\bibitem{xie-hpca14}
M.~Xie, D.~Tong, K.~Huang, and X.~Cheng.
\newblock {Improving System Throughput and Fairness Simultaneously in Shared
  Memory CMP Systems via Dynamic Bank Partitioning}.
\newblock In {\em HPCA}, 2014.

\bibitem{xie-iccad13}
X.~Xie, Y.~Liang, G.~Sun, and D.~Chen.
\newblock {An Efficient Compiler Framework for Cache Bypassing on {GPUs}}.
\newblock In {\em ICCAD}, 2013.

\bibitem{xie-hpca15}
X.~Xie, Y.~Liang, Y.~Wang, G.~Sun, and T.~Wang.
\newblock {Coordinated Static and Dynamic Cache Bypassing for {GPUs}}.
\newblock In {\em HPCA}, 2015.

\bibitem{xiong-taco16}
D.~Xiong, K.~Huang, X.~Jiang, and X.~Yan.
\newblock {Memory Access Scheduling Based on Dynamic Multilevel Priority in
  Shared DRAM Systems}.
\newblock {\em ACM TACO}, 13(4), Dec. 2016.

\bibitem{warp-slicer}
Q.~Xu, H.~Jeon, K.~Kim, W.~W. Ro, and M.~Annavaram.
\newblock {Warped-Slicer: Efficient Intra-SM Slicing through Dynamic Resource
  Partitioning for GPU Multiprogramming}.
\newblock In {\em ISCA}, 2016.

\bibitem{yoon-iccd2012}
H.~Yoon, J.~Meza, R.~Ausavarungnirun, R.~Harding, and O.~Mutlu.
\newblock {Row Buffer Locality Aware Caching Policies for Hybrid Memories}.
\newblock In {\em ICCD}, 2012.

\bibitem{yu-dasc}
B.~Yu, J.~Ma, T.~Chen, and M.~Wu.
\newblock {Global Priority Table for Last-Level Caches}.
\newblock In {\em DASC}, 2011.

\bibitem{banshee}
X.~Yu, C.~J. Hughes, N.~Satish, O.~Mutlu, and S.~Devadas.
\newblock {Banshee: Bandwidth-Efficient DRAM Caching via Software/Hardware
  Cooperation}.
\newblock In {\em MICRO}, 2017.

\bibitem{complexity}
G.~Yuan, A.~Bakhoda, and T.~Aamodt.
\newblock {Complexity Effective Memory Access Scheduling for Many-Core
  Accelerator Architectures}.
\newblock In {\em MICRO}, 2009.

\bibitem{zhang-ispled14}
C.~Zhang, G.~Sun, P.~Li, T.~Wang, D.~Niu, and Y.~Chen.
\newblock {SBAC: A Statistics Based Cache Bypassing Method for
  Asymmetric-access Caches}.
\newblock In {\em ISPLED}, 2014.

\bibitem{zhang-2014}
D.~Zhang, N.~Jayasena, A.~Lyashevsky, J.~L. Greathouse, L.~Xu, and
  M.~Ignatowski.
\newblock {TOP-PIM: Throughput-oriented Programmable Processing in Memory}.
\newblock In {\em HPDC}, 2014.

\bibitem{zhang-ieee2001}
L.~Zhang, Z.~Fang, M.~Parker, B.~K. Mathew, L.~Schaelicke, J.~B. Carter, W.~C.
  Hsieh, and S.~A. McKee.
\newblock {The Impulse Memory Controller}.
\newblock {\em IEEE TC}, 50(11):1117--1132, 2001.

\bibitem{xiangdong95}
X.~Zhang and Y.~Yan.
\newblock {Comparative Modeling and Evaluation of CC-NUMA and COMA on
  Hierarchical Ring Architectures}.
\newblock {\em IEEE TPDS}, 1995.

\bibitem{jishen-firm}
J.~Zhao, O.~Mutlu, and Y.~Xie.
\newblock {FIRM: Fair and High-Performance Memory Control for Persistent Memory
  Systems}.
\newblock In {\em MICRO}, 2014.

\bibitem{zhao-iccd2005}
L.~Zhao, R.~Iyer, S.~Makineni, L.~Bhuyan, and D.~Newell.
\newblock {Hardware Support for Bulk Data Movement in Server Platforms}.
\newblock In {\em ICCD}, 2005.

\bibitem{zheng-micro2008}
H.~Zheng, J.~Lin, Z.~Zhang, E.~Gorbatov, H.~David, and Z.~Zhu.
\newblock {Mini-rank: Adaptive DRAM Architecture for Improving Memory Power
  Efficiency}.
\newblock In {\em MICRO}, 2008.

\bibitem{tianhao-hpca16}
T.~Zheng, D.~Nellans, A.~Zulfiqar, M.~Stephenson, and S.~W. Keckler.
\newblock {Towards High Performance Paged Memory for GPUs}.
\newblock In {\em HPCA}, 2016.

\bibitem{frfcfs-patent}
W.~K. Zuravleff and T.~Robinson.
\newblock {Controller for a Synchronous DRAM That Maximizes Throughput by
  Allowing Memory Requests and Commands to Be Issued Out of Order}.
\newblock In {\em US Patent Number 5,630,096}, 1997.

\end{thebibliography}

\end{document}